\documentclass[a4paper]{article}
\usepackage{amssymb,amsmath}
\usepackage {bsymb,mybcode}
\usepackage{graphicx}
\usepackage {bsymb,b2latex,mybcode,times}
\usepackage[all]{xy}
\usepackage{url,psboxit}
\usepackage{tabularx}
\usepackage{b2latex}

\urldef{\mailsa}\path|{mery, singhnne}@loria.fr|

\newcommand{\event}[1]{\textsf{#1}}
\newcommand{\model}[1]{\textsc{#1}}
\newcommand{\aircraft}[3]{
\boxed{
\begin{array}{c}	
\xymatrixcolsep{1pc}\xymatrix{
& \boxed{\scriptstyle\textsc{A},#1}  &  \\
\boxed{\scriptstyle\textsc{C},#3}\ar@{-}[ur] \ar@{-}[rr]	&  & \boxed{\scriptstyle\textsc{B},#2} \ar@{-}[ul]\\
	}
\end{array}}
 }

\newcommand{\of}{\left(\begin{array}{l}}
\newcommand{\ff}{\end{array}\right)}

\newcommand{\req}[2]{
\noindent
\fbox{
\begin{minipage}{1.0\linewidth}
#1:{\it  #2}    
\end{minipage}
}
}

\begin{document}

\title{Modelling an Aircraft Landing System in Event-B\thanks{The current report is  the companion paper of the paper~\cite{ABZ2014ls} accepted for publication in the volume 433 of the serie Communications in Computer Information Science. The Event-B models are available at the link http://eb2all.loria.fr. Processed on \today. }}

\author{Dominique M\'ery\\ Universit\'e de Lorraine, LORIA\\ BP 239, Nancy, France\\ {Dominique.Mery@loria.fr} \and Neeraj Kumar Singh\\
McMaster Centre for Software Certification\\ Hamilton, ON, Canada\\ {singhn10@mcmaster.ca}
}

\maketitle

\begin{abstract}

  The failure of hardware or software in a critical system can lead to
  loss of  lives. The design errors can  be main source of the failures
  that  can be introduced   during system development  process. Formal
  techniques are an alternative approach to  verify the correctness of
  critical  systems,  overcoming   limitations  of   the   traditional
  validation techniques such as simulation and testing. The increasing
  complexity and  failure rate brings  new  challenges in the  area of
  verification   and   validation  of   avionic   systems.  Since  the
  reliability  of  the   software     cannot  be  quantified,      the
  \textit{correct by  construction} approach can  implement a reliable
  system.  Refinement plays   a  major role   to build a  large system
  incrementally from an abstract specification to a concrete system.
  This paper  contributes as  a   stepwise formal development   of the
  landing system of an aircraft. The formal models include the complex
  behaviour, temporal   behaviour  and  sequence of   operations of  the
  landing gear system. The  models are formalized in Event-B  modelling
  language,  which supports stepwise refinement.   This case study  is
  considered as a benchmark for techniques  and tools dedicated to the
  verification of behavioural properties of systems.
\end{abstract}

\subsection*{Keywords} {Abstract model, Event-B, Event-driven approach, Proof-based development, Refinement, Landing Gear System}

\section{Introduction} 

In the  cutting edge  technology  of  aircraft, the  requirements  for
avionic systems become  increasingly complex. The failure of  hardware
or software in such a  complex system can lead  to  loss of lives.  The
increasing  complexity and failure  rate  brings new challenges in the
area  of verification and validation of  avionic  systems. The Federal
Aviation Administration   (FAA) ensures  that  aircraft meets  highest
safety standards. The FAA recommends  the catastrophic failures of the
aircraft and  suggests probabilities of  failure  on the order  of per
flight hour~\cite{FAA1988}.

Hardware component failures and design errors  are two main reasonable
factors to  major the reliability of the  avionics. There  are several
techniques like redundancy and voting are  used to handle the hardware
failures.   However, the  design  errors   can be  introduced  at  the
development  phase,  which   may    include  errors in    the   system
specification,  and  errors  made during   the  implementation of  the
software or hardware~\cite{Johnson2001}.

The complexity  of    software has been  tremendously  increased.  Our
experience, intuition  and   developed methodologies is  reliable  for
building the   continuous system, but software  exhibits discontinuous
behaviour.   To verify the correctness  of    the system, it is  highly
desirable  to reason about millions   of  sequences of discrete  state
transitions. Traditional  techniques like testing  and simulations are
infeasible to  test the  correctness of  a system~\cite{Butler1993}.
Since the   reliability of  the  software  cannot be  quantified,  the
avionic  software  must       be   developed  using    \textit{correct      by
construction}~\cite{Leavens:roadmap} approach    that can  produce  the
correct       design     and    implementation      of  the      final
system~\cite{singh2013}.

This paper describes  how  rigorous analysis employing formal  methods
can be applied to the software  development process. Formal methods is
considered as an alternative approach for certification in the DO-178B
standard for avionics  software development. We propose the refinement
based correct by construction approach to develop a critical system.
The  nature   of  the   refinement    that    we verify   using    the
RODIN~\cite{rodin} proof  tools  is  a safety  refinement.   Thus, the
behaviour of  final resulting system  is preserved by an abstract model
as  well as in the correctly  refined models.  Proof-based development
methods~\cite{abrial2010}  integrate formal  proof  techniques in  the
development of  software systems.  The main   idea is to start  with a
very abstract model  of   the system under development.    Details are
gradually  added  to this first model  by  building a sequence of more
concrete events.   The relationship between  two successive  models in
this sequence is  \textit{refinement}~\cite{abrial2010,back79a}.  Here
we  present  stepwise     development  to  model   and  verify    such
interdisciplinary  requirements  in   Event-B~\cite{losl-b,abrial2010}
modelling language.  The correctness of each step is proved in order to
achieve a reliable system.

In this paper, we present the stepwise  formalization of the benchmark
case study landing system of an aircraft.  The current work intends to
explore those problems related to modelling  the sequence of operations
of landing  gears and doors associated  with hydraulic cylinders under
the real-time constraints and to evaluate the refinement process.

The outline of the remaining  paper is as  follows. Section 2 presents
selection of  the  case  study related   to the  landing  system of an
aircraft for formalization. In  Section 3,     we  explore the   incremental   proof-based  formal
development of   the  landing system. Finally, in Section   4, we conclude   the  paper.

\section{Basic Overview of Landing Gear System}

The landing gear is  an essential system  that allows an  aircraft to
land safely,  and supports the  entire  weight of  an aircraft  during
landing and ground operations.  The basic engineering and  operational
behaviors behind a  landing gear system  are very complex. There are
several  types of gears,  which depend on the  aircraft design and its
intended use.  Most landing gears have  wheels to facilitate operation
to    and     form        hard   surfaces,     such   as       airport
runways~\cite{Aviation2012}.
  
Three basic arrangements of landing gear are used: tail wheel type
landing gear, tandem landing gear, and tricycle-type landing gear. The
most commonly used landing gear arrangement is the tricycle-type
landing gear. All these aircraft landing gears are further classified
into fixed and retractable categories. Single engine and light weight
aircrafts use fixed landing gear while the retractable gear is used in
heavy aircrafts~\cite{Aviation2012}.

\begin{figure}
\centering
\includegraphics[scale=.35]{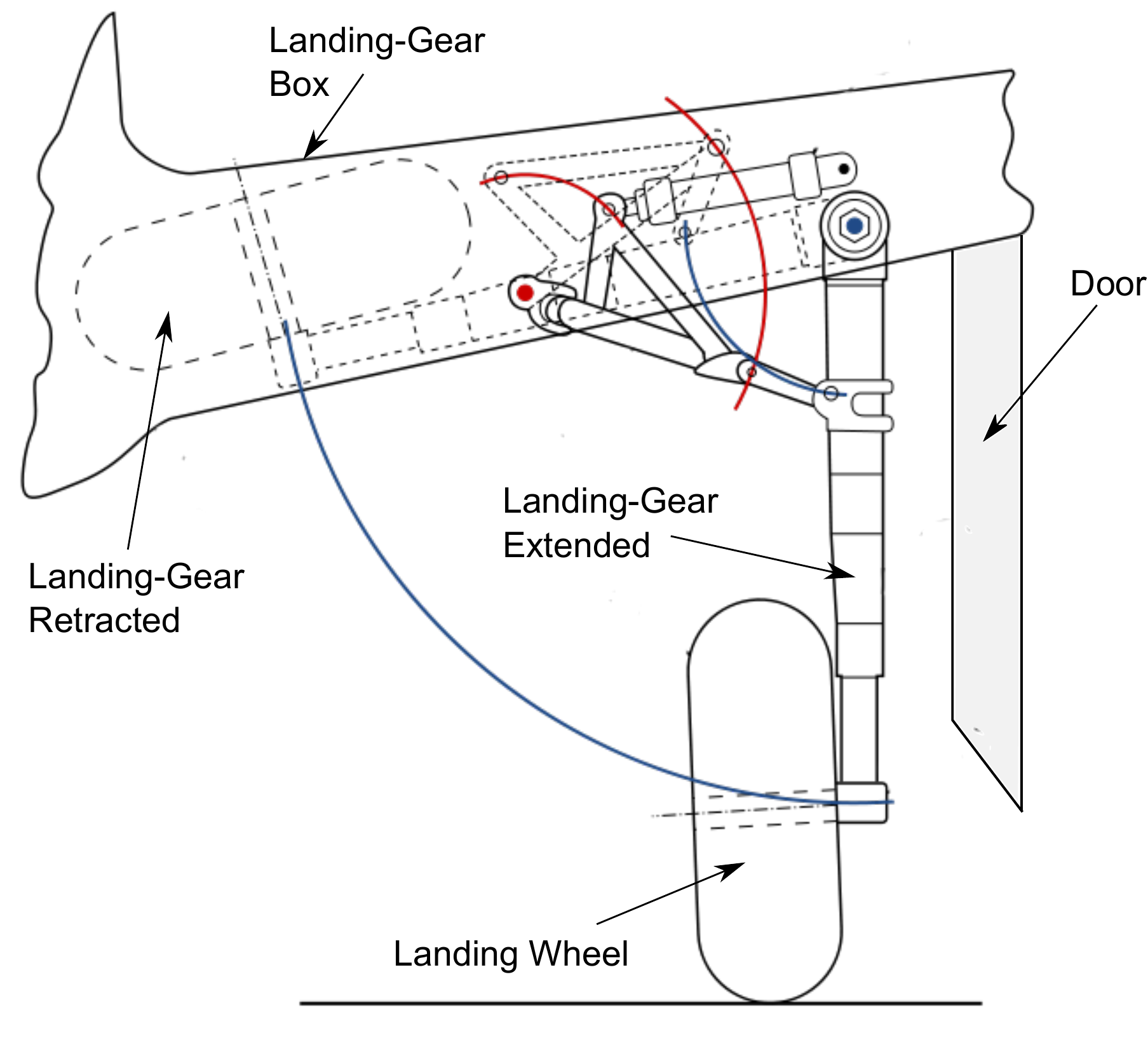}
\caption{Landing Gear System}
\label{fig1}
\end{figure}

The  landing   system controls  the   maneuvering  landing  gears  and
associated  doors.  Fig.~\ref{fig1} depicts basic components  of a
landing system. The landing system  is made of three different landing
sets, which corresponds to front, left and  right. The main components
of a landing system are doors, landing gears and hydraulic cylinders.

The landing  gear system   is controlled  by the  software  in nominal
operating mode, and an    emergency  mode is  handled    analogically.
Generally, landing  system always use nominal  mode. In case of system
failure,  the pilot can   activate the  emergency mode.  However,  the
landing  system can   be activated in  emergency   mode only when  any
anomaly is detected in the system.

There  are sequential   operations  of the landing    gear system. The
sequential operations for extending gears are: open  the doors of gear
boxes, extend  the landing gears, and  close the  doors. Similarly the
sequential operations for retraction gears are: open the door, retract
the landing  gears,  and close   the doors.  During  these  sequential
operations there are several parameters and conditions, which can be used
to assess the health of a landing system~\cite{Boniol2013}.

There are three main components of the landing gear system: 1)
mechanical system, 2) digital system, and 3) pilot interface. The
mechanical system is composed of three landing sets, where each set
contains landing gear box, and a door with latching boxes. The landing
gears and doors motions are performed with the help of cylinders. The
cylinder position is used to identify the various states of the door
or landing gear positions. Hydraulic power is used to control the
cylinders with the help of electro-valves. These electro-valves are
activated by a digital system.  The digital system is composed of two
identical computing modules, which execute parallel. The digital
system is only the responsible for controlling mechanical parts like
gears and doors, and for detecting anomalies. The pilot interface has
an Up/Down handle and a set of indicators. The handle is used by pilot
for extending or retracting the landing gear sequence, and a set of
indicators is the different type of lights for giving the actual
position of gears and doors, and the system state. A detailed
description of the landing gear system is given
in~\cite{Aviation2012,Boniol2013}.

The landing gear system is a critical embedded system, where all the
operations are based on the state of a physical device, and required temporal 
behaviour. The main challenge is to model the system behaviour
of the landing gear system, and to prove the safety requirements under
the consideration of physical behaviour of hydraulic devices.

\section{Formal Development of the Landing System }

The development is   progressively designing  the  landing system   by
integrating observations and elements of the  document. The first model
is specific as  abstract  as possible  and  it  captures the different possible \textit{big} steps of the system by defining the synchronous atomic events. For example, the sequence of door opening, door closing, gear extension and gear retraction etc.

\subsection{\model{M1}: Moving Up and Down}


When  the system is  moving up  (resp.  \textit{down}) till retraction
(resp.  extension),  it will  be    in a position  \textit{halt}   and
\textit{up} (resp.     \textit{down}),  namely  \textit{haltup} (resp.
\textit{haltdown)}. The first model observes the positions of the
global  state which considers that the  landing system is either
moving   \textit{down}  from a  \textit{haltup}  position,  or  moving
\textit{up}  from a  \textit{haltdown}   position.  The   global state
expresses the state of handle at  an initialization in a \textit{down}
state  ($button   := DOWN$)  and  the   gear system  is  halted  in  a
\textit{haltdown} position ($phase     := haltdown$). It   means  that
initially the gear system is  extended and locked. Two state variables
record  these informations namely $button$   and $phase$. Events model
the possible observable    actions and modifications over the   global
system:

\begin{itemize}
\item \event{PressDOWN} is  enabled, when the  gear system is halted up and retracted; the  system is  in a new state corresponding  to  the \textit{movingup} action. The intention is to extend the gear system.
\item \event{PressUP} is  enabled, when the  gear system is halted down and extended; the  system is  in a new state corresponding  to  the \textit{movingdown} action. The intention if to retract the gear system.
  \end{itemize}

\begin{figure}
\tiny  \centering
  
\begin{center}
\fbox{  
\begin{minipage}{0.7\linewidth}
\xymatrixcolsep{6pc}
\xymatrix{
\boxed{\boxed{{UP,haltup}}} \ar@{>}[r]_ {\event{Press\_Down}} & \boxed{{DOWN,movingdown}} \ar@{-->}[d]_ {\event{movingdown}} {\ar@{>}@/^/[dl]^{\ \ \event{PressUP}}}  \\
  \boxed{{UP,movingup}}  \ar@{-->}[u]_ {\event{movingup}} \ar@{>}@/^/[ur]^{\ \ \event{PressDOWN}} &  \boxed{{DOWN,haltdown}} \ar@{>}[l]_{\event{Press\_Up}}  \\
}
\end{minipage}
}  
\end{center}
\caption{State-based automaton for the model \model{M1}} \label{fig:m1}

\end{figure}
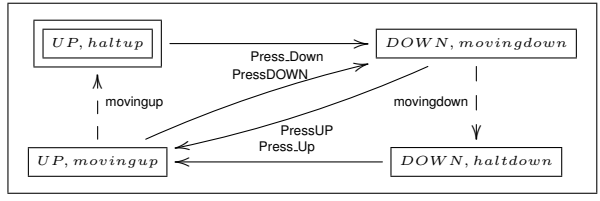

  Moreover,  when one of  events  \event{PressDOWN} or \event{PressUP}
  (solid  labelled transitions  in Fig.~\ref{fig:m1})   is  observed, the
  system    should provide  a  \textit{service}   corresponding  to an
  effective action (dashed  labelled transitions in Fig.~\ref{fig:m1}) of
  the landing system and physically moving  gears.  The landing system
  \textit{reacts}   (dashed labelled transitions in Fig.~\ref{fig:m1}) to
  the  orders of    the   pilot    (solid   labelled      transitions in
  Fig.~\ref{fig:m1}).

  \begin{itemize}
  \item \event{movingup} is  an action supported by  engine which helps to move the landing system into the state \textit{haltup} and to the retracted state.

  \item \event{movingdopwn} is  an action supported by  engine which helps to move the landing system into the state \textit{haltdown} and to the extended state.

\end{itemize}

Events express that, when the  button remains UP (resp. DOWN), the
reaction of the system is to reach  the state \textit{retracted} (resp.
\textit{extended}).   The current diagram assumes  that  the system is
operating in normal mode.  The detection of anomalies is left for the next
refinements.  The diagram contains the  main goals of the system which
is operating in a cyclic mode. The  requirements R11bis and R12bis are
clearly satisfied, as well as R12 and R11. Other requirements are not
considered since  they are   related  to  features  that are  not  yet
defined.

\subsection{\model{M2}: Opening and Closing Doors}

The model \model{M2} is considering different possible steps in the
moving up or in the moving down phases. However, the different steps
are possibly victims of counters orders. The pilot decides to press UP
and  then to press  DOWN or reciprocally.   These movements affect the
classical cycle of the  system starting from  a locked closed position
to another one without interrupt.  First observation leads to consider
that we identify   that doors are   alternatively \textit{opening} and
\textit{closing}.   We add  a  detail on the fact  that  the doors are
opened when they  are unlocked  and when  they  are  closed, they  are
locked. A new state is enriching the previous  one by a state variable
for  doors states ($dstate$) and a  variable for expressing when doors
are   locked ($lstate$).  Three  variables are    used  to control the
possible change of  decisions and expressing the  sequentialisation of
\textit{extension} scenario  or  \textit{retraction} scenario:  $p$,
$l$, $i$.

The next invariant states that when the doors are  opened, the doors are
unlocked ($M2\_inv5$); when one door is opened, all  the doors are opened ($M2\_inv3$)
and when a door is closed, all the doors are closed ($M2\_inv4$). 

\begin{center}
\scriptsize
$
\begin{Bcode}
{M2\_inv1 }:{ dstate \in  DOORS \tfun  SDOORS }\\
{M2\_ inv2 }:{ lstate \in  DOORS \tfun  SLOCKS }\\
{M2\_ inv3 }:{ dstate^{-1} [\{ OPEN\} ] \neq  \emptyset  \limp  dstate^{-1} [\{ OPEN\} ]=DOORS }\\
{M2\_ inv4 }:{ dstate^{-1} [\{ CLOSED\} ] \neq  \emptyset  \limp  dstate^{-1} [\{ CLOSED\} ]=DOORS }\\
{M2\_ inv5 }:{ dstate[DOORS]=\{ OPEN\}  \limp  lstate[DOORS]=\{ UNLOCKED\}  }\\
{M2\_ inv6 }:{ l=E \land  p=R \limp  lstate[DOORS]=\{ UNLOCKED\}  }\\
{M2\_ inv7 }:{ l=R \land  p=E \limp  lstate[DOORS]=\{ UNLOCKED\}  }
\end{Bcode}
$
\end{center}

\begin{figure}
\centering
\includegraphics[scale=.80]{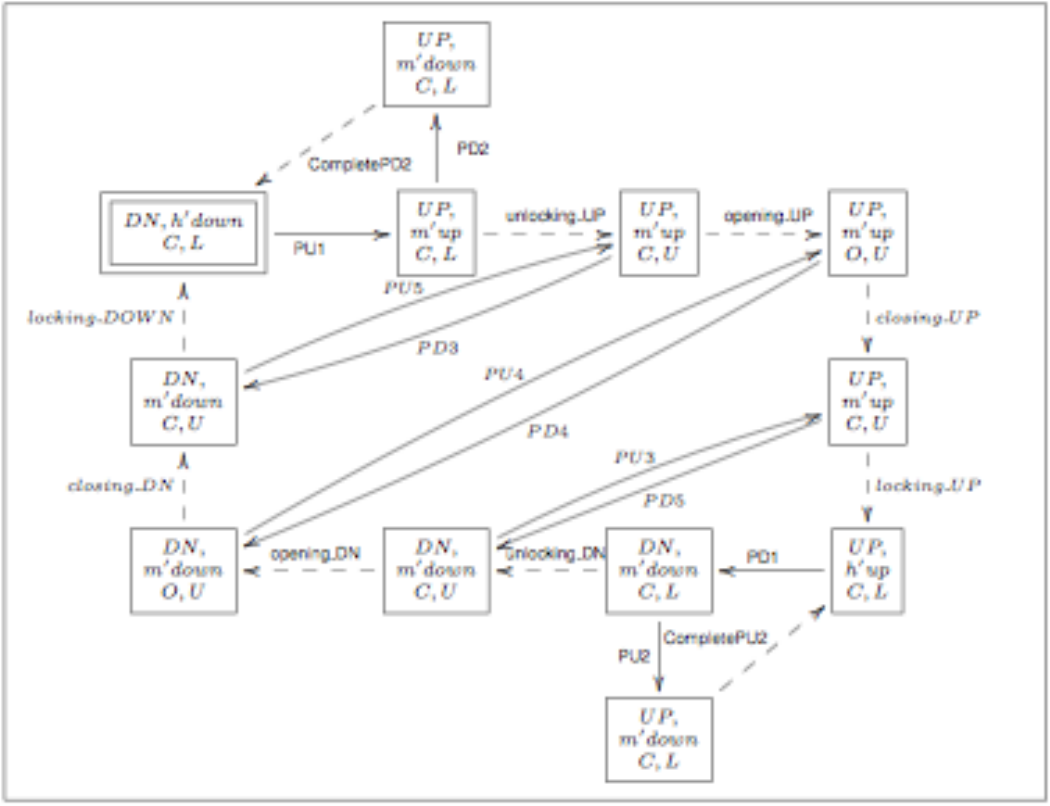}
\caption{State-based Automaton for Events in model \model{M2}}
\label{fig:m2}
\end{figure}

Events are now capturing the observation of opening and closing with
possible counter orders by the pilot.  We have not yet considered the
state of \textit{ flying }or \textit{grounding}.  Initially,  doors are closed and the
state is \textit{haltdown}.  It means that the landing system is
corresponding to a state on ground and should be obviously extended.
The three auxiliary variables ($p$,$l$,$i$) are set to $R$ to mean that the
system is ready to retract whenever the pilot wants. We do not
consider the case when a \textit{crazy} pilot would try to retract
when the aircraft is on the ground but we may consider that we observe
a safe situation. Further refinements will forbid these kinds of
possible behaviours. Events are refining the previous four events and
we refine the two events \event{PressDown} and \event{PressUp} by
events that can interrupt the initial scenario and switch to the other
scenario. Fig.~\ref{fig:m2} describes
the state-based automaton for the model \model{M2} and we use the
following notations $UP$ for $button=UP$, $DN$ for $button=DOWN$, $C$
for $dstate[DOORSQ]=\{CLOSED\}$, $O$ for $dstate[DOORSQ]=\{OPEN\}$,
$L$ for $lstate[DOORSQ]=\{LOCKED\}$, $U$ for
$lstate[DOORSQ]=\{UNLOCKED\}$, $m'down$ for $phase=movingdown", $m'up$
for $phase=movingup$, $h'down$ for $phase=haltdown", $h'up$ for
$phase=haltup$. The dashed and plain arrows present the distinction between two different types of actions. Dashed arrows show that it is an action of the system, and plain arrows show that it is an action of the pilot. 


%
%

The diagram   Fig.~\ref{fig:m2} confirms the  requirements.  The
model  is validated  using  ProB and the  sequences  of retraction and
extension are observed according to the requirements.

\subsection{\model{M3}: Observing the gears}

The next observation leads us to  consider the full mechanical system.
In fact, doors are opened and  closed but we  have the time to see that
gears are either   moving  out (extension   scenario) or   moving  in
(retraction scenario). The next model is  refining the previous one by
adding gears and observing different states of the  gears ($gstate \in GEARS
\tfun SGEARS$).  $SGEARS$ is defined as enumerated set $partition(SGEARS$, $\{ RETRACTED\}$, $\{ EXTENDED\} $, $\{ RETRACTING\}$ ,   $\{  EXTENDING\}$) to capture the multiple states of gears.  There are obvious    invariant properties that express  that the  doors are opened  when  the gears are
moving.  The invariants are listed  as follow:

\begin{center}
\scriptsize
$
\begin{Bcode}
 { M3\_inv1 }:{ gstate \in  GEARS \tfun  SGEARS }\\
 { M3\_inv2 }:{ \forall  door\qdot  
\left(\begin{array}{l} 
door \in  DOORS \land  dstate(door)=CLOSED\\
  \land  ran(gstate)\neq \{ RETRACTED\}\\  
\limp\\  
ran(gstate)=\{ EXTENDED\} 
\end{array}\right) }\\
{ M3\_inv3 }:{ \forall  door\qdot  
\left(
\begin{array}{l} 
door \in  DOORS \land  dstate(door)=CLOSED\\   
\land  ran(gstate)\neq \{ EXTENDED\}\\
\limp\\  
ran(gstate)=\{ RETRACTED\}\\ 
\end{array}\right) }
\end{Bcode}
$
\end{center}

\begin{center}
\scriptsize
$
\begin{Bcode}
{M3\_inv4 }:{ 
\left(\begin{array}{l} 
ran(gstate)\neq \{ RETRACTED\}  \land  ran(gstate)\neq \{ EXTENDED\}\\   
\limp\\  ran(dstate)=\{ OPEN\}
\end{array}\right) }\\
{M3\_inv5 }:{ 
\left(\begin{array}{l} 
ran(dstate)=\{ CLOSED\}\\  \limp\\
ran(gstate)\binter \{ RETRACTING,EXTENDING\}  =\emptyset
\end{array}\right) }
\end{Bcode}
$
\end{center}

$M3\_inv2$ and  $M3\_inv3$ express that  when doors are opened,  either the
gears  are extended or the  gears  are retracted.  When  the doors are
closed, the gears  are not in moving state  ($M3\_inv4$ and $M3\_inv5$). When the gears
are moving,   the   doors are  opened.    The  expression of the
simultaneaous  state of the doors either  closed or opened, as well as
the gears either extended  or  retracted, prepare the conditions  of
the   synchronisation    over      the  doors    and     the    gears.
Fig.~\ref{fig:m2} is  now  detailed by  splitting  the two  states
$(DN,  m'down, O, U)$ and $(UP, m'up, O, U)$ and by  considering that the new
variable $gstate$ is  modified at this  stage. We are introducing four
new events corresponding to the  retraction of gears and to the extension of gears.

The \textit{retraction} phase  is decomposed into two events   \event {retracting\_gears} and 
\event {retraction}  and the gears are transiting from a state \textit{EXTENDED} into the state \textit{RETRACTING} and finally the state \textit{RETRACTED}.

\begin{center}
\scriptsize
$
\begin{Bcode}
  \Bevent {retracting\_gears}\\
		\quad \Bkeyword{WHEN}\\
			\quad\quad { grd1 }:{ dstate[DOORS]=\{ OPEN\}  }\\
			\quad\quad { grd2 }:{ gstate[GEARS]=\{ EXTENDED\}  }\\
			\quad\quad { grd3 }:{ p=R }\\
		\quad \Bkeyword{THEN}\\
			\quad\quad { act1 }:{ gstate :=  \{ a\mapsto b| a \in  GEARS \land  b=RETRACTING\}  }\\
\quad \Bkeyword{END}\\
\Bevent {retraction}\\
		\quad \Bkeyword{WHEN}\\
			\quad\quad { grd1 }:{ dstate[DOORS]=\{ OPEN\}  }\\
			\quad\quad { grd2 }:{ gstate[GEARS]=\{ RETRACTING\}  }\\
		\quad \Bkeyword{THEN}\\
			\quad\quad { act1 }:{ gstate:=   \{ a\mapsto b|  a \in  GEARS \land  b= RETRACTED\}  }\\
		\quad \Bkeyword{END}
                \end{Bcode}
$
\end{center}

The \textit{extension} phase   is decomposed  into two events   \event
{extending\_gears} and \event {extension} and the gears are transiting
from a state \textit{RETRACTED} into the state \textit{EXTENDING} and
finally the state \textit{EXTENDED}.

\begin{center}
\scriptsize
$
\begin{Bcode}
\Bevent {extending\_gears}\\
		\quad \Bkeyword{WHEN}\\
			\quad\quad { grd1 }:{ dstate[DOORS]=\{ OPEN\}  }\\
			\quad\quad { grd2 }:{ gstate[GEARS]=\{ RETRACTED\}  }\\
			\quad\quad { grd3 }:{ p=E }\\
		\quad \Bkeyword{THEN}\\
			\quad\quad { act1 }:{ gstate :=  \{ a\mapsto b| a \in  GEARS \land  b=EXTENDING\}  }\\
		\quad \Bkeyword{END}\\
\Bevent {extension}\\
		\quad \Bkeyword{WHEN}\\
			\quad\quad { grd1 }:{ dstate[DOORS]=\{ OPEN\}  }\\
			\quad\quad { grd2 }:{ gstate[GEARS]=\{ EXTENDING\}  }\\
		\quad \Bkeyword{THEN}\\
			\quad\quad { act1 }:{ gstate :=  \{ a\mapsto b|  a \in  GEARS \land  b=EXTENDED\}  }
\\			
		\quad \Bkeyword{END}
\end{Bcode}
$
\end{center}

\begin{figure}
\centering
\includegraphics[scale=.80]{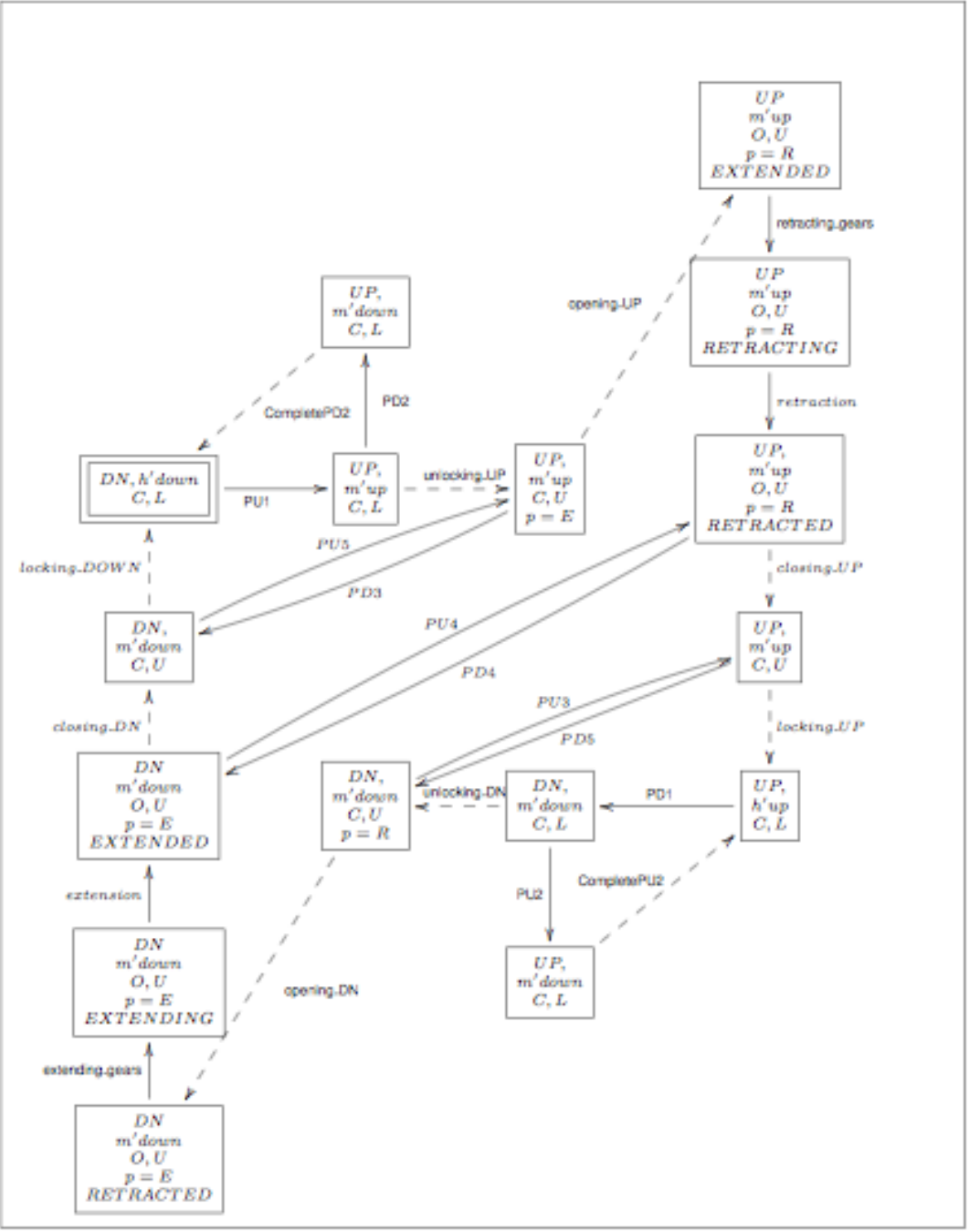}
\caption{State-based Automaton for Events in model  \model{M3}}
\label{fig:m3}
\end{figure}

The events \event{PU4} and   \event{PD4} are both refined into   three
events which  are controlling or  switching from the  retraction to the
extension and vice-versa. The two   possible scenarios (extension  and
retraction) have a meaning and we can address the requirements R21 and R22.

The model \model{M3} is refined into a new model called \model{M30}
which is \textit{forbidden} the use of buttons.  The model is clearly
satisfying the requirement over the successive actions. ProB is also used to validate the behaviour of system. 


\subsection{\model{M4}: Sensors and Actuators}

 In this refinement, we address the problem of sensors   and 
actuators. We introduce the management of sensors and actuators by
considering the collection of values of  sensors and an abstract
expression of computing modules  for analysing  the  sensed
values.  We introduce a list of new  variables according to the Fig.~\ref{fig:cm}:

\begin{figure}
  \centering
  \includegraphics[scale=0.40]{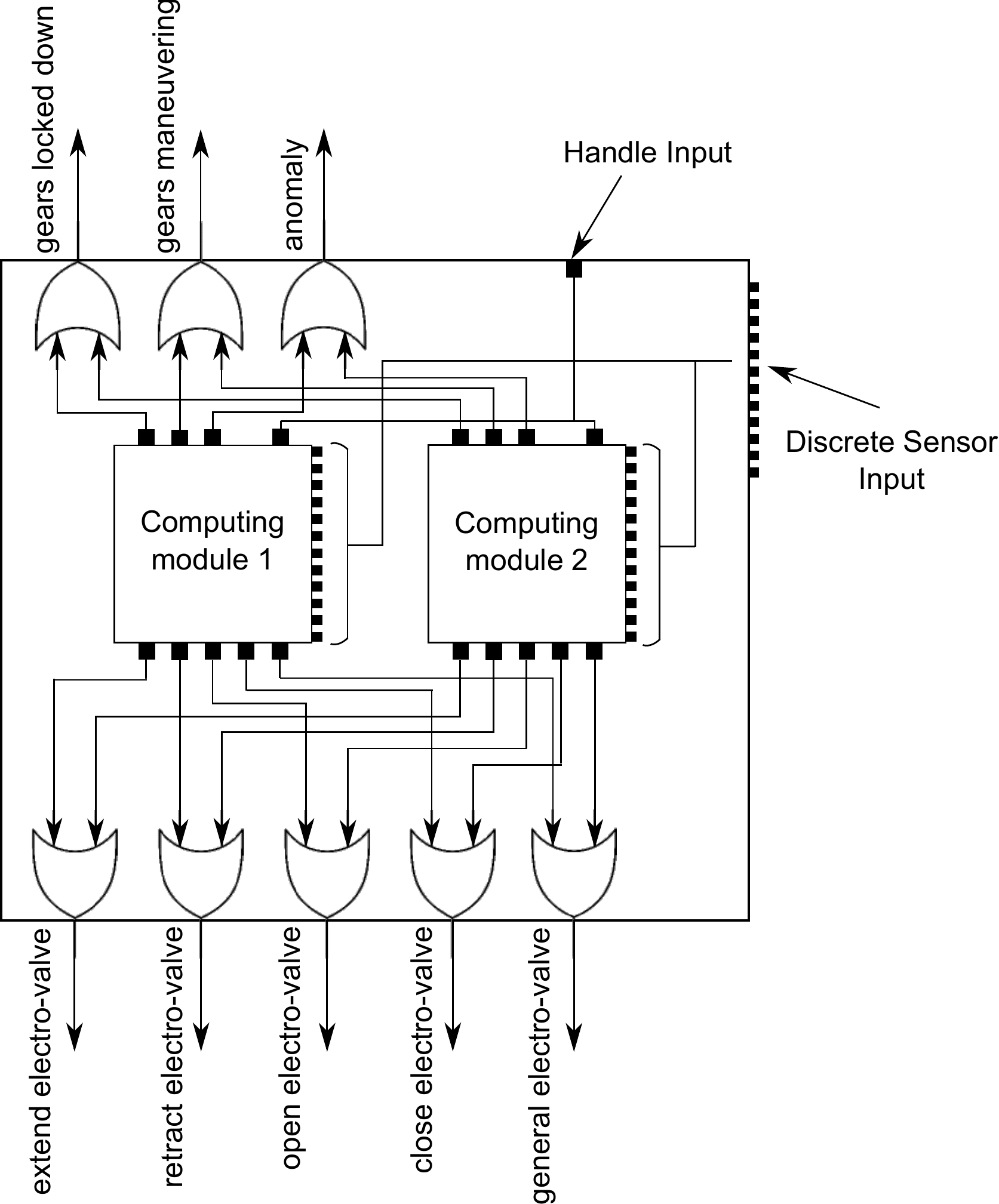}
  \caption{Architecture of the computing modules}
  \label{fig:cm}
\end{figure}

\begin{itemize}
\item Variables for expressing the sensors states: \textit{handle} (for the pilot
  interface), \textit{analogical\_switch},   \textit{gear\_extended}  (sensor  for
  detecting  an extension  activity), \textit{gear\_retracted} (sensor for
  detecting a retraction  activity), \textit{gear\_shock\_absorber}  (sensor
  for detecting the flight or ground mode), \textit{door\_closed} (sensor for
  stating when the doors are closed), \textit{door\_open} (sensor for stating
  when the doors are opened),  \textit{circuit\_pressurized} (sensor for  the
  pressure control).

\item Variables for getting the evaluation of the sensed states of the
  system   by  the  computing    modules:  \textit{general\_EV}, \textit{close\_EV},
  \textit{retract\_EV}, \textit{extend\_EV}, \textit{open\_EV}.

  \item Variables for  modelling the computation  of the sensed  state
    from  the  collected    sensed  values:   \textit{general\_EV\_func},
    \textit{close\_EV\_func},  \textit{retract\_EV\_func}, \textit{extend\_EV\_func},
  \textit{open\_EV\_func}.

\item  Variables   for   collecting  output of      computing modules:
  \textit{gears\_locked\_down}, \textit{gears\_man}, \textit{anomaly}.

\end{itemize}

New variables are  used  for  adding some  constraints  over  guards of
previous events:

\begin{itemize}
\item \event{HPD1} and \event{HPU1} are  two events corresponding to
  the order by  the  pilot interface  to  extend (HPD1) or  to retract
  (HPU1) the  gears.  For instance,  the  guard  $\forall x\qdot  x\in
  1\upto 3  \limp handle(x)=UP $ senses that  the handle is UP and then
  it  moves to  DOWN  ($handle  :\in 1\upto  3 \tfun  \{  DOWN\}$. The
  sensors are triplicated and we define each sensor value by a function
  from $1..3$ into the sensors values.

\item     \event  {Analogical\_switch\_closed}        and       \event
  {Analogical\_switch\_open} are two events  for updating  the general
  switch  for protecting the system agiants  abnormal behaviour of the
  digital part.

\item 	\event {Circuit\_pressurized} manages  the sensor of the pressure control.

\item \event {Computing\_Module\_1\_2} models in a very abstract
  way for computing and updating of $EV$ variables using sensors values.

\item \event {Failure\_Detection} detects any failure in the system.
	
\end{itemize}

The  model introduces  sensors   and values synthesized from   sensors
values.   We  have used a   very abstract way to  state  the values of
sensors.  The  model  \model{M4} is not   analyse-able with ProB.  The
previous requirements R11, R11bis, R12, R12bis, R22, R21 are remaining
satisfied  by   the   model  \model{M4}  by   refinement.   We need to
strengthening  the guards  of events  ($\forall  x\qdot x\in  1\upto 3
\limp handle(x)=button$).  The reader will notice  that the two events
\event{HPU1}  and   \event{HPD1}  are   the  external   interfaces  for
controlling the  events to associate the  functionality of handle with
old variable $button$.  The guard $gear\_shock\_absorber = \{ a\mapsto
b | a \in 1\upto 3 \land b = ground\}$ indicates that now we know that
either we are  on the ground  or  not: it  means  that we assume  that
sensors   are trusted and this  assumption  is valid.   The state of $
gear\_shock\_absorber$ is modified according to the figure 11, page 12
of \cite{Boniol2013} and  it is the reason for updating in two
events \event{extension} and \event{retraction}.

In this refinement,  the number of proof obligations is very high (247) but  it is possible to add intermediate models  for progressive development.

\begin{itemize}
\item The two events \event{HPU1} and \event{HPD1} are adding a small amount of  new proof obligations.

\item The  unproved proof obligations appearing in the summary are mainly  typing properties and   they are discharged either using the SMT solver  or  a procedure. We consider that they are   pseudo-automatic proof obligations.
\end{itemize}

\subsection{\model{M5}: Managing electro-valves}

The  model \model{M5} takes into  account  the management  of electro-valves
used for moving the  gears from a  position  to another one. Four  new
variables  are modelling pressure   states  (page 10, subsection  3.2,
Electro-valves) and they model the hydraulical features of the
system:  ${  general\_EV\_Hout   }$, ${   close\_EV\_Hout  }$,    $  {
  retract\_EV\_Hout }$, $ { extend\_EV\_Hout }$, ${ open\_EV\_Hout }$,
${ A\_Switch\_Out  }$. The   invariant   is stating that  either   the
pressure is $on$ or $off$ by the two possible values: $0$ or $Hin$:

\begin{center}
\scriptsize
$
\begin{Bcode}
{ inv1 }:{ general\_EV\_Hout \in  \{ 0, Hin\}  }\\
{ inv2 }:{ close\_EV\_Hout \in  \{ 0, Hin\}  }\\
{ inv3 }:{ retract\_EV\_Hout \in  \{ 0, Hin\}  }\\
{ inv4 }:{ extend\_EV\_Hout \in  \{ 0, Hin\}  }\\
{ inv5 }:{ open\_EV\_Hout \in  \{ 0, Hin\}  }
\end{Bcode}
$
\end{center}

The summary of   new proof obligations is  simply  that  19 new  proof
obligations are generated and automatically  discharged. In the previous
development, the values were less precise and  we got a problem in the
next  refinements  with some proof obligations  to  discharge.   A new
event \event {Update\_Hout}  is   introduced to update the  values  of
sensors for the hyraulic part:

\begin{center}
\scriptsize
$
\begin{Bcode}
  \Bevent {Update\_Hout}\\		
\quad \Bkeyword{BEGIN}\\
\quad\quad { act1 }:{ general\_EV\_Hout :|  \left(
\begin{array}{l} (general\_EV = TRUE \land   general\_EV\_Hout' = Hin)\\
 \lor  (general\_EV = FALSE \land  general\_EV\_Hout' = 0)\\
\lor  (A\_Switch\_Out = TRUE \land   general\_EV\_Hout' = Hin)\\
 \lor  (A\_Switch\_Out = FALSE \land   general\_EV\_Hout' = 0)
\end{array}\right) }\\
\quad\quad { act2 }:{ close\_EV\_Hout :| 
\left(
\begin{array}{l}
(close\_EV = TRUE \land  close\_EV\_Hout' = Hin)\\ 
\lor  (close\_EV = FALSE \land  close\_EV\_Hout' = 0)
\end{array}
\right)
}\\
\quad\quad { act3 }:{ open\_EV\_Hout :|  
\left(
\begin{array}{l}
(open\_EV = TRUE \land  open\_EV\_Hout' = Hin)\\ 
\lor  (open\_EV = FALSE \land  open\_EV\_Hout' = 0)
\end{array}
\right)}\\
\quad\quad { act4 }:{ extend\_EV\_Hout :|  
\left(
\begin{array}{l}
(extend\_EV = TRUE \land  extend\_EV\_Hout' = Hin)\\ \lor  (extend\_EV = FALSE \land  extend\_EV\_Hout' = 0)
\end{array}
\right) }\\
\quad\quad { act5 }:{ retract\_EV\_Hout :|  
\left(
\begin{array}{l}
(retract\_EV = TRUE \land  retract\_EV\_Hout' = Hin)\\ 
\lor  (retract\_EV = FALSE \land  retract\_EV\_Hout' = 0)
\end{array}
\right) }\\
\quad \Bkeyword{END}
\end{Bcode}
$
\end{center}

The  event  \event {Circuit\_pressurized}   is  refined by two  events
considering  that the sensing  is OK or not; it   assigns the value of
$Hout$.

\begin{center}
\scriptsize
$
\begin{Bcode}
	 \Bevent {Circuit\_pressurized}\\
	\quad \BRevent {Circuit\_pressurized}\\
		\quad \Bkeyword{WHEN}\\
			
			\quad\quad { grd1 }:{ general\_EV\_Hout = Hin }\\
		\quad \Bkeyword{THEN}\\
			\quad\quad { act9 }:{ circuit\_pressurized :\in  1\upto 3 \tfun  \{ TRUE\}  }\\
			
		\quad \Bkeyword{END}
                \end{Bcode}
$
\end{center}

\begin{center}
\scriptsize

$
                \begin{Bcode}
\Bevent {Circuit\_pressurized\_notOK}\\
	\quad \BRevent {Circuit\_pressurized}\\
		\quad \Bkeyword{WHEN}\\
			\quad\quad { grd1 }:{ general\_EV\_Hout = 0 }\\
		\quad \Bkeyword{THEN}\\
			\quad\quad { act9 }:{ circuit\_pressurized :\in  1\upto 3 \tfun  \{ FALSE\}  }\\
		\quad \Bkeyword{END}
\end{Bcode}
$
\end{center}

\subsection{\model{M6}:  Integrating Cylinders Behaviours}

The next step is to integrate the cylinders behaviour according to the
electro-valves circuit and to control the process, which is computing
from sensors values the global state of the system. It leads to
strengthen guards of events opening and closing doors and gears using
cyliders sensors and hydraulic pressure (\event
{opening\_doors\_DOWN}, \event {opening\_doors\_UP}, \event
{closing\_doors\_UP}, \event {closing\_doors\_DOWN}, \event
{unlocking\_UP}, \event {locking\_UP}, \event {unlocking\_DOWN},
\event {locking\_DOWN}, \event {retracting\_gears}, \event
{retraction}, \event {extending\_gears}, \event {extension}).  The
event \event {CylinderMovingOrStop} models the change of the cylinders
according to the pressure, when the value of $state$ is $cylinder$. It
leads to a next state which \textit{activates} the computing
modules.

\begin{center}
\scriptsize
$
\begin{Bcode}
\Bevent {CylinderMovingOrStop}\\
\textit{ Cylinder Moving or Stop according to the output of hydraulic circuit }\\
\quad \Bkeyword{WHEN}\\
\quad\quad { grd1 }:{ state = cylinder }\\
	\quad \Bkeyword{THEN}\\
\quad\quad { act1 }: SGCylinder :|\\   \left(\begin{array}{l}
\ \left(\begin{array}{l}
SGCylinder' = \{ a\mapsto b |  a\in  GEARS\cprod \{ GCYF,GCYR,GCYL\} \land  b=MOVING  \}\\
 \land\   extend\_EV\_Hout = Hin 
\end{array}\right)\\ 
\lor \left(\begin{array}{l}
  SGCylinder' = \{ a\mapsto b |  a\in  GEARS\cprod \{ GCYF,GCYR,GCYL\}  \land  b=STOP\}\\   \land\   extend\_EV\_Hout = 0 
\end{array}\right)\\ 
\lor
\left(\begin{array}{l}
SGCylinder' = \{ a\mapsto b |  a\in  GEARS\cprod \{ GCYF,GCYR,GCYL\}  \land  b=MOVING\}\\  \land\   retract\_EV\_Hout = Hin 
\end{array}\right)\\ 
\lor \left(\begin{array}{l}
SGCylinder' = \{ a\mapsto b |  a\in  GEARS\cprod \{ GCYF,GCYR,GCYL\}  \land  b=STOP\}\\  \land\   retract\_EV\_Hout = 0
\end{array}\right)\\ 
\end{array}
\right)\\
\quad\quad { act2 }: SDCylinder :|\\  \left(\begin{array}{l}
\of 
SDCylinder' = \{ a\mapsto b |  a\in  DOORS\cprod \{ DCYF,DCYR,DCYL\}  \land  b=MOVING\}\\
  \land\   open\_EV\_Hout = Hin
\ff\\
\lor \of SDCylinder' = \{ a\mapsto b |  a\in  DOORS\cprod \{ DCYF,DCYR,DCYL\}  \land  b=STOP\}\\   \land\   open\_EV\_Hout = 0
\ff\\
\lor \of SDCylinder' = \{ a\mapsto b |  a\in  DOORS\cprod \{ DCYF,DCYR,DCYL\}  \land  b=MOVING\}\\  
\land\   close\_EV\_Hout = Hin
\ff\\
\lor \of SDCylinder' = \{ a\mapsto b |  a\in  DOORS\cprod \{ DCYF,DCYR,DCYL\} \land  b=STOP\}\\  \land\   close\_EV\_Hout = 0
\ff
\end{array}
\right) \\
\quad\quad { act3 }:{ state :=  computing }\\
		
		\quad \Bkeyword{END}
\end{Bcode}
$
\end{center}

More than 50 \% of the proof obligations are manually discharged.
However, it appears that the disjunction of actions allows us to have
a unique view of the cylinders behaviours.  The proofs to discharge
are not complex and are mainly discharged by several clicks on
procedures buttons.

\subsection{\model{M7}: Failure Detection}
The model \model{M7} is modelling  the detection of different possible
failures.  Page 16 and page 17 of the case study have given a list of conditions for  detecting anomalies: \textit{Analogical  switch   monitoring, Pressure sensor
  monitoring,   Doors   motion  monitoring,  Gears  motion monitoring,
  Expected behavior in case of anomaly.} The decision is to refine the
event \event{Failure\_Detection} into  six events which are  modelling
the    different      cases    for     failure   detection:    \event
{Failure\_Detection\_Generic\_Monitoring},                      \event
{Failure\_Detection\_Ana-logical\_Switch},                       \event
{Failure\_Detection\_Pressure\_Sensor},                         \event
{Failure\_Detection\_Doors},     \event   {Failure\_Detection\_Gears},
\event  {Failure\_Detection\_Generic\_Monitoring}.  The decision is to
postpone  the introduction of timing   constraints  in the last  model.
However,  we  have     to  strengthen    the guards  of   events    \event
{opening\_doors\_DOWN},       \event    {opening\_doors\_UP},   \event
{closing\_doors\_UP},   \event       {closing\_doors\_DOWN},    \event
{unlocking\_UP},    \event {locking\_UP}, \event    {unlocking\_DOWN},
\event {locking\_DOWN} by adding a condition $anomaly=FALSE$.

\subsection{\model{M8}: Timing Requirements}

The   time pattern~\cite{cansell06}   provides  a way    to add timing
properties. The pattern adds an event \event {tic\_tock} simulating the
progression of time. Timing  properties are derived from the document.
We agree with  possible  discussions on the  modelling of  time but it
appears that further works are required to get a better integration of
a \textit{more real time} approach. However, we think that the current
model \model{M8} is an abstraction of another automaton with real time
features~\cite{DBLP:journals/tcs/AlurD94}.

\begin{center}
\scriptsize
$
\begin{Bcode}
\Bevent {tic\_tock}\\
\quad \Bkeyword{ANY}\\
			\quad\quad {tm }\\
		\quad \Bkeyword{WHERE}\\
			\quad\quad { grd1 }:{ tm \in  \nat }\\
			\quad\quad { grd2 }:{ tm >  time }\\
			\quad\quad { grd3 }:{ ran(at) \neq  \emptyset  \limp  tm \leq  min(ran(at)) }\\
		\quad \Bkeyword{THEN}\\
			\quad\quad { act1 }:{ time :=  tm }\\
		\quad \Bkeyword{END}
\end{Bcode}
$
\end{center}

The pilot uses the handle and the handle is taking  some time to change
the value of the sensors. 

\begin{center}
\scriptsize
$
\begin{Bcode}
\Bevent {HPD1}\\
	\quad \BRevent {HPD1}\\
		\quad \Bkeyword{WHEN}\\
			\quad\quad { grd3 }:{ \forall x\qdot x\in 1\upto 3 \limp  handle(x)=UP }\\
		\quad \Bkeyword{THEN}\\
			\quad\quad { act2 }:{ handle :\in  1\upto 3 \tfun  \{ DOWN\}  }\\
			\quad\quad { act3 }:{ at :=  at \bunion  \{ (index+1) \mapsto  (time+160)\}  }\\
			\quad\quad { act4 }:{ handleDown\_interval :=  time + 40000 }\\
			\quad\quad { act5 }:{ handleUp\_interval :=  0 }\\
			\quad\quad { act6 }:{ index :=  index +1 }\\
		\quad \Bkeyword{END}\\
		
\Bevent {HPU1}\\
\quad \BRevent {HPU1}\\
		\quad \Bkeyword{WHEN}\\
			\quad\quad { grd3 }:{ \forall x\qdot x\in 1\upto 3 \limp  handle(x)=DOWN }
			
		\quad \Bkeyword{THEN}\\
			
			\quad\quad { act2 }:{ handle :\in  1\upto 3 \tfun  \{ UP\}  }\\
			\quad\quad { act3 }:{ at :=  at \bunion  \{ (index+1) \mapsto  (time+160)\}   }\\
			\quad\quad { act4 }:{ handleUp\_interval :=  time + 40000 }\\
			\quad\quad { act5 }:{ handleDown\_interval :=  0 }\\
			\quad\quad { act6 }:{ index :=  index + 1 }\\			
		\quad \Bkeyword{END}
\end{Bcode}
$
\end{center}

The proof   assistant   is not efficient    on  this new  refinement.
However, now we can cover  requirements with timing aspects.


\subsection{\model{M9}: Adding Lights}
The last refinement of our development introduces the interface of
the pilot:   \textit{the lights}. These  lights  are modelled  by a variable as $pilot\_interface\_light \in colorSet \tfun lightState $. Initially,
p$ilot\_interface\_light$ is set to $\{ Green\mapsto Off,Orange\mapsto
Off,Red\mapsto Off\}$.  The following  events are informing  the pilot
by interpreting the results of the computing modules and they are extracted from the document:

\begin{itemize}
\item \event {pilot\_interface\_Green\_light\_On}:   green light is on; when gears locked down is true.
\item  \event {pilot\_interface\_Orange\_light\_On}:   orange light is on, when gears maneuvering is true.
\item  \event {pilot\_interface\_Red\_light\_On}:   red light is on, when anomaly is detected (true).
\item  \event {pilot\_interface\_Green\_light\_Off}:  green light is off, when gears locked down is false.
\item \event {pilot\_interface\_Orange\_light\_Off}:  orange light is off, when gears maneuvering is false.
\item  \event {pilot\_interface\_Red\_light\_Off}:   red light is off, when anomaly is detected (false).
\end{itemize}

\section{Conclusion}

Validation  and verification   are   processed  by  using   the   ProB
tool~\cite{prob03} and Proof Statistics. \textit{Validation} refers to
gaining confidence that the  developed  formal models are   consistent
with the requirements,    which are  expressed in  the    requirements
document~\cite{Boniol2013}.  The  landing    system   specification is
developed and formally proved by  the Event-B tool support prover. The
developed formal models are also  validated  by the ProB tool  through
animation   and  model  checker  tool   support of   the abstract  and
successive refined models  under some constraints  of the tool.    These
constraints  are the selection of   parameters  for testing the  given
model, and avoiding the failure of the  tool during animation or model
checking.  However, we use  this tool on abstract  and all the refined
models to check that the developed specification is deadlock free from
an initial model to the concrete model. Due to
features of ProB,    we have used  ProB  for  the models   \model{M1},
\model{M2} and \model{M3}. 

The Table-Fig\ref{fig:table} is expressing the proof statistics of the
development in the RODIN tool.  These statistics measure the size of
the model, the proof obligations are generated and discharged by the
Rodin platform, and those are interactively proved.  The complete
development of landing system results in 529(100\%) proof obligations,
in which 448(84,68\%) are proved completely automatically by the RODIN
tool.  The remaining 81(15,31\%) proof obligations are proved
interactively using RODIN tool. In the models, many proof obligations
are generated due to introduction of new functional and temporal
behaviors.  In order to guarantee the correctness of these functional
and temporal behaviors, we have established various invariants in
stepwise refinement. Most of the proofs are automatically discharged
and the interactively discharged proof obligations are discharged by
simple sequence of using automatic procedures of Rodin.

\begin{figure}
  \centering
\fontsize{8}{8}\selectfont
\newcolumntype{Z}{@{\hspace{2pt}}X}
\setlength\extrarowheight{4pt}
\noindent
\begin{tabularx}{1.0\textwidth}
  {|p{1.5cm}|p{6cm}|Z|Z|Z|}
    \hline
     {\bf Model} &	{\bf Requirements}	& {\bf Total PO}	 & {\bf Auto}	& {\bf Man}\\ \hline
      M1	        &R11, R11bis,R12, R12bis  	      &  10 &	10	 &0 \\ \hline
      M2	        &  R11, R11bis,R12, R12bis     	      & 33 &	33	 & 0 \\ \hline
      M3	        &  R11, R11bis,R12, R12bis,    R22, R21 	      &44  &	44	 & 0\\ \hline
      M4	        &   R11, R11bis,R12, R12bis,    R22, R21 	      &264 &	252	 & 12\\ \hline 
      M5	        &   R11, R11bis,R12, R12bis,    R22, R21 	      &19 &	19	 & 0\\ \hline 
      M6	        &   R11, R11bis,R12, R12bis,    R22, R21 	      &49&	20 	 & 29\\ \hline
      M7	        &   R11, R11bis,R12, R12bis,    R22, R21 	      &1&	0 	 & 1\\ \hline 
      M8	        &   R11, R11bis,R12, R12bis,    R22, R21 	      &56&23 & 33\\ \hline 
      M9	        &   R11, R11bis,R12, R12bis,    R22, R21 	      &9&3 &6\\ \hline 
Total	        &   R11, R11bis,R12, R12bis,    R22, R21 	      &529&448 &81\\ \hline 
\end{tabularx}
\caption{Table  of  requirements satisfied by models and proof statistics}
\label{fig:table}
\end{figure}

The  current version of the development  is the n$^{th}$ version.  The
document describes a concrete   system with sensors,  mechanical parts
and digital  part.   A first attempt  by  one of  the authors  was  to
propose  a   sequence  of  refined   models too   much  close  of this
description.  Then we try  to have a global  view of the system and to
provide   a very abstract  initial   model.   In  a  second round   of
derivation of models, we got a wrong model, since we did not take into
account  the  counter   orders.     Finally,   the diagram    of   the
Fig.~\ref{fig:m3}   summarizes main steps   of  the system.  From this
model, we decide   to make elements more   concrete  and we  introduce
sensors,  computing modules.   Timing requirements   are added in  the
pre-last model  \model{M8}  which is then  equipped by  lights  in the
model \model{M9}. Our models are still too abstract and we have to get
comments and feedbacks from the domain experts.

\bibliographystyle{plain}
\bibliography{citation}


\clearpage
\appendix

\section{Requirements}
\label{sec:req}


\req{R11}{When the  command line   is  working (normal mode),   if the
  landing gear command button  has  been pushed  DOWN and stays  DOWN,
  then the gears will be locked down and the doors will be seen closed
  less than 15 seconds after the button has been pushed.  }

\req{R12}{ When  the command  line  is working  (normal mode),  if the
  landing gear command button has been  pushed UP  and stays UP,  then
  the gears will be locked retracted and the doors will be seen closed
  less than 15 seconds  after the button has  been pushed. Note that a
  weaker  version  of   these two requirements    could  be considered
  as well. This weaker version does not take into account quantitative
  time.}

\req{ (R11bis)}{  When the command line is  working (normal  mode), if
  the landing gear command button has been pushed DOWN and stays DOWN,
  then eventually the gears will be locked down and  the doors will be
  seen closed.  }

\req{(R12bis)}{ When the command line is working (normal mode), if the
landing gear command button has been pushed UP and stays UP, then
eventually the gears will be locked retracted and the doors will be
seen closed.}



\req{  (R21)}{ When the command line is working (normal mode), if the landing gear command button remains in the DOWN position, then retraction sequence is not observed.
}

\req{(R22)}{ When the command line is working (normal mode), if the
  landing gear command button remains in the UP position, then
  outgoing sequence is not observed.  }


\req{ (R31)}{ When the command line is working (normal mode), the
  stimulation of the gears outgoing or the retraction electro-valves
  can only happen when the three doors are locked open.}

\req{  (R32)}{ When the command line is working (normal mode), the stimulation of the doors opening or closure electro-valves can only happen when the three gears are locked down or up.
}

\req{(R41)}{ When the command line is working (normal mode), opening and closure doors electro-valves are not stimulated simultaneously ; outgoing and retraction gears electro-valves are not stimulated simultaneously.
}

\req{  (R42)}{ When the command line is working (normal mode), opening doors electro-valve and closure doors electro-valve are not stimulated simultaneously outgoing gears electro-valve and retraction gears electro-valve are not stimulated simultaneously
}

\req{(R51)}{

When the command line is working (normal mode), it is not possible to stimulate the maneuvering electro-valve (opening, closure, outgoing or retraction) without stimulating the general electro-valve.

}

\req{(R61)}{ If one of the three doors is still seen locked in the closed position more than 0.5 second after stimulating the opening electro-valve, then the boolean output  normal mode is set to  false.
}

\req{  (R62)}{ If one of the three doors is still seen locked in the open position more than 0.5 second after stimulating the closure electro-valve, then the boolean output  normal mode is set to  false. 
}

\req{  (R63)}{ If one of the three gears is still seen locked in the down position more than 0.5 second after stimulating the retraction electro-valve, then the boolean output  normal mode is set to  false. 
}

\req{  (R64)}{ If one of the three gears is still seen locked in the up position more than 0.5 second after stimulating the outgoing electro-valve, then the boolean output  normal mode is set to  false.
}

\req{ (R71)}{ If one of the three doors is not seen locked in the open
position more than 2 seconds after stimulating the opening
electro-valve, then the boolean output normal mode is set to false.  }

\req{ (R72)}{ If one of the three doors is not seen locked in the closed position more than 2 seconds after stimulating the closure electro-valve, then the boolean output  normal mode is set to  false.
}

\req{  (R73)}{ If one of the three gears is not seen locked in the up position more than 10 seconds after stimulating the retraction electro-valve, then the boolean output  normal mode is set to  false.
}

\req{  (R74)}{ If one of the three gears is not seen locked in the down position more than 10 seconds after stimulating the outgoing electro-valve, then the !boolean output  normal mode is set to  false.
}

\req{  (R81)}{ When at least one computing module is working, if the landing gear command button has been DOWN for 15 seconds, and if the gears are not locked down after 15 seconds, then the red light Ólanding gear system failure is on.
}

\req{  (R82)}{ When at least one computing module is working, if the landing gear command button has been UP for 15 seconds, and if the gears are not locked retracted after 15 seconds, then the red light landing gear system failure is on.
}


\section{Introduction of the Modeling Framework}

We  summarize the concepts of the \textsc{Event~B} modeling language
developed by Abrial~\cite{abrial2010} and  indicate
the links with the tool called \textsc{RODIN}~\cite{rodin}. The   modeling
process deals with various languages, as seen by considering the
\textit{triptych\footnote{The term 'triptych' covers the three phases of software development: domain description, requirements prescription and software design.}} of Bjoerner~\cite{dines}: ${\cal
D},{\cal S} \longrightarrow {\cal R}$. Here, the domain ${\cal D}$
deals with properties, axioms, sets, constants, functions, relations,
and theories. The system model ${\cal S}$ expresses a model or
a refinement-based chain of models of the system. Finally, ${\cal R}$ expresses
requirements for the system to be designed. Considering the \textsc{Event~B}
modeling language, we notice that the language can express
\textit{safety} properties, which are either \textit{invariants} or
\textit{theorems} in a machine corresponding to the system. Recall that two main structures are available in \textsc{Event~B}:

\begin{itemize}
\item Contexts express static informations about the model.
\item Machines express dynamic informations about the model, invariants,
safety properties, and events.
\end{itemize}

A \textsc{Event~B} model is defined either as a context or as a machine.  The
triptych of Bjoerner~\cite{dines,losl-eatcs} ${\cal
  D},{\cal S} \longrightarrow {\cal R}$ is translated as follows:
${\cal C},{\cal M} \longrightarrow {\cal R}$, where ${\cal C}$ is a
context, ${\cal M}$ is a machine and ${\cal R}$ are the requirements.
The relation $ \longrightarrow$ is defined to be a logical
satisfaction relation with respect to an underlying
logico-mathematical theory. The satisfaction relation is supported by
the RODIN platform.  A machine is organizing events modifying state
variables and it uses static informations defined in a context.  These
basic structure mechanisms are extended by the refinement mechanism
which provides a mechanism for relating an abstract model and a
concrete model by adding new events or by adding new variables.  This
mechanism allows us to develop gradually \textsc{Event~B} models and to
validate each decision step using the proof tool.  The refinement
relationship should be expressed as follows: a model $M$ is refined by
a model $P$, when $P$ is simulating $M$. The final concrete model is
close to the behavior of real system that is executing events using
real source code. We give details now on the definition of events,
refinement and guidelines for developing complex system models.

\subsection{Modeling Actions Over States}

\textsc{Event~B}~\cite{abrial2010} is based
on the B notation. It extends the methodological scope of basic
concepts to take into account the idea of \textit{formal reactive 
models}. Briefly, a formal reactive  model is characterized by a
(finite) list $x$ of \textit{state variables} possibly modified by a
(finite) list of \textit{events}, where an invariant $I(x)$ states
properties that must always be satisfied by the variables $x$ and
\textit{maintained} by the activation of the events. In the following,
we summarize the definitions and principles of formal models and
explain how they can be managed by tools~\cite{rodin}.

Generalized substitutions are borrowed from the B notation, which  express changes in the value of state variables. An event has three main parts, namely a list of local parameters, a guard and a relation over values denotes \textit{pre} values of variables and \textit{post} values of variables. The most common event representation is ($\Bkeyword{any}\; t\; \Bkeyword{where}\; G(t,x) \;$ $\Bkeyword{then} \; x:|(R(x,x',t))\; \Bkeyword{end}$). The \textit{before-after} predicate $BA(e)(x,x')$, associated with each event, describes the event as a logical predicate for expressing the relationship linking values of the state variables just before ($x$) and just after ($x'$) the \textit{execution} of event \textit{e}. The form is semantically equivalent to $\exists \, t \bcdot\, (G(t,x) \band R(x,x',t))$.

\begin{center}
\fbox{
\begin{minipage}{.88\linewidth}
\small 
\noindent
\textsc{Proof obligations}
\noindent
\begin{itemize}
\item{\sf (INV1)}   $Init(x) \ \Rightarrow \ I(x)$
\item{\sf (INV2)}  $I(x)\ \band \  BA(e)(x,x') \ \Rightarrow \ I(x')$
\item{\sf (FIS)}  $I(x)\ \band\ {\sf grd}(e)(x)\    \Rightarrow  \exists y. BA(e)(x,y)$
\end{itemize}
\end{minipage}
}

\textbf{Table-1 }\textsc{Event~B} proof obligations  
\end{center}

Proof obligations (\textsf{INV 1} and \textsf{INV 2}) are produced
by the RODIN tool~\cite{rodin} from events to state that an
invariant condition $I(x)$ is preserved. Their general form follows
immediately from the definition of the before--after predicate
$BA(e)(x,x')$ of each event $e$ (see Table-1). Note
that it follows from the two guarded forms of the events that this
obligation is trivially discharged when the guard of the event is
false. Whenever this is the case, the event is said to be \textit{disabled}.
The proof obligation \textsf{FIS} expresses the feasibility of the
event $e$ with respect to the invariant $I$. By proving feasibility, we achieve that $BA(e)(x,y)$ provides an after state whenever $grd(e)(x)$ holds. This means that the guard indeed represents the enabling condition of the event.

The intention of specifying a guard of an event is that the event may always occur when a given guard is true. There is, however, some interaction between guards and nondeterministic assignments, namely  $x :| BA(e)(x,x')$. The predicate $BA(e)(x,x')$ of an action $x :| BA(e)(x,x')$ is not satisfiable or a set ($S$) is empty in an action predicate ($v :\in S$). Both cases show violations of the event feasibility proof obligation. We say that an assignment is feasible if there is an after-state satisfying the corresponding before-after predicate. For each event, its feasibility must be proved. Note, that for deterministic assignments the proof of feasibility is trivial. Also note, that feasibility of the initialization of a machine yields the existence of an initial state of the machine. It is not necessary to require an extra initialization.

\subsection{Model Refinement}

The refinement of a formal model allows us to enrich the model via a
\textit{step-by-step} approach and is the foundation of our
correct-by-construction approach~\cite{Leavens:roadmap}. Refinement
provides a way to strengthen invariants and to add details to a model.
It is also used to transform an abstract model to a more concrete
version by modifying the state description. This is done by extending
the list of state variables (possibly suppressing some of them), by
refining each abstract event to a set of possible concrete version,
and by adding new events. The abstract ($x$) and
concrete ($y$) state variables are linked by means of a
\textit{gluing invariant} $J(x,y)$. A number of proof obligations
ensure that (1) each abstract event is correctly refined by its
corresponding concrete version, (2) each new event refines $skip$, (3)
no new event takes control for ever, and (4) relative
deadlock freedom is preserved. Details of the formulation of these
proofs follows.

We suppose that an abstract model $AM$ with variables $x$ and invariant $I(x)$ is refined by a concrete model $CM$ with variables $y$ and gluing invariant $J(x,y)$. Event $e$ is in abstract model $AM$ and event $f$ is in concrete model $CM$. Event $f$ refines event $e$. $BA(e)(x,x')$ and $BA(f)(y,y')$ are predicates of events $e$ and $f$ respectively, we have to prove the following statement, corresponding to proof obligation (1):

\begin{center}
\scriptsize 
$
\begin{Bcode}
I(x) \band J(x,y) \band BA(f)(y,y') \
\Rightarrow \ \exists x' \,\bcdot\,(BA(e)(x,x') \band J(x',y'))
\end{Bcode}
$

\end{center}

The new events introduced in a refinement step can be viewed as hidden
events not visible to the environment of a system and are thus outside
the control of the environment. In \textsc{Event~B}, requiring a new event to
refine $skip$ means that the effect of  the new  event
is not observable in the abstract model. Any number of executions of an internal action may
occur in between each execution of a visible action. Now, proof
obligation (2) states that $BA(f)(y,y')$ must refine $skip$ ($x'=x$),
generating the following simple statement to prove (2):

\begin{center}
\scriptsize
$
\begin{Bcode}
I(x) \band J(x,y) \band BA(f)(y,y') \
\Rightarrow \ J(x,y')
\end{Bcode}
$
\end{center}

In refining a model,  an existing event can be refined by
strengthening the guard and/or the before--after predicate (effectively
reducing the degree of nondeterminism), or a new event can be added
to refine the skip event. The feasibility condition is crucial to
avoiding possible states that have no successor, such as
division by zero. Furthermore, this refinement guarantees that the set
of traces of the refined model contains (up to stuttering) the traces
of the resulting model.   The refinement   of  an event $e$ by an event $f$ means that the event $f$ simulates the event $e$.

The \textsc{Event~B} modeling language is supported
by the RODIN platform~\cite{rodin} and has been introduced in
publications~\cite{abrial2010}, where the many case
studies and discussions about the language itself and the foundations of the
\textsc{Event~B} approach. The language of \textit{generalized substitutions}
is very rich, enabling the expression of any relation between states in a
set-theoretical context. The expressive power of the language leads to a
requirement for help in writing relational specifications, which is why we
should provide guidelines for assisting the development of
\textsc{Event~B} models.

\subsection{Time-Based Pattern in Event-B}

The purpose of a design pattern~\cite{gamma} is to capture structures and to make decisions within a design that are common to similar modeling and analysis tasks. They can be re-applied when undertaking similar tasks in order to reduce the duplication of effort. The design pattern approach is the possibility to reuse solutions from earlier developments in the current project. This will lead to a correct refinement in the chain of models, without arising proof obligations. Since the correctness (i.e proof obligations are proved) of the pattern has been proved during its development, nothing is to be proved again when using this pattern.

The landing gear system is characterized by  their functions, which can  be expressed by  analyzing the real-time patterns. Sequence of operations related to doors and gears, are performed under the real-time constraints. D.   Cansell et. all~\cite{cansell06} have introduced  the time  constraint pattern. In this case study, we use the same time pattern to solve the timing requirements of the landing system. This time pattern is fully based on timed automaton. The  timed automaton is a finite state machine that is useful to model the components  of real-time systems. In a model,  the timed automata interacts with  each other  and defines a timed transition system. Besides  ordinary action   transitions  that can represent  input, output  and  internal  actions. A   timed transition system has time  progress transitions. Such  time progress transitions result in synchronous progress of   all clock variables in the  model. Here we apply the time pattern to model the sequential operations of doors and gears of  the landing system in continuous
progressive time constraint. In  the model every events are  controlled under time constraints, which means action of any event activates only when time constraint satisfies on  specific time. The time progress is
also an event,   so there  is   no modification of  the  underlying  B language. It  is only  a  modeling technique instead  of a specialized formal  system. The timed variable is in $\nat$ $(natural\ numbers)$
but  the time  constraint  can   be written  in   terms involving  unknown constants  or expressions between  different times. Finally, the timed event observations can be constrained  by other events which determine future activations. 

\subsection{Tools Environments for \textsc{Event~B}}

The     \textsc{Event~B} modeling  language    is    supported by the   Atelier
B~\cite{atelierb} environment  and by the RODIN platform~\cite{rodin}.
Both   environments    provide facilities   for    editing   machines,
refinements, contexts and  projects, for generating proof  obligations
corresponding to a given property, for proving proof obligations in an
automatic or/and  interactive process and for  animating  models.  The
internal prover is shared by the two environments  and there are hints
generated by the prover  interface for helping the interactive proofs.
However, the refinement process of machines should be progressive when
adding new  elements  to a given  current   model and the  goal  is to
distribute  the     complexity  of  proofs   through  the  proof-based
refinement.  These tools are  based on logical and semantical concepts
of \textsc{Event~B} models (machines,  contexts,refinement) and our methodology
for modeling medical protocol or guidelines can be built from them.

\section{M1}
\label{sec:M1}

\begin{description}
\BTitle{M1}{27Jan2014}{10:44:59 AM}
\MACHINE{M1}
\SEES{C0}
\VARIABLES
	\begin{description}
		\Item{ button }
		\Item{ phase }
	\end{description}
\INVARIANTS
	\begin{description}
		\nItem{ inv1 }{ button \in  POSITIONS }
		\nItem{ inv2 }{ phase \in  PHASES }
		\nItem{ inv3 }{ phase=movingup \limp  button=UP }
		\nItem{ inv4 }{ phase=movingdown \limp  button = DOWN }
		\nItem{ inv5 }{ button=UP \limp  phase\notin \{ movingdown,haltdown\}  }
		\nItem{ inv6 }{ button=DOWN \limp  phase\notin \{ movingup,haltup\}  }
	\end{description}
\EVENTS
	\INITIALISATION
		\begin{description}
		\BeginAct
			\begin{description}
			\nItem{ act1 }{ button :=  DOWN }
			\nItem{ act2 }{ phase :=  haltdown }
			\end{description}
		\EndAct
		\end{description}
	\EVT {PressDOWN}
		\begin{description}
		\WhenGrd
			\begin{description}
			\nItem{ grd1 }{ button=UP }
			\end{description}
		\ThenAct
			\begin{description}
			\nItem{ act1 }{ phase:= movingdown }
			\nItem{ act2 }{ button:= DOWN }
			\end{description}
		\EndAct
		\end{description}
	\EVT {PressUP}
		\begin{description}
		\WhenGrd
			\begin{description}
			\nItem{ grd1 }{ button=DOWN }
			\end{description}
		\ThenAct
			\begin{description}
			\nItem{ act1 }{ phase:= movingup }
			\nItem{ act2 }{ button:= UP }
			\end{description}
		\EndAct
		\end{description}
	\EVT {movingup}
		\begin{description}
		\WhenGrd
			\begin{description}
			\nItem{ grd1 }{ phase=movingup  }
			\end{description}
		\ThenAct
			\begin{description}
			\nItem{ act1 }{ phase:= haltup }
			\end{description}
		\EndAct
		\end{description}
	\EVT {movingdown}
		\begin{description}
		\WhenGrd
			\begin{description}
			\nItem{ grd1 }{ phase=movingdown }
			\end{description}
		\ThenAct
			\begin{description}
			\nItem{ act1 }{ phase:= haltdown }
			\end{description}
		\EndAct
		\end{description}
\END
\end{description}

\section{M2}
\label{sec:M2}

\begin{description}
\BTitle{M2}{27Jan2014}{10:44:59 AM}
\MACHINE{M2}
\REFINES{M1}
\SEES{C0}
\VARIABLES
	\begin{description}
		\Item{ dstate }
		\Item{ lstate }
		\Item{ phase }
		\Item{ button }
		\Item{ p }
		\Item{ l }
		\Item{ i }
	\end{description}
\INVARIANTS
	\begin{description}
		\nItem{ inv1 }{ dstate \in  DOORS \tfun  SDOORS }
		\nItemY{ inv2 }{ dstate^{-1} [\{ OPEN\} ] \neq  \emptyset  \limp  dstate^{-1} [\{ OPEN\} ]=DOORS }{ 		\\\hspace*{1,4 cm}  when one door is open, each door is open. }
		\nItemY{ inv3 }{ dstate^{-1} [\{ CLOSED\} ] \neq  \emptyset  \limp  dstate^{-1} [\{ CLOSED\} ]=DOORS }{ 		\\\hspace*{1,4 cm}  when a door is closed, t each door is closed }
		\nItem{ inv6 }{ lstate \in  DOORS \tfun  SLOCKS }
		\nItem{ inv7 }{ dstate[DOORS]=\{ OPEN\}  \limp  lstate[DOORS]=\{ UNLOCKED\}  }
		\nItem{ inv12 }{ p \in  P }
		\nItem{ inv13 }{ l \in  P }
		\nItem{ inv14 }{ i \in  P }
		\nItem{ inv15 }{ l=E \land  p=R \limp  lstate[DOORS]=\{ UNLOCKED\}  }
		\nItem{ inv16 }{ l=R \land  p=E \limp  lstate[DOORS]=\{ UNLOCKED\}  }
	\end{description}
\EVENTS
	\INITIALISATION
		\\\textit{extended}
		\begin{description}
		\BeginAct
			\begin{description}
			\nItemX{ act1 }{ button :=  DOWN }
			\nItemX{ act2 }{ phase :=  haltdown }
			\nItemY{ act3 }{ dstate :|  (dstate'\in  DOORS \tfun  SDOORS \land  dstate'=\{ a\mapsto b|  a \in  DOORS \land  b=CLOSED\} ) }{ 		\\\hspace*{1,4 cm}  missing elements of the invariant }
			\nItem{ act4 }{ lstate :=  \{ a\mapsto b| a\in DOORS\land  b=LOCKED\}  }
			\nItem{ act5 }{ p :=  R }
			\nItem{ act6 }{ l :=  R }
			\nItem{ act7 }{ i :=  R }
			\end{description}
		\EndAct
		\end{description}
	\EVT {opening\_doors\_DOWN}
		\begin{description}
		\WhenGrd
			\begin{description}
			\nItem{ grd1 }{ dstate[DOORS]= \{ CLOSED\}  }
			\nItem{ grd5 }{ lstate[DOORS]=\{ UNLOCKED\}  }
			\nItem{ grd7 }{ phase=movingdown }
			\nItem{ grd8 }{ p=R }
			\nItem{ grd9 }{ l=R }
			\end{description}
		\ThenAct
			\begin{description}
			\nItem{ act1 }{ dstate :=  \{ a\mapsto b|  a \in  DOORS \land  b=OPEN\}  }
			\nItem{ act2 }{ p:= E }
			\end{description}
		\EndAct
		\end{description}
	\EVT {opening\_doors\_UP}
		\begin{description}
		\WhenGrd
			\begin{description}
			\nItem{ grd1 }{ dstate[DOORS]= \{ CLOSED\}  }
			\nItem{ grd4 }{ lstate[DOORS]=\{ UNLOCKED\}  }
			\nItem{ grd5 }{ phase= movingup }
			\nItem{ grd6 }{ p=E }
			\nItem{ grd7 }{ l=E }
			\end{description}
		\ThenAct
			\begin{description}
			\nItem{ act1 }{ dstate :=  \{ a\mapsto b|  a \in  DOORS \land  b=OPEN\}  }
			\nItem{ act2 }{ p:= R }
			\end{description}
		\EndAct
		\end{description}
	\EVT {closing\_doors\_UP}
		\begin{description}
		\AnyPrm
			\begin{description}
			\Item{f }
			\end{description}
		\WhereGrd
			\begin{description}
			\nItem{ grd1 }{ dstate[DOORS]=\{ OPEN\}  }
			\nItem{ grd3 }{ f \in  DOORS \tfun  SDOORS }
			\nItem{ grd4 }{ \forall e\qdot  e \in  DOORS \limp  f(e)=CLOSED }
			\nItem{ grd5 }{ phase=movingup }
			\nItem{ grd6 }{ p=R }
			\end{description}
		\ThenAct
			\begin{description}
			\nItem{ act1 }{ dstate:= f }
			\end{description}
		\EndAct
		\end{description}
	\EVT {closing\_doors\_DOWN}
		\begin{description}
		\AnyPrm
			\begin{description}
			\Item{f }
			\end{description}
		\WhereGrd
			\begin{description}
			\nItem{ grd1 }{ dstate[DOORS]=\{ OPEN\}  }
			\nItem{ grd3 }{ f \in  DOORS \tfun  SDOORS }
			\nItem{ grd4 }{ \forall e\qdot  e \in  DOORS \limp  f(e)=CLOSED }
			\nItem{ grd5 }{ phase=movingdown }
			\nItem{ grd6 }{ p=E }
			\end{description}
		\ThenAct
			\begin{description}
			\nItem{ act1 }{ dstate:= f }
			\end{description}
		\EndAct
		\end{description}
	\EVT {unlocking\_UP}
		\begin{description}
		\WhenGrd
			\begin{description}
			\nItem{ grd3 }{ lstate[DOORS]=\{ LOCKED\}  }
			\nItem{ grd4 }{ phase=movingup }
			\nItem{ grd5 }{ l=E }
			\nItem{ grd6 }{ p=E }
			\nItem{ grd7 }{ i=E }
			\end{description}
		\ThenAct
			\begin{description}
			\nItem{ act1 }{ lstate:= \{ a\mapsto b| a\in DOORS \land  b=UNLOCKED\}  }
			\end{description}
		\EndAct
		\end{description}
	\EVT {locking\_UP}
	\REF {movingup}
		\begin{description}
		\WhenGrd
			\begin{description}
			\nItem{ grd3 }{ dstate[DOORS]=\{ CLOSED\}  }
			\nItem{ grd4 }{ phase=movingup }
			\nItem{ grd5 }{ lstate[DOORS]=\{ UNLOCKED\}  }
			\nItem{ grd6 }{ p=R }
			\nItem{ grd7 }{ l=E }
			\end{description}
		\ThenAct
			\begin{description}
			\nItem{ act1 }{ lstate:= \{ a\mapsto b| a\in DOORS \land  b=LOCKED\}  }
			\nItem{ act3 }{ phase:= haltup }
			\nItemY{ act4 }{ l:= R }{ 		\\\hspace*{1,4 cm}  added by D Mery }
			\end{description}
		\EndAct
		\end{description}
	\EVT {unlocking\_DOWN}
		\begin{description}
		\WhenGrd
			\begin{description}
			\nItem{ grd3 }{ lstate[DOORS]=\{ LOCKED\}  }
			\nItem{ grd4 }{ phase=movingdown }
			\nItem{ grd5 }{ l=R }
			\nItem{ grd6 }{ p=R }
			\nItem{ grd7 }{ i=R }
			\end{description}
		\ThenAct
			\begin{description}
			\nItem{ act1 }{ lstate:= \{ a\mapsto b| a\in DOORS \land  b=UNLOCKED\}  }
			\end{description}
		\EndAct
		\end{description}
	\EVT {locking\_DOWN}
	\REF {movingdown}
		\begin{description}
		\WhenGrd
			\begin{description}
			\nItem{ grd1 }{ dstate[DOORS]=\{ CLOSED\}  }
			\nItem{ grd2 }{ phase=movingdown }
			\nItem{ grd3 }{ lstate[DOORS]=\{ UNLOCKED\}  }
			\nItem{ grd4 }{ p=E }
			\nItem{ grd5 }{ l=R }
			\end{description}
		\ThenAct
			\begin{description}
			\nItem{ act1 }{ lstate:= \{ a\mapsto b| a\in DOORS \land  b = LOCKED\}  }
			\nItem{ act3 }{ phase:= haltdown }
			\nItem{ act4 }{ l:= E }
			\end{description}
		\EndAct
		\end{description}
	\EVT {PD1}
	\REF {PressDOWN}
		\begin{description}
		\WhenGrd
			\begin{description}
			\nItem{ grd1 }{ button=UP }
			\nItem{ grd2 }{ phase=haltup }
			\end{description}
		\ThenAct
			\begin{description}
			\nItem{ act1 }{ phase:= movingdown }
			\nItem{ act2 }{ button:= DOWN }
			\nItem{ act3 }{ l:= R }
			\nItem{ act4 }{ p:= R }
			\nItem{ act5 }{ i:= R }
			\end{description}
		\EndAct
		\end{description}
	\EVT {PU1}
	\REF {PressUP}
		\begin{description}
		\WhenGrd
			\begin{description}
			\nItem{ grd1 }{ button=DOWN }
			\nItem{ grd2 }{ phase=haltdown }
			\end{description}
		\ThenAct
			\begin{description}
			\nItem{ act1 }{ phase:= movingup }
			\nItem{ act2 }{ button:= UP }
			\nItem{ act3 }{ l:= E }
			\nItem{ act4 }{ p:= E }
			\nItem{ act5 }{ i:= E }
			\end{description}
		\EndAct
		\end{description}
	\EVT {PU2}
	\REF {PressUP}
		\begin{description}
		\WhenGrd
			\begin{description}
			\nItem{ grd1 }{ l=R }
			\nItem{ grd2 }{ p=R }
			\nItem{ grd3 }{ phase=movingdown }
			\nItem{ grd4 }{ button=DOWN }
			\nItem{ grd5 }{ i=R }
			\nItem{ grd6 }{ lstate[DOORS]=\{ LOCKED\}  }
			\end{description}
		\ThenAct
			\begin{description}
			\nItem{ act1 }{ phase:= movingup }
			\nItem{ act4 }{ button:=   UP }
			\nItem{ act5 }{ l:= E }
			\nItem{ act6 }{ p:= E }
			\nItem{ act7 }{ i:= R }
			\end{description}
		\EndAct
		\end{description}
	\EVT {CompletePU2}
	\REF {movingup}
		\begin{description}
		\WhenGrd
			\begin{description}
			\nItem{ grd1 }{ phase=movingup }
			\nItem{ grd2 }{ button=UP }
			\nItem{ grd3 }{ l=E }
			\nItem{ grd4 }{ p=E }
			\nItem{ grd5 }{ i=R }
			\end{description}
		\ThenAct
			\begin{description}
			\nItem{ act1 }{ phase:= haltup }
			\end{description}
		\EndAct
		\end{description}
	\EVT {PU3}
	\REF {PressUP}
		\begin{description}
		\WhenGrd
			\begin{description}
			\nItem{ grd1 }{ dstate[DOORS]=\{ CLOSED\}  }
			\nItem{ grd2 }{ lstate[DOORS]=\{ UNLOCKED\}  }
			\nItem{ grd3 }{ phase = movingdown }
			\nItem{ grd4 }{ p=R }
			\nItem{ grd5 }{ l=R }
			\nItem{ grd6 }{ button=DOWN }
			\end{description}
		\ThenAct
			\begin{description}
			\nItem{ act1 }{ phase:= movingup }
			\nItem{ act2 }{ p:= R }
			\nItem{ act3 }{ l:= E }
			\nItem{ act4 }{ button:= UP }
			\end{description}
		\EndAct
		\end{description}
	\EVT {PU4}
	\REF {PressUP}
		\begin{description}
		\WhenGrd
			\begin{description}
			\nItem{ grd1 }{ dstate[DOORS]=\{ OPEN\}  }
			\nItem{ grd2 }{ phase=movingdown }
			\nItem{ grd3 }{ p=E }
			\nItem{ grd4 }{ button=DOWN }
			\end{description}
		\ThenAct
			\begin{description}
			\nItem{ act1 }{ phase:= movingup }
			\nItem{ act2 }{ p:= R }
			\nItem{ act3 }{ button:= UP }
			\nItem{ act4 }{ i:= E }
			\nItem{ act5 }{ l:= E }
			\end{description}
		\EndAct
		\end{description}
	\EVT {PU5}
	\REF {PressUP}
		\begin{description}
		\WhenGrd
			\begin{description}
			\nItem{ grd1 }{ dstate[DOORS]=\{ CLOSED\}  }
			\nItem{ grd2 }{ phase=movingdown }
			\nItem{ grd3 }{ p=E }
			\nItem{ grd4 }{ button=DOWN }
			\nItem{ grd5 }{ lstate[DOORS]=\{ UNLOCKED\}  }
			\end{description}
		\ThenAct
			\begin{description}
			\nItem{ act1 }{ phase:= movingup }
			\nItem{ act3 }{ button:= UP }
			\nItem{ act4 }{ i:= E }
			\nItem{ act5 }{ l:= E }
			\end{description}
		\EndAct
		\end{description}
	\EVT {PD2}
	\REF {PressDOWN}
		\begin{description}
		\WhenGrd
			\begin{description}
			\nItem{ grd1 }{ l=E }
			\nItem{ grd2 }{ p=E }
			\nItem{ grd3 }{ phase=movingup }
			\nItem{ grd4 }{ i=E }
			\nItem{ grd5 }{ lstate[DOORS]=\{ LOCKED\}  }
			\end{description}
		\ThenAct
			\begin{description}
			\nItem{ act1 }{ phase:= movingdown }
			\nItem{ act2 }{ button:= DOWN }
			\nItem{ act3 }{ l:= R }
			\nItem{ act4 }{ p:= R }
			\nItem{ act5 }{ i:= E }
			\end{description}
		\EndAct
		\end{description}
	\EVT {CompletePD2}
	\REF {movingdown}
		\begin{description}
		\WhenGrd
			\begin{description}
			\nItem{ grd1 }{ phase=movingdown }
			\nItem{ grd2 }{ button=DOWN }
			\nItem{ grd3 }{ l=R }
			\nItem{ grd4 }{ p=R }
			\nItem{ grd5 }{ i=E }
			\end{description}
		\ThenAct
			\begin{description}
			\nItem{ act1 }{ phase:= haltdown }
			\end{description}
		\EndAct
		\end{description}
	\EVT {PD3}
	\REF {PressDOWN}
		\begin{description}
		\WhenGrd
			\begin{description}
			\nItem{ grd1 }{ dstate[DOORS]=\{ CLOSED\}  }
			\nItem{ grd2 }{ lstate[DOORS]=\{ UNLOCKED\}  }
			\nItem{ grd3 }{ phase=movingup }
			\nItem{ grd4 }{ p=E }
			\nItem{ grd5 }{ l=E }
			\nItem{ grd6 }{ button=UP }
			\end{description}
		\ThenAct
			\begin{description}
			\nItem{ act1 }{ phase:= movingdown }
			\nItem{ act2 }{ p:= E }
			\nItem{ act3 }{ l:= R }
			\nItem{ act4 }{ button:= DOWN }
			\end{description}
		\EndAct
		\end{description}
	\EVT {PD4}
	\REF {PressDOWN}
		\begin{description}
		\WhenGrd
			\begin{description}
			\nItem{ grd1 }{ dstate[DOORS]=\{ OPEN\}  }
			\nItem{ grd2 }{ phase=movingup }
			\nItem{ grd3 }{ p=R }
			\nItem{ grd4 }{ button=UP }
			\end{description}
		\ThenAct
			\begin{description}
			\nItem{ act1 }{ phase:= movingdown }
			\nItem{ act2 }{ p:= E }
			\nItem{ act3 }{ button:= DOWN }
			\nItem{ act4 }{ i:= R }
			\nItem{ act5 }{ l:= R }
			\end{description}
		\EndAct
		\end{description}
	\EVT {PD5}
	\REF {PressDOWN}
		\begin{description}
		\WhenGrd
			\begin{description}
			\nItem{ grd1 }{ dstate[DOORS]=\{ CLOSED\}  }
			\nItem{ grd2 }{ phase=movingup }
			\nItem{ grd3 }{ p=R }
			\nItem{ grd4 }{ button=UP }
			\nItem{ grd5 }{ lstate[DOORS]=\{ UNLOCKED\}  }
			\end{description}
		\ThenAct
			\begin{description}
			\nItem{ act1 }{ phase :=  movingdown }
			\nItem{ act2 }{ button:= DOWN }
			\nItem{ act3 }{ i:= R }
			\nItem{ act4 }{ l:= R }
			\end{description}
		\EndAct
		\end{description}
\END
\end{description}

\section{M3}
\label{sec:M3}

\begin{description}
\BTitle{M3}{27Jan2014}{10:44:59 AM}
\MACHINE{M3}
\REFINES{M2}
\SEES{C0}
\VARIABLES
	\begin{description}
		\Item{ dstate }
		\Item{ lstate }
		\Item{ phase }
		\Item{ button }
		\Item{ p }
		\Item{ l }
		\Item{ i }
		\Item{ gstate }
	\end{description}
\INVARIANTS
	\begin{description}
		\nItem{ M3\_inv1 }{ gstate \in  GEARS \tfun  SGEARS }
		\nItemY{ M3\_inv3 }{ \forall  door\qdot  door \in  DOORS \land  dstate(door)=CLOSED  \land  ran(gstate)\neq \{ RETRACTED\}  \limp  ran(gstate)=\{ EXTENDED\}  }{ 		\\\hspace*{2 cm}  gears can not be out or moving in this case. }
		\nItem{ M3\_inv6 }{ \forall  door\qdot  door \in  DOORS \land  dstate(door)=CLOSED  \land  ran(gstate)\neq \{ EXTENDED\}   \limp  ran(gstate)=\{ RETRACTED\}  }
		\nItem{ M3\_inv7 }{ ran(gstate)\neq \{ RETRACTED\}  \land  ran(gstate)\neq \{ EXTENDED\}   \limp  ran(dstate)=\{ OPEN\}  }
		\nItem{ M3\_inv11 }{ ran(dstate)=\{ CLOSED\}  \limp  ran(gstate)\binter \{ RETRACTING,EXTENDING\}  =\emptyset  }
	\end{description}
\EVENTS
	\INITIALISATION
		\\\textit{extended}
		\begin{description}
		\BeginAct
			\begin{description}
			\nItemX{ act1 }{ button :=  DOWN }
			\nItemX{ act2 }{ phase :=  haltdown }
			\nItemXY{ act3 }{ dstate :|  (dstate'\in  DOORS \tfun  SDOORS \land  dstate'=\{ a\mapsto b|  a \in  DOORS \land  b=CLOSED\} ) }{ 		\\\hspace*{1,4 cm}  missing elements of the invariant }
			\nItemX{ act4 }{ lstate :=  \{ a\mapsto b| a\in DOORS\land  b=LOCKED\}  }
			\nItemX{ act5 }{ p :=  R }
			\nItemX{ act6 }{ l :=  R }
			\nItemX{ act7 }{ i :=  R }
			\nItem{ act8 }{ gstate :|  (gstate' \in  GEARS \tfun  SGEARS \land  gstate'=\{ a\mapsto b |  a \in  GEARS  \land  b=EXTENDED\} ) }
			\end{description}
		\EndAct
		\end{description}
	\EVT {opening\_doors\_DOWN}
	\EXTD {opening\_doors\_DOWN}
		\begin{description}
		\WhenGrd
			\begin{description}
			\nItemX{ grd1 }{ dstate[DOORS]= \{ CLOSED\}  }
			\nItemX{ grd5 }{ lstate[DOORS]=\{ UNLOCKED\}  }
			\nItemX{ grd7 }{ phase=movingdown }
			\nItemX{ grd8 }{ p=R }
			\nItemX{ grd9 }{ l=R }
			\end{description}
		\ThenAct
			\begin{description}
			\nItemX{ act1 }{ dstate :=  \{ a\mapsto b|  a \in  DOORS \land  b=OPEN\}  }
			\nItemX{ act2 }{ p:= E }
			\end{description}
		\EndAct
		\end{description}
	\EVT {opening\_doors\_UP}
	\EXTD {opening\_doors\_UP}
		\begin{description}
		\WhenGrd
			\begin{description}
			\nItemX{ grd1 }{ dstate[DOORS]= \{ CLOSED\}  }
			\nItemX{ grd4 }{ lstate[DOORS]=\{ UNLOCKED\}  }
			\nItemX{ grd5 }{ phase= movingup }
			\nItemX{ grd6 }{ p=E }
			\nItemX{ grd7 }{ l=E }
			\end{description}
		\ThenAct
			\begin{description}
			\nItemX{ act1 }{ dstate :=  \{ a\mapsto b|  a \in  DOORS \land  b=OPEN\}  }
			\nItemX{ act2 }{ p:= R }
			\end{description}
		\EndAct
		\end{description}
	\EVT {closing\_doors\_UP}
	\REF {closing\_doors\_UP}
		\begin{description}
		\AnyPrm
			\begin{description}
			\Item{f }
			\end{description}
		\WhereGrd
			\begin{description}
			\nItem{ grd1 }{ dstate[DOORS]=\{ OPEN\}  }
			\nItem{ grd3 }{ f \in  DOORS \tfun  SDOORS }
			\nItem{ grd4 }{ \forall e\qdot  e \in  DOORS \limp  f(e)=CLOSED }
			\nItem{ grd5 }{ phase=movingup }
			\nItem{ grd6 }{ p=R }
			\nItem{ grd7 }{ gstate[GEARS]=\{ RETRACTED\}  }
			\end{description}
		\ThenAct
			\begin{description}
			\nItem{ act1 }{ dstate:= f }
			\end{description}
		\EndAct
		\end{description}
	\EVT {closing\_doors\_DOWN}
	\REF {closing\_doors\_DOWN}
		\begin{description}
		\AnyPrm
			\begin{description}
			\Item{f }
			\end{description}
		\WhereGrd
			\begin{description}
			\nItem{ grd1 }{ dstate[DOORS]=\{ OPEN\}  }
			\nItem{ grd3 }{ f \in  DOORS \tfun  SDOORS }
			\nItem{ grd4 }{ \forall e\qdot  e \in  DOORS \limp  f(e)=CLOSED }
			\nItem{ grd5 }{ phase=movingdown }
			\nItem{ grd6 }{ p=E }
			\nItem{ grd7 }{ gstate[GEARS]=\{ EXTENDED\}  }
			\end{description}
		\ThenAct
			\begin{description}
			\nItem{ act1 }{ dstate:= f }
			\end{description}
		\EndAct
		\end{description}
	\EVT {unlocking\_UP}
	\EXTD {unlocking\_UP}
		\begin{description}
		\WhenGrd
			\begin{description}
			\nItemX{ grd3 }{ lstate[DOORS]=\{ LOCKED\}  }
			\nItemX{ grd4 }{ phase=movingup }
			\nItemX{ grd5 }{ l=E }
			\nItemX{ grd6 }{ p=E }
			\nItemX{ grd7 }{ i=E }
			\end{description}
		\ThenAct
			\begin{description}
			\nItemX{ act1 }{ lstate:= \{ a\mapsto b| a\in DOORS \land  b=UNLOCKED\}  }
			\end{description}
		\EndAct
		\end{description}
	\EVT {locking\_UP}
	\EXTD {locking\_UP}
		\begin{description}
		\WhenGrd
			\begin{description}
			\nItemX{ grd3 }{ dstate[DOORS]=\{ CLOSED\}  }
			\nItemX{ grd4 }{ phase=movingup }
			\nItemX{ grd5 }{ lstate[DOORS]=\{ UNLOCKED\}  }
			\nItemX{ grd6 }{ p=R }
			\nItemX{ grd7 }{ l=E }
			\end{description}
		\ThenAct
			\begin{description}
			\nItemX{ act1 }{ lstate:= \{ a\mapsto b| a\in DOORS \land  b=LOCKED\}  }
			\nItemX{ act3 }{ phase:= haltup }
			\nItemXY{ act4 }{ l:= R }{ 		\\\hspace*{1,4 cm}  added by D Mery }
			\end{description}
		\EndAct
		\end{description}
	\EVT {unlocking\_DOWN}
	\EXTD {unlocking\_DOWN}
		\begin{description}
		\WhenGrd
			\begin{description}
			\nItemX{ grd3 }{ lstate[DOORS]=\{ LOCKED\}  }
			\nItemX{ grd4 }{ phase=movingdown }
			\nItemX{ grd5 }{ l=R }
			\nItemX{ grd6 }{ p=R }
			\nItemX{ grd7 }{ i=R }
			\end{description}
		\ThenAct
			\begin{description}
			\nItemX{ act1 }{ lstate:= \{ a\mapsto b| a\in DOORS \land  b=UNLOCKED\}  }
			\end{description}
		\EndAct
		\end{description}
	\EVT {locking\_DOWN}
	\EXTD {locking\_DOWN}
		\begin{description}
		\WhenGrd
			\begin{description}
			\nItemX{ grd1 }{ dstate[DOORS]=\{ CLOSED\}  }
			\nItemX{ grd2 }{ phase=movingdown }
			\nItemX{ grd3 }{ lstate[DOORS]=\{ UNLOCKED\}  }
			\nItemX{ grd4 }{ p=E }
			\nItemX{ grd5 }{ l=R }
			\end{description}
		\ThenAct
			\begin{description}
			\nItemX{ act1 }{ lstate:= \{ a\mapsto b| a\in DOORS \land  b = LOCKED\}  }
			\nItemX{ act3 }{ phase:= haltdown }
			\nItemX{ act4 }{ l:= E }
			\end{description}
		\EndAct
		\end{description}
	\EVT {PD1}
	\EXTD {PD1}
		\begin{description}
		\WhenGrd
			\begin{description}
			\nItemX{ grd1 }{ button=UP }
			\nItemX{ grd2 }{ phase=haltup }
			\end{description}
		\ThenAct
			\begin{description}
			\nItemX{ act1 }{ phase:= movingdown }
			\nItemX{ act2 }{ button:= DOWN }
			\nItemX{ act3 }{ l:= R }
			\nItemX{ act4 }{ p:= R }
			\nItemX{ act5 }{ i:= R }
			\end{description}
		\EndAct
		\end{description}
	\EVT {PU1}
	\EXTD {PU1}
		\begin{description}
		\WhenGrd
			\begin{description}
			\nItemX{ grd1 }{ button=DOWN }
			\nItemX{ grd2 }{ phase=haltdown }
			\end{description}
		\ThenAct
			\begin{description}
			\nItemX{ act1 }{ phase:= movingup }
			\nItemX{ act2 }{ button:= UP }
			\nItemX{ act3 }{ l:= E }
			\nItemX{ act4 }{ p:= E }
			\nItemX{ act5 }{ i:= E }
			\end{description}
		\EndAct
		\end{description}
	\EVT {PU2}
	\EXTD {PU2}
		\begin{description}
		\WhenGrd
			\begin{description}
			\nItemX{ grd1 }{ l=R }
			\nItemX{ grd2 }{ p=R }
			\nItemX{ grd3 }{ phase=movingdown }
			\nItemX{ grd4 }{ button=DOWN }
			\nItemX{ grd5 }{ i=R }
			\nItemX{ grd6 }{ lstate[DOORS]=\{ LOCKED\}  }
			\end{description}
		\ThenAct
			\begin{description}
			\nItemX{ act1 }{ phase:= movingup }
			\nItemX{ act4 }{ button:=   UP }
			\nItemX{ act5 }{ l:= E }
			\nItemX{ act6 }{ p:= E }
			\nItemX{ act7 }{ i:= R }
			\end{description}
		\EndAct
		\end{description}
	\EVT {CompletePU2}
	\EXTD {CompletePU2}
		\begin{description}
		\WhenGrd
			\begin{description}
			\nItemX{ grd1 }{ phase=movingup }
			\nItemX{ grd2 }{ button=UP }
			\nItemX{ grd3 }{ l=E }
			\nItemX{ grd4 }{ p=E }
			\nItemX{ grd5 }{ i=R }
			\end{description}
		\ThenAct
			\begin{description}
			\nItemX{ act1 }{ phase:= haltup }
			\end{description}
		\EndAct
		\end{description}
	\EVT {PU3}
	\EXTD {PU3}
		\begin{description}
		\WhenGrd
			\begin{description}
			\nItemX{ grd1 }{ dstate[DOORS]=\{ CLOSED\}  }
			\nItemX{ grd2 }{ lstate[DOORS]=\{ UNLOCKED\}  }
			\nItemX{ grd3 }{ phase = movingdown }
			\nItemX{ grd4 }{ p=R }
			\nItemX{ grd5 }{ l=R }
			\nItemX{ grd6 }{ button=DOWN }
			\end{description}
		\ThenAct
			\begin{description}
			\nItemX{ act1 }{ phase:= movingup }
			\nItemX{ act2 }{ p:= R }
			\nItemX{ act3 }{ l:= E }
			\nItemX{ act4 }{ button:= UP }
			\end{description}
		\EndAct
		\end{description}
	\EVT {PU4}
	\EXTD {PU4}
		\begin{description}
		\WhenGrd
			\begin{description}
			\nItemX{ grd1 }{ dstate[DOORS]=\{ OPEN\}  }
			\nItemX{ grd2 }{ phase=movingdown }
			\nItemX{ grd3 }{ p=E }
			\nItemX{ grd4 }{ button=DOWN }
			\end{description}
		\ThenAct
			\begin{description}
			\nItemX{ act1 }{ phase:= movingup }
			\nItemX{ act2 }{ p:= R }
			\nItemX{ act3 }{ button:= UP }
			\nItemX{ act4 }{ i:= E }
			\nItemX{ act5 }{ l:= E }
			\end{description}
		\EndAct
		\end{description}
	\EVT {PU5}
	\EXTD {PU5}
		\begin{description}
		\WhenGrd
			\begin{description}
			\nItemX{ grd1 }{ dstate[DOORS]=\{ CLOSED\}  }
			\nItemX{ grd2 }{ phase=movingdown }
			\nItemX{ grd3 }{ p=E }
			\nItemX{ grd4 }{ button=DOWN }
			\nItemX{ grd5 }{ lstate[DOORS]=\{ UNLOCKED\}  }
			\end{description}
		\ThenAct
			\begin{description}
			\nItemX{ act1 }{ phase:= movingup }
			\nItemX{ act3 }{ button:= UP }
			\nItemX{ act4 }{ i:= E }
			\nItemX{ act5 }{ l:= E }
			\end{description}
		\EndAct
		\end{description}
	\EVT {PD2}
	\EXTD {PD2}
		\begin{description}
		\WhenGrd
			\begin{description}
			\nItemX{ grd1 }{ l=E }
			\nItemX{ grd2 }{ p=E }
			\nItemX{ grd3 }{ phase=movingup }
			\nItemX{ grd4 }{ i=E }
			\nItemX{ grd5 }{ lstate[DOORS]=\{ LOCKED\}  }
			\end{description}
		\ThenAct
			\begin{description}
			\nItemX{ act1 }{ phase:= movingdown }
			\nItemX{ act2 }{ button:= DOWN }
			\nItemX{ act3 }{ l:= R }
			\nItemX{ act4 }{ p:= R }
			\nItemX{ act5 }{ i:= E }
			\end{description}
		\EndAct
		\end{description}
	\EVT {CompletePD2}
	\EXTD {CompletePD2}
		\begin{description}
		\WhenGrd
			\begin{description}
			\nItemX{ grd1 }{ phase=movingdown }
			\nItemX{ grd2 }{ button=DOWN }
			\nItemX{ grd3 }{ l=R }
			\nItemX{ grd4 }{ p=R }
			\nItemX{ grd5 }{ i=E }
			\end{description}
		\ThenAct
			\begin{description}
			\nItemX{ act1 }{ phase:= haltdown }
			\end{description}
		\EndAct
		\end{description}
	\EVT {PD3}
	\EXTD {PD3}
		\begin{description}
		\WhenGrd
			\begin{description}
			\nItemX{ grd1 }{ dstate[DOORS]=\{ CLOSED\}  }
			\nItemX{ grd2 }{ lstate[DOORS]=\{ UNLOCKED\}  }
			\nItemX{ grd3 }{ phase=movingup }
			\nItemX{ grd4 }{ p=E }
			\nItemX{ grd5 }{ l=E }
			\nItemX{ grd6 }{ button=UP }
			\end{description}
		\ThenAct
			\begin{description}
			\nItemX{ act1 }{ phase:= movingdown }
			\nItemX{ act2 }{ p:= E }
			\nItemX{ act3 }{ l:= R }
			\nItemX{ act4 }{ button:= DOWN }
			\end{description}
		\EndAct
		\end{description}
	\EVT {PD4}
	\EXTD {PD4}
		\begin{description}
		\WhenGrd
			\begin{description}
			\nItemX{ grd1 }{ dstate[DOORS]=\{ OPEN\}  }
			\nItemX{ grd2 }{ phase=movingup }
			\nItemX{ grd3 }{ p=R }
			\nItemX{ grd4 }{ button=UP }
			\end{description}
		\ThenAct
			\begin{description}
			\nItemX{ act1 }{ phase:= movingdown }
			\nItemX{ act2 }{ p:= E }
			\nItemX{ act3 }{ button:= DOWN }
			\nItemX{ act4 }{ i:= R }
			\nItemX{ act5 }{ l:= R }
			\end{description}
		\EndAct
		\end{description}
	\EVT {PD5}
	\EXTD {PD5}
		\begin{description}
		\WhenGrd
			\begin{description}
			\nItemX{ grd1 }{ dstate[DOORS]=\{ CLOSED\}  }
			\nItemX{ grd2 }{ phase=movingup }
			\nItemX{ grd3 }{ p=R }
			\nItemX{ grd4 }{ button=UP }
			\nItemX{ grd5 }{ lstate[DOORS]=\{ UNLOCKED\}  }
			\end{description}
		\ThenAct
			\begin{description}
			\nItemX{ act1 }{ phase :=  movingdown }
			\nItemX{ act2 }{ button:= DOWN }
			\nItemX{ act3 }{ i:= R }
			\nItemX{ act4 }{ l:= R }
			\end{description}
		\EndAct
		\end{description}
	\EVT {retracting\_gears}
		\begin{description}
		\WhenGrd
			\begin{description}
			\nItem{ grd1 }{ dstate[DOORS]=\{ OPEN\}  }
			\nItem{ grd2 }{ gstate[GEARS]=\{ EXTENDED\}  }
			\nItem{ grd3 }{ p=R }
			\end{description}
		\ThenAct
			\begin{description}
			\nItem{ act1 }{ gstate :=  \{ a\mapsto b| a \in  GEARS \land  b=RETRACTING\}  }
			\end{description}
		\EndAct
		\end{description}
	\EVT {retraction}
		\begin{description}
		\WhenGrd
			\begin{description}
			\nItem{ grd1 }{ dstate[DOORS]=\{ OPEN\}  }
			\nItem{ grd2 }{ gstate[GEARS]=\{ RETRACTING\}  }
			\end{description}
		\ThenAct
			\begin{description}
			\nItem{ act1 }{ gstate:=   \{ a\mapsto b|  a \in  GEARS \land  b= RETRACTED\}  }
			\end{description}
		\EndAct
		\end{description}
	\EVT {extending\_gears}
		\begin{description}
		\WhenGrd
			\begin{description}
			\nItem{ grd1 }{ dstate[DOORS]=\{ OPEN\}  }
			\nItem{ grd2 }{ gstate[GEARS]=\{ RETRACTED\}  }
			\nItem{ grd3 }{ p=E }
			\end{description}
		\ThenAct
			\begin{description}
			\nItem{ act1 }{ gstate :=  \{ a\mapsto b| a \in  GEARS \land  b=EXTENDING\}  }
			\end{description}
		\EndAct
		\end{description}
	\EVT {extension}
		\begin{description}
		\WhenGrd
			\begin{description}
			\nItem{ grd1 }{ dstate[DOORS]=\{ OPEN\}  }
			\nItem{ grd2 }{ gstate[GEARS]=\{ EXTENDING\}  }
			\end{description}
		\ThenAct
			\begin{description}
			\nItem{ act1 }{ gstate :=  \{ a\mapsto b|  a \in  GEARS \land  b=EXTENDED\}  }
			\end{description}
		\EndAct
		\end{description}
\END
\end{description}

\section{M4}
\label{sec:M4}

\begin{description}
\BTitle{M4}{27Jan2014}{10:44:59 AM}
\MACHINE{M4}\cmt{		\\\hspace*{1 cm}  Reading Sensor		\\\hspace*{0,8 cm}  Computing Module  }
\REFINES{M3}
\SEES{C1}
\VARIABLES
	\begin{description}
		\Item{ dstate }
		\Item{ lstate }
		\Item{ phase }
		\Item{ button }
		\Item{ p }
		\Item{ l }
		\Item{ i }
		\Item{ gstate }
		\Item{ handle }
		\Item{ analogical\_switch }
		\Item{ gear\_extended }
		\Item{ gear\_retracted }
		\Item{ gear\_shock\_absorber }
		\Item{ door\_closed }
		\Item{ door\_open }
		\Item{ circuit\_pressurized }
		\Item{ general\_EV }
		\Item{ close\_EV }
		\Item{ retract\_EV }
		\Item{ extend\_EV }
		\Item{ open\_EV }
		\Item{ gears\_locked\_down }
		\Item{ gears\_man }
		\Item{ anomaly }
		\Item{ general\_EV\_func }
		\Item{ close\_EV\_func }
		\Item{ retract\_EV\_func }
		\Item{ extend\_EV\_func }
		\Item{ open\_EV\_func }
		\Item{ gears\_locked\_down\_func }
		\Item{ gears\_man\_func }
		\Item{ anomaly\_func }
		\Item{ A\_Switch\_Out }
	\end{description}
\INVARIANTS
	\begin{description}
		\nItem{ inv3 }{ handle \in  1\upto 3 \tfun  POSITIONS }
		\nItem{ inv4 }{ analogical\_switch \in  1\upto 3 \tfun  A\_Switch }
		\nItem{ inv5 }{ gear\_extended \in  1\upto 3 \tfun  (GEARS \tfun  BOOL) }
		\nItem{ inv6 }{ gear\_retracted \in  1\upto 3 \tfun  (GEARS \tfun  BOOL) }
		\nItem{ inv7 }{ gear\_shock\_absorber \in  1\upto 3 \tfun  GEAR\_ABSORBER }
		\nItem{ inv8 }{ door\_closed \in  1\upto 3 \tfun  (DOORS \tfun  BOOL) }
		\nItem{ inv9 }{ door\_open \in  1\upto 3 \tfun  (DOORS \tfun  BOOL) }
		\nItem{ inv10 }{ circuit\_pressurized \in  1\upto 3 \tfun  BOOL }
		\nItem{ inv13 }{ general\_EV \in  BOOL }
		\nItem{ inv14 }{ close\_EV \in  BOOL }
		\nItem{ inv15 }{ retract\_EV \in  BOOL }
		\nItem{ inv16 }{ extend\_EV \in  BOOL }
		\nItem{ inv18 }{ open\_EV \in  BOOL }
		\nItem{ inv19 }{ gears\_locked\_down \in  BOOL }
		\nItem{ inv20 }{ gears\_man \in  BOOL }
		\nItem{ inv21 }{ anomaly \in  BOOL }
		\nItem{ inv22 }{ general\_EV\_func \in  (1\upto 3 \tfun  POSITIONS) \cprod  (1\upto 3 \tfun  A\_Switch) \cprod  (1\upto 3 \tfun  (GEARS \tfun  BOOL)) \cprod  (1\upto 3 \tfun  (GEARS \tfun  BOOL)) \cprod  (1\upto 3 \tfun  GEAR\_ABSORBER) \cprod  (1\upto 3 \tfun  (DOORS \tfun  BOOL)) \cprod  (1\upto 3 \tfun  (DOORS \tfun  BOOL)) \cprod  (1\upto 3 \tfun  BOOL) \tfun  BOOL }
		\nItem{ inv23 }{ close\_EV\_func \in  (1\upto 3 \tfun  POSITIONS) \cprod  (1\upto 3 \tfun  A\_Switch) \cprod  (1\upto 3 \tfun  (GEARS \tfun  BOOL)) \cprod  (1\upto 3 \tfun  (GEARS \tfun  BOOL)) \cprod  (1\upto 3 \tfun  GEAR\_ABSORBER) \cprod  (1\upto 3 \tfun  (DOORS \tfun  BOOL)) \cprod  (1\upto 3 \tfun  (DOORS \tfun  BOOL)) \cprod  (1\upto 3 \tfun  BOOL) \tfun  BOOL }
		\nItem{ inv24 }{ retract\_EV\_func \in  (1\upto 3 \tfun  POSITIONS) \cprod  (1\upto 3 \tfun  A\_Switch) \cprod  (1\upto 3 \tfun  (GEARS \tfun  BOOL)) \cprod  (1\upto 3 \tfun  (GEARS \tfun  BOOL)) \cprod  (1\upto 3 \tfun  GEAR\_ABSORBER) \cprod  (1\upto 3 \tfun  (DOORS \tfun  BOOL)) \cprod  (1\upto 3 \tfun  (DOORS \tfun  BOOL)) \cprod  (1\upto 3 \tfun  BOOL) \tfun  BOOL }
		\nItem{ inv25 }{ extend\_EV\_func \in  (1\upto 3 \tfun  POSITIONS) \cprod  (1\upto 3 \tfun  A\_Switch) \cprod  (1\upto 3 \tfun  (GEARS \tfun  BOOL)) \cprod  (1\upto 3 \tfun  (GEARS \tfun  BOOL)) \cprod  (1\upto 3 \tfun  GEAR\_ABSORBER) \cprod  (1\upto 3 \tfun  (DOORS \tfun  BOOL)) \cprod  (1\upto 3 \tfun  (DOORS \tfun  BOOL)) \cprod  (1\upto 3 \tfun  BOOL) \tfun  BOOL }
		\nItem{ inv26 }{ open\_EV\_func \in  (1\upto 3 \tfun  POSITIONS) \cprod  (1\upto 3 \tfun  A\_Switch) \cprod  (1\upto 3 \tfun  (GEARS \tfun  BOOL)) \cprod  (1\upto 3 \tfun  (GEARS \tfun  BOOL)) \cprod  (1\upto 3 \tfun  GEAR\_ABSORBER) \cprod  (1\upto 3 \tfun  (DOORS \tfun  BOOL)) \cprod  (1\upto 3 \tfun  (DOORS \tfun  BOOL)) \cprod  (1\upto 3 \tfun  BOOL) \tfun  BOOL }
		\nItem{ inv27 }{ gears\_locked\_down\_func \in  (1\upto 3 \tfun  POSITIONS) \cprod  (1\upto 3 \tfun  A\_Switch) \cprod  (1\upto 3 \tfun  (GEARS \tfun  BOOL)) \cprod  (1\upto 3 \tfun  (GEARS \tfun  BOOL)) \cprod  (1\upto 3 \tfun  GEAR\_ABSORBER) \cprod  (1\upto 3 \tfun  (DOORS \tfun  BOOL)) \cprod  (1\upto 3 \tfun  (DOORS \tfun  BOOL)) \cprod  (1\upto 3 \tfun  BOOL) \tfun  BOOL }
		\nItem{ inv28 }{ gears\_man\_func \in  (1\upto 3 \tfun  POSITIONS) \cprod  (1\upto 3 \tfun  A\_Switch) \cprod  (1\upto 3 \tfun  (GEARS \tfun  BOOL)) \cprod  (1\upto 3 \tfun  (GEARS \tfun  BOOL)) \cprod  (1\upto 3 \tfun  GEAR\_ABSORBER) \cprod  (1\upto 3 \tfun  (DOORS \tfun  BOOL)) \cprod  (1\upto 3 \tfun  (DOORS \tfun  BOOL)) \cprod  (1\upto 3 \tfun  BOOL) \tfun  BOOL }
		\nItem{ inv29 }{ anomaly\_func \in  (1\upto 3 \tfun  POSITIONS) \cprod  (1\upto 3 \tfun  A\_Switch) \cprod  (1\upto 3 \tfun  (GEARS \tfun  BOOL)) \cprod  (1\upto 3 \tfun  (GEARS \tfun  BOOL)) \cprod  (1\upto 3 \tfun  GEAR\_ABSORBER) \cprod  (1\upto 3 \tfun  (DOORS \tfun  BOOL)) \cprod  (1\upto 3 \tfun  (DOORS \tfun  BOOL)) \cprod  (1\upto 3 \tfun  BOOL) \tfun  BOOL }
		\nItem{ inv30 }{ A\_Switch\_Out \in  BOOL }
		\nItem{ M1\_inv1 }{ button \in  POSITIONS }
		\nItem{ M1\_inv2 }{ phase \in  PHASES }
		\nItem{ M1\_inv3 }{ phase=movingup \limp  button=UP }
		\nItem{ M1\_inv4 }{ phase=movingdown \limp  button = DOWN }
		\nItem{ M1\_inv5 }{ button=UP \limp  phase\notin \{ movingdown,haltdown\}  }
		\nItem{ M1\_inv6 }{ button=DOWN \limp  phase\notin \{ movingup,haltup\}  }
		\nItem{ M2\_inv1 }{ dstate \in  DOORS \tfun  SDOORS }
		\nItemY{ M2\_inv2 }{ dstate^{-1} [\{ OPEN\} ] \neq  \emptyset  \limp  dstate^{-1} [\{ OPEN\} ]=DOORS }{ 		\\\hspace*{2 cm}  when one door is open, each door is open. }
		\nItemY{ M2\_inv3 }{ dstate^{-1} [\{ CLOSED\} ] \neq  \emptyset  \limp  dstate^{-1} [\{ CLOSED\} ]=DOORS }{ 		\\\hspace*{2 cm}  when a door is closed, t each door is closed }
		\nItem{ M2\_inv6 }{ lstate \in  DOORS \tfun  SLOCKS }
		\nItem{ M2\_inv7 }{ dstate[DOORS]=\{ OPEN\}  \limp  lstate[DOORS]=\{ UNLOCKED\}  }
		\nItem{ M2\_inv12 }{ p \in  P }
		\nItem{ M2\_inv13 }{ l \in  P }
		\nItem{ M2\_inv14 }{ i \in  P }
		\nItem{ M2\_inv15 }{ l=E \land  p=R \limp  lstate[DOORS]=\{ UNLOCKED\}  }
		\nItem{ M2\_inv16 }{ l=R \land  p=E \limp  lstate[DOORS]=\{ UNLOCKED\}  }
		\nItem{ M3\_inv1 }{ gstate \in  GEARS \tfun  SGEARS }
		\nItemY{ M3\_inv3 }{ \forall  door\qdot  door \in  DOORS \land  dstate(door)=CLOSED  \land  ran(gstate)\neq \{ RETRACTED\}  \limp  ran(gstate)=\{ EXTENDED\}  }{ 		\\\hspace*{2 cm}  gears can not be out or moving in this case. }
		\nItem{ M3\_inv6 }{ \forall  door\qdot  door \in  DOORS \land  dstate(door)=CLOSED  \land  ran(gstate)\neq \{ EXTENDED\}   \limp  ran(gstate)=\{ RETRACTED\}  }
		\nItem{ M3\_inv7 }{ ran(gstate)\neq \{ RETRACTED\}  \land  ran(gstate)\neq \{ EXTENDED\}   \limp  ran(dstate)=\{ OPEN\}  }
		\nItem{ M3\_inv11 }{ ran(dstate)=\{ CLOSED\}  \limp  ran(gstate)\binter \{ RETRACTING,EXTENDING\}  =\emptyset  }
	\end{description}
\EVENTS
	\INITIALISATION
		\begin{description}
		\BeginAct
			\begin{description}
			\nItem{ act1 }{ button :=  DOWN }
			\nItem{ act2 }{ phase :=  haltdown }
			\nItemY{ act3 }{ dstate :|  (dstate'\in  DOORS \tfun  SDOORS \land  dstate'=\{ a\mapsto b|  a \in  DOORS \land  b=CLOSED\} ) }{ 		\\\hspace*{1,4 cm}  missing elements of the invariant }
			\nItem{ act4 }{ lstate :=  \{ a\mapsto b| a\in DOORS\land  b=LOCKED\}  }
			\nItem{ act5 }{ p :=  R }
			\nItem{ act6 }{ l :=  R }
			\nItem{ act7 }{ i :=  R }
			\nItem{ act8 }{ gstate :|  (gstate' \in  GEARS \tfun  SGEARS \land  gstate'=\{ a\mapsto b |  a \in  GEARS  \land  b=EXTENDED\} ) }
			\nItem{ act14 }{ handle :\in  1\upto 3 \tfun  \{ DOWN\}  }
			\nItem{ act15 }{ analogical\_switch :\in  1\upto 3 \tfun  \{ open\}  }
			\nItem{ act16 }{ gear\_extended :\in  1\upto 3 \tfun  (GEARS \tfun  \{ TRUE\} ) }
			\nItem{ act17 }{ gear\_retracted :\in  1\upto 3 \tfun  (GEARS \tfun  \{ FALSE\} ) }
			\nItem{ act18 }{ gear\_shock\_absorber :\in  1\upto 3 \tfun  \{ ground\}  }
			\nItem{ act19 }{ door\_closed :\in  1\upto 3 \tfun  (DOORS \tfun  \{ TRUE\} ) }
			\nItem{ act20 }{ door\_open :\in  1\upto 3 \tfun  (DOORS \tfun  \{ FALSE\} ) }
			\nItem{ act21 }{ circuit\_pressurized :\in  1\upto 3 \tfun  \{ FALSE\}  }
			\nItem{ act22 }{ general\_EV :=  FALSE }
			\nItem{ act23 }{ close\_EV :=  TRUE }
			\nItem{ act24 }{ retract\_EV :=  FALSE }
			\nItem{ act25 }{ extend\_EV :=  TRUE }
			\nItem{ act27 }{ open\_EV :=  FALSE }
			\nItem{ act28 }{ gears\_locked\_down :=  TRUE }
			\nItem{ act29 }{ gears\_man :=  FALSE }
			\nItem{ act30 }{ anomaly :=  FALSE }
			\nItem{ act31 }{ general\_EV\_func :\in  (1\upto 3 \tfun  POSITIONS) \cprod  (1\upto 3 \tfun  A\_Switch) \cprod  (1\upto 3 \tfun  (GEARS \tfun  BOOL)) \cprod  (1\upto 3 \tfun  (GEARS \tfun  BOOL)) \cprod  (1\upto 3 \tfun  GEAR\_ABSORBER) \cprod  (1\upto 3 \tfun  (DOORS \tfun  BOOL)) \cprod  (1\upto 3 \tfun  (DOORS \tfun  BOOL)) \cprod  (1\upto 3 \tfun  BOOL) \tfun  BOOL }
			\nItem{ act32 }{ close\_EV\_func :\in  (1\upto 3 \tfun  POSITIONS) \cprod  (1\upto 3 \tfun  A\_Switch) \cprod  (1\upto 3 \tfun  (GEARS \tfun  BOOL)) \cprod  (1\upto 3 \tfun  (GEARS \tfun  BOOL)) \cprod  (1\upto 3 \tfun  GEAR\_ABSORBER) \cprod  (1\upto 3 \tfun  (DOORS \tfun  BOOL)) \cprod  (1\upto 3 \tfun  (DOORS \tfun  BOOL)) \cprod  (1\upto 3 \tfun  BOOL) \tfun  BOOL }
			\nItem{ act33 }{ retract\_EV\_func :\in  (1\upto 3 \tfun  POSITIONS) \cprod  (1\upto 3 \tfun  A\_Switch) \cprod  (1\upto 3 \tfun  (GEARS \tfun  BOOL)) \cprod  (1\upto 3 \tfun  (GEARS \tfun  BOOL)) \cprod  (1\upto 3 \tfun  GEAR\_ABSORBER) \cprod  (1\upto 3 \tfun  (DOORS \tfun  BOOL)) \cprod  (1\upto 3 \tfun  (DOORS \tfun  BOOL)) \cprod  (1\upto 3 \tfun  BOOL) \tfun  BOOL }
			\nItem{ act34 }{ extend\_EV\_func :\in  (1\upto 3 \tfun  POSITIONS) \cprod  (1\upto 3 \tfun  A\_Switch) \cprod  (1\upto 3 \tfun  (GEARS \tfun  BOOL)) \cprod  (1\upto 3 \tfun  (GEARS \tfun  BOOL)) \cprod  (1\upto 3 \tfun  GEAR\_ABSORBER) \cprod  (1\upto 3 \tfun  (DOORS \tfun  BOOL)) \cprod  (1\upto 3 \tfun  (DOORS \tfun  BOOL)) \cprod  (1\upto 3 \tfun  BOOL) \tfun  BOOL }
			\nItem{ act35 }{ open\_EV\_func :\in  (1\upto 3 \tfun  POSITIONS) \cprod  (1\upto 3 \tfun  A\_Switch) \cprod  (1\upto 3 \tfun  (GEARS \tfun  BOOL)) \cprod  (1\upto 3 \tfun  (GEARS \tfun  BOOL)) \cprod  (1\upto 3 \tfun  GEAR\_ABSORBER) \cprod  (1\upto 3 \tfun  (DOORS \tfun  BOOL)) \cprod  (1\upto 3 \tfun  (DOORS \tfun  BOOL)) \cprod  (1\upto 3 \tfun  BOOL) \tfun  BOOL }
			\nItem{ act36 }{ gears\_locked\_down\_func :\in  (1\upto 3 \tfun  POSITIONS) \cprod  (1\upto 3 \tfun  A\_Switch) \cprod  (1\upto 3 \tfun  (GEARS \tfun  BOOL)) \cprod  (1\upto 3 \tfun  (GEARS \tfun  BOOL)) \cprod  (1\upto 3 \tfun  GEAR\_ABSORBER) \cprod  (1\upto 3 \tfun  (DOORS \tfun  BOOL)) \cprod  (1\upto 3 \tfun  (DOORS \tfun  BOOL)) \cprod  (1\upto 3 \tfun  BOOL) \tfun  BOOL }
			\nItem{ act37 }{ gears\_man\_func :\in  (1\upto 3 \tfun  POSITIONS) \cprod  (1\upto 3 \tfun  A\_Switch) \cprod  (1\upto 3 \tfun  (GEARS \tfun  BOOL)) \cprod  (1\upto 3 \tfun  (GEARS \tfun  BOOL)) \cprod  (1\upto 3 \tfun  GEAR\_ABSORBER) \cprod  (1\upto 3 \tfun  (DOORS \tfun  BOOL)) \cprod  (1\upto 3 \tfun  (DOORS \tfun  BOOL)) \cprod  (1\upto 3 \tfun  BOOL) \tfun  BOOL }
			\nItem{ act38 }{ anomaly\_func :\in  (1\upto 3 \tfun  POSITIONS) \cprod  (1\upto 3 \tfun  A\_Switch) \cprod  (1\upto 3 \tfun  (GEARS \tfun  BOOL)) \cprod  (1\upto 3 \tfun  (GEARS \tfun  BOOL)) \cprod  (1\upto 3 \tfun  GEAR\_ABSORBER) \cprod  (1\upto 3 \tfun  (DOORS \tfun  BOOL)) \cprod  (1\upto 3 \tfun  (DOORS \tfun  BOOL)) \cprod  (1\upto 3 \tfun  BOOL) \tfun  BOOL }
			\nItem{ act39 }{ A\_Switch\_Out :=  FALSE }
			\end{description}
		\EndAct
		\end{description}
	\EVT {opening\_doors\_DOWN}
	\REF {opening\_doors\_DOWN}
		\begin{description}
		\WhenGrd
			\begin{description}
			\nItem{ grd1 }{ dstate[DOORS]= \{ CLOSED\}  }
			\nItem{ grd5 }{ lstate[DOORS]=\{ UNLOCKED\}  }
			\nItem{ grd7 }{ phase=movingdown }
			\nItem{ grd8 }{ p=R }
			\nItem{ grd9 }{ l=R }
			\nItem{ grd10 }{ door\_open = \{ a\mapsto b |  a \in  1\upto 3 \land  b \in  DOORS \tfun  \{ FALSE\} \}   }
			\nItem{ grd11 }{ door\_closed = \{ a\mapsto b |  a \in  1\upto 3 \land  b \in  DOORS \tfun  \{ FALSE\} \}   }
			\nItem{ grd12 }{ \forall x\qdot x\in 1\upto 3 \limp  handle(x)=button }
			\end{description}
		\ThenAct
			\begin{description}
			\nItem{ act1 }{ dstate :=  \{ a\mapsto b|  a \in  DOORS \land  b=OPEN\}  }
			\nItem{ act2 }{ p:= E }
			\nItem{ act3 }{ door\_open :\in  1\upto 3 \tfun  (DOORS \tfun  \{ TRUE\} ) }
			\end{description}
		\EndAct
		\end{description}
	\EVT {opening\_doors\_UP}
	\REF {opening\_doors\_UP}
		\begin{description}
		\WhenGrd
			\begin{description}
			\nItem{ grd1 }{ dstate[DOORS]= \{ CLOSED\}  }
			\nItem{ grd4 }{ lstate[DOORS]=\{ UNLOCKED\}  }
			\nItem{ grd5 }{ phase= movingup }
			\nItem{ grd6 }{ p=E }
			\nItem{ grd7 }{ l=E }
			\nItem{ grd8 }{ door\_open = \{ a\mapsto b |  a \in  1\upto 3 \land  b \in  DOORS \tfun  \{ FALSE\} \}   }
			\nItem{ grd9 }{ door\_closed = \{ a\mapsto b |  a \in  1\upto 3 \land  b \in  DOORS \tfun  \{ FALSE\} \}   }
			\nItem{ grd10 }{ \forall x\qdot x\in 1\upto 3 \limp  handle(x)=button }
			\end{description}
		\ThenAct
			\begin{description}
			\nItem{ act1 }{ dstate :=  \{ a\mapsto b|  a \in  DOORS \land  b=OPEN\}  }
			\nItem{ act2 }{ p:= R }
			\nItem{ act3 }{ door\_open:\in  1\upto 3 \tfun  (DOORS \tfun  \{ TRUE\} )  }
			\end{description}
		\EndAct
		\end{description}
	\EVT {closing\_doors\_UP}
	\REF {closing\_doors\_UP}
		\begin{description}
		\AnyPrm
			\begin{description}
			\Item{f }
			\end{description}
		\WhereGrd
			\begin{description}
			\nItem{ grd1 }{ dstate[DOORS]=\{ OPEN\}  }
			\nItem{ grd3 }{ f \in  DOORS \tfun  SDOORS }
			\nItem{ grd4 }{ \forall e\qdot  e \in  DOORS \limp  f(e)=CLOSED }
			\nItem{ grd5 }{ phase=movingup }
			\nItem{ grd6 }{ p=R }
			\nItem{ grd7 }{ gstate[GEARS]=\{ RETRACTED\}  }
			\nItem{ grd8 }{ \forall x\qdot x\in 1\upto 3 \limp  handle(x)=button }
			\end{description}
		\ThenAct
			\begin{description}
			\nItem{ act1 }{ dstate:= f }
			\end{description}
		\EndAct
		\end{description}
	\EVT {closing\_doors\_DOWN}
	\REF {closing\_doors\_DOWN}
		\begin{description}
		\AnyPrm
			\begin{description}
			\Item{f }
			\end{description}
		\WhereGrd
			\begin{description}
			\nItem{ grd1 }{ dstate[DOORS]=\{ OPEN\}  }
			\nItem{ grd3 }{ f \in  DOORS \tfun  SDOORS }
			\nItem{ grd4 }{ \forall e\qdot  e \in  DOORS \limp  f(e)=CLOSED }
			\nItem{ grd5 }{ phase=movingdown }
			\nItem{ grd6 }{ p=E }
			\nItem{ grd7 }{ gstate[GEARS]=\{ EXTENDED\}  }
			\nItem{ grd8 }{ \forall x\qdot x\in 1\upto 3 \limp  handle(x)=button }
			\end{description}
		\ThenAct
			\begin{description}
			\nItem{ act1 }{ dstate:= f }
			\end{description}
		\EndAct
		\end{description}
	\EVT {unlocking\_UP}
	\REF {unlocking\_UP}
		\begin{description}
		\WhenGrd
			\begin{description}
			\nItem{ grd3 }{ lstate[DOORS]=\{ LOCKED\}  }
			\nItem{ grd4 }{ phase=movingup }
			\nItem{ grd5 }{ l=E }
			\nItem{ grd6 }{ p=E }
			\nItem{ grd7 }{ i=E }
			\nItem{ grd8 }{ door\_open = \{ a\mapsto b |  a \in  1\upto 3 \land  b \in  DOORS \tfun  \{ FALSE\} \}   }
			\nItem{ grd9 }{ door\_closed = \{ a\mapsto b |  a \in  1\upto 3 \land  b \in  DOORS \tfun  \{ TRUE\} \}   }
			\nItem{ grd10 }{ \forall x\qdot x\in 1\upto 3 \limp  handle(x)=button }
			\end{description}
		\ThenAct
			\begin{description}
			\nItem{ act1 }{ lstate:= \{ a\mapsto b| a\in DOORS \land  b=UNLOCKED\}  }
			\nItem{ act2 }{ door\_closed :\in  1\upto 3 \tfun  (DOORS \tfun  \{ FALSE\} )  }
			\end{description}
		\EndAct
		\end{description}
	\EVT {locking\_UP}
	\REF {locking\_UP}
		\begin{description}
		\WhenGrd
			\begin{description}
			\nItem{ grd3 }{ dstate[DOORS]=\{ CLOSED\}  }
			\nItem{ grd4 }{ phase=movingup }
			\nItem{ grd5 }{ lstate[DOORS]=\{ UNLOCKED\}  }
			\nItem{ grd6 }{ p=R }
			\nItem{ grd7 }{ l=E }
			\nItem{ grd9 }{ door\_open = \{ a\mapsto b |  a \in  1\upto 3 \land  b \in  DOORS \tfun  \{ FALSE\} \}   }
			\nItem{ grd10 }{ door\_closed = \{ a\mapsto b |  a \in  1\upto 3 \land  b \in  DOORS \tfun  \{ FALSE\} \}   }
			\nItem{ grd11 }{ \forall x\qdot x\in 1\upto 3 \limp  handle(x)=button }
			\end{description}
		\ThenAct
			\begin{description}
			\nItem{ act1 }{ lstate:= \{ a\mapsto b| a\in DOORS \land  b=LOCKED\}  }
			\nItem{ act3 }{ phase:= haltup }
			\nItemY{ act4 }{ l:= R }{ 		\\\hspace*{1,4 cm}  added by D Mery }
			\nItem{ act44 }{ door\_closed :\in  1\upto 3 \tfun  (DOORS \tfun  \{ TRUE\} )  }
			\end{description}
		\EndAct
		\end{description}
	\EVT {unlocking\_DOWN}
	\REF {unlocking\_DOWN}
		\begin{description}
		\WhenGrd
			\begin{description}
			\nItem{ grd3 }{ lstate[DOORS]=\{ LOCKED\}  }
			\nItem{ grd4 }{ phase=movingdown }
			\nItem{ grd5 }{ l=R }
			\nItem{ grd6 }{ p=R }
			\nItem{ grd7 }{ i=R }
			\nItem{ grd8 }{ door\_open = \{ a\mapsto b |  a \in  1\upto 3 \land  b \in  DOORS \tfun  \{ FALSE\} \}   }
			\nItem{ grd9 }{ door\_closed = \{ a\mapsto b |  a \in  1\upto 3 \land  b \in  DOORS \tfun  \{ TRUE\} \}   }
			\nItem{ grd10 }{ \forall x\qdot x\in 1\upto 3 \limp  handle(x)=button }
			\end{description}
		\ThenAct
			\begin{description}
			\nItem{ act1 }{ lstate:= \{ a\mapsto b| a\in DOORS \land  b=UNLOCKED\}  }
			\nItem{ act2 }{ door\_closed :\in  1\upto 3 \tfun  (DOORS \tfun  \{ FALSE\} ) }
			\end{description}
		\EndAct
		\end{description}
	\EVT {locking\_DOWN}
	\REF {locking\_DOWN}
		\begin{description}
		\WhenGrd
			\begin{description}
			\nItem{ grd1 }{ dstate[DOORS]=\{ CLOSED\}  }
			\nItem{ grd2 }{ phase=movingdown }
			\nItem{ grd3 }{ lstate[DOORS]=\{ UNLOCKED\}  }
			\nItem{ grd4 }{ p=E }
			\nItem{ grd5 }{ l=R }
			\nItem{ grd7 }{ door\_open = \{ a\mapsto b |  a \in  1\upto 3 \land  b \in  DOORS \tfun  \{ FALSE\} \}   }
			\nItem{ grd8 }{ door\_closed = \{ a\mapsto b |  a \in  1\upto 3 \land  b \in  DOORS \tfun  \{ FALSE\} \}   }
			\nItem{ grd9 }{ \forall x\qdot x\in 1\upto 3 \limp  handle(x)=button }
			\end{description}
		\ThenAct
			\begin{description}
			\nItem{ act1 }{ lstate:= \{ a\mapsto b| a\in DOORS \land  b = LOCKED\}  }
			\nItem{ act3 }{ phase:= haltdown }
			\nItem{ act4 }{ l:= E }
			\nItem{ act5 }{ door\_closed :\in  1\upto 3 \tfun  (DOORS \tfun  \{ TRUE\} )  }
			\end{description}
		\EndAct
		\end{description}
	\EVT {PD1}
	\REF {PD1}
		\begin{description}
		\WhenGrd
			\begin{description}
			\nItem{ grd1 }{ button=UP }
			\nItem{ grd2 }{ phase=haltup }
			\nItem{ grd3 }{ \forall x\qdot x\in 1\upto 3 \limp  handle(x)=DOWN }
			\end{description}
		\ThenAct
			\begin{description}
			\nItem{ act1 }{ phase:= movingdown }
			\nItem{ act2 }{ button:= DOWN }
			\nItem{ act3 }{ l:= R }
			\nItem{ act4 }{ p:= R }
			\nItem{ act5 }{ i:= R }
			\end{description}
		\EndAct
		\end{description}
	\EVT {PU1}
	\REF {PU1}
		\begin{description}
		\WhenGrd
			\begin{description}
			\nItem{ grd1 }{ button=DOWN }
			\nItem{ grd2 }{ phase=haltdown }
			\nItem{ grd3 }{ \forall x\qdot x\in 1\upto 3 \limp  handle(x)=UP }
			\end{description}
		\ThenAct
			\begin{description}
			\nItem{ act1 }{ phase:= movingup }
			\nItem{ act2 }{ button:= UP }
			\nItem{ act3 }{ l:= E }
			\nItem{ act4 }{ p:= E }
			\nItem{ act5 }{ i:= E }
			\end{description}
		\EndAct
		\end{description}
	\EVT {PU2}
	\REF {PU2}
		\begin{description}
		\WhenGrd
			\begin{description}
			\nItem{ grd1 }{ l=R }
			\nItem{ grd2 }{ p=R }
			\nItem{ grd3 }{ phase=movingdown }
			\nItem{ grd4 }{ button=DOWN }
			\nItem{ grd5 }{ i=R }
			\nItem{ grd6 }{ lstate[DOORS]=\{ LOCKED\}  }
			\nItem{ grd7 }{ door\_open = \{ a\mapsto b |  a \in  1\upto 3 \land  b \in  DOORS \tfun  \{ FALSE\} \}   }
			\nItem{ grd8 }{ door\_closed = \{ a\mapsto b |  a \in  1\upto 3 \land  b \in  DOORS \tfun  \{ TRUE\} \}   }
			\nItem{ grd9 }{ \forall x\qdot x\in 1\upto 3 \limp  handle(x)=UP }
			\end{description}
		\ThenAct
			\begin{description}
			\nItem{ act1 }{ phase:= movingup }
			\nItem{ act4 }{ button:=   UP }
			\nItem{ act5 }{ l:= E }
			\nItem{ act6 }{ p:= E }
			\nItem{ act7 }{ i:= R }
			\end{description}
		\EndAct
		\end{description}
	\EVT {CompletePU2}
	\REF {CompletePU2}
		\begin{description}
		\WhenGrd
			\begin{description}
			\nItem{ grd1 }{ phase=movingup }
			\nItem{ grd2 }{ button=UP }
			\nItem{ grd3 }{ l=E }
			\nItem{ grd4 }{ p=E }
			\nItem{ grd5 }{ i=R }
			\end{description}
		\ThenAct
			\begin{description}
			\nItem{ act1 }{ phase:= haltup }
			\end{description}
		\EndAct
		\end{description}
	\EVT {PU3}
	\REF {PU3}
		\begin{description}
		\WhenGrd
			\begin{description}
			\nItem{ grd1 }{ dstate[DOORS]=\{ CLOSED\}  }
			\nItem{ grd2 }{ lstate[DOORS]=\{ UNLOCKED\}  }
			\nItem{ grd3 }{ phase = movingdown }
			\nItem{ grd4 }{ p=R }
			\nItem{ grd5 }{ l=R }
			\nItem{ grd6 }{ button=DOWN }
			\nItem{ grd7 }{ door\_open = \{ a\mapsto b |  a \in  1\upto 3 \land  b \in  DOORS \tfun  \{ FALSE\} \}   }
			\nItem{ grd8 }{ door\_closed = \{ a\mapsto b |  a \in  1\upto 3 \land  b \in  DOORS \tfun  \{ FALSE\} \}   }
			\nItem{ grd9 }{ \forall x\qdot x\in 1\upto 3 \limp  handle(x)=UP }
			\end{description}
		\ThenAct
			\begin{description}
			\nItem{ act1 }{ phase:= movingup }
			\nItem{ act2 }{ p:= R }
			\nItem{ act3 }{ l:= E }
			\nItem{ act4 }{ button:= UP }
			\end{description}
		\EndAct
		\end{description}
	\EVT {PU4}
	\REF {PU4}
		\begin{description}
		\WhenGrd
			\begin{description}
			\nItem{ grd1 }{ dstate[DOORS]=\{ OPEN\}  }
			\nItem{ grd2 }{ phase=movingdown }
			\nItem{ grd3 }{ p=E }
			\nItem{ grd4 }{ button=DOWN }
			\nItem{ grd7 }{ \forall x\qdot x\in 1\upto 3 \limp  handle(x)=UP }
			\end{description}
		\ThenAct
			\begin{description}
			\nItem{ act1 }{ phase:= movingup }
			\nItem{ act2 }{ p:= R }
			\nItem{ act3 }{ button:= UP }
			\nItem{ act4 }{ i:= E }
			\nItem{ act5 }{ l:= E }
			\end{description}
		\EndAct
		\end{description}
	\EVT {PU5}
	\REF {PU5}
		\begin{description}
		\WhenGrd
			\begin{description}
			\nItem{ grd1 }{ dstate[DOORS]=\{ CLOSED\}  }
			\nItem{ grd2 }{ phase=movingdown }
			\nItem{ grd3 }{ p=E }
			\nItem{ grd4 }{ button=DOWN }
			\nItem{ grd5 }{ lstate[DOORS]=\{ UNLOCKED\}  }
			\nItem{ grd6 }{ door\_open = \{ a\mapsto b |  a \in  1\upto 3 \land  b \in  DOORS \tfun  \{ FALSE\} \}   }
			\nItem{ grd7 }{ door\_closed = \{ a\mapsto b |  a \in  1\upto 3 \land  b \in  DOORS \tfun  \{ FALSE\} \}   }
			\nItem{ grd8 }{ \forall x\qdot x\in 1\upto 3 \limp  handle(x)=UP }
			\end{description}
		\ThenAct
			\begin{description}
			\nItem{ act1 }{ phase:= movingup }
			\nItem{ act3 }{ button:= UP }
			\nItem{ act4 }{ i:= E }
			\nItem{ act5 }{ l:= E }
			\end{description}
		\EndAct
		\end{description}
	\EVT {PD2}
	\REF {PD2}
		\begin{description}
		\WhenGrd
			\begin{description}
			\nItem{ grd1 }{ l=E }
			\nItem{ grd2 }{ p=E }
			\nItem{ grd3 }{ phase=movingup }
			\nItem{ grd4 }{ i=E }
			\nItem{ grd5 }{ lstate[DOORS]=\{ LOCKED\}  }
			\nItem{ grd6 }{ \forall x\qdot x\in 1\upto 3 \limp  handle(x)=DOWN }
			\end{description}
		\ThenAct
			\begin{description}
			\nItem{ act1 }{ phase:= movingdown }
			\nItem{ act2 }{ button:= DOWN }
			\nItem{ act3 }{ l:= R }
			\nItem{ act4 }{ p:= R }
			\nItem{ act5 }{ i:= E }
			\end{description}
		\EndAct
		\end{description}
	\EVT {CompletePD2}
	\REF {CompletePD2}
		\begin{description}
		\WhenGrd
			\begin{description}
			\nItem{ grd1 }{ phase=movingdown }
			\nItem{ grd2 }{ button=DOWN }
			\nItem{ grd3 }{ l=R }
			\nItem{ grd4 }{ p=R }
			\nItem{ grd5 }{ i=E }
			\end{description}
		\ThenAct
			\begin{description}
			\nItem{ act1 }{ phase:= haltdown }
			\end{description}
		\EndAct
		\end{description}
	\EVT {PD3}
	\REF {PD3}
		\begin{description}
		\WhenGrd
			\begin{description}
			\nItem{ grd1 }{ dstate[DOORS]=\{ CLOSED\}  }
			\nItem{ grd2 }{ lstate[DOORS]=\{ UNLOCKED\}  }
			\nItem{ grd3 }{ phase=movingup }
			\nItem{ grd4 }{ p=E }
			\nItem{ grd5 }{ l=E }
			\nItem{ grd6 }{ button=UP }
			\nItem{ grd7 }{ door\_open = \{ a\mapsto b |  a \in  1\upto 3 \land  b \in  DOORS \tfun  \{ FALSE\} \}   }
			\nItem{ grd8 }{ door\_closed = \{ a\mapsto b |  a \in  1\upto 3 \land  b \in  DOORS \tfun  \{ FALSE\} \}   }
			\nItem{ grd9 }{ \forall x\qdot x\in 1\upto 3 \limp  handle(x)=DOWN }
			\end{description}
		\ThenAct
			\begin{description}
			\nItem{ act1 }{ phase:= movingdown }
			\nItem{ act2 }{ p:= E }
			\nItem{ act3 }{ l:= R }
			\nItem{ act4 }{ button:= DOWN }
			\end{description}
		\EndAct
		\end{description}
	\EVT {PD4}
	\REF {PD4}
		\begin{description}
		\WhenGrd
			\begin{description}
			\nItem{ grd1 }{ dstate[DOORS]=\{ OPEN\}  }
			\nItem{ grd2 }{ phase=movingup }
			\nItem{ grd3 }{ p=R }
			\nItem{ grd4 }{ button=UP }
			\nItem{ grd6 }{ \forall x\qdot x\in 1\upto 3 \limp  handle(x)=DOWN }
			\end{description}
		\ThenAct
			\begin{description}
			\nItem{ act1 }{ phase:= movingdown }
			\nItem{ act2 }{ p:= E }
			\nItem{ act3 }{ button:= DOWN }
			\nItem{ act4 }{ i:= R }
			\nItem{ act5 }{ l:= R }
			\end{description}
		\EndAct
		\end{description}
	\EVT {PD5}
	\REF {PD5}
		\begin{description}
		\WhenGrd
			\begin{description}
			\nItem{ grd1 }{ dstate[DOORS]=\{ CLOSED\}  }
			\nItem{ grd2 }{ phase=movingup }
			\nItem{ grd3 }{ p=R }
			\nItem{ grd4 }{ button=UP }
			\nItem{ grd5 }{ lstate[DOORS]=\{ UNLOCKED\}  }
			\nItem{ grd6 }{ door\_open = \{ a\mapsto b |  a \in  1\upto 3 \land  b \in  DOORS \tfun  \{ FALSE\} \}   }
			\nItem{ grd7 }{ door\_closed = \{ a\mapsto b |  a \in  1\upto 3 \land  b \in  DOORS \tfun  \{ FALSE\} \}   }
			\nItem{ grd8 }{ \forall x\qdot x\in 1\upto 3 \limp  handle(x)=DOWN }
			\end{description}
		\ThenAct
			\begin{description}
			\nItem{ act1 }{ phase :=  movingdown }
			\nItem{ act2 }{ button:= DOWN }
			\nItem{ act3 }{ i:= R }
			\nItem{ act4 }{ l:= R }
			\end{description}
		\EndAct
		\end{description}
	\EVT {retracting\_gears}
	\REF {retracting\_gears}
		\begin{description}
		\WhenGrd
			\begin{description}
			\nItem{ grd1 }{ dstate[DOORS]=\{ OPEN\}  }
			\nItem{ grd2 }{ gstate[GEARS]=\{ EXTENDED\}  }
			\nItem{ grd3 }{ p=R }
			\nItem{ grd6 }{ gear\_extended = \{ a\mapsto b |  a \in  1\upto 3 \land  b \in  GEARS \tfun  \{ TRUE\} \}   }
			\nItem{ grd7 }{ gear\_retracted = \{ a\mapsto b |  a \in  1\upto 3 \land  b \in  GEARS \tfun  \{ FALSE\} \}   }
			\nItem{ grd8 }{ gear\_shock\_absorber = \{ a\mapsto b |  a \in  1\upto 3 \land  b = ground\}   }
			\nItem{ grd9 }{ \forall x\qdot x\in 1\upto 3 \limp  handle(x)=button }
			\end{description}
		\ThenAct
			\begin{description}
			\nItem{ act1 }{ gstate :=  \{ a\mapsto b| a \in  GEARS \land  b=RETRACTING\}  }
			\nItem{ act2 }{ gear\_extended :\in  1\upto 3 \tfun  (GEARS \tfun  \{ FALSE\} )  }
			\nItem{ act3 }{ gear\_shock\_absorber :=  \{ a\mapsto b |  a \in  1\upto 3 \land  b = flight\}   }
			\end{description}
		\EndAct
		\end{description}
	\EVT {retraction}
	\REF {retraction}
		\begin{description}
		\WhenGrd
			\begin{description}
			\nItem{ grd1 }{ dstate[DOORS]=\{ OPEN\}  }
			\nItem{ grd2 }{ gstate[GEARS]=\{ RETRACTING\}  }
			\nItem{ grd4 }{ gear\_extended = \{ a\mapsto b |  a \in  1\upto 3 \land  b \in  GEARS \tfun  \{ FALSE\} \}   }
			\nItem{ grd5 }{ gear\_retracted = \{ a\mapsto b |  a \in  1\upto 3 \land  b \in  GEARS \tfun  \{ FALSE\} \}   }
			\nItem{ grd6 }{ gear\_shock\_absorber = \{ a\mapsto b |  a \in  1\upto 3 \land  b = flight\}   }
			\nItem{ grd7 }{ \forall x\qdot x\in 1\upto 3 \limp  handle(x)=button }
			\end{description}
		\ThenAct
			\begin{description}
			\nItem{ act1 }{ gstate:=   \{ a\mapsto b|  a \in  GEARS \land  b= RETRACTED\}  }
			\nItem{ act2 }{ gear\_retracted :\in  1\upto 3 \tfun  (GEARS \tfun  \{ TRUE\} )  }
			\end{description}
		\EndAct
		\end{description}
	\EVT {extending\_gears}
	\REF {extending\_gears}
		\begin{description}
		\WhenGrd
			\begin{description}
			\nItem{ grd1 }{ dstate[DOORS]=\{ OPEN\}  }
			\nItem{ grd2 }{ gstate[GEARS]=\{ RETRACTED\}  }
			\nItem{ grd3 }{ p=E }
			\nItem{ grd5 }{ gear\_retracted = \{ a\mapsto b |  a \in  1\upto 3 \land  b \in  GEARS \tfun  \{ TRUE\} \}   }
			\nItem{ grd6 }{ gear\_extended = \{ a\mapsto b |  a \in  1\upto 3 \land  b \in  GEARS \tfun  \{ FALSE\} \}   }
			\nItem{ grd7 }{ gear\_shock\_absorber = \{ a\mapsto b |  a \in  1\upto 3 \land  b = flight\}   }
			\nItem{ grd8 }{ \forall x\qdot x\in 1\upto 3 \limp  handle(x)=button }
			\end{description}
		\ThenAct
			\begin{description}
			\nItem{ act1 }{ gstate :=  \{ a\mapsto b| a \in  GEARS \land  b=EXTENDING\}  }
			\nItem{ act2 }{ gear\_retracted :\in  1\upto 3 \tfun  (GEARS \tfun  \{ FALSE\} ) }
			\end{description}
		\EndAct
		\end{description}
	\EVT {extension}
	\REF {extension}
		\begin{description}
		\WhenGrd
			\begin{description}
			\nItem{ grd1 }{ dstate[DOORS]=\{ OPEN\}  }
			\nItem{ grd2 }{ gstate[GEARS]=\{ EXTENDING\}  }
			\nItem{ grd4 }{ gear\_retracted = \{ a\mapsto b |  a \in  1\upto 3 \land  b \in  GEARS \tfun  \{ FALSE\} \}   }
			\nItem{ grd5 }{ gear\_extended = \{ a\mapsto b |  a \in  1\upto 3 \land  b \in  GEARS \tfun  \{ FALSE\} \}   }
			\nItem{ grd6 }{ gear\_shock\_absorber = \{ a\mapsto b |  a \in  1\upto 3 \land  b = flight\}   }
			\nItem{ grd7 }{ \forall x\qdot x\in 1\upto 3 \limp  handle(x)=button }
			\end{description}
		\ThenAct
			\begin{description}
			\nItem{ act1 }{ gstate :=  \{ a\mapsto b|  a \in  GEARS \land  b=EXTENDED\}  }
			\nItem{ act2 }{ gear\_extended :\in  1\upto 3 \tfun  (GEARS \tfun  \{ TRUE\} )  }
			\nItem{ act3 }{ gear\_shock\_absorber :=  \{ a\mapsto b |  a \in  1\upto 3 \land  b = ground\}   }
			\end{description}
		\EndAct
		\end{description}
	\EVT {HPD1}
		\begin{description}
		\WhenGrd
			\begin{description}
			\nItem{ grd3 }{ \forall x\qdot x\in 1\upto 3 \limp  handle(x)=UP }
			\end{description}
		\ThenAct
			\begin{description}
			\nItem{ act2 }{ handle :\in  1\upto 3 \tfun  \{ DOWN\}  }
			\end{description}
		\EndAct
		\end{description}
	\EVT {HPU1}
		\begin{description}
		\WhenGrd
			\begin{description}
			\nItem{ grd3 }{ \forall x\qdot x\in 1\upto 3 \limp  handle(x)=DOWN }
			\end{description}
		\ThenAct
			\begin{description}
			\nItem{ act2 }{ handle :\in  1\upto 3 \tfun  \{ UP\}  }
			\end{description}
		\EndAct
		\end{description}
	\EVT {Analogical\_switch\_closed}
		\begin{description}
		\AnyPrm
			\begin{description}
			\ItemY{in }{in port }
			\end{description}
		\WhereGrd
			\begin{description}
			\nItem{ grd1 }{ in = general\_EV }
			\nItem{ grd2 }{ \forall x\qdot x\in 1\upto 3 \limp  (handle(x)=UP \lor  handle(x)=DOWN) }
			\end{description}
		\ThenAct
			\begin{description}
			\nItem{ act3 }{ analogical\_switch :\in  1\upto 3 \tfun  \{ closed\}  }
			\nItem{ act4 }{ A\_Switch\_Out :=  TRUE }
			\end{description}
		\EndAct
		\end{description}
	\EVT {Analogical\_switch\_open}
		\begin{description}
		\AnyPrm
			\begin{description}
			\ItemY{in }{in port }
			\end{description}
		\WhereGrd
			\begin{description}
			\nItem{ grd1 }{ in = general\_EV }
			\nItem{ grd2 }{ \forall x\qdot x\in 1\upto 3 \limp  (handle(x)=UP \lor  handle(x)=DOWN) }
			\end{description}
		\ThenAct
			\begin{description}
			\nItem{ act3 }{ analogical\_switch :\in  1\upto 3 \tfun  \{ open\}  }
			\nItem{ act4 }{ A\_Switch\_Out :=  FALSE }
			\end{description}
		\EndAct
		\end{description}
	\EVT {Circuit\_pressurized}
		\begin{description}
		\BeginAct
			\begin{description}
			\nItem{ act9 }{ circuit\_pressurized :\in  1\upto 3 \tfun  BOOL }
			\end{description}
		\EndAct
		\end{description}
	\EVT {Computing\_Module\_1\_2}
		\begin{description}
		\BeginAct
			\begin{description}
			\nItem{ act1 }{ general\_EV :=  general\_EV\_func(handle \mapsto  analogical\_switch \mapsto  gear\_extended \mapsto  gear\_retracted \mapsto  gear\_shock\_absorber \mapsto  door\_open \mapsto  door\_closed \mapsto  circuit\_pressurized) }
			\nItem{ act2 }{ close\_EV :=  close\_EV\_func(handle \mapsto  analogical\_switch \mapsto  gear\_extended \mapsto  gear\_retracted \mapsto  gear\_shock\_absorber \mapsto  door\_open \mapsto  door\_closed \mapsto  circuit\_pressurized) }
			\nItem{ act3 }{ retract\_EV :=  retract\_EV\_func(handle \mapsto  analogical\_switch \mapsto  gear\_extended \mapsto  gear\_retracted \mapsto  gear\_shock\_absorber \mapsto  door\_open \mapsto  door\_closed \mapsto  circuit\_pressurized) }
			\nItem{ act4 }{ extend\_EV :=  extend\_EV\_func(handle \mapsto  analogical\_switch \mapsto  gear\_extended \mapsto  gear\_retracted \mapsto  gear\_shock\_absorber \mapsto  door\_open \mapsto  door\_closed \mapsto  circuit\_pressurized) }
			\nItem{ act5 }{ open\_EV :=  open\_EV\_func(handle \mapsto  analogical\_switch \mapsto  gear\_extended \mapsto  gear\_retracted \mapsto  gear\_shock\_absorber \mapsto  door\_open \mapsto  door\_closed \mapsto  circuit\_pressurized) }
			\nItem{ act6 }{ gears\_locked\_down :=  gears\_locked\_down\_func(handle \mapsto  analogical\_switch \mapsto  gear\_extended \mapsto  gear\_retracted \mapsto  gear\_shock\_absorber \mapsto  door\_open \mapsto  door\_closed \mapsto  circuit\_pressurized) }
			\nItem{ act7 }{ gears\_man :=  gears\_man\_func(handle \mapsto  analogical\_switch \mapsto  gear\_extended \mapsto  gear\_retracted \mapsto  gear\_shock\_absorber \mapsto  door\_open \mapsto  door\_closed \mapsto  circuit\_pressurized) }
			\nItem{ act8 }{ anomaly :=  anomaly\_func(handle \mapsto  analogical\_switch \mapsto  gear\_extended \mapsto  gear\_retracted \mapsto  gear\_shock\_absorber \mapsto  door\_open \mapsto  door\_closed \mapsto  circuit\_pressurized) }
			\end{description}
		\EndAct
		\end{description}
	\EVT {Failure\_Detection}
		\begin{description}
		\BeginAct
			\begin{description}
			\nItem{ act1 }{ anomaly :=  TRUE }
			\end{description}
		\EndAct
		\end{description}
\END
\end{description}

\section{M5}
\label{sec:M5}

\begin{description}
\BTitle{M5}{27Jan2014}{10:44:59 AM}
\MACHINE{M5}\cmt{		\\\hspace*{1 cm}  Hydraulic circuit output for Electro-valves. }
\REFINES{M4}
\SEES{C1}
\VARIABLES
	\begin{description}
		\Item{ dstate }
		\Item{ lstate }
		\Item{ phase }
		\Item{ button }
		\Item{ p }
		\Item{ l }
		\Item{ i }
		\Item{ gstate }
		\Item{ handle }
		\Item{ analogical\_switch }
		\Item{ gear\_extended }
		\Item{ gear\_retracted }
		\Item{ gear\_shock\_absorber }
		\Item{ door\_closed }
		\Item{ door\_open }
		\Item{ circuit\_pressurized }
		\Item{ general\_EV }
		\Item{ close\_EV }
		\Item{ retract\_EV }
		\Item{ extend\_EV }
		\Item{ open\_EV }
		\Item{ gears\_locked\_down }
		\Item{ gears\_man }
		\Item{ anomaly }
		\Item{ general\_EV\_func }
		\Item{ close\_EV\_func }
		\Item{ retract\_EV\_func }
		\Item{ extend\_EV\_func }
		\Item{ open\_EV\_func }
		\Item{ gears\_locked\_down\_func }
		\Item{ gears\_man\_func }
		\Item{ anomaly\_func }
		\Item{ general\_EV\_Hout }
		\Item{ close\_EV\_Hout }
		\Item{ retract\_EV\_Hout }
		\Item{ extend\_EV\_Hout }
		\Item{ open\_EV\_Hout }
		\Item{ A\_Switch\_Out }
	\end{description}
\INVARIANTS
	\begin{description}
		\nItem{ inv1 }{ general\_EV\_Hout \in  \{ 0, Hin\}  }
		\nItem{ inv2 }{ close\_EV\_Hout \in  \{ 0, Hin\}  }
		\nItem{ inv3 }{ retract\_EV\_Hout \in  \{ 0, Hin\}  }
		\nItem{ inv4 }{ extend\_EV\_Hout \in  \{ 0, Hin\}  }
		\nItem{ inv5 }{ open\_EV\_Hout \in  \{ 0, Hin\}  }
	\end{description}
\EVENTS
	\INITIALISATION
		\\\textit{extended}
		\begin{description}
		\BeginAct
			\begin{description}
			\nItemX{ act1 }{ button :=  DOWN }
			\nItemX{ act2 }{ phase :=  haltdown }
			\nItemXY{ act3 }{ dstate :|  (dstate'\in  DOORS \tfun  SDOORS \land  dstate'=\{ a\mapsto b|  a \in  DOORS \land  b=CLOSED\} ) }{ 		\\\hspace*{1,4 cm}  missing elements of the invariant }
			\nItemX{ act4 }{ lstate :=  \{ a\mapsto b| a\in DOORS\land  b=LOCKED\}  }
			\nItemX{ act5 }{ p :=  R }
			\nItemX{ act6 }{ l :=  R }
			\nItemX{ act7 }{ i :=  R }
			\nItemX{ act8 }{ gstate :|  (gstate' \in  GEARS \tfun  SGEARS \land  gstate'=\{ a\mapsto b |  a \in  GEARS  \land  b=EXTENDED\} ) }
			\nItemX{ act14 }{ handle :\in  1\upto 3 \tfun  \{ DOWN\}  }
			\nItemX{ act15 }{ analogical\_switch :\in  1\upto 3 \tfun  \{ open\}  }
			\nItemX{ act16 }{ gear\_extended :\in  1\upto 3 \tfun  (GEARS \tfun  \{ TRUE\} ) }
			\nItemX{ act17 }{ gear\_retracted :\in  1\upto 3 \tfun  (GEARS \tfun  \{ FALSE\} ) }
			\nItemX{ act18 }{ gear\_shock\_absorber :\in  1\upto 3 \tfun  \{ ground\}  }
			\nItemX{ act19 }{ door\_closed :\in  1\upto 3 \tfun  (DOORS \tfun  \{ TRUE\} ) }
			\nItemX{ act20 }{ door\_open :\in  1\upto 3 \tfun  (DOORS \tfun  \{ FALSE\} ) }
			\nItemX{ act21 }{ circuit\_pressurized :\in  1\upto 3 \tfun  \{ FALSE\}  }
			\nItemX{ act22 }{ general\_EV :=  FALSE }
			\nItemX{ act23 }{ close\_EV :=  TRUE }
			\nItemX{ act24 }{ retract\_EV :=  FALSE }
			\nItemX{ act25 }{ extend\_EV :=  TRUE }
			\nItemX{ act27 }{ open\_EV :=  FALSE }
			\nItemX{ act28 }{ gears\_locked\_down :=  TRUE }
			\nItemX{ act29 }{ gears\_man :=  FALSE }
			\nItemX{ act30 }{ anomaly :=  FALSE }
			\nItemX{ act31 }{ general\_EV\_func :\in  (1\upto 3 \tfun  POSITIONS) \cprod  (1\upto 3 \tfun  A\_Switch) \cprod  (1\upto 3 \tfun  (GEARS \tfun  BOOL)) \cprod  (1\upto 3 \tfun  (GEARS \tfun  BOOL)) \cprod  (1\upto 3 \tfun  GEAR\_ABSORBER) \cprod  (1\upto 3 \tfun  (DOORS \tfun  BOOL)) \cprod  (1\upto 3 \tfun  (DOORS \tfun  BOOL)) \cprod  (1\upto 3 \tfun  BOOL) \tfun  BOOL }
			\nItemX{ act32 }{ close\_EV\_func :\in  (1\upto 3 \tfun  POSITIONS) \cprod  (1\upto 3 \tfun  A\_Switch) \cprod  (1\upto 3 \tfun  (GEARS \tfun  BOOL)) \cprod  (1\upto 3 \tfun  (GEARS \tfun  BOOL)) \cprod  (1\upto 3 \tfun  GEAR\_ABSORBER) \cprod  (1\upto 3 \tfun  (DOORS \tfun  BOOL)) \cprod  (1\upto 3 \tfun  (DOORS \tfun  BOOL)) \cprod  (1\upto 3 \tfun  BOOL) \tfun  BOOL }
			\nItemX{ act33 }{ retract\_EV\_func :\in  (1\upto 3 \tfun  POSITIONS) \cprod  (1\upto 3 \tfun  A\_Switch) \cprod  (1\upto 3 \tfun  (GEARS \tfun  BOOL)) \cprod  (1\upto 3 \tfun  (GEARS \tfun  BOOL)) \cprod  (1\upto 3 \tfun  GEAR\_ABSORBER) \cprod  (1\upto 3 \tfun  (DOORS \tfun  BOOL)) \cprod  (1\upto 3 \tfun  (DOORS \tfun  BOOL)) \cprod  (1\upto 3 \tfun  BOOL) \tfun  BOOL }
			\nItemX{ act34 }{ extend\_EV\_func :\in  (1\upto 3 \tfun  POSITIONS) \cprod  (1\upto 3 \tfun  A\_Switch) \cprod  (1\upto 3 \tfun  (GEARS \tfun  BOOL)) \cprod  (1\upto 3 \tfun  (GEARS \tfun  BOOL)) \cprod  (1\upto 3 \tfun  GEAR\_ABSORBER) \cprod  (1\upto 3 \tfun  (DOORS \tfun  BOOL)) \cprod  (1\upto 3 \tfun  (DOORS \tfun  BOOL)) \cprod  (1\upto 3 \tfun  BOOL) \tfun  BOOL }
			\nItemX{ act35 }{ open\_EV\_func :\in  (1\upto 3 \tfun  POSITIONS) \cprod  (1\upto 3 \tfun  A\_Switch) \cprod  (1\upto 3 \tfun  (GEARS \tfun  BOOL)) \cprod  (1\upto 3 \tfun  (GEARS \tfun  BOOL)) \cprod  (1\upto 3 \tfun  GEAR\_ABSORBER) \cprod  (1\upto 3 \tfun  (DOORS \tfun  BOOL)) \cprod  (1\upto 3 \tfun  (DOORS \tfun  BOOL)) \cprod  (1\upto 3 \tfun  BOOL) \tfun  BOOL }
			\nItemX{ act36 }{ gears\_locked\_down\_func :\in  (1\upto 3 \tfun  POSITIONS) \cprod  (1\upto 3 \tfun  A\_Switch) \cprod  (1\upto 3 \tfun  (GEARS \tfun  BOOL)) \cprod  (1\upto 3 \tfun  (GEARS \tfun  BOOL)) \cprod  (1\upto 3 \tfun  GEAR\_ABSORBER) \cprod  (1\upto 3 \tfun  (DOORS \tfun  BOOL)) \cprod  (1\upto 3 \tfun  (DOORS \tfun  BOOL)) \cprod  (1\upto 3 \tfun  BOOL) \tfun  BOOL }
			\nItemX{ act37 }{ gears\_man\_func :\in  (1\upto 3 \tfun  POSITIONS) \cprod  (1\upto 3 \tfun  A\_Switch) \cprod  (1\upto 3 \tfun  (GEARS \tfun  BOOL)) \cprod  (1\upto 3 \tfun  (GEARS \tfun  BOOL)) \cprod  (1\upto 3 \tfun  GEAR\_ABSORBER) \cprod  (1\upto 3 \tfun  (DOORS \tfun  BOOL)) \cprod  (1\upto 3 \tfun  (DOORS \tfun  BOOL)) \cprod  (1\upto 3 \tfun  BOOL) \tfun  BOOL }
			\nItemX{ act38 }{ anomaly\_func :\in  (1\upto 3 \tfun  POSITIONS) \cprod  (1\upto 3 \tfun  A\_Switch) \cprod  (1\upto 3 \tfun  (GEARS \tfun  BOOL)) \cprod  (1\upto 3 \tfun  (GEARS \tfun  BOOL)) \cprod  (1\upto 3 \tfun  GEAR\_ABSORBER) \cprod  (1\upto 3 \tfun  (DOORS \tfun  BOOL)) \cprod  (1\upto 3 \tfun  (DOORS \tfun  BOOL)) \cprod  (1\upto 3 \tfun  BOOL) \tfun  BOOL }
			\nItemX{ act39 }{ A\_Switch\_Out :=  FALSE }
			\nItem{ act40 }{ close\_EV\_Hout :=  0 }
			\nItem{ act41 }{ retract\_EV\_Hout :=  0 }
			\nItem{ act42 }{ extend\_EV\_Hout :=  0 }
			\nItem{ act43 }{ open\_EV\_Hout :=  0 }
			\nItem{ act44 }{ general\_EV\_Hout :=  0 }
			\end{description}
		\EndAct
		\end{description}
	\EVT {opening\_doors\_DOWN}
	\EXTD {opening\_doors\_DOWN}
		\begin{description}
		\WhenGrd
			\begin{description}
			\nItemX{ grd1 }{ dstate[DOORS]= \{ CLOSED\}  }
			\nItemX{ grd5 }{ lstate[DOORS]=\{ UNLOCKED\}  }
			\nItemX{ grd7 }{ phase=movingdown }
			\nItemX{ grd8 }{ p=R }
			\nItemX{ grd9 }{ l=R }
			\nItemX{ grd10 }{ door\_open = \{ a\mapsto b |  a \in  1\upto 3 \land  b \in  DOORS \tfun  \{ FALSE\} \}   }
			\nItemX{ grd11 }{ door\_closed = \{ a\mapsto b |  a \in  1\upto 3 \land  b \in  DOORS \tfun  \{ FALSE\} \}   }
			\nItemX{ grd12 }{ \forall x\qdot x\in 1\upto 3 \limp  handle(x)=button }
			\end{description}
		\ThenAct
			\begin{description}
			\nItemX{ act1 }{ dstate :=  \{ a\mapsto b|  a \in  DOORS \land  b=OPEN\}  }
			\nItemX{ act2 }{ p:= E }
			\nItemX{ act3 }{ door\_open :\in  1\upto 3 \tfun  (DOORS \tfun  \{ TRUE\} ) }
			\end{description}
		\EndAct
		\end{description}
	\EVT {opening\_doors\_UP}
	\EXTD {opening\_doors\_UP}
		\begin{description}
		\WhenGrd
			\begin{description}
			\nItemX{ grd1 }{ dstate[DOORS]= \{ CLOSED\}  }
			\nItemX{ grd4 }{ lstate[DOORS]=\{ UNLOCKED\}  }
			\nItemX{ grd5 }{ phase= movingup }
			\nItemX{ grd6 }{ p=E }
			\nItemX{ grd7 }{ l=E }
			\nItemX{ grd8 }{ door\_open = \{ a\mapsto b |  a \in  1\upto 3 \land  b \in  DOORS \tfun  \{ FALSE\} \}   }
			\nItemX{ grd9 }{ door\_closed = \{ a\mapsto b |  a \in  1\upto 3 \land  b \in  DOORS \tfun  \{ FALSE\} \}   }
			\nItemX{ grd10 }{ \forall x\qdot x\in 1\upto 3 \limp  handle(x)=button }
			\end{description}
		\ThenAct
			\begin{description}
			\nItemX{ act1 }{ dstate :=  \{ a\mapsto b|  a \in  DOORS \land  b=OPEN\}  }
			\nItemX{ act2 }{ p:= R }
			\nItemX{ act3 }{ door\_open:\in  1\upto 3 \tfun  (DOORS \tfun  \{ TRUE\} )  }
			\end{description}
		\EndAct
		\end{description}
	\EVT {closing\_doors\_UP}
	\EXTD {closing\_doors\_UP}
		\begin{description}
		\AnyPrm
			\begin{description}
			\ItemX{f }
			\end{description}
		\WhereGrd
			\begin{description}
			\nItemX{ grd1 }{ dstate[DOORS]=\{ OPEN\}  }
			\nItemX{ grd3 }{ f \in  DOORS \tfun  SDOORS }
			\nItemX{ grd4 }{ \forall e\qdot  e \in  DOORS \limp  f(e)=CLOSED }
			\nItemX{ grd5 }{ phase=movingup }
			\nItemX{ grd6 }{ p=R }
			\nItemX{ grd7 }{ gstate[GEARS]=\{ RETRACTED\}  }
			\nItemX{ grd8 }{ \forall x\qdot x\in 1\upto 3 \limp  handle(x)=button }
			\end{description}
		\ThenAct
			\begin{description}
			\nItemX{ act1 }{ dstate:= f }
			\end{description}
		\EndAct
		\end{description}
	\EVT {closing\_doors\_DOWN}
	\EXTD {closing\_doors\_DOWN}
		\begin{description}
		\AnyPrm
			\begin{description}
			\ItemX{f }
			\end{description}
		\WhereGrd
			\begin{description}
			\nItemX{ grd1 }{ dstate[DOORS]=\{ OPEN\}  }
			\nItemX{ grd3 }{ f \in  DOORS \tfun  SDOORS }
			\nItemX{ grd4 }{ \forall e\qdot  e \in  DOORS \limp  f(e)=CLOSED }
			\nItemX{ grd5 }{ phase=movingdown }
			\nItemX{ grd6 }{ p=E }
			\nItemX{ grd7 }{ gstate[GEARS]=\{ EXTENDED\}  }
			\nItemX{ grd8 }{ \forall x\qdot x\in 1\upto 3 \limp  handle(x)=button }
			\end{description}
		\ThenAct
			\begin{description}
			\nItemX{ act1 }{ dstate:= f }
			\end{description}
		\EndAct
		\end{description}
	\EVT {unlocking\_UP}
	\EXTD {unlocking\_UP}
		\begin{description}
		\WhenGrd
			\begin{description}
			\nItemX{ grd3 }{ lstate[DOORS]=\{ LOCKED\}  }
			\nItemX{ grd4 }{ phase=movingup }
			\nItemX{ grd5 }{ l=E }
			\nItemX{ grd6 }{ p=E }
			\nItemX{ grd7 }{ i=E }
			\nItemX{ grd8 }{ door\_open = \{ a\mapsto b |  a \in  1\upto 3 \land  b \in  DOORS \tfun  \{ FALSE\} \}   }
			\nItemX{ grd9 }{ door\_closed = \{ a\mapsto b |  a \in  1\upto 3 \land  b \in  DOORS \tfun  \{ TRUE\} \}   }
			\nItemX{ grd10 }{ \forall x\qdot x\in 1\upto 3 \limp  handle(x)=button }
			\end{description}
		\ThenAct
			\begin{description}
			\nItemX{ act1 }{ lstate:= \{ a\mapsto b| a\in DOORS \land  b=UNLOCKED\}  }
			\nItemX{ act2 }{ door\_closed :\in  1\upto 3 \tfun  (DOORS \tfun  \{ FALSE\} )  }
			\end{description}
		\EndAct
		\end{description}
	\EVT {locking\_UP}
	\EXTD {locking\_UP}
		\begin{description}
		\WhenGrd
			\begin{description}
			\nItemX{ grd3 }{ dstate[DOORS]=\{ CLOSED\}  }
			\nItemX{ grd4 }{ phase=movingup }
			\nItemX{ grd5 }{ lstate[DOORS]=\{ UNLOCKED\}  }
			\nItemX{ grd6 }{ p=R }
			\nItemX{ grd7 }{ l=E }
			\nItemX{ grd9 }{ door\_open = \{ a\mapsto b |  a \in  1\upto 3 \land  b \in  DOORS \tfun  \{ FALSE\} \}   }
			\nItemX{ grd10 }{ door\_closed = \{ a\mapsto b |  a \in  1\upto 3 \land  b \in  DOORS \tfun  \{ FALSE\} \}   }
			\nItemX{ grd11 }{ \forall x\qdot x\in 1\upto 3 \limp  handle(x)=button }
			\end{description}
		\ThenAct
			\begin{description}
			\nItemX{ act1 }{ lstate:= \{ a\mapsto b| a\in DOORS \land  b=LOCKED\}  }
			\nItemX{ act3 }{ phase:= haltup }
			\nItemXY{ act4 }{ l:= R }{ 		\\\hspace*{1,4 cm}  added by D Mery }
			\nItemX{ act44 }{ door\_closed :\in  1\upto 3 \tfun  (DOORS \tfun  \{ TRUE\} )  }
			\end{description}
		\EndAct
		\end{description}
	\EVT {unlocking\_DOWN}
	\EXTD {unlocking\_DOWN}
		\begin{description}
		\WhenGrd
			\begin{description}
			\nItemX{ grd3 }{ lstate[DOORS]=\{ LOCKED\}  }
			\nItemX{ grd4 }{ phase=movingdown }
			\nItemX{ grd5 }{ l=R }
			\nItemX{ grd6 }{ p=R }
			\nItemX{ grd7 }{ i=R }
			\nItemX{ grd8 }{ door\_open = \{ a\mapsto b |  a \in  1\upto 3 \land  b \in  DOORS \tfun  \{ FALSE\} \}   }
			\nItemX{ grd9 }{ door\_closed = \{ a\mapsto b |  a \in  1\upto 3 \land  b \in  DOORS \tfun  \{ TRUE\} \}   }
			\nItemX{ grd10 }{ \forall x\qdot x\in 1\upto 3 \limp  handle(x)=button }
			\end{description}
		\ThenAct
			\begin{description}
			\nItemX{ act1 }{ lstate:= \{ a\mapsto b| a\in DOORS \land  b=UNLOCKED\}  }
			\nItemX{ act2 }{ door\_closed :\in  1\upto 3 \tfun  (DOORS \tfun  \{ FALSE\} ) }
			\end{description}
		\EndAct
		\end{description}
	\EVT {locking\_DOWN}
	\EXTD {locking\_DOWN}
		\begin{description}
		\WhenGrd
			\begin{description}
			\nItemX{ grd1 }{ dstate[DOORS]=\{ CLOSED\}  }
			\nItemX{ grd2 }{ phase=movingdown }
			\nItemX{ grd3 }{ lstate[DOORS]=\{ UNLOCKED\}  }
			\nItemX{ grd4 }{ p=E }
			\nItemX{ grd5 }{ l=R }
			\nItemX{ grd7 }{ door\_open = \{ a\mapsto b |  a \in  1\upto 3 \land  b \in  DOORS \tfun  \{ FALSE\} \}   }
			\nItemX{ grd8 }{ door\_closed = \{ a\mapsto b |  a \in  1\upto 3 \land  b \in  DOORS \tfun  \{ FALSE\} \}   }
			\nItemX{ grd9 }{ \forall x\qdot x\in 1\upto 3 \limp  handle(x)=button }
			\end{description}
		\ThenAct
			\begin{description}
			\nItemX{ act1 }{ lstate:= \{ a\mapsto b| a\in DOORS \land  b = LOCKED\}  }
			\nItemX{ act3 }{ phase:= haltdown }
			\nItemX{ act4 }{ l:= E }
			\nItemX{ act5 }{ door\_closed :\in  1\upto 3 \tfun  (DOORS \tfun  \{ TRUE\} )  }
			\end{description}
		\EndAct
		\end{description}
	\EVT {PD1}
	\EXTD {PD1}
		\begin{description}
		\WhenGrd
			\begin{description}
			\nItemX{ grd1 }{ button=UP }
			\nItemX{ grd2 }{ phase=haltup }
			\nItemX{ grd3 }{ \forall x\qdot x\in 1\upto 3 \limp  handle(x)=DOWN }
			\end{description}
		\ThenAct
			\begin{description}
			\nItemX{ act1 }{ phase:= movingdown }
			\nItemX{ act2 }{ button:= DOWN }
			\nItemX{ act3 }{ l:= R }
			\nItemX{ act4 }{ p:= R }
			\nItemX{ act5 }{ i:= R }
			\end{description}
		\EndAct
		\end{description}
	\EVT {PU1}
	\EXTD {PU1}
		\begin{description}
		\WhenGrd
			\begin{description}
			\nItemX{ grd1 }{ button=DOWN }
			\nItemX{ grd2 }{ phase=haltdown }
			\nItemX{ grd3 }{ \forall x\qdot x\in 1\upto 3 \limp  handle(x)=UP }
			\end{description}
		\ThenAct
			\begin{description}
			\nItemX{ act1 }{ phase:= movingup }
			\nItemX{ act2 }{ button:= UP }
			\nItemX{ act3 }{ l:= E }
			\nItemX{ act4 }{ p:= E }
			\nItemX{ act5 }{ i:= E }
			\end{description}
		\EndAct
		\end{description}
	\EVT {PU2}
	\EXTD {PU2}
		\begin{description}
		\WhenGrd
			\begin{description}
			\nItemX{ grd1 }{ l=R }
			\nItemX{ grd2 }{ p=R }
			\nItemX{ grd3 }{ phase=movingdown }
			\nItemX{ grd4 }{ button=DOWN }
			\nItemX{ grd5 }{ i=R }
			\nItemX{ grd6 }{ lstate[DOORS]=\{ LOCKED\}  }
			\nItemX{ grd7 }{ door\_open = \{ a\mapsto b |  a \in  1\upto 3 \land  b \in  DOORS \tfun  \{ FALSE\} \}   }
			\nItemX{ grd8 }{ door\_closed = \{ a\mapsto b |  a \in  1\upto 3 \land  b \in  DOORS \tfun  \{ TRUE\} \}   }
			\nItemX{ grd9 }{ \forall x\qdot x\in 1\upto 3 \limp  handle(x)=UP }
			\end{description}
		\ThenAct
			\begin{description}
			\nItemX{ act1 }{ phase:= movingup }
			\nItemX{ act4 }{ button:=   UP }
			\nItemX{ act5 }{ l:= E }
			\nItemX{ act6 }{ p:= E }
			\nItemX{ act7 }{ i:= R }
			\end{description}
		\EndAct
		\end{description}
	\EVT {CompletePU2}
	\EXTD {CompletePU2}
		\begin{description}
		\WhenGrd
			\begin{description}
			\nItemX{ grd1 }{ phase=movingup }
			\nItemX{ grd2 }{ button=UP }
			\nItemX{ grd3 }{ l=E }
			\nItemX{ grd4 }{ p=E }
			\nItemX{ grd5 }{ i=R }
			\end{description}
		\ThenAct
			\begin{description}
			\nItemX{ act1 }{ phase:= haltup }
			\end{description}
		\EndAct
		\end{description}
	\EVT {PU3}
	\EXTD {PU3}
		\begin{description}
		\WhenGrd
			\begin{description}
			\nItemX{ grd1 }{ dstate[DOORS]=\{ CLOSED\}  }
			\nItemX{ grd2 }{ lstate[DOORS]=\{ UNLOCKED\}  }
			\nItemX{ grd3 }{ phase = movingdown }
			\nItemX{ grd4 }{ p=R }
			\nItemX{ grd5 }{ l=R }
			\nItemX{ grd6 }{ button=DOWN }
			\nItemX{ grd7 }{ door\_open = \{ a\mapsto b |  a \in  1\upto 3 \land  b \in  DOORS \tfun  \{ FALSE\} \}   }
			\nItemX{ grd8 }{ door\_closed = \{ a\mapsto b |  a \in  1\upto 3 \land  b \in  DOORS \tfun  \{ FALSE\} \}   }
			\nItemX{ grd9 }{ \forall x\qdot x\in 1\upto 3 \limp  handle(x)=UP }
			\end{description}
		\ThenAct
			\begin{description}
			\nItemX{ act1 }{ phase:= movingup }
			\nItemX{ act2 }{ p:= R }
			\nItemX{ act3 }{ l:= E }
			\nItemX{ act4 }{ button:= UP }
			\end{description}
		\EndAct
		\end{description}
	\EVT {PU4}
	\EXTD {PU4}
		\begin{description}
		\WhenGrd
			\begin{description}
			\nItemX{ grd1 }{ dstate[DOORS]=\{ OPEN\}  }
			\nItemX{ grd2 }{ phase=movingdown }
			\nItemX{ grd3 }{ p=E }
			\nItemX{ grd4 }{ button=DOWN }
			\nItemX{ grd7 }{ \forall x\qdot x\in 1\upto 3 \limp  handle(x)=UP }
			\end{description}
		\ThenAct
			\begin{description}
			\nItemX{ act1 }{ phase:= movingup }
			\nItemX{ act2 }{ p:= R }
			\nItemX{ act3 }{ button:= UP }
			\nItemX{ act4 }{ i:= E }
			\nItemX{ act5 }{ l:= E }
			\end{description}
		\EndAct
		\end{description}
	\EVT {PU5}
	\EXTD {PU5}
		\begin{description}
		\WhenGrd
			\begin{description}
			\nItemX{ grd1 }{ dstate[DOORS]=\{ CLOSED\}  }
			\nItemX{ grd2 }{ phase=movingdown }
			\nItemX{ grd3 }{ p=E }
			\nItemX{ grd4 }{ button=DOWN }
			\nItemX{ grd5 }{ lstate[DOORS]=\{ UNLOCKED\}  }
			\nItemX{ grd6 }{ door\_open = \{ a\mapsto b |  a \in  1\upto 3 \land  b \in  DOORS \tfun  \{ FALSE\} \}   }
			\nItemX{ grd7 }{ door\_closed = \{ a\mapsto b |  a \in  1\upto 3 \land  b \in  DOORS \tfun  \{ FALSE\} \}   }
			\nItemX{ grd8 }{ \forall x\qdot x\in 1\upto 3 \limp  handle(x)=UP }
			\end{description}
		\ThenAct
			\begin{description}
			\nItemX{ act1 }{ phase:= movingup }
			\nItemX{ act3 }{ button:= UP }
			\nItemX{ act4 }{ i:= E }
			\nItemX{ act5 }{ l:= E }
			\end{description}
		\EndAct
		\end{description}
	\EVT {PD2}
	\EXTD {PD2}
		\begin{description}
		\WhenGrd
			\begin{description}
			\nItemX{ grd1 }{ l=E }
			\nItemX{ grd2 }{ p=E }
			\nItemX{ grd3 }{ phase=movingup }
			\nItemX{ grd4 }{ i=E }
			\nItemX{ grd5 }{ lstate[DOORS]=\{ LOCKED\}  }
			\nItemX{ grd6 }{ \forall x\qdot x\in 1\upto 3 \limp  handle(x)=DOWN }
			\end{description}
		\ThenAct
			\begin{description}
			\nItemX{ act1 }{ phase:= movingdown }
			\nItemX{ act2 }{ button:= DOWN }
			\nItemX{ act3 }{ l:= R }
			\nItemX{ act4 }{ p:= R }
			\nItemX{ act5 }{ i:= E }
			\end{description}
		\EndAct
		\end{description}
	\EVT {CompletePD2}
	\EXTD {CompletePD2}
		\begin{description}
		\WhenGrd
			\begin{description}
			\nItemX{ grd1 }{ phase=movingdown }
			\nItemX{ grd2 }{ button=DOWN }
			\nItemX{ grd3 }{ l=R }
			\nItemX{ grd4 }{ p=R }
			\nItemX{ grd5 }{ i=E }
			\end{description}
		\ThenAct
			\begin{description}
			\nItemX{ act1 }{ phase:= haltdown }
			\end{description}
		\EndAct
		\end{description}
	\EVT {PD3}
	\EXTD {PD3}
		\begin{description}
		\WhenGrd
			\begin{description}
			\nItemX{ grd1 }{ dstate[DOORS]=\{ CLOSED\}  }
			\nItemX{ grd2 }{ lstate[DOORS]=\{ UNLOCKED\}  }
			\nItemX{ grd3 }{ phase=movingup }
			\nItemX{ grd4 }{ p=E }
			\nItemX{ grd5 }{ l=E }
			\nItemX{ grd6 }{ button=UP }
			\nItemX{ grd7 }{ door\_open = \{ a\mapsto b |  a \in  1\upto 3 \land  b \in  DOORS \tfun  \{ FALSE\} \}   }
			\nItemX{ grd8 }{ door\_closed = \{ a\mapsto b |  a \in  1\upto 3 \land  b \in  DOORS \tfun  \{ FALSE\} \}   }
			\nItemX{ grd9 }{ \forall x\qdot x\in 1\upto 3 \limp  handle(x)=DOWN }
			\end{description}
		\ThenAct
			\begin{description}
			\nItemX{ act1 }{ phase:= movingdown }
			\nItemX{ act2 }{ p:= E }
			\nItemX{ act3 }{ l:= R }
			\nItemX{ act4 }{ button:= DOWN }
			\end{description}
		\EndAct
		\end{description}
	\EVT {PD4}
	\EXTD {PD4}
		\begin{description}
		\WhenGrd
			\begin{description}
			\nItemX{ grd1 }{ dstate[DOORS]=\{ OPEN\}  }
			\nItemX{ grd2 }{ phase=movingup }
			\nItemX{ grd3 }{ p=R }
			\nItemX{ grd4 }{ button=UP }
			\nItemX{ grd6 }{ \forall x\qdot x\in 1\upto 3 \limp  handle(x)=DOWN }
			\end{description}
		\ThenAct
			\begin{description}
			\nItemX{ act1 }{ phase:= movingdown }
			\nItemX{ act2 }{ p:= E }
			\nItemX{ act3 }{ button:= DOWN }
			\nItemX{ act4 }{ i:= R }
			\nItemX{ act5 }{ l:= R }
			\end{description}
		\EndAct
		\end{description}
	\EVT {PD5}
	\EXTD {PD5}
		\begin{description}
		\WhenGrd
			\begin{description}
			\nItemX{ grd1 }{ dstate[DOORS]=\{ CLOSED\}  }
			\nItemX{ grd2 }{ phase=movingup }
			\nItemX{ grd3 }{ p=R }
			\nItemX{ grd4 }{ button=UP }
			\nItemX{ grd5 }{ lstate[DOORS]=\{ UNLOCKED\}  }
			\nItemX{ grd6 }{ door\_open = \{ a\mapsto b |  a \in  1\upto 3 \land  b \in  DOORS \tfun  \{ FALSE\} \}   }
			\nItemX{ grd7 }{ door\_closed = \{ a\mapsto b |  a \in  1\upto 3 \land  b \in  DOORS \tfun  \{ FALSE\} \}   }
			\nItemX{ grd8 }{ \forall x\qdot x\in 1\upto 3 \limp  handle(x)=DOWN }
			\end{description}
		\ThenAct
			\begin{description}
			\nItemX{ act1 }{ phase :=  movingdown }
			\nItemX{ act2 }{ button:= DOWN }
			\nItemX{ act3 }{ i:= R }
			\nItemX{ act4 }{ l:= R }
			\end{description}
		\EndAct
		\end{description}
	\EVT {retracting\_gears}
	\EXTD {retracting\_gears}
		\begin{description}
		\WhenGrd
			\begin{description}
			\nItemX{ grd1 }{ dstate[DOORS]=\{ OPEN\}  }
			\nItemX{ grd2 }{ gstate[GEARS]=\{ EXTENDED\}  }
			\nItemX{ grd3 }{ p=R }
			\nItemX{ grd6 }{ gear\_extended = \{ a\mapsto b |  a \in  1\upto 3 \land  b \in  GEARS \tfun  \{ TRUE\} \}   }
			\nItemX{ grd7 }{ gear\_retracted = \{ a\mapsto b |  a \in  1\upto 3 \land  b \in  GEARS \tfun  \{ FALSE\} \}   }
			\nItemX{ grd8 }{ gear\_shock\_absorber = \{ a\mapsto b |  a \in  1\upto 3 \land  b = ground\}   }
			\nItemX{ grd9 }{ \forall x\qdot x\in 1\upto 3 \limp  handle(x)=button }
			\end{description}
		\ThenAct
			\begin{description}
			\nItemX{ act1 }{ gstate :=  \{ a\mapsto b| a \in  GEARS \land  b=RETRACTING\}  }
			\nItemX{ act2 }{ gear\_extended :\in  1\upto 3 \tfun  (GEARS \tfun  \{ FALSE\} )  }
			\nItemX{ act3 }{ gear\_shock\_absorber :=  \{ a\mapsto b |  a \in  1\upto 3 \land  b = flight\}   }
			\end{description}
		\EndAct
		\end{description}
	\EVT {retraction}
	\EXTD {retraction}
		\begin{description}
		\WhenGrd
			\begin{description}
			\nItemX{ grd1 }{ dstate[DOORS]=\{ OPEN\}  }
			\nItemX{ grd2 }{ gstate[GEARS]=\{ RETRACTING\}  }
			\nItemX{ grd4 }{ gear\_extended = \{ a\mapsto b |  a \in  1\upto 3 \land  b \in  GEARS \tfun  \{ FALSE\} \}   }
			\nItemX{ grd5 }{ gear\_retracted = \{ a\mapsto b |  a \in  1\upto 3 \land  b \in  GEARS \tfun  \{ FALSE\} \}   }
			\nItemX{ grd6 }{ gear\_shock\_absorber = \{ a\mapsto b |  a \in  1\upto 3 \land  b = flight\}   }
			\nItemX{ grd7 }{ \forall x\qdot x\in 1\upto 3 \limp  handle(x)=button }
			\end{description}
		\ThenAct
			\begin{description}
			\nItemX{ act1 }{ gstate:=   \{ a\mapsto b|  a \in  GEARS \land  b= RETRACTED\}  }
			\nItemX{ act2 }{ gear\_retracted :\in  1\upto 3 \tfun  (GEARS \tfun  \{ TRUE\} )  }
			\end{description}
		\EndAct
		\end{description}
	\EVT {extending\_gears}
	\EXTD {extending\_gears}
		\begin{description}
		\WhenGrd
			\begin{description}
			\nItemX{ grd1 }{ dstate[DOORS]=\{ OPEN\}  }
			\nItemX{ grd2 }{ gstate[GEARS]=\{ RETRACTED\}  }
			\nItemX{ grd3 }{ p=E }
			\nItemX{ grd5 }{ gear\_retracted = \{ a\mapsto b |  a \in  1\upto 3 \land  b \in  GEARS \tfun  \{ TRUE\} \}   }
			\nItemX{ grd6 }{ gear\_extended = \{ a\mapsto b |  a \in  1\upto 3 \land  b \in  GEARS \tfun  \{ FALSE\} \}   }
			\nItemX{ grd7 }{ gear\_shock\_absorber = \{ a\mapsto b |  a \in  1\upto 3 \land  b = flight\}   }
			\nItemX{ grd8 }{ \forall x\qdot x\in 1\upto 3 \limp  handle(x)=button }
			\end{description}
		\ThenAct
			\begin{description}
			\nItemX{ act1 }{ gstate :=  \{ a\mapsto b| a \in  GEARS \land  b=EXTENDING\}  }
			\nItemX{ act2 }{ gear\_retracted :\in  1\upto 3 \tfun  (GEARS \tfun  \{ FALSE\} ) }
			\end{description}
		\EndAct
		\end{description}
	\EVT {extension}
	\EXTD {extension}
		\begin{description}
		\WhenGrd
			\begin{description}
			\nItemX{ grd1 }{ dstate[DOORS]=\{ OPEN\}  }
			\nItemX{ grd2 }{ gstate[GEARS]=\{ EXTENDING\}  }
			\nItemX{ grd4 }{ gear\_retracted = \{ a\mapsto b |  a \in  1\upto 3 \land  b \in  GEARS \tfun  \{ FALSE\} \}   }
			\nItemX{ grd5 }{ gear\_extended = \{ a\mapsto b |  a \in  1\upto 3 \land  b \in  GEARS \tfun  \{ FALSE\} \}   }
			\nItemX{ grd6 }{ gear\_shock\_absorber = \{ a\mapsto b |  a \in  1\upto 3 \land  b = flight\}   }
			\nItemX{ grd7 }{ \forall x\qdot x\in 1\upto 3 \limp  handle(x)=button }
			\end{description}
		\ThenAct
			\begin{description}
			\nItemX{ act1 }{ gstate :=  \{ a\mapsto b|  a \in  GEARS \land  b=EXTENDED\}  }
			\nItemX{ act2 }{ gear\_extended :\in  1\upto 3 \tfun  (GEARS \tfun  \{ TRUE\} )  }
			\nItemX{ act3 }{ gear\_shock\_absorber :=  \{ a\mapsto b |  a \in  1\upto 3 \land  b = ground\}   }
			\end{description}
		\EndAct
		\end{description}
	\EVT {HPD1}
	\EXTD {HPD1}
		\begin{description}
		\WhenGrd
			\begin{description}
			\nItemX{ grd3 }{ \forall x\qdot x\in 1\upto 3 \limp  handle(x)=UP }
			\end{description}
		\ThenAct
			\begin{description}
			\nItemX{ act2 }{ handle :\in  1\upto 3 \tfun  \{ DOWN\}  }
			\end{description}
		\EndAct
		\end{description}
	\EVT {HPU1}
	\EXTD {HPU1}
		\begin{description}
		\WhenGrd
			\begin{description}
			\nItemX{ grd3 }{ \forall x\qdot x\in 1\upto 3 \limp  handle(x)=DOWN }
			\end{description}
		\ThenAct
			\begin{description}
			\nItemX{ act2 }{ handle :\in  1\upto 3 \tfun  \{ UP\}  }
			\end{description}
		\EndAct
		\end{description}
	\EVT {Analogical\_switch\_closed}
	\EXTD {Analogical\_switch\_closed}
		\begin{description}
		\AnyPrm
			\begin{description}
			\ItemXY{in }{in port }
			\end{description}
		\WhereGrd
			\begin{description}
			\nItemX{ grd1 }{ in = general\_EV }
			\nItemX{ grd2 }{ \forall x\qdot x\in 1\upto 3 \limp  (handle(x)=UP \lor  handle(x)=DOWN) }
			\end{description}
		\ThenAct
			\begin{description}
			\nItemX{ act3 }{ analogical\_switch :\in  1\upto 3 \tfun  \{ closed\}  }
			\nItemX{ act4 }{ A\_Switch\_Out :=  TRUE }
			\end{description}
		\EndAct
		\end{description}
	\EVT {Analogical\_switch\_open}
	\EXTD {Analogical\_switch\_open}
		\begin{description}
		\AnyPrm
			\begin{description}
			\ItemXY{in }{in port }
			\end{description}
		\WhereGrd
			\begin{description}
			\nItemX{ grd1 }{ in = general\_EV }
			\nItemX{ grd2 }{ \forall x\qdot x\in 1\upto 3 \limp  (handle(x)=UP \lor  handle(x)=DOWN) }
			\end{description}
		\ThenAct
			\begin{description}
			\nItemX{ act3 }{ analogical\_switch :\in  1\upto 3 \tfun  \{ open\}  }
			\nItemX{ act4 }{ A\_Switch\_Out :=  FALSE }
			\end{description}
		\EndAct
		\end{description}
	\EVT {Circuit\_pressurized\_OK}
	\REF {Circuit\_pressurized}
		\begin{description}
		\WhenGrd
			\begin{description}
			\nItem{ grd1 }{ general\_EV\_Hout = Hin }
			\end{description}
		\ThenAct
			\begin{description}
			\nItem{ act9 }{ circuit\_pressurized :\in  1\upto 3 \tfun  \{ TRUE\}  }
			\end{description}
		\EndAct
		\end{description}
	\EVT {Circuit\_pressurized\_notOK}
	\REF {Circuit\_pressurized}
		\begin{description}
		\WhenGrd
			\begin{description}
			\nItem{ grd1 }{ general\_EV\_Hout = 0 }
			\end{description}
		\ThenAct
			\begin{description}
			\nItem{ act9 }{ circuit\_pressurized :\in  1\upto 3 \tfun  \{ FALSE\}  }
			\end{description}
		\EndAct
		\end{description}
	\EVT {Computing\_Module\_1\_2}
	\EXTD {Computing\_Module\_1\_2}
		\begin{description}
		\BeginAct
			\begin{description}
			\nItemX{ act1 }{ general\_EV :=  general\_EV\_func(handle \mapsto  analogical\_switch \mapsto  gear\_extended \mapsto  gear\_retracted \mapsto  gear\_shock\_absorber \mapsto  door\_open \mapsto  door\_closed \mapsto  circuit\_pressurized) }
			\nItemX{ act2 }{ close\_EV :=  close\_EV\_func(handle \mapsto  analogical\_switch \mapsto  gear\_extended \mapsto  gear\_retracted \mapsto  gear\_shock\_absorber \mapsto  door\_open \mapsto  door\_closed \mapsto  circuit\_pressurized) }
			\nItemX{ act3 }{ retract\_EV :=  retract\_EV\_func(handle \mapsto  analogical\_switch \mapsto  gear\_extended \mapsto  gear\_retracted \mapsto  gear\_shock\_absorber \mapsto  door\_open \mapsto  door\_closed \mapsto  circuit\_pressurized) }
			\nItemX{ act4 }{ extend\_EV :=  extend\_EV\_func(handle \mapsto  analogical\_switch \mapsto  gear\_extended \mapsto  gear\_retracted \mapsto  gear\_shock\_absorber \mapsto  door\_open \mapsto  door\_closed \mapsto  circuit\_pressurized) }
			\nItemX{ act5 }{ open\_EV :=  open\_EV\_func(handle \mapsto  analogical\_switch \mapsto  gear\_extended \mapsto  gear\_retracted \mapsto  gear\_shock\_absorber \mapsto  door\_open \mapsto  door\_closed \mapsto  circuit\_pressurized) }
			\nItemX{ act6 }{ gears\_locked\_down :=  gears\_locked\_down\_func(handle \mapsto  analogical\_switch \mapsto  gear\_extended \mapsto  gear\_retracted \mapsto  gear\_shock\_absorber \mapsto  door\_open \mapsto  door\_closed \mapsto  circuit\_pressurized) }
			\nItemX{ act7 }{ gears\_man :=  gears\_man\_func(handle \mapsto  analogical\_switch \mapsto  gear\_extended \mapsto  gear\_retracted \mapsto  gear\_shock\_absorber \mapsto  door\_open \mapsto  door\_closed \mapsto  circuit\_pressurized) }
			\nItemX{ act8 }{ anomaly :=  anomaly\_func(handle \mapsto  analogical\_switch \mapsto  gear\_extended \mapsto  gear\_retracted \mapsto  gear\_shock\_absorber \mapsto  door\_open \mapsto  door\_closed \mapsto  circuit\_pressurized) }
			\end{description}
		\EndAct
		\end{description}
	\EVT {Update\_Hout}\cmt{		\\\hspace*{2,8 cm}  Assign the value of Hout  }
		\begin{description}
		\BeginAct
			\begin{description}
			\nItemY{ act1 }{ general\_EV\_Hout :|  ((general\_EV = TRUE \land   general\_EV\_Hout' = Hin) \lor  (general\_EV = FALSE \land  general\_EV\_Hout' = 0)		\\\hspace*{1,2 cm}  \lor  (A\_Switch\_Out = TRUE \land   general\_EV\_Hout' = Hin) \lor  (A\_Switch\_Out = FALSE \land   general\_EV\_Hout' = 0)) }{ 		\\\hspace*{1,4 cm}  pass the current value of hydraulic input port (Hin) to hydraulic output port (Hout) }
			\nItem{ act2 }{ close\_EV\_Hout :|  ((close\_EV = TRUE \land  close\_EV\_Hout' = Hin) \lor  (close\_EV = FALSE \land  close\_EV\_Hout' = 0)) }
			\nItem{ act3 }{ open\_EV\_Hout :|  ((open\_EV = TRUE \land  open\_EV\_Hout' = Hin) \lor  (open\_EV = FALSE \land  open\_EV\_Hout' = 0)) }
			\nItem{ act4 }{ extend\_EV\_Hout :|  ((extend\_EV = TRUE \land  extend\_EV\_Hout' = Hin) \lor  (extend\_EV = FALSE \land  extend\_EV\_Hout' = 0)) }
			\nItem{ act5 }{ retract\_EV\_Hout :|  ((retract\_EV = TRUE \land  retract\_EV\_Hout' = Hin) \lor  (retract\_EV = FALSE \land  retract\_EV\_Hout' = 0)) }
			\end{description}
		\EndAct
		\end{description}
	\EVT {Failure\_Detection}
	\EXTD {Failure\_Detection}
		\begin{description}
		\BeginAct
			\begin{description}
			\nItemX{ act1 }{ anomaly :=  TRUE }
			\end{description}
		\EndAct
		\end{description}
\END
\end{description}

\section{M6}
\label{sec:M6}

\begin{description}
\BTitle{M6}{27Jan2014}{10:44:59 AM}
\MACHINE{M6}\cmt{		\\\hspace*{1 cm}  Integration of Cylinder bhavior according to the Electro-valves circuit		\\\hspace*{0,8 cm}  Strengthing guards of opening and closing doors and gears using cyliders  sensors, and haudrlic pressure.  }
\REFINES{M5}
\SEES{C1}
\VARIABLES
	\begin{description}
		\Item{ dstate }
		\Item{ lstate }
		\Item{ phase }
		\Item{ button }
		\Item{ p }
		\Item{ l }
		\Item{ i }
		\Item{ gstate }
		\Item{ handle }
		\Item{ analogical\_switch }
		\Item{ gear\_extended }
		\Item{ gear\_retracted }
		\Item{ gear\_shock\_absorber }
		\Item{ door\_closed }
		\Item{ door\_open }
		\Item{ circuit\_pressurized }
		\Item{ general\_EV }
		\Item{ close\_EV }
		\Item{ retract\_EV }
		\Item{ extend\_EV }
		\Item{ open\_EV }
		\Item{ gears\_locked\_down }
		\Item{ gears\_man }
		\Item{ anomaly }
		\Item{ general\_EV\_func }
		\Item{ close\_EV\_func }
		\Item{ retract\_EV\_func }
		\Item{ extend\_EV\_func }
		\Item{ open\_EV\_func }
		\Item{ gears\_locked\_down\_func }
		\Item{ gears\_man\_func }
		\Item{ anomaly\_func }
		\Item{ general\_EV\_Hout }
		\Item{ close\_EV\_Hout }
		\Item{ retract\_EV\_Hout }
		\Item{ extend\_EV\_Hout }
		\Item{ open\_EV\_Hout }
		\Item{ state }
		\ItemY{ SDCylinder }{State of Door Cylinder}
		\ItemY{ SGCylinder }{State of Gear Cylinder}
		\Item{ A\_Switch\_Out }
	\end{description}
\INVARIANTS
	\begin{description}
		\nItem{ inv1 }{ SDCylinder \in  DOORS \cprod  \{ DCYF,DCYR,DCYL\}  \tfun  S\_CYLINDER }
		\nItem{ inv2 }{ SGCylinder \in  GEARS \cprod  \{ GCYF,GCYR,GCYL\}   \tfun  S\_CYLINDER }
		\nItem{ inv17 }{ state \in  SPHASES }
		\nItem{ inv4 }{  SDCylinder = \{ a\mapsto b |  a\in  DOORS\cprod CYLINDER \land  b=STOP\}  \land  dstate^{-1} [\{ CLOSED\} ] \neq  \emptyset  \limp  dstate^{-1} [\{ CLOSED\} ]= DOORS }
		\nItem{ inv5 }{  SGCylinder = \{ a\mapsto b |  a\in  GEARS\cprod CYLINDER \land  b=STOP\}  \land  		\\\hspace*{1,2 cm}  (\forall door\qdot door \in  DOORS \land  dstate(door)=CLOSED)  \land  ran(gstate)\neq \{ RETRACTED\}  \land  ran(gstate)\neq \{ RETRACTING\}  \land  ran(gstate)\neq \{ EXTENDING\}  \limp  ran(gstate)=\{ EXTENDED\}  }
		\nItem{ inv6 }{  SGCylinder = \{ a\mapsto b |  a\in  GEARS\cprod CYLINDER \land  b=STOP\}  \land  		\\\hspace*{1,2 cm}  (\forall door\qdot door \in  DOORS \land  dstate(door)=CLOSED)  \land  ran(gstate)\neq \{ EXTENDED\}  \land  ran(gstate)\neq \{ RETRACTING\}  \land  ran(gstate)\neq \{ EXTENDING\}  \limp  ran(gstate)=\{ RETRACTED\}  }
		\nItem{ inv7 }{ SGCylinder = \{ a\mapsto b |  a\in  GEARS\cprod CYLINDER \land  b=MOVING\}  \land  		\\\hspace*{1,2 cm}  (\forall door\qdot door \in  DOORS \land  dstate(door)=CLOSED)  \land  ran(gstate)\neq \{ RETRACTED\}  \land  ran(gstate)\neq \{ EXTENDED\}  \land  ran(gstate)\neq \{ EXTENDING\}  \limp  ran(gstate)=\{ RETRACTING\}  }
		\nItem{ inv8 }{ SGCylinder = \{ a\mapsto b |  a\in  GEARS\cprod CYLINDER \land  b=MOVING\}  \land  		\\\hspace*{1,2 cm}  (\forall door\qdot door \in  DOORS \land  dstate(door)=CLOSED)  \land  ran(gstate)\neq \{ RETRACTED\}  \land  ran(gstate)\neq \{ EXTENDED\}  \land  ran(gstate)\neq \{ RETRACTING\}  \limp  ran(gstate)=\{ EXTENDING\}  }
	\end{description}
\EVENTS
	\INITIALISATION
		\\\textit{extended}
		\begin{description}
		\BeginAct
			\begin{description}
			\nItemX{ act1 }{ button :=  DOWN }
			\nItemX{ act2 }{ phase :=  haltdown }
			\nItemXY{ act3 }{ dstate :|  (dstate'\in  DOORS \tfun  SDOORS \land  dstate'=\{ a\mapsto b|  a \in  DOORS \land  b=CLOSED\} ) }{ 		\\\hspace*{1,4 cm}  missing elements of the invariant }
			\nItemX{ act4 }{ lstate :=  \{ a\mapsto b| a\in DOORS\land  b=LOCKED\}  }
			\nItemX{ act5 }{ p :=  R }
			\nItemX{ act6 }{ l :=  R }
			\nItemX{ act7 }{ i :=  R }
			\nItemX{ act8 }{ gstate :|  (gstate' \in  GEARS \tfun  SGEARS \land  gstate'=\{ a\mapsto b |  a \in  GEARS  \land  b=EXTENDED\} ) }
			\nItemX{ act14 }{ handle :\in  1\upto 3 \tfun  \{ DOWN\}  }
			\nItemX{ act15 }{ analogical\_switch :\in  1\upto 3 \tfun  \{ open\}  }
			\nItemX{ act16 }{ gear\_extended :\in  1\upto 3 \tfun  (GEARS \tfun  \{ TRUE\} ) }
			\nItemX{ act17 }{ gear\_retracted :\in  1\upto 3 \tfun  (GEARS \tfun  \{ FALSE\} ) }
			\nItemX{ act18 }{ gear\_shock\_absorber :\in  1\upto 3 \tfun  \{ ground\}  }
			\nItemX{ act19 }{ door\_closed :\in  1\upto 3 \tfun  (DOORS \tfun  \{ TRUE\} ) }
			\nItemX{ act20 }{ door\_open :\in  1\upto 3 \tfun  (DOORS \tfun  \{ FALSE\} ) }
			\nItemX{ act21 }{ circuit\_pressurized :\in  1\upto 3 \tfun  \{ FALSE\}  }
			\nItemX{ act22 }{ general\_EV :=  FALSE }
			\nItemX{ act23 }{ close\_EV :=  TRUE }
			\nItemX{ act24 }{ retract\_EV :=  FALSE }
			\nItemX{ act25 }{ extend\_EV :=  TRUE }
			\nItemX{ act27 }{ open\_EV :=  FALSE }
			\nItemX{ act28 }{ gears\_locked\_down :=  TRUE }
			\nItemX{ act29 }{ gears\_man :=  FALSE }
			\nItemX{ act30 }{ anomaly :=  FALSE }
			\nItemX{ act31 }{ general\_EV\_func :\in  (1\upto 3 \tfun  POSITIONS) \cprod  (1\upto 3 \tfun  A\_Switch) \cprod  (1\upto 3 \tfun  (GEARS \tfun  BOOL)) \cprod  (1\upto 3 \tfun  (GEARS \tfun  BOOL)) \cprod  (1\upto 3 \tfun  GEAR\_ABSORBER) \cprod  (1\upto 3 \tfun  (DOORS \tfun  BOOL)) \cprod  (1\upto 3 \tfun  (DOORS \tfun  BOOL)) \cprod  (1\upto 3 \tfun  BOOL) \tfun  BOOL }
			\nItemX{ act32 }{ close\_EV\_func :\in  (1\upto 3 \tfun  POSITIONS) \cprod  (1\upto 3 \tfun  A\_Switch) \cprod  (1\upto 3 \tfun  (GEARS \tfun  BOOL)) \cprod  (1\upto 3 \tfun  (GEARS \tfun  BOOL)) \cprod  (1\upto 3 \tfun  GEAR\_ABSORBER) \cprod  (1\upto 3 \tfun  (DOORS \tfun  BOOL)) \cprod  (1\upto 3 \tfun  (DOORS \tfun  BOOL)) \cprod  (1\upto 3 \tfun  BOOL) \tfun  BOOL }
			\nItemX{ act33 }{ retract\_EV\_func :\in  (1\upto 3 \tfun  POSITIONS) \cprod  (1\upto 3 \tfun  A\_Switch) \cprod  (1\upto 3 \tfun  (GEARS \tfun  BOOL)) \cprod  (1\upto 3 \tfun  (GEARS \tfun  BOOL)) \cprod  (1\upto 3 \tfun  GEAR\_ABSORBER) \cprod  (1\upto 3 \tfun  (DOORS \tfun  BOOL)) \cprod  (1\upto 3 \tfun  (DOORS \tfun  BOOL)) \cprod  (1\upto 3 \tfun  BOOL) \tfun  BOOL }
			\nItemX{ act34 }{ extend\_EV\_func :\in  (1\upto 3 \tfun  POSITIONS) \cprod  (1\upto 3 \tfun  A\_Switch) \cprod  (1\upto 3 \tfun  (GEARS \tfun  BOOL)) \cprod  (1\upto 3 \tfun  (GEARS \tfun  BOOL)) \cprod  (1\upto 3 \tfun  GEAR\_ABSORBER) \cprod  (1\upto 3 \tfun  (DOORS \tfun  BOOL)) \cprod  (1\upto 3 \tfun  (DOORS \tfun  BOOL)) \cprod  (1\upto 3 \tfun  BOOL) \tfun  BOOL }
			\nItemX{ act35 }{ open\_EV\_func :\in  (1\upto 3 \tfun  POSITIONS) \cprod  (1\upto 3 \tfun  A\_Switch) \cprod  (1\upto 3 \tfun  (GEARS \tfun  BOOL)) \cprod  (1\upto 3 \tfun  (GEARS \tfun  BOOL)) \cprod  (1\upto 3 \tfun  GEAR\_ABSORBER) \cprod  (1\upto 3 \tfun  (DOORS \tfun  BOOL)) \cprod  (1\upto 3 \tfun  (DOORS \tfun  BOOL)) \cprod  (1\upto 3 \tfun  BOOL) \tfun  BOOL }
			\nItemX{ act36 }{ gears\_locked\_down\_func :\in  (1\upto 3 \tfun  POSITIONS) \cprod  (1\upto 3 \tfun  A\_Switch) \cprod  (1\upto 3 \tfun  (GEARS \tfun  BOOL)) \cprod  (1\upto 3 \tfun  (GEARS \tfun  BOOL)) \cprod  (1\upto 3 \tfun  GEAR\_ABSORBER) \cprod  (1\upto 3 \tfun  (DOORS \tfun  BOOL)) \cprod  (1\upto 3 \tfun  (DOORS \tfun  BOOL)) \cprod  (1\upto 3 \tfun  BOOL) \tfun  BOOL }
			\nItemX{ act37 }{ gears\_man\_func :\in  (1\upto 3 \tfun  POSITIONS) \cprod  (1\upto 3 \tfun  A\_Switch) \cprod  (1\upto 3 \tfun  (GEARS \tfun  BOOL)) \cprod  (1\upto 3 \tfun  (GEARS \tfun  BOOL)) \cprod  (1\upto 3 \tfun  GEAR\_ABSORBER) \cprod  (1\upto 3 \tfun  (DOORS \tfun  BOOL)) \cprod  (1\upto 3 \tfun  (DOORS \tfun  BOOL)) \cprod  (1\upto 3 \tfun  BOOL) \tfun  BOOL }
			\nItemX{ act38 }{ anomaly\_func :\in  (1\upto 3 \tfun  POSITIONS) \cprod  (1\upto 3 \tfun  A\_Switch) \cprod  (1\upto 3 \tfun  (GEARS \tfun  BOOL)) \cprod  (1\upto 3 \tfun  (GEARS \tfun  BOOL)) \cprod  (1\upto 3 \tfun  GEAR\_ABSORBER) \cprod  (1\upto 3 \tfun  (DOORS \tfun  BOOL)) \cprod  (1\upto 3 \tfun  (DOORS \tfun  BOOL)) \cprod  (1\upto 3 \tfun  BOOL) \tfun  BOOL }
			\nItemX{ act39 }{ A\_Switch\_Out :=  FALSE }
			\nItemX{ act40 }{ close\_EV\_Hout :=  0 }
			\nItemX{ act41 }{ retract\_EV\_Hout :=  0 }
			\nItemX{ act42 }{ extend\_EV\_Hout :=  0 }
			\nItemX{ act43 }{ open\_EV\_Hout :=  0 }
			\nItemX{ act44 }{ general\_EV\_Hout :=  0 }
			\nItem{ act45 }{ SDCylinder :\in  DOORS \cprod  \{ DCYF,DCYR,DCYL\}  \tfun   \{ STOP\}  }
			\nItem{ act46 }{ SGCylinder :\in  GEARS \cprod  \{ GCYF,GCYR,GCYL\}  \tfun  \{ STOP\}   }
			\nItem{ act26 }{ state :=  computing }
			\end{description}
		\EndAct
		\end{description}
	\EVT {opening\_doors\_DOWN}
	\EXTD {opening\_doors\_DOWN}
		\begin{description}
		\WhenGrd
			\begin{description}
			\nItemX{ grd1 }{ dstate[DOORS]= \{ CLOSED\}  }
			\nItemX{ grd5 }{ lstate[DOORS]=\{ UNLOCKED\}  }
			\nItemX{ grd7 }{ phase=movingdown }
			\nItemX{ grd8 }{ p=R }
			\nItemX{ grd9 }{ l=R }
			\nItemX{ grd10 }{ door\_open = \{ a\mapsto b |  a \in  1\upto 3 \land  b \in  DOORS \tfun  \{ FALSE\} \}   }
			\nItemX{ grd11 }{ door\_closed = \{ a\mapsto b |  a \in  1\upto 3 \land  b \in  DOORS \tfun  \{ FALSE\} \}   }
			\nItemX{ grd12 }{ \forall x\qdot x\in 1\upto 3 \limp  handle(x)=button }
			\nItem{ grd3 }{ SDCylinder = \{ a\mapsto b |  a\in  DOORS\cprod CYLINDER \land  b=MOVING\}  }
			\end{description}
		\ThenAct
			\begin{description}
			\nItemX{ act1 }{ dstate :=  \{ a\mapsto b|  a \in  DOORS \land  b=OPEN\}  }
			\nItemX{ act2 }{ p:= E }
			\nItemX{ act3 }{ door\_open :\in  1\upto 3 \tfun  (DOORS \tfun  \{ TRUE\} ) }
			\end{description}
		\EndAct
		\end{description}
	\EVT {opening\_doors\_UP}
	\EXTD {opening\_doors\_UP}
		\begin{description}
		\WhenGrd
			\begin{description}
			\nItemX{ grd1 }{ dstate[DOORS]= \{ CLOSED\}  }
			\nItemX{ grd4 }{ lstate[DOORS]=\{ UNLOCKED\}  }
			\nItemX{ grd5 }{ phase= movingup }
			\nItemX{ grd6 }{ p=E }
			\nItemX{ grd7 }{ l=E }
			\nItemX{ grd8 }{ door\_open = \{ a\mapsto b |  a \in  1\upto 3 \land  b \in  DOORS \tfun  \{ FALSE\} \}   }
			\nItemX{ grd9 }{ door\_closed = \{ a\mapsto b |  a \in  1\upto 3 \land  b \in  DOORS \tfun  \{ FALSE\} \}   }
			\nItemX{ grd10 }{ \forall x\qdot x\in 1\upto 3 \limp  handle(x)=button }
			\nItem{ grd3 }{ SDCylinder = \{ a\mapsto b |  a\in  DOORS\cprod CYLINDER \land  b=MOVING\}  }
			\end{description}
		\ThenAct
			\begin{description}
			\nItemX{ act1 }{ dstate :=  \{ a\mapsto b|  a \in  DOORS \land  b=OPEN\}  }
			\nItemX{ act2 }{ p:= R }
			\nItemX{ act3 }{ door\_open:\in  1\upto 3 \tfun  (DOORS \tfun  \{ TRUE\} )  }
			\end{description}
		\EndAct
		\end{description}
	\EVT {closing\_doors\_UP}
	\EXTD {closing\_doors\_UP}
		\begin{description}
		\AnyPrm
			\begin{description}
			\ItemX{f }
			\end{description}
		\WhereGrd
			\begin{description}
			\nItemX{ grd1 }{ dstate[DOORS]=\{ OPEN\}  }
			\nItemX{ grd3 }{ f \in  DOORS \tfun  SDOORS }
			\nItemX{ grd4 }{ \forall e\qdot  e \in  DOORS \limp  f(e)=CLOSED }
			\nItemX{ grd5 }{ phase=movingup }
			\nItemX{ grd6 }{ p=R }
			\nItemX{ grd7 }{ gstate[GEARS]=\{ RETRACTED\}  }
			\nItemX{ grd8 }{ \forall x\qdot x\in 1\upto 3 \limp  handle(x)=button }
			\end{description}
		\ThenAct
			\begin{description}
			\nItemX{ act1 }{ dstate:= f }
			\end{description}
		\EndAct
		\end{description}
	\EVT {closing\_doors\_DOWN}
	\EXTD {closing\_doors\_DOWN}
		\begin{description}
		\AnyPrm
			\begin{description}
			\ItemX{f }
			\end{description}
		\WhereGrd
			\begin{description}
			\nItemX{ grd1 }{ dstate[DOORS]=\{ OPEN\}  }
			\nItemX{ grd3 }{ f \in  DOORS \tfun  SDOORS }
			\nItemX{ grd4 }{ \forall e\qdot  e \in  DOORS \limp  f(e)=CLOSED }
			\nItemX{ grd5 }{ phase=movingdown }
			\nItemX{ grd6 }{ p=E }
			\nItemX{ grd7 }{ gstate[GEARS]=\{ EXTENDED\}  }
			\nItemX{ grd8 }{ \forall x\qdot x\in 1\upto 3 \limp  handle(x)=button }
			\end{description}
		\ThenAct
			\begin{description}
			\nItemX{ act1 }{ dstate:= f }
			\end{description}
		\EndAct
		\end{description}
	\EVT {unlocking\_UP}
	\EXTD {unlocking\_UP}
		\begin{description}
		\WhenGrd
			\begin{description}
			\nItemX{ grd3 }{ lstate[DOORS]=\{ LOCKED\}  }
			\nItemX{ grd4 }{ phase=movingup }
			\nItemX{ grd5 }{ l=E }
			\nItemX{ grd6 }{ p=E }
			\nItemX{ grd7 }{ i=E }
			\nItemX{ grd8 }{ door\_open = \{ a\mapsto b |  a \in  1\upto 3 \land  b \in  DOORS \tfun  \{ FALSE\} \}   }
			\nItemX{ grd9 }{ door\_closed = \{ a\mapsto b |  a \in  1\upto 3 \land  b \in  DOORS \tfun  \{ TRUE\} \}   }
			\nItemX{ grd10 }{ \forall x\qdot x\in 1\upto 3 \limp  handle(x)=button }
			\end{description}
		\ThenAct
			\begin{description}
			\nItemX{ act1 }{ lstate:= \{ a\mapsto b| a\in DOORS \land  b=UNLOCKED\}  }
			\nItemX{ act2 }{ door\_closed :\in  1\upto 3 \tfun  (DOORS \tfun  \{ FALSE\} )  }
			\end{description}
		\EndAct
		\end{description}
	\EVT {locking\_UP}
	\EXTD {locking\_UP}
		\begin{description}
		\WhenGrd
			\begin{description}
			\nItemX{ grd3 }{ dstate[DOORS]=\{ CLOSED\}  }
			\nItemX{ grd4 }{ phase=movingup }
			\nItemX{ grd5 }{ lstate[DOORS]=\{ UNLOCKED\}  }
			\nItemX{ grd6 }{ p=R }
			\nItemX{ grd7 }{ l=E }
			\nItemX{ grd9 }{ door\_open = \{ a\mapsto b |  a \in  1\upto 3 \land  b \in  DOORS \tfun  \{ FALSE\} \}   }
			\nItemX{ grd10 }{ door\_closed = \{ a\mapsto b |  a \in  1\upto 3 \land  b \in  DOORS \tfun  \{ FALSE\} \}   }
			\nItemX{ grd11 }{ \forall x\qdot x\in 1\upto 3 \limp  handle(x)=button }
			\nItem{ grd8 }{ SDCylinder = \{ a\mapsto b |  a\in  DOORS\cprod CYLINDER \land  b=STOP\}  }
			\end{description}
		\ThenAct
			\begin{description}
			\nItemX{ act1 }{ lstate:= \{ a\mapsto b| a\in DOORS \land  b=LOCKED\}  }
			\nItemX{ act3 }{ phase:= haltup }
			\nItemXY{ act4 }{ l:= R }{ 		\\\hspace*{1,4 cm}  added by D Mery }
			\nItemX{ act44 }{ door\_closed :\in  1\upto 3 \tfun  (DOORS \tfun  \{ TRUE\} )  }
			\end{description}
		\EndAct
		\end{description}
	\EVT {unlocking\_DOWN}
	\EXTD {unlocking\_DOWN}
		\begin{description}
		\WhenGrd
			\begin{description}
			\nItemX{ grd3 }{ lstate[DOORS]=\{ LOCKED\}  }
			\nItemX{ grd4 }{ phase=movingdown }
			\nItemX{ grd5 }{ l=R }
			\nItemX{ grd6 }{ p=R }
			\nItemX{ grd7 }{ i=R }
			\nItemX{ grd8 }{ door\_open = \{ a\mapsto b |  a \in  1\upto 3 \land  b \in  DOORS \tfun  \{ FALSE\} \}   }
			\nItemX{ grd9 }{ door\_closed = \{ a\mapsto b |  a \in  1\upto 3 \land  b \in  DOORS \tfun  \{ TRUE\} \}   }
			\nItemX{ grd10 }{ \forall x\qdot x\in 1\upto 3 \limp  handle(x)=button }
			\end{description}
		\ThenAct
			\begin{description}
			\nItemX{ act1 }{ lstate:= \{ a\mapsto b| a\in DOORS \land  b=UNLOCKED\}  }
			\nItemX{ act2 }{ door\_closed :\in  1\upto 3 \tfun  (DOORS \tfun  \{ FALSE\} ) }
			\end{description}
		\EndAct
		\end{description}
	\EVT {locking\_DOWN}
	\EXTD {locking\_DOWN}
		\begin{description}
		\WhenGrd
			\begin{description}
			\nItemX{ grd1 }{ dstate[DOORS]=\{ CLOSED\}  }
			\nItemX{ grd2 }{ phase=movingdown }
			\nItemX{ grd3 }{ lstate[DOORS]=\{ UNLOCKED\}  }
			\nItemX{ grd4 }{ p=E }
			\nItemX{ grd5 }{ l=R }
			\nItemX{ grd7 }{ door\_open = \{ a\mapsto b |  a \in  1\upto 3 \land  b \in  DOORS \tfun  \{ FALSE\} \}   }
			\nItemX{ grd8 }{ door\_closed = \{ a\mapsto b |  a \in  1\upto 3 \land  b \in  DOORS \tfun  \{ FALSE\} \}   }
			\nItemX{ grd9 }{ \forall x\qdot x\in 1\upto 3 \limp  handle(x)=button }
			\nItem{ grd6 }{ SDCylinder = \{ a\mapsto b |  a\in  DOORS\cprod CYLINDER \land  b=STOP\}  }
			\end{description}
		\ThenAct
			\begin{description}
			\nItemX{ act1 }{ lstate:= \{ a\mapsto b| a\in DOORS \land  b = LOCKED\}  }
			\nItemX{ act3 }{ phase:= haltdown }
			\nItemX{ act4 }{ l:= E }
			\nItemX{ act5 }{ door\_closed :\in  1\upto 3 \tfun  (DOORS \tfun  \{ TRUE\} )  }
			\end{description}
		\EndAct
		\end{description}
	\EVT {PD1}
	\EXTD {PD1}
		\begin{description}
		\WhenGrd
			\begin{description}
			\nItemX{ grd1 }{ button=UP }
			\nItemX{ grd2 }{ phase=haltup }
			\nItemX{ grd3 }{ \forall x\qdot x\in 1\upto 3 \limp  handle(x)=DOWN }
			\end{description}
		\ThenAct
			\begin{description}
			\nItemX{ act1 }{ phase:= movingdown }
			\nItemX{ act2 }{ button:= DOWN }
			\nItemX{ act3 }{ l:= R }
			\nItemX{ act4 }{ p:= R }
			\nItemX{ act5 }{ i:= R }
			\end{description}
		\EndAct
		\end{description}
	\EVT {PU1}
	\EXTD {PU1}
		\begin{description}
		\WhenGrd
			\begin{description}
			\nItemX{ grd1 }{ button=DOWN }
			\nItemX{ grd2 }{ phase=haltdown }
			\nItemX{ grd3 }{ \forall x\qdot x\in 1\upto 3 \limp  handle(x)=UP }
			\end{description}
		\ThenAct
			\begin{description}
			\nItemX{ act1 }{ phase:= movingup }
			\nItemX{ act2 }{ button:= UP }
			\nItemX{ act3 }{ l:= E }
			\nItemX{ act4 }{ p:= E }
			\nItemX{ act5 }{ i:= E }
			\end{description}
		\EndAct
		\end{description}
	\EVT {PU2}
	\EXTD {PU2}
		\begin{description}
		\WhenGrd
			\begin{description}
			\nItemX{ grd1 }{ l=R }
			\nItemX{ grd2 }{ p=R }
			\nItemX{ grd3 }{ phase=movingdown }
			\nItemX{ grd4 }{ button=DOWN }
			\nItemX{ grd5 }{ i=R }
			\nItemX{ grd6 }{ lstate[DOORS]=\{ LOCKED\}  }
			\nItemX{ grd7 }{ door\_open = \{ a\mapsto b |  a \in  1\upto 3 \land  b \in  DOORS \tfun  \{ FALSE\} \}   }
			\nItemX{ grd8 }{ door\_closed = \{ a\mapsto b |  a \in  1\upto 3 \land  b \in  DOORS \tfun  \{ TRUE\} \}   }
			\nItemX{ grd9 }{ \forall x\qdot x\in 1\upto 3 \limp  handle(x)=UP }
			\end{description}
		\ThenAct
			\begin{description}
			\nItemX{ act1 }{ phase:= movingup }
			\nItemX{ act4 }{ button:=   UP }
			\nItemX{ act5 }{ l:= E }
			\nItemX{ act6 }{ p:= E }
			\nItemX{ act7 }{ i:= R }
			\end{description}
		\EndAct
		\end{description}
	\EVT {CompletePU2}
	\EXTD {CompletePU2}
		\begin{description}
		\WhenGrd
			\begin{description}
			\nItemX{ grd1 }{ phase=movingup }
			\nItemX{ grd2 }{ button=UP }
			\nItemX{ grd3 }{ l=E }
			\nItemX{ grd4 }{ p=E }
			\nItemX{ grd5 }{ i=R }
			\end{description}
		\ThenAct
			\begin{description}
			\nItemX{ act1 }{ phase:= haltup }
			\end{description}
		\EndAct
		\end{description}
	\EVT {PU3}
	\EXTD {PU3}
		\begin{description}
		\WhenGrd
			\begin{description}
			\nItemX{ grd1 }{ dstate[DOORS]=\{ CLOSED\}  }
			\nItemX{ grd2 }{ lstate[DOORS]=\{ UNLOCKED\}  }
			\nItemX{ grd3 }{ phase = movingdown }
			\nItemX{ grd4 }{ p=R }
			\nItemX{ grd5 }{ l=R }
			\nItemX{ grd6 }{ button=DOWN }
			\nItemX{ grd7 }{ door\_open = \{ a\mapsto b |  a \in  1\upto 3 \land  b \in  DOORS \tfun  \{ FALSE\} \}   }
			\nItemX{ grd8 }{ door\_closed = \{ a\mapsto b |  a \in  1\upto 3 \land  b \in  DOORS \tfun  \{ FALSE\} \}   }
			\nItemX{ grd9 }{ \forall x\qdot x\in 1\upto 3 \limp  handle(x)=UP }
			\end{description}
		\ThenAct
			\begin{description}
			\nItemX{ act1 }{ phase:= movingup }
			\nItemX{ act2 }{ p:= R }
			\nItemX{ act3 }{ l:= E }
			\nItemX{ act4 }{ button:= UP }
			\end{description}
		\EndAct
		\end{description}
	\EVT {PU4}
	\EXTD {PU4}
		\begin{description}
		\WhenGrd
			\begin{description}
			\nItemX{ grd1 }{ dstate[DOORS]=\{ OPEN\}  }
			\nItemX{ grd2 }{ phase=movingdown }
			\nItemX{ grd3 }{ p=E }
			\nItemX{ grd4 }{ button=DOWN }
			\nItemX{ grd7 }{ \forall x\qdot x\in 1\upto 3 \limp  handle(x)=UP }
			\end{description}
		\ThenAct
			\begin{description}
			\nItemX{ act1 }{ phase:= movingup }
			\nItemX{ act2 }{ p:= R }
			\nItemX{ act3 }{ button:= UP }
			\nItemX{ act4 }{ i:= E }
			\nItemX{ act5 }{ l:= E }
			\end{description}
		\EndAct
		\end{description}
	\EVT {PU5}
	\EXTD {PU5}
		\begin{description}
		\WhenGrd
			\begin{description}
			\nItemX{ grd1 }{ dstate[DOORS]=\{ CLOSED\}  }
			\nItemX{ grd2 }{ phase=movingdown }
			\nItemX{ grd3 }{ p=E }
			\nItemX{ grd4 }{ button=DOWN }
			\nItemX{ grd5 }{ lstate[DOORS]=\{ UNLOCKED\}  }
			\nItemX{ grd6 }{ door\_open = \{ a\mapsto b |  a \in  1\upto 3 \land  b \in  DOORS \tfun  \{ FALSE\} \}   }
			\nItemX{ grd7 }{ door\_closed = \{ a\mapsto b |  a \in  1\upto 3 \land  b \in  DOORS \tfun  \{ FALSE\} \}   }
			\nItemX{ grd8 }{ \forall x\qdot x\in 1\upto 3 \limp  handle(x)=UP }
			\end{description}
		\ThenAct
			\begin{description}
			\nItemX{ act1 }{ phase:= movingup }
			\nItemX{ act3 }{ button:= UP }
			\nItemX{ act4 }{ i:= E }
			\nItemX{ act5 }{ l:= E }
			\end{description}
		\EndAct
		\end{description}
	\EVT {PD2}
	\EXTD {PD2}
		\begin{description}
		\WhenGrd
			\begin{description}
			\nItemX{ grd1 }{ l=E }
			\nItemX{ grd2 }{ p=E }
			\nItemX{ grd3 }{ phase=movingup }
			\nItemX{ grd4 }{ i=E }
			\nItemX{ grd5 }{ lstate[DOORS]=\{ LOCKED\}  }
			\nItemX{ grd6 }{ \forall x\qdot x\in 1\upto 3 \limp  handle(x)=DOWN }
			\end{description}
		\ThenAct
			\begin{description}
			\nItemX{ act1 }{ phase:= movingdown }
			\nItemX{ act2 }{ button:= DOWN }
			\nItemX{ act3 }{ l:= R }
			\nItemX{ act4 }{ p:= R }
			\nItemX{ act5 }{ i:= E }
			\end{description}
		\EndAct
		\end{description}
	\EVT {CompletePD2}
	\EXTD {CompletePD2}
		\begin{description}
		\WhenGrd
			\begin{description}
			\nItemX{ grd1 }{ phase=movingdown }
			\nItemX{ grd2 }{ button=DOWN }
			\nItemX{ grd3 }{ l=R }
			\nItemX{ grd4 }{ p=R }
			\nItemX{ grd5 }{ i=E }
			\end{description}
		\ThenAct
			\begin{description}
			\nItemX{ act1 }{ phase:= haltdown }
			\end{description}
		\EndAct
		\end{description}
	\EVT {PD3}
	\EXTD {PD3}
		\begin{description}
		\WhenGrd
			\begin{description}
			\nItemX{ grd1 }{ dstate[DOORS]=\{ CLOSED\}  }
			\nItemX{ grd2 }{ lstate[DOORS]=\{ UNLOCKED\}  }
			\nItemX{ grd3 }{ phase=movingup }
			\nItemX{ grd4 }{ p=E }
			\nItemX{ grd5 }{ l=E }
			\nItemX{ grd6 }{ button=UP }
			\nItemX{ grd7 }{ door\_open = \{ a\mapsto b |  a \in  1\upto 3 \land  b \in  DOORS \tfun  \{ FALSE\} \}   }
			\nItemX{ grd8 }{ door\_closed = \{ a\mapsto b |  a \in  1\upto 3 \land  b \in  DOORS \tfun  \{ FALSE\} \}   }
			\nItemX{ grd9 }{ \forall x\qdot x\in 1\upto 3 \limp  handle(x)=DOWN }
			\end{description}
		\ThenAct
			\begin{description}
			\nItemX{ act1 }{ phase:= movingdown }
			\nItemX{ act2 }{ p:= E }
			\nItemX{ act3 }{ l:= R }
			\nItemX{ act4 }{ button:= DOWN }
			\end{description}
		\EndAct
		\end{description}
	\EVT {PD4}
	\EXTD {PD4}
		\begin{description}
		\WhenGrd
			\begin{description}
			\nItemX{ grd1 }{ dstate[DOORS]=\{ OPEN\}  }
			\nItemX{ grd2 }{ phase=movingup }
			\nItemX{ grd3 }{ p=R }
			\nItemX{ grd4 }{ button=UP }
			\nItemX{ grd6 }{ \forall x\qdot x\in 1\upto 3 \limp  handle(x)=DOWN }
			\end{description}
		\ThenAct
			\begin{description}
			\nItemX{ act1 }{ phase:= movingdown }
			\nItemX{ act2 }{ p:= E }
			\nItemX{ act3 }{ button:= DOWN }
			\nItemX{ act4 }{ i:= R }
			\nItemX{ act5 }{ l:= R }
			\end{description}
		\EndAct
		\end{description}
	\EVT {PD5}
	\EXTD {PD5}
		\begin{description}
		\WhenGrd
			\begin{description}
			\nItemX{ grd1 }{ dstate[DOORS]=\{ CLOSED\}  }
			\nItemX{ grd2 }{ phase=movingup }
			\nItemX{ grd3 }{ p=R }
			\nItemX{ grd4 }{ button=UP }
			\nItemX{ grd5 }{ lstate[DOORS]=\{ UNLOCKED\}  }
			\nItemX{ grd6 }{ door\_open = \{ a\mapsto b |  a \in  1\upto 3 \land  b \in  DOORS \tfun  \{ FALSE\} \}   }
			\nItemX{ grd7 }{ door\_closed = \{ a\mapsto b |  a \in  1\upto 3 \land  b \in  DOORS \tfun  \{ FALSE\} \}   }
			\nItemX{ grd8 }{ \forall x\qdot x\in 1\upto 3 \limp  handle(x)=DOWN }
			\end{description}
		\ThenAct
			\begin{description}
			\nItemX{ act1 }{ phase :=  movingdown }
			\nItemX{ act2 }{ button:= DOWN }
			\nItemX{ act3 }{ i:= R }
			\nItemX{ act4 }{ l:= R }
			\end{description}
		\EndAct
		\end{description}
	\EVT {retracting\_gears}
	\EXTD {retracting\_gears}
		\begin{description}
		\WhenGrd
			\begin{description}
			\nItemX{ grd1 }{ dstate[DOORS]=\{ OPEN\}  }
			\nItemX{ grd2 }{ gstate[GEARS]=\{ EXTENDED\}  }
			\nItemX{ grd3 }{ p=R }
			\nItemX{ grd6 }{ gear\_extended = \{ a\mapsto b |  a \in  1\upto 3 \land  b \in  GEARS \tfun  \{ TRUE\} \}   }
			\nItemX{ grd7 }{ gear\_retracted = \{ a\mapsto b |  a \in  1\upto 3 \land  b \in  GEARS \tfun  \{ FALSE\} \}   }
			\nItemX{ grd8 }{ gear\_shock\_absorber = \{ a\mapsto b |  a \in  1\upto 3 \land  b = ground\}   }
			\nItemX{ grd9 }{ \forall x\qdot x\in 1\upto 3 \limp  handle(x)=button }
			\nItem{ grd5 }{ SGCylinder = \{ a\mapsto b |  a\in  GEARS\cprod CYLINDER \land  b=MOVING\}  }
			\end{description}
		\ThenAct
			\begin{description}
			\nItemX{ act1 }{ gstate :=  \{ a\mapsto b| a \in  GEARS \land  b=RETRACTING\}  }
			\nItemX{ act2 }{ gear\_extended :\in  1\upto 3 \tfun  (GEARS \tfun  \{ FALSE\} )  }
			\nItemX{ act3 }{ gear\_shock\_absorber :=  \{ a\mapsto b |  a \in  1\upto 3 \land  b = flight\}   }
			\end{description}
		\EndAct
		\end{description}
	\EVT {retraction}
	\EXTD {retraction}
		\begin{description}
		\WhenGrd
			\begin{description}
			\nItemX{ grd1 }{ dstate[DOORS]=\{ OPEN\}  }
			\nItemX{ grd2 }{ gstate[GEARS]=\{ RETRACTING\}  }
			\nItemX{ grd4 }{ gear\_extended = \{ a\mapsto b |  a \in  1\upto 3 \land  b \in  GEARS \tfun  \{ FALSE\} \}   }
			\nItemX{ grd5 }{ gear\_retracted = \{ a\mapsto b |  a \in  1\upto 3 \land  b \in  GEARS \tfun  \{ FALSE\} \}   }
			\nItemX{ grd6 }{ gear\_shock\_absorber = \{ a\mapsto b |  a \in  1\upto 3 \land  b = flight\}   }
			\nItemX{ grd7 }{ \forall x\qdot x\in 1\upto 3 \limp  handle(x)=button }
			\nItem{ grd3 }{ SGCylinder = \{ a\mapsto b |  a\in  GEARS\cprod CYLINDER \land  b=STOP\}  }
			\end{description}
		\ThenAct
			\begin{description}
			\nItemX{ act1 }{ gstate:=   \{ a\mapsto b|  a \in  GEARS \land  b= RETRACTED\}  }
			\nItemX{ act2 }{ gear\_retracted :\in  1\upto 3 \tfun  (GEARS \tfun  \{ TRUE\} )  }
			\end{description}
		\EndAct
		\end{description}
	\EVT {extending\_gears}
	\EXTD {extending\_gears}
		\begin{description}
		\WhenGrd
			\begin{description}
			\nItemX{ grd1 }{ dstate[DOORS]=\{ OPEN\}  }
			\nItemX{ grd2 }{ gstate[GEARS]=\{ RETRACTED\}  }
			\nItemX{ grd3 }{ p=E }
			\nItemX{ grd5 }{ gear\_retracted = \{ a\mapsto b |  a \in  1\upto 3 \land  b \in  GEARS \tfun  \{ TRUE\} \}   }
			\nItemX{ grd6 }{ gear\_extended = \{ a\mapsto b |  a \in  1\upto 3 \land  b \in  GEARS \tfun  \{ FALSE\} \}   }
			\nItemX{ grd7 }{ gear\_shock\_absorber = \{ a\mapsto b |  a \in  1\upto 3 \land  b = flight\}   }
			\nItemX{ grd8 }{ \forall x\qdot x\in 1\upto 3 \limp  handle(x)=button }
			\nItem{ grd4 }{ SGCylinder = \{ a\mapsto b |  a\in  GEARS\cprod CYLINDER \land  b=MOVING\}  }
			\end{description}
		\ThenAct
			\begin{description}
			\nItemX{ act1 }{ gstate :=  \{ a\mapsto b| a \in  GEARS \land  b=EXTENDING\}  }
			\nItemX{ act2 }{ gear\_retracted :\in  1\upto 3 \tfun  (GEARS \tfun  \{ FALSE\} ) }
			\end{description}
		\EndAct
		\end{description}
	\EVT {extension}
	\EXTD {extension}
		\begin{description}
		\WhenGrd
			\begin{description}
			\nItemX{ grd1 }{ dstate[DOORS]=\{ OPEN\}  }
			\nItemX{ grd2 }{ gstate[GEARS]=\{ EXTENDING\}  }
			\nItemX{ grd4 }{ gear\_retracted = \{ a\mapsto b |  a \in  1\upto 3 \land  b \in  GEARS \tfun  \{ FALSE\} \}   }
			\nItemX{ grd5 }{ gear\_extended = \{ a\mapsto b |  a \in  1\upto 3 \land  b \in  GEARS \tfun  \{ FALSE\} \}   }
			\nItemX{ grd6 }{ gear\_shock\_absorber = \{ a\mapsto b |  a \in  1\upto 3 \land  b = flight\}   }
			\nItemX{ grd7 }{ \forall x\qdot x\in 1\upto 3 \limp  handle(x)=button }
			\nItem{ grd3 }{ SGCylinder = \{ a\mapsto b |  a\in  GEARS\cprod CYLINDER \land  b=STOP\}  }
			\end{description}
		\ThenAct
			\begin{description}
			\nItemX{ act1 }{ gstate :=  \{ a\mapsto b|  a \in  GEARS \land  b=EXTENDED\}  }
			\nItemX{ act2 }{ gear\_extended :\in  1\upto 3 \tfun  (GEARS \tfun  \{ TRUE\} )  }
			\nItemX{ act3 }{ gear\_shock\_absorber :=  \{ a\mapsto b |  a \in  1\upto 3 \land  b = ground\}   }
			\end{description}
		\EndAct
		\end{description}
	\EVT {HPD1}
	\EXTD {HPD1}
		\begin{description}
		\WhenGrd
			\begin{description}
			\nItemX{ grd3 }{ \forall x\qdot x\in 1\upto 3 \limp  handle(x)=UP }
			\end{description}
		\ThenAct
			\begin{description}
			\nItemX{ act2 }{ handle :\in  1\upto 3 \tfun  \{ DOWN\}  }
			\end{description}
		\EndAct
		\end{description}
	\EVT {HPU1}
	\EXTD {HPU1}
		\begin{description}
		\WhenGrd
			\begin{description}
			\nItemX{ grd3 }{ \forall x\qdot x\in 1\upto 3 \limp  handle(x)=DOWN }
			\end{description}
		\ThenAct
			\begin{description}
			\nItemX{ act2 }{ handle :\in  1\upto 3 \tfun  \{ UP\}  }
			\end{description}
		\EndAct
		\end{description}
	\EVT {Analogical\_switch\_closed}
	\EXTD {Analogical\_switch\_closed}
		\begin{description}
		\AnyPrm
			\begin{description}
			\ItemXY{in }{in port }
			\end{description}
		\WhereGrd
			\begin{description}
			\nItemX{ grd1 }{ in = general\_EV }
			\nItemX{ grd2 }{ \forall x\qdot x\in 1\upto 3 \limp  (handle(x)=UP \lor  handle(x)=DOWN) }
			\end{description}
		\ThenAct
			\begin{description}
			\nItemX{ act3 }{ analogical\_switch :\in  1\upto 3 \tfun  \{ closed\}  }
			\nItemX{ act4 }{ A\_Switch\_Out :=  TRUE }
			\end{description}
		\EndAct
		\end{description}
	\EVT {Analogical\_switch\_open}
	\EXTD {Analogical\_switch\_open}
		\begin{description}
		\AnyPrm
			\begin{description}
			\ItemXY{in }{in port }
			\end{description}
		\WhereGrd
			\begin{description}
			\nItemX{ grd1 }{ in = general\_EV }
			\nItemX{ grd2 }{ \forall x\qdot x\in 1\upto 3 \limp  (handle(x)=UP \lor  handle(x)=DOWN) }
			\end{description}
		\ThenAct
			\begin{description}
			\nItemX{ act3 }{ analogical\_switch :\in  1\upto 3 \tfun  \{ open\}  }
			\nItemX{ act4 }{ A\_Switch\_Out :=  FALSE }
			\end{description}
		\EndAct
		\end{description}
	\EVT {Circuit\_pressurized\_OK}
	\EXTD {Circuit\_pressurized\_OK}
		\begin{description}
		\WhenGrd
			\begin{description}
			\nItemX{ grd1 }{ general\_EV\_Hout = Hin }
			\end{description}
		\ThenAct
			\begin{description}
			\nItemX{ act9 }{ circuit\_pressurized :\in  1\upto 3 \tfun  \{ TRUE\}  }
			\end{description}
		\EndAct
		\end{description}
	\EVT {Circuit\_pressurized\_notOK}
	\EXTD {Circuit\_pressurized\_notOK}
		\begin{description}
		\WhenGrd
			\begin{description}
			\nItemX{ grd1 }{ general\_EV\_Hout = 0 }
			\end{description}
		\ThenAct
			\begin{description}
			\nItemX{ act9 }{ circuit\_pressurized :\in  1\upto 3 \tfun  \{ FALSE\}  }
			\end{description}
		\EndAct
		\end{description}
	\EVT {Computing\_Module\_1\_2}
	\EXTD {Computing\_Module\_1\_2}
		\begin{description}
		\WhenGrd
			\begin{description}
			\nItem{ grd1 }{ state=computing }
			\end{description}
		\ThenAct
			\begin{description}
			\nItemX{ act1 }{ general\_EV :=  general\_EV\_func(handle \mapsto  analogical\_switch \mapsto  gear\_extended \mapsto  gear\_retracted \mapsto  gear\_shock\_absorber \mapsto  door\_open \mapsto  door\_closed \mapsto  circuit\_pressurized) }
			\nItemX{ act2 }{ close\_EV :=  close\_EV\_func(handle \mapsto  analogical\_switch \mapsto  gear\_extended \mapsto  gear\_retracted \mapsto  gear\_shock\_absorber \mapsto  door\_open \mapsto  door\_closed \mapsto  circuit\_pressurized) }
			\nItemX{ act3 }{ retract\_EV :=  retract\_EV\_func(handle \mapsto  analogical\_switch \mapsto  gear\_extended \mapsto  gear\_retracted \mapsto  gear\_shock\_absorber \mapsto  door\_open \mapsto  door\_closed \mapsto  circuit\_pressurized) }
			\nItemX{ act4 }{ extend\_EV :=  extend\_EV\_func(handle \mapsto  analogical\_switch \mapsto  gear\_extended \mapsto  gear\_retracted \mapsto  gear\_shock\_absorber \mapsto  door\_open \mapsto  door\_closed \mapsto  circuit\_pressurized) }
			\nItemX{ act5 }{ open\_EV :=  open\_EV\_func(handle \mapsto  analogical\_switch \mapsto  gear\_extended \mapsto  gear\_retracted \mapsto  gear\_shock\_absorber \mapsto  door\_open \mapsto  door\_closed \mapsto  circuit\_pressurized) }
			\nItemX{ act6 }{ gears\_locked\_down :=  gears\_locked\_down\_func(handle \mapsto  analogical\_switch \mapsto  gear\_extended \mapsto  gear\_retracted \mapsto  gear\_shock\_absorber \mapsto  door\_open \mapsto  door\_closed \mapsto  circuit\_pressurized) }
			\nItemX{ act7 }{ gears\_man :=  gears\_man\_func(handle \mapsto  analogical\_switch \mapsto  gear\_extended \mapsto  gear\_retracted \mapsto  gear\_shock\_absorber \mapsto  door\_open \mapsto  door\_closed \mapsto  circuit\_pressurized) }
			\nItemX{ act8 }{ anomaly :=  anomaly\_func(handle \mapsto  analogical\_switch \mapsto  gear\_extended \mapsto  gear\_retracted \mapsto  gear\_shock\_absorber \mapsto  door\_open \mapsto  door\_closed \mapsto  circuit\_pressurized) }
			\nItem{ act9 }{ state:= electroValve }
			\end{description}
		\EndAct
		\end{description}
	\EVT {Update\_Hout}\cmt{		\\\hspace*{2,8 cm}  Assign the value of Hout  }
	\EXTD {Update\_Hout}
		\begin{description}
		\WhenGrd
			\begin{description}
			\nItem{ grd1 }{ state = electroValve }
			\end{description}
		\ThenAct
			\begin{description}
			\nItemXY{ act1 }{ general\_EV\_Hout :|  ((general\_EV = TRUE \land   general\_EV\_Hout' = Hin) \lor  (general\_EV = FALSE \land  general\_EV\_Hout' = 0)		\\\hspace*{1,2 cm}  \lor  (A\_Switch\_Out = TRUE \land   general\_EV\_Hout' = Hin) \lor  (A\_Switch\_Out = FALSE \land   general\_EV\_Hout' = 0)) }{ 		\\\hspace*{1,4 cm}  pass the current value of hydraulic input port (Hin) to hydraulic output port (Hout) }
			\nItemX{ act2 }{ close\_EV\_Hout :|  ((close\_EV = TRUE \land  close\_EV\_Hout' = Hin) \lor  (close\_EV = FALSE \land  close\_EV\_Hout' = 0)) }
			\nItemX{ act3 }{ open\_EV\_Hout :|  ((open\_EV = TRUE \land  open\_EV\_Hout' = Hin) \lor  (open\_EV = FALSE \land  open\_EV\_Hout' = 0)) }
			\nItemX{ act4 }{ extend\_EV\_Hout :|  ((extend\_EV = TRUE \land  extend\_EV\_Hout' = Hin) \lor  (extend\_EV = FALSE \land  extend\_EV\_Hout' = 0)) }
			\nItemX{ act5 }{ retract\_EV\_Hout :|  ((retract\_EV = TRUE \land  retract\_EV\_Hout' = Hin) \lor  (retract\_EV = FALSE \land  retract\_EV\_Hout' = 0)) }
			\nItem{ act6 }{ state :=  cylinder }
			\end{description}
		\EndAct
		\end{description}
	\EVT {CylinderMovingOrStop}\cmt{		\\\hspace*{4,6 cm}  Cylinder Moving or Stop according to the output of hydraulic circuit }
		\begin{description}
		\WhenGrd
			\begin{description}
			\nItem{ grd1 }{ state = cylinder }
			\end{description}
		\ThenAct
			\begin{description}
			\nItem{ act1 }{ SGCylinder :|  ((SGCylinder' = \{ a\mapsto b |  a\in  GEARS\cprod \{ GCYF,GCYR,GCYL\}  \land  b=MOVING\}  \land  extend\_EV\_Hout = Hin ) \lor  		\\\hspace*{1,4 cm}  (SGCylinder' = \{ a\mapsto b |  a\in  GEARS\cprod \{ GCYF,GCYR,GCYL\}  \land  b=STOP\}  \land  extend\_EV\_Hout = 0 ) \lor 		\\\hspace*{1,4 cm}  (SGCylinder' = \{ a\mapsto b |  a\in  GEARS\cprod \{ GCYF,GCYR,GCYL\}  \land  b=MOVING\}  \land  retract\_EV\_Hout = Hin ) \lor 		\\\hspace*{1,4 cm}  (SGCylinder' = \{ a\mapsto b |  a\in  GEARS\cprod \{ GCYF,GCYR,GCYL\}  \land  b=STOP\}  \land  retract\_EV\_Hout = 0 )) }
			\nItem{ act2 }{ SDCylinder :|  ((SDCylinder' = \{ a\mapsto b |  a\in  DOORS\cprod \{ DCYF,DCYR,DCYL\}  \land  b=MOVING\}  \land  open\_EV\_Hout = Hin) \lor  		\\\hspace*{1,4 cm}  (SDCylinder' = \{ a\mapsto b |  a\in  DOORS\cprod \{ DCYF,DCYR,DCYL\}  \land  b=STOP\}  \land  open\_EV\_Hout = 0) \lor 		\\\hspace*{1,4 cm}  (SDCylinder' = \{ a\mapsto b |  a\in  DOORS\cprod \{ DCYF,DCYR,DCYL\}  \land  b=MOVING\}  \land  close\_EV\_Hout = Hin) \lor 		\\\hspace*{1,4 cm}  (SDCylinder' = \{ a\mapsto b |  a\in  DOORS\cprod \{ DCYF,DCYR,DCYL\} \land  b=STOP\}  \land  close\_EV\_Hout = 0)) }
			\nItem{ act3 }{ state :=  computing }
			\end{description}
		\EndAct
		\end{description}
	\EVT {Failure\_Detection}
	\EXTD {Failure\_Detection}
		\begin{description}
		\BeginAct
			\begin{description}
			\nItemX{ act1 }{ anomaly :=  TRUE }
			\end{description}
		\EndAct
		\end{description}
\END
\end{description}

\section{M7}
\label{sec:M7}

\begin{description}
\BTitle{M7}{27Jan2014}{10:44:59 AM}
\MACHINE{M7}\cmt{		\\\hspace*{1 cm}  Failure Modelling		\\\hspace*{0,8 cm}  Generic Monitoring failure		\\\hspace*{0,8 cm}  Failure detection is added for doors and gears motion monitioring (Page 17)		\\\hspace*{0,8 cm}  Analogical Switch Monitioring failure (Page 16)		\\\hspace*{0,8 cm}  Pressure Sensor Monitioring faliure (Page 16)		\\\hspace*{0,8 cm}  But timing requirements can be added only in last. }
\REFINES{M6}
\SEES{C1}
\VARIABLES
	\begin{description}
		\Item{ dstate }
		\Item{ lstate }
		\Item{ phase }
		\Item{ button }
		\Item{ p }
		\Item{ l }
		\Item{ i }
		\Item{ gstate }
		\Item{ handle }
		\Item{ analogical\_switch }
		\Item{ gear\_extended }
		\Item{ gear\_retracted }
		\Item{ gear\_shock\_absorber }
		\Item{ door\_closed }
		\Item{ door\_open }
		\Item{ circuit\_pressurized }
		\Item{ general\_EV }
		\Item{ close\_EV }
		\Item{ retract\_EV }
		\Item{ extend\_EV }
		\Item{ open\_EV }
		\Item{ gears\_locked\_down }
		\Item{ gears\_man }
		\Item{ anomaly }
		\Item{ general\_EV\_func }
		\Item{ close\_EV\_func }
		\Item{ retract\_EV\_func }
		\Item{ extend\_EV\_func }
		\Item{ open\_EV\_func }
		\Item{ gears\_locked\_down\_func }
		\Item{ gears\_man\_func }
		\Item{ anomaly\_func }
		\Item{ general\_EV\_Hout }
		\Item{ close\_EV\_Hout }
		\Item{ retract\_EV\_Hout }
		\Item{ extend\_EV\_Hout }
		\Item{ open\_EV\_Hout }
		\Item{ A\_Switch\_Out }
		\ItemY{ SDCylinder }{State of Door Cylinder}
		\ItemY{ SGCylinder }{State of Gear Cylinder}
		\Item{ state }
	\end{description}
\EVENTS
	\INITIALISATION
		\\\textit{extended}
		\begin{description}
		\BeginAct
			\begin{description}
			\nItemX{ act1 }{ button :=  DOWN }
			\nItemX{ act2 }{ phase :=  haltdown }
			\nItemXY{ act3 }{ dstate :|  (dstate'\in  DOORS \tfun  SDOORS \land  dstate'=\{ a\mapsto b|  a \in  DOORS \land  b=CLOSED\} ) }{ 		\\\hspace*{1,4 cm}  missing elements of the invariant }
			\nItemX{ act4 }{ lstate :=  \{ a\mapsto b| a\in DOORS\land  b=LOCKED\}  }
			\nItemX{ act5 }{ p :=  R }
			\nItemX{ act6 }{ l :=  R }
			\nItemX{ act7 }{ i :=  R }
			\nItemX{ act8 }{ gstate :|  (gstate' \in  GEARS \tfun  SGEARS \land  gstate'=\{ a\mapsto b |  a \in  GEARS  \land  b=EXTENDED\} ) }
			\nItemX{ act14 }{ handle :\in  1\upto 3 \tfun  \{ DOWN\}  }
			\nItemX{ act15 }{ analogical\_switch :\in  1\upto 3 \tfun  \{ open\}  }
			\nItemX{ act16 }{ gear\_extended :\in  1\upto 3 \tfun  (GEARS \tfun  \{ TRUE\} ) }
			\nItemX{ act17 }{ gear\_retracted :\in  1\upto 3 \tfun  (GEARS \tfun  \{ FALSE\} ) }
			\nItemX{ act18 }{ gear\_shock\_absorber :\in  1\upto 3 \tfun  \{ ground\}  }
			\nItemX{ act19 }{ door\_closed :\in  1\upto 3 \tfun  (DOORS \tfun  \{ TRUE\} ) }
			\nItemX{ act20 }{ door\_open :\in  1\upto 3 \tfun  (DOORS \tfun  \{ FALSE\} ) }
			\nItemX{ act21 }{ circuit\_pressurized :\in  1\upto 3 \tfun  \{ FALSE\}  }
			\nItemX{ act22 }{ general\_EV :=  FALSE }
			\nItemX{ act23 }{ close\_EV :=  TRUE }
			\nItemX{ act24 }{ retract\_EV :=  FALSE }
			\nItemX{ act25 }{ extend\_EV :=  TRUE }
			\nItemX{ act27 }{ open\_EV :=  FALSE }
			\nItemX{ act28 }{ gears\_locked\_down :=  TRUE }
			\nItemX{ act29 }{ gears\_man :=  FALSE }
			\nItemX{ act30 }{ anomaly :=  FALSE }
			\nItemX{ act31 }{ general\_EV\_func :\in  (1\upto 3 \tfun  POSITIONS) \cprod  (1\upto 3 \tfun  A\_Switch) \cprod  (1\upto 3 \tfun  (GEARS \tfun  BOOL)) \cprod  (1\upto 3 \tfun  (GEARS \tfun  BOOL)) \cprod  (1\upto 3 \tfun  GEAR\_ABSORBER) \cprod  (1\upto 3 \tfun  (DOORS \tfun  BOOL)) \cprod  (1\upto 3 \tfun  (DOORS \tfun  BOOL)) \cprod  (1\upto 3 \tfun  BOOL) \tfun  BOOL }
			\nItemX{ act32 }{ close\_EV\_func :\in  (1\upto 3 \tfun  POSITIONS) \cprod  (1\upto 3 \tfun  A\_Switch) \cprod  (1\upto 3 \tfun  (GEARS \tfun  BOOL)) \cprod  (1\upto 3 \tfun  (GEARS \tfun  BOOL)) \cprod  (1\upto 3 \tfun  GEAR\_ABSORBER) \cprod  (1\upto 3 \tfun  (DOORS \tfun  BOOL)) \cprod  (1\upto 3 \tfun  (DOORS \tfun  BOOL)) \cprod  (1\upto 3 \tfun  BOOL) \tfun  BOOL }
			\nItemX{ act33 }{ retract\_EV\_func :\in  (1\upto 3 \tfun  POSITIONS) \cprod  (1\upto 3 \tfun  A\_Switch) \cprod  (1\upto 3 \tfun  (GEARS \tfun  BOOL)) \cprod  (1\upto 3 \tfun  (GEARS \tfun  BOOL)) \cprod  (1\upto 3 \tfun  GEAR\_ABSORBER) \cprod  (1\upto 3 \tfun  (DOORS \tfun  BOOL)) \cprod  (1\upto 3 \tfun  (DOORS \tfun  BOOL)) \cprod  (1\upto 3 \tfun  BOOL) \tfun  BOOL }
			\nItemX{ act34 }{ extend\_EV\_func :\in  (1\upto 3 \tfun  POSITIONS) \cprod  (1\upto 3 \tfun  A\_Switch) \cprod  (1\upto 3 \tfun  (GEARS \tfun  BOOL)) \cprod  (1\upto 3 \tfun  (GEARS \tfun  BOOL)) \cprod  (1\upto 3 \tfun  GEAR\_ABSORBER) \cprod  (1\upto 3 \tfun  (DOORS \tfun  BOOL)) \cprod  (1\upto 3 \tfun  (DOORS \tfun  BOOL)) \cprod  (1\upto 3 \tfun  BOOL) \tfun  BOOL }
			\nItemX{ act35 }{ open\_EV\_func :\in  (1\upto 3 \tfun  POSITIONS) \cprod  (1\upto 3 \tfun  A\_Switch) \cprod  (1\upto 3 \tfun  (GEARS \tfun  BOOL)) \cprod  (1\upto 3 \tfun  (GEARS \tfun  BOOL)) \cprod  (1\upto 3 \tfun  GEAR\_ABSORBER) \cprod  (1\upto 3 \tfun  (DOORS \tfun  BOOL)) \cprod  (1\upto 3 \tfun  (DOORS \tfun  BOOL)) \cprod  (1\upto 3 \tfun  BOOL) \tfun  BOOL }
			\nItemX{ act36 }{ gears\_locked\_down\_func :\in  (1\upto 3 \tfun  POSITIONS) \cprod  (1\upto 3 \tfun  A\_Switch) \cprod  (1\upto 3 \tfun  (GEARS \tfun  BOOL)) \cprod  (1\upto 3 \tfun  (GEARS \tfun  BOOL)) \cprod  (1\upto 3 \tfun  GEAR\_ABSORBER) \cprod  (1\upto 3 \tfun  (DOORS \tfun  BOOL)) \cprod  (1\upto 3 \tfun  (DOORS \tfun  BOOL)) \cprod  (1\upto 3 \tfun  BOOL) \tfun  BOOL }
			\nItemX{ act37 }{ gears\_man\_func :\in  (1\upto 3 \tfun  POSITIONS) \cprod  (1\upto 3 \tfun  A\_Switch) \cprod  (1\upto 3 \tfun  (GEARS \tfun  BOOL)) \cprod  (1\upto 3 \tfun  (GEARS \tfun  BOOL)) \cprod  (1\upto 3 \tfun  GEAR\_ABSORBER) \cprod  (1\upto 3 \tfun  (DOORS \tfun  BOOL)) \cprod  (1\upto 3 \tfun  (DOORS \tfun  BOOL)) \cprod  (1\upto 3 \tfun  BOOL) \tfun  BOOL }
			\nItemX{ act38 }{ anomaly\_func :\in  (1\upto 3 \tfun  POSITIONS) \cprod  (1\upto 3 \tfun  A\_Switch) \cprod  (1\upto 3 \tfun  (GEARS \tfun  BOOL)) \cprod  (1\upto 3 \tfun  (GEARS \tfun  BOOL)) \cprod  (1\upto 3 \tfun  GEAR\_ABSORBER) \cprod  (1\upto 3 \tfun  (DOORS \tfun  BOOL)) \cprod  (1\upto 3 \tfun  (DOORS \tfun  BOOL)) \cprod  (1\upto 3 \tfun  BOOL) \tfun  BOOL }
			\nItemX{ act39 }{ A\_Switch\_Out :=  FALSE }
			\nItemX{ act40 }{ close\_EV\_Hout :=  0 }
			\nItemX{ act41 }{ retract\_EV\_Hout :=  0 }
			\nItemX{ act42 }{ extend\_EV\_Hout :=  0 }
			\nItemX{ act43 }{ open\_EV\_Hout :=  0 }
			\nItemX{ act44 }{ general\_EV\_Hout :=  0 }
			\nItemX{ act45 }{ SDCylinder :\in  DOORS \cprod  \{ DCYF,DCYR,DCYL\}  \tfun   \{ STOP\}  }
			\nItemX{ act46 }{ SGCylinder :\in  GEARS \cprod  \{ GCYF,GCYR,GCYL\}  \tfun  \{ STOP\}   }
			\nItemX{ act26 }{ state :=  computing }
			\end{description}
		\EndAct
		\end{description}
	\EVT {opening\_doors\_DOWN}
	\EXTD {opening\_doors\_DOWN}
		\begin{description}
		\WhenGrd
			\begin{description}
			\nItemX{ grd1 }{ dstate[DOORS]= \{ CLOSED\}  }
			\nItemX{ grd5 }{ lstate[DOORS]=\{ UNLOCKED\}  }
			\nItemX{ grd7 }{ phase=movingdown }
			\nItemX{ grd8 }{ p=R }
			\nItemX{ grd9 }{ l=R }
			\nItemX{ grd10 }{ door\_open = \{ a\mapsto b |  a \in  1\upto 3 \land  b \in  DOORS \tfun  \{ FALSE\} \}   }
			\nItemX{ grd11 }{ door\_closed = \{ a\mapsto b |  a \in  1\upto 3 \land  b \in  DOORS \tfun  \{ FALSE\} \}   }
			\nItemX{ grd12 }{ \forall x\qdot x\in 1\upto 3 \limp  handle(x)=button }
			\nItemX{ grd3 }{ SDCylinder = \{ a\mapsto b |  a\in  DOORS\cprod CYLINDER \land  b=MOVING\}  }
			\nItem{ grd13 }{ anomaly = FALSE }
			\end{description}
		\ThenAct
			\begin{description}
			\nItemX{ act1 }{ dstate :=  \{ a\mapsto b|  a \in  DOORS \land  b=OPEN\}  }
			\nItemX{ act2 }{ p:= E }
			\nItemX{ act3 }{ door\_open :\in  1\upto 3 \tfun  (DOORS \tfun  \{ TRUE\} ) }
			\end{description}
		\EndAct
		\end{description}
	\EVT {opening\_doors\_UP}
	\EXTD {opening\_doors\_UP}
		\begin{description}
		\WhenGrd
			\begin{description}
			\nItemX{ grd1 }{ dstate[DOORS]= \{ CLOSED\}  }
			\nItemX{ grd4 }{ lstate[DOORS]=\{ UNLOCKED\}  }
			\nItemX{ grd5 }{ phase= movingup }
			\nItemX{ grd6 }{ p=E }
			\nItemX{ grd7 }{ l=E }
			\nItemX{ grd8 }{ door\_open = \{ a\mapsto b |  a \in  1\upto 3 \land  b \in  DOORS \tfun  \{ FALSE\} \}   }
			\nItemX{ grd9 }{ door\_closed = \{ a\mapsto b |  a \in  1\upto 3 \land  b \in  DOORS \tfun  \{ FALSE\} \}   }
			\nItemX{ grd10 }{ \forall x\qdot x\in 1\upto 3 \limp  handle(x)=button }
			\nItemX{ grd3 }{ SDCylinder = \{ a\mapsto b |  a\in  DOORS\cprod CYLINDER \land  b=MOVING\}  }
			\nItem{ grd11 }{ anomaly = FALSE }
			\end{description}
		\ThenAct
			\begin{description}
			\nItemX{ act1 }{ dstate :=  \{ a\mapsto b|  a \in  DOORS \land  b=OPEN\}  }
			\nItemX{ act2 }{ p:= R }
			\nItemX{ act3 }{ door\_open:\in  1\upto 3 \tfun  (DOORS \tfun  \{ TRUE\} )  }
			\end{description}
		\EndAct
		\end{description}
	\EVT {closing\_doors\_UP}
	\EXTD {closing\_doors\_UP}
		\begin{description}
		\AnyPrm
			\begin{description}
			\ItemX{f }
			\end{description}
		\WhereGrd
			\begin{description}
			\nItemX{ grd1 }{ dstate[DOORS]=\{ OPEN\}  }
			\nItemX{ grd3 }{ f \in  DOORS \tfun  SDOORS }
			\nItemX{ grd4 }{ \forall e\qdot  e \in  DOORS \limp  f(e)=CLOSED }
			\nItemX{ grd5 }{ phase=movingup }
			\nItemX{ grd6 }{ p=R }
			\nItemX{ grd7 }{ gstate[GEARS]=\{ RETRACTED\}  }
			\nItemX{ grd8 }{ \forall x\qdot x\in 1\upto 3 \limp  handle(x)=button }
			\nItem{ grd9 }{ anomaly = FALSE }
			\end{description}
		\ThenAct
			\begin{description}
			\nItemX{ act1 }{ dstate:= f }
			\end{description}
		\EndAct
		\end{description}
	\EVT {closing\_doors\_DOWN}
	\EXTD {closing\_doors\_DOWN}
		\begin{description}
		\AnyPrm
			\begin{description}
			\ItemX{f }
			\end{description}
		\WhereGrd
			\begin{description}
			\nItemX{ grd1 }{ dstate[DOORS]=\{ OPEN\}  }
			\nItemX{ grd3 }{ f \in  DOORS \tfun  SDOORS }
			\nItemX{ grd4 }{ \forall e\qdot  e \in  DOORS \limp  f(e)=CLOSED }
			\nItemX{ grd5 }{ phase=movingdown }
			\nItemX{ grd6 }{ p=E }
			\nItemX{ grd7 }{ gstate[GEARS]=\{ EXTENDED\}  }
			\nItemX{ grd8 }{ \forall x\qdot x\in 1\upto 3 \limp  handle(x)=button }
			\nItem{ grd9 }{ anomaly = FALSE }
			\end{description}
		\ThenAct
			\begin{description}
			\nItemX{ act1 }{ dstate:= f }
			\end{description}
		\EndAct
		\end{description}
	\EVT {unlocking\_UP}
	\EXTD {unlocking\_UP}
		\begin{description}
		\WhenGrd
			\begin{description}
			\nItemX{ grd3 }{ lstate[DOORS]=\{ LOCKED\}  }
			\nItemX{ grd4 }{ phase=movingup }
			\nItemX{ grd5 }{ l=E }
			\nItemX{ grd6 }{ p=E }
			\nItemX{ grd7 }{ i=E }
			\nItemX{ grd8 }{ door\_open = \{ a\mapsto b |  a \in  1\upto 3 \land  b \in  DOORS \tfun  \{ FALSE\} \}   }
			\nItemX{ grd9 }{ door\_closed = \{ a\mapsto b |  a \in  1\upto 3 \land  b \in  DOORS \tfun  \{ TRUE\} \}   }
			\nItemX{ grd10 }{ \forall x\qdot x\in 1\upto 3 \limp  handle(x)=button }
			\nItem{ grd11 }{ anomaly = FALSE }
			\end{description}
		\ThenAct
			\begin{description}
			\nItemX{ act1 }{ lstate:= \{ a\mapsto b| a\in DOORS \land  b=UNLOCKED\}  }
			\nItemX{ act2 }{ door\_closed :\in  1\upto 3 \tfun  (DOORS \tfun  \{ FALSE\} )  }
			\end{description}
		\EndAct
		\end{description}
	\EVT {locking\_UP}
	\EXTD {locking\_UP}
		\begin{description}
		\WhenGrd
			\begin{description}
			\nItemX{ grd3 }{ dstate[DOORS]=\{ CLOSED\}  }
			\nItemX{ grd4 }{ phase=movingup }
			\nItemX{ grd5 }{ lstate[DOORS]=\{ UNLOCKED\}  }
			\nItemX{ grd6 }{ p=R }
			\nItemX{ grd7 }{ l=E }
			\nItemX{ grd9 }{ door\_open = \{ a\mapsto b |  a \in  1\upto 3 \land  b \in  DOORS \tfun  \{ FALSE\} \}   }
			\nItemX{ grd10 }{ door\_closed = \{ a\mapsto b |  a \in  1\upto 3 \land  b \in  DOORS \tfun  \{ FALSE\} \}   }
			\nItemX{ grd11 }{ \forall x\qdot x\in 1\upto 3 \limp  handle(x)=button }
			\nItemX{ grd8 }{ SDCylinder = \{ a\mapsto b |  a\in  DOORS\cprod CYLINDER \land  b=STOP\}  }
			\nItem{ grd12 }{ anomaly = FALSE }
			\end{description}
		\ThenAct
			\begin{description}
			\nItemX{ act1 }{ lstate:= \{ a\mapsto b| a\in DOORS \land  b=LOCKED\}  }
			\nItemX{ act3 }{ phase:= haltup }
			\nItemXY{ act4 }{ l:= R }{ 		\\\hspace*{1,4 cm}  added by D Mery }
			\nItemX{ act44 }{ door\_closed :\in  1\upto 3 \tfun  (DOORS \tfun  \{ TRUE\} )  }
			\end{description}
		\EndAct
		\end{description}
	\EVT {unlocking\_DOWN}
	\EXTD {unlocking\_DOWN}
		\begin{description}
		\WhenGrd
			\begin{description}
			\nItemX{ grd3 }{ lstate[DOORS]=\{ LOCKED\}  }
			\nItemX{ grd4 }{ phase=movingdown }
			\nItemX{ grd5 }{ l=R }
			\nItemX{ grd6 }{ p=R }
			\nItemX{ grd7 }{ i=R }
			\nItemX{ grd8 }{ door\_open = \{ a\mapsto b |  a \in  1\upto 3 \land  b \in  DOORS \tfun  \{ FALSE\} \}   }
			\nItemX{ grd9 }{ door\_closed = \{ a\mapsto b |  a \in  1\upto 3 \land  b \in  DOORS \tfun  \{ TRUE\} \}   }
			\nItemX{ grd10 }{ \forall x\qdot x\in 1\upto 3 \limp  handle(x)=button }
			\nItem{ grd11 }{ anomaly = FALSE }
			\end{description}
		\ThenAct
			\begin{description}
			\nItemX{ act1 }{ lstate:= \{ a\mapsto b| a\in DOORS \land  b=UNLOCKED\}  }
			\nItemX{ act2 }{ door\_closed :\in  1\upto 3 \tfun  (DOORS \tfun  \{ FALSE\} ) }
			\end{description}
		\EndAct
		\end{description}
	\EVT {locking\_DOWN}
	\EXTD {locking\_DOWN}
		\begin{description}
		\WhenGrd
			\begin{description}
			\nItemX{ grd1 }{ dstate[DOORS]=\{ CLOSED\}  }
			\nItemX{ grd2 }{ phase=movingdown }
			\nItemX{ grd3 }{ lstate[DOORS]=\{ UNLOCKED\}  }
			\nItemX{ grd4 }{ p=E }
			\nItemX{ grd5 }{ l=R }
			\nItemX{ grd7 }{ door\_open = \{ a\mapsto b |  a \in  1\upto 3 \land  b \in  DOORS \tfun  \{ FALSE\} \}   }
			\nItemX{ grd8 }{ door\_closed = \{ a\mapsto b |  a \in  1\upto 3 \land  b \in  DOORS \tfun  \{ FALSE\} \}   }
			\nItemX{ grd9 }{ \forall x\qdot x\in 1\upto 3 \limp  handle(x)=button }
			\nItemX{ grd6 }{ SDCylinder = \{ a\mapsto b |  a\in  DOORS\cprod CYLINDER \land  b=STOP\}  }
			\nItem{ grd10 }{ anomaly = FALSE }
			\end{description}
		\ThenAct
			\begin{description}
			\nItemX{ act1 }{ lstate:= \{ a\mapsto b| a\in DOORS \land  b = LOCKED\}  }
			\nItemX{ act3 }{ phase:= haltdown }
			\nItemX{ act4 }{ l:= E }
			\nItemX{ act5 }{ door\_closed :\in  1\upto 3 \tfun  (DOORS \tfun  \{ TRUE\} )  }
			\end{description}
		\EndAct
		\end{description}
	\EVT {PD1}
	\EXTD {PD1}
		\begin{description}
		\WhenGrd
			\begin{description}
			\nItemX{ grd1 }{ button=UP }
			\nItemX{ grd2 }{ phase=haltup }
			\nItemX{ grd3 }{ \forall x\qdot x\in 1\upto 3 \limp  handle(x)=DOWN }
			\end{description}
		\ThenAct
			\begin{description}
			\nItemX{ act1 }{ phase:= movingdown }
			\nItemX{ act2 }{ button:= DOWN }
			\nItemX{ act3 }{ l:= R }
			\nItemX{ act4 }{ p:= R }
			\nItemX{ act5 }{ i:= R }
			\end{description}
		\EndAct
		\end{description}
	\EVT {PU1}
	\EXTD {PU1}
		\begin{description}
		\WhenGrd
			\begin{description}
			\nItemX{ grd1 }{ button=DOWN }
			\nItemX{ grd2 }{ phase=haltdown }
			\nItemX{ grd3 }{ \forall x\qdot x\in 1\upto 3 \limp  handle(x)=UP }
			\end{description}
		\ThenAct
			\begin{description}
			\nItemX{ act1 }{ phase:= movingup }
			\nItemX{ act2 }{ button:= UP }
			\nItemX{ act3 }{ l:= E }
			\nItemX{ act4 }{ p:= E }
			\nItemX{ act5 }{ i:= E }
			\end{description}
		\EndAct
		\end{description}
	\EVT {PU2}
	\EXTD {PU2}
		\begin{description}
		\WhenGrd
			\begin{description}
			\nItemX{ grd1 }{ l=R }
			\nItemX{ grd2 }{ p=R }
			\nItemX{ grd3 }{ phase=movingdown }
			\nItemX{ grd4 }{ button=DOWN }
			\nItemX{ grd5 }{ i=R }
			\nItemX{ grd6 }{ lstate[DOORS]=\{ LOCKED\}  }
			\nItemX{ grd7 }{ door\_open = \{ a\mapsto b |  a \in  1\upto 3 \land  b \in  DOORS \tfun  \{ FALSE\} \}   }
			\nItemX{ grd8 }{ door\_closed = \{ a\mapsto b |  a \in  1\upto 3 \land  b \in  DOORS \tfun  \{ TRUE\} \}   }
			\nItemX{ grd9 }{ \forall x\qdot x\in 1\upto 3 \limp  handle(x)=UP }
			\end{description}
		\ThenAct
			\begin{description}
			\nItemX{ act1 }{ phase:= movingup }
			\nItemX{ act4 }{ button:=   UP }
			\nItemX{ act5 }{ l:= E }
			\nItemX{ act6 }{ p:= E }
			\nItemX{ act7 }{ i:= R }
			\end{description}
		\EndAct
		\end{description}
	\EVT {CompletePU2}
	\EXTD {CompletePU2}
		\begin{description}
		\WhenGrd
			\begin{description}
			\nItemX{ grd1 }{ phase=movingup }
			\nItemX{ grd2 }{ button=UP }
			\nItemX{ grd3 }{ l=E }
			\nItemX{ grd4 }{ p=E }
			\nItemX{ grd5 }{ i=R }
			\end{description}
		\ThenAct
			\begin{description}
			\nItemX{ act1 }{ phase:= haltup }
			\end{description}
		\EndAct
		\end{description}
	\EVT {PU3}
	\EXTD {PU3}
		\begin{description}
		\WhenGrd
			\begin{description}
			\nItemX{ grd1 }{ dstate[DOORS]=\{ CLOSED\}  }
			\nItemX{ grd2 }{ lstate[DOORS]=\{ UNLOCKED\}  }
			\nItemX{ grd3 }{ phase = movingdown }
			\nItemX{ grd4 }{ p=R }
			\nItemX{ grd5 }{ l=R }
			\nItemX{ grd6 }{ button=DOWN }
			\nItemX{ grd7 }{ door\_open = \{ a\mapsto b |  a \in  1\upto 3 \land  b \in  DOORS \tfun  \{ FALSE\} \}   }
			\nItemX{ grd8 }{ door\_closed = \{ a\mapsto b |  a \in  1\upto 3 \land  b \in  DOORS \tfun  \{ FALSE\} \}   }
			\nItemX{ grd9 }{ \forall x\qdot x\in 1\upto 3 \limp  handle(x)=UP }
			\end{description}
		\ThenAct
			\begin{description}
			\nItemX{ act1 }{ phase:= movingup }
			\nItemX{ act2 }{ p:= R }
			\nItemX{ act3 }{ l:= E }
			\nItemX{ act4 }{ button:= UP }
			\end{description}
		\EndAct
		\end{description}
	\EVT {PU4}
	\EXTD {PU4}
		\begin{description}
		\WhenGrd
			\begin{description}
			\nItemX{ grd1 }{ dstate[DOORS]=\{ OPEN\}  }
			\nItemX{ grd2 }{ phase=movingdown }
			\nItemX{ grd3 }{ p=E }
			\nItemX{ grd4 }{ button=DOWN }
			\nItemX{ grd7 }{ \forall x\qdot x\in 1\upto 3 \limp  handle(x)=UP }
			\end{description}
		\ThenAct
			\begin{description}
			\nItemX{ act1 }{ phase:= movingup }
			\nItemX{ act2 }{ p:= R }
			\nItemX{ act3 }{ button:= UP }
			\nItemX{ act4 }{ i:= E }
			\nItemX{ act5 }{ l:= E }
			\end{description}
		\EndAct
		\end{description}
	\EVT {PU5}
	\EXTD {PU5}
		\begin{description}
		\WhenGrd
			\begin{description}
			\nItemX{ grd1 }{ dstate[DOORS]=\{ CLOSED\}  }
			\nItemX{ grd2 }{ phase=movingdown }
			\nItemX{ grd3 }{ p=E }
			\nItemX{ grd4 }{ button=DOWN }
			\nItemX{ grd5 }{ lstate[DOORS]=\{ UNLOCKED\}  }
			\nItemX{ grd6 }{ door\_open = \{ a\mapsto b |  a \in  1\upto 3 \land  b \in  DOORS \tfun  \{ FALSE\} \}   }
			\nItemX{ grd7 }{ door\_closed = \{ a\mapsto b |  a \in  1\upto 3 \land  b \in  DOORS \tfun  \{ FALSE\} \}   }
			\nItemX{ grd8 }{ \forall x\qdot x\in 1\upto 3 \limp  handle(x)=UP }
			\end{description}
		\ThenAct
			\begin{description}
			\nItemX{ act1 }{ phase:= movingup }
			\nItemX{ act3 }{ button:= UP }
			\nItemX{ act4 }{ i:= E }
			\nItemX{ act5 }{ l:= E }
			\end{description}
		\EndAct
		\end{description}
	\EVT {PD2}
	\EXTD {PD2}
		\begin{description}
		\WhenGrd
			\begin{description}
			\nItemX{ grd1 }{ l=E }
			\nItemX{ grd2 }{ p=E }
			\nItemX{ grd3 }{ phase=movingup }
			\nItemX{ grd4 }{ i=E }
			\nItemX{ grd5 }{ lstate[DOORS]=\{ LOCKED\}  }
			\nItemX{ grd6 }{ \forall x\qdot x\in 1\upto 3 \limp  handle(x)=DOWN }
			\end{description}
		\ThenAct
			\begin{description}
			\nItemX{ act1 }{ phase:= movingdown }
			\nItemX{ act2 }{ button:= DOWN }
			\nItemX{ act3 }{ l:= R }
			\nItemX{ act4 }{ p:= R }
			\nItemX{ act5 }{ i:= E }
			\end{description}
		\EndAct
		\end{description}
	\EVT {CompletePD2}
	\EXTD {CompletePD2}
		\begin{description}
		\WhenGrd
			\begin{description}
			\nItemX{ grd1 }{ phase=movingdown }
			\nItemX{ grd2 }{ button=DOWN }
			\nItemX{ grd3 }{ l=R }
			\nItemX{ grd4 }{ p=R }
			\nItemX{ grd5 }{ i=E }
			\end{description}
		\ThenAct
			\begin{description}
			\nItemX{ act1 }{ phase:= haltdown }
			\end{description}
		\EndAct
		\end{description}
	\EVT {PD3}
	\EXTD {PD3}
		\begin{description}
		\WhenGrd
			\begin{description}
			\nItemX{ grd1 }{ dstate[DOORS]=\{ CLOSED\}  }
			\nItemX{ grd2 }{ lstate[DOORS]=\{ UNLOCKED\}  }
			\nItemX{ grd3 }{ phase=movingup }
			\nItemX{ grd4 }{ p=E }
			\nItemX{ grd5 }{ l=E }
			\nItemX{ grd6 }{ button=UP }
			\nItemX{ grd7 }{ door\_open = \{ a\mapsto b |  a \in  1\upto 3 \land  b \in  DOORS \tfun  \{ FALSE\} \}   }
			\nItemX{ grd8 }{ door\_closed = \{ a\mapsto b |  a \in  1\upto 3 \land  b \in  DOORS \tfun  \{ FALSE\} \}   }
			\nItemX{ grd9 }{ \forall x\qdot x\in 1\upto 3 \limp  handle(x)=DOWN }
			\end{description}
		\ThenAct
			\begin{description}
			\nItemX{ act1 }{ phase:= movingdown }
			\nItemX{ act2 }{ p:= E }
			\nItemX{ act3 }{ l:= R }
			\nItemX{ act4 }{ button:= DOWN }
			\end{description}
		\EndAct
		\end{description}
	\EVT {PD4}
	\EXTD {PD4}
		\begin{description}
		\WhenGrd
			\begin{description}
			\nItemX{ grd1 }{ dstate[DOORS]=\{ OPEN\}  }
			\nItemX{ grd2 }{ phase=movingup }
			\nItemX{ grd3 }{ p=R }
			\nItemX{ grd4 }{ button=UP }
			\nItemX{ grd6 }{ \forall x\qdot x\in 1\upto 3 \limp  handle(x)=DOWN }
			\end{description}
		\ThenAct
			\begin{description}
			\nItemX{ act1 }{ phase:= movingdown }
			\nItemX{ act2 }{ p:= E }
			\nItemX{ act3 }{ button:= DOWN }
			\nItemX{ act4 }{ i:= R }
			\nItemX{ act5 }{ l:= R }
			\end{description}
		\EndAct
		\end{description}
	\EVT {PD5}
	\EXTD {PD5}
		\begin{description}
		\WhenGrd
			\begin{description}
			\nItemX{ grd1 }{ dstate[DOORS]=\{ CLOSED\}  }
			\nItemX{ grd2 }{ phase=movingup }
			\nItemX{ grd3 }{ p=R }
			\nItemX{ grd4 }{ button=UP }
			\nItemX{ grd5 }{ lstate[DOORS]=\{ UNLOCKED\}  }
			\nItemX{ grd6 }{ door\_open = \{ a\mapsto b |  a \in  1\upto 3 \land  b \in  DOORS \tfun  \{ FALSE\} \}   }
			\nItemX{ grd7 }{ door\_closed = \{ a\mapsto b |  a \in  1\upto 3 \land  b \in  DOORS \tfun  \{ FALSE\} \}   }
			\nItemX{ grd8 }{ \forall x\qdot x\in 1\upto 3 \limp  handle(x)=DOWN }
			\end{description}
		\ThenAct
			\begin{description}
			\nItemX{ act1 }{ phase :=  movingdown }
			\nItemX{ act2 }{ button:= DOWN }
			\nItemX{ act3 }{ i:= R }
			\nItemX{ act4 }{ l:= R }
			\end{description}
		\EndAct
		\end{description}
	\EVT {retracting\_gears}
	\EXTD {retracting\_gears}
		\begin{description}
		\WhenGrd
			\begin{description}
			\nItemX{ grd1 }{ dstate[DOORS]=\{ OPEN\}  }
			\nItemX{ grd2 }{ gstate[GEARS]=\{ EXTENDED\}  }
			\nItemX{ grd3 }{ p=R }
			\nItemX{ grd6 }{ gear\_extended = \{ a\mapsto b |  a \in  1\upto 3 \land  b \in  GEARS \tfun  \{ TRUE\} \}   }
			\nItemX{ grd7 }{ gear\_retracted = \{ a\mapsto b |  a \in  1\upto 3 \land  b \in  GEARS \tfun  \{ FALSE\} \}   }
			\nItemX{ grd8 }{ gear\_shock\_absorber = \{ a\mapsto b |  a \in  1\upto 3 \land  b = ground\}   }
			\nItemX{ grd9 }{ \forall x\qdot x\in 1\upto 3 \limp  handle(x)=button }
			\nItemX{ grd5 }{ SGCylinder = \{ a\mapsto b |  a\in  GEARS\cprod CYLINDER \land  b=MOVING\}  }
			\end{description}
		\ThenAct
			\begin{description}
			\nItemX{ act1 }{ gstate :=  \{ a\mapsto b| a \in  GEARS \land  b=RETRACTING\}  }
			\nItemX{ act2 }{ gear\_extended :\in  1\upto 3 \tfun  (GEARS \tfun  \{ FALSE\} )  }
			\nItemX{ act3 }{ gear\_shock\_absorber :=  \{ a\mapsto b |  a \in  1\upto 3 \land  b = flight\}   }
			\end{description}
		\EndAct
		\end{description}
	\EVT {retraction}
	\EXTD {retraction}
		\begin{description}
		\WhenGrd
			\begin{description}
			\nItemX{ grd1 }{ dstate[DOORS]=\{ OPEN\}  }
			\nItemX{ grd2 }{ gstate[GEARS]=\{ RETRACTING\}  }
			\nItemX{ grd4 }{ gear\_extended = \{ a\mapsto b |  a \in  1\upto 3 \land  b \in  GEARS \tfun  \{ FALSE\} \}   }
			\nItemX{ grd5 }{ gear\_retracted = \{ a\mapsto b |  a \in  1\upto 3 \land  b \in  GEARS \tfun  \{ FALSE\} \}   }
			\nItemX{ grd6 }{ gear\_shock\_absorber = \{ a\mapsto b |  a \in  1\upto 3 \land  b = flight\}   }
			\nItemX{ grd7 }{ \forall x\qdot x\in 1\upto 3 \limp  handle(x)=button }
			\nItemX{ grd3 }{ SGCylinder = \{ a\mapsto b |  a\in  GEARS\cprod CYLINDER \land  b=STOP\}  }
			\end{description}
		\ThenAct
			\begin{description}
			\nItemX{ act1 }{ gstate:=   \{ a\mapsto b|  a \in  GEARS \land  b= RETRACTED\}  }
			\nItemX{ act2 }{ gear\_retracted :\in  1\upto 3 \tfun  (GEARS \tfun  \{ TRUE\} )  }
			\end{description}
		\EndAct
		\end{description}
	\EVT {extending\_gears}
	\EXTD {extending\_gears}
		\begin{description}
		\WhenGrd
			\begin{description}
			\nItemX{ grd1 }{ dstate[DOORS]=\{ OPEN\}  }
			\nItemX{ grd2 }{ gstate[GEARS]=\{ RETRACTED\}  }
			\nItemX{ grd3 }{ p=E }
			\nItemX{ grd5 }{ gear\_retracted = \{ a\mapsto b |  a \in  1\upto 3 \land  b \in  GEARS \tfun  \{ TRUE\} \}   }
			\nItemX{ grd6 }{ gear\_extended = \{ a\mapsto b |  a \in  1\upto 3 \land  b \in  GEARS \tfun  \{ FALSE\} \}   }
			\nItemX{ grd7 }{ gear\_shock\_absorber = \{ a\mapsto b |  a \in  1\upto 3 \land  b = flight\}   }
			\nItemX{ grd8 }{ \forall x\qdot x\in 1\upto 3 \limp  handle(x)=button }
			\nItemX{ grd4 }{ SGCylinder = \{ a\mapsto b |  a\in  GEARS\cprod CYLINDER \land  b=MOVING\}  }
			\end{description}
		\ThenAct
			\begin{description}
			\nItemX{ act1 }{ gstate :=  \{ a\mapsto b| a \in  GEARS \land  b=EXTENDING\}  }
			\nItemX{ act2 }{ gear\_retracted :\in  1\upto 3 \tfun  (GEARS \tfun  \{ FALSE\} ) }
			\end{description}
		\EndAct
		\end{description}
	\EVT {extension}
	\EXTD {extension}
		\begin{description}
		\WhenGrd
			\begin{description}
			\nItemX{ grd1 }{ dstate[DOORS]=\{ OPEN\}  }
			\nItemX{ grd2 }{ gstate[GEARS]=\{ EXTENDING\}  }
			\nItemX{ grd4 }{ gear\_retracted = \{ a\mapsto b |  a \in  1\upto 3 \land  b \in  GEARS \tfun  \{ FALSE\} \}   }
			\nItemX{ grd5 }{ gear\_extended = \{ a\mapsto b |  a \in  1\upto 3 \land  b \in  GEARS \tfun  \{ FALSE\} \}   }
			\nItemX{ grd6 }{ gear\_shock\_absorber = \{ a\mapsto b |  a \in  1\upto 3 \land  b = flight\}   }
			\nItemX{ grd7 }{ \forall x\qdot x\in 1\upto 3 \limp  handle(x)=button }
			\nItemX{ grd3 }{ SGCylinder = \{ a\mapsto b |  a\in  GEARS\cprod CYLINDER \land  b=STOP\}  }
			\end{description}
		\ThenAct
			\begin{description}
			\nItemX{ act1 }{ gstate :=  \{ a\mapsto b|  a \in  GEARS \land  b=EXTENDED\}  }
			\nItemX{ act2 }{ gear\_extended :\in  1\upto 3 \tfun  (GEARS \tfun  \{ TRUE\} )  }
			\nItemX{ act3 }{ gear\_shock\_absorber :=  \{ a\mapsto b |  a \in  1\upto 3 \land  b = ground\}   }
			\end{description}
		\EndAct
		\end{description}
	\EVT {HPD1}
	\EXTD {HPD1}
		\begin{description}
		\WhenGrd
			\begin{description}
			\nItemX{ grd3 }{ \forall x\qdot x\in 1\upto 3 \limp  handle(x)=UP }
			\end{description}
		\ThenAct
			\begin{description}
			\nItemX{ act2 }{ handle :\in  1\upto 3 \tfun  \{ DOWN\}  }
			\end{description}
		\EndAct
		\end{description}
	\EVT {HPU1}
	\EXTD {HPU1}
		\begin{description}
		\WhenGrd
			\begin{description}
			\nItemX{ grd3 }{ \forall x\qdot x\in 1\upto 3 \limp  handle(x)=DOWN }
			\end{description}
		\ThenAct
			\begin{description}
			\nItemX{ act2 }{ handle :\in  1\upto 3 \tfun  \{ UP\}  }
			\end{description}
		\EndAct
		\end{description}
	\EVT {Analogical\_switch\_closed}
	\EXTD {Analogical\_switch\_closed}
		\begin{description}
		\AnyPrm
			\begin{description}
			\ItemXY{in }{in port }
			\end{description}
		\WhereGrd
			\begin{description}
			\nItemX{ grd1 }{ in = general\_EV }
			\nItemX{ grd2 }{ \forall x\qdot x\in 1\upto 3 \limp  (handle(x)=UP \lor  handle(x)=DOWN) }
			\end{description}
		\ThenAct
			\begin{description}
			\nItemX{ act3 }{ analogical\_switch :\in  1\upto 3 \tfun  \{ closed\}  }
			\nItemX{ act4 }{ A\_Switch\_Out :=  TRUE }
			\end{description}
		\EndAct
		\end{description}
	\EVT {Analogical\_switch\_open}
	\EXTD {Analogical\_switch\_open}
		\begin{description}
		\AnyPrm
			\begin{description}
			\ItemXY{in }{in port }
			\end{description}
		\WhereGrd
			\begin{description}
			\nItemX{ grd1 }{ in = general\_EV }
			\nItemX{ grd2 }{ \forall x\qdot x\in 1\upto 3 \limp  (handle(x)=UP \lor  handle(x)=DOWN) }
			\end{description}
		\ThenAct
			\begin{description}
			\nItemX{ act3 }{ analogical\_switch :\in  1\upto 3 \tfun  \{ open\}  }
			\nItemX{ act4 }{ A\_Switch\_Out :=  FALSE }
			\end{description}
		\EndAct
		\end{description}
	\EVT {Circuit\_pressurized\_OK}
	\EXTD {Circuit\_pressurized\_OK}
		\begin{description}
		\WhenGrd
			\begin{description}
			\nItemX{ grd1 }{ general\_EV\_Hout = Hin }
			\end{description}
		\ThenAct
			\begin{description}
			\nItemX{ act9 }{ circuit\_pressurized :\in  1\upto 3 \tfun  \{ TRUE\}  }
			\end{description}
		\EndAct
		\end{description}
	\EVT {Circuit\_pressurized\_notOK}
	\EXTD {Circuit\_pressurized\_notOK}
		\begin{description}
		\WhenGrd
			\begin{description}
			\nItemX{ grd1 }{ general\_EV\_Hout = 0 }
			\end{description}
		\ThenAct
			\begin{description}
			\nItemX{ act9 }{ circuit\_pressurized :\in  1\upto 3 \tfun  \{ FALSE\}  }
			\end{description}
		\EndAct
		\end{description}
	\EVT {Computing\_Module\_1\_2}
	\EXTD {Computing\_Module\_1\_2}
		\begin{description}
		\WhenGrd
			\begin{description}
			\nItemX{ grd1 }{ state=computing }
			\end{description}
		\ThenAct
			\begin{description}
			\nItemX{ act1 }{ general\_EV :=  general\_EV\_func(handle \mapsto  analogical\_switch \mapsto  gear\_extended \mapsto  gear\_retracted \mapsto  gear\_shock\_absorber \mapsto  door\_open \mapsto  door\_closed \mapsto  circuit\_pressurized) }
			\nItemX{ act2 }{ close\_EV :=  close\_EV\_func(handle \mapsto  analogical\_switch \mapsto  gear\_extended \mapsto  gear\_retracted \mapsto  gear\_shock\_absorber \mapsto  door\_open \mapsto  door\_closed \mapsto  circuit\_pressurized) }
			\nItemX{ act3 }{ retract\_EV :=  retract\_EV\_func(handle \mapsto  analogical\_switch \mapsto  gear\_extended \mapsto  gear\_retracted \mapsto  gear\_shock\_absorber \mapsto  door\_open \mapsto  door\_closed \mapsto  circuit\_pressurized) }
			\nItemX{ act4 }{ extend\_EV :=  extend\_EV\_func(handle \mapsto  analogical\_switch \mapsto  gear\_extended \mapsto  gear\_retracted \mapsto  gear\_shock\_absorber \mapsto  door\_open \mapsto  door\_closed \mapsto  circuit\_pressurized) }
			\nItemX{ act5 }{ open\_EV :=  open\_EV\_func(handle \mapsto  analogical\_switch \mapsto  gear\_extended \mapsto  gear\_retracted \mapsto  gear\_shock\_absorber \mapsto  door\_open \mapsto  door\_closed \mapsto  circuit\_pressurized) }
			\nItemX{ act6 }{ gears\_locked\_down :=  gears\_locked\_down\_func(handle \mapsto  analogical\_switch \mapsto  gear\_extended \mapsto  gear\_retracted \mapsto  gear\_shock\_absorber \mapsto  door\_open \mapsto  door\_closed \mapsto  circuit\_pressurized) }
			\nItemX{ act7 }{ gears\_man :=  gears\_man\_func(handle \mapsto  analogical\_switch \mapsto  gear\_extended \mapsto  gear\_retracted \mapsto  gear\_shock\_absorber \mapsto  door\_open \mapsto  door\_closed \mapsto  circuit\_pressurized) }
			\nItemX{ act8 }{ anomaly :=  anomaly\_func(handle \mapsto  analogical\_switch \mapsto  gear\_extended \mapsto  gear\_retracted \mapsto  gear\_shock\_absorber \mapsto  door\_open \mapsto  door\_closed \mapsto  circuit\_pressurized) }
			\nItemX{ act9 }{ state:= electroValve }
			\end{description}
		\EndAct
		\end{description}
	\EVT {Update\_Hout}\cmt{		\\\hspace*{2,8 cm}  Assign the value of Hout  }
	\EXTD {Update\_Hout}
		\begin{description}
		\WhenGrd
			\begin{description}
			\nItemX{ grd1 }{ state = electroValve }
			\end{description}
		\ThenAct
			\begin{description}
			\nItemXY{ act1 }{ general\_EV\_Hout :|  ((general\_EV = TRUE \land   general\_EV\_Hout' = Hin) \lor  (general\_EV = FALSE \land  general\_EV\_Hout' = 0)		\\\hspace*{1,2 cm}  \lor  (A\_Switch\_Out = TRUE \land   general\_EV\_Hout' = Hin) \lor  (A\_Switch\_Out = FALSE \land   general\_EV\_Hout' = 0)) }{ 		\\\hspace*{1,4 cm}  pass the current value of hydraulic input port (Hin) to hydraulic output port (Hout) }
			\nItemX{ act2 }{ close\_EV\_Hout :|  ((close\_EV = TRUE \land  close\_EV\_Hout' = Hin) \lor  (close\_EV = FALSE \land  close\_EV\_Hout' = 0)) }
			\nItemX{ act3 }{ open\_EV\_Hout :|  ((open\_EV = TRUE \land  open\_EV\_Hout' = Hin) \lor  (open\_EV = FALSE \land  open\_EV\_Hout' = 0)) }
			\nItemX{ act4 }{ extend\_EV\_Hout :|  ((extend\_EV = TRUE \land  extend\_EV\_Hout' = Hin) \lor  (extend\_EV = FALSE \land  extend\_EV\_Hout' = 0)) }
			\nItemX{ act5 }{ retract\_EV\_Hout :|  ((retract\_EV = TRUE \land  retract\_EV\_Hout' = Hin) \lor  (retract\_EV = FALSE \land  retract\_EV\_Hout' = 0)) }
			\nItemX{ act6 }{ state :=  cylinder }
			\end{description}
		\EndAct
		\end{description}
	\EVT {CylinderMovingOrStop}\cmt{		\\\hspace*{4,6 cm}  Cylinder Moving or Stop according to the output of hydraulic circuit }
	\EXTD {CylinderMovingOrStop}
		\begin{description}
		\WhenGrd
			\begin{description}
			\nItemX{ grd1 }{ state = cylinder }
			\end{description}
		\ThenAct
			\begin{description}
			\nItemX{ act1 }{ SGCylinder :|  ((SGCylinder' = \{ a\mapsto b |  a\in  GEARS\cprod \{ GCYF,GCYR,GCYL\}  \land  b=MOVING\}  \land  extend\_EV\_Hout = Hin ) \lor  		\\\hspace*{1,4 cm}  (SGCylinder' = \{ a\mapsto b |  a\in  GEARS\cprod \{ GCYF,GCYR,GCYL\}  \land  b=STOP\}  \land  extend\_EV\_Hout = 0 ) \lor 		\\\hspace*{1,4 cm}  (SGCylinder' = \{ a\mapsto b |  a\in  GEARS\cprod \{ GCYF,GCYR,GCYL\}  \land  b=MOVING\}  \land  retract\_EV\_Hout = Hin ) \lor 		\\\hspace*{1,4 cm}  (SGCylinder' = \{ a\mapsto b |  a\in  GEARS\cprod \{ GCYF,GCYR,GCYL\}  \land  b=STOP\}  \land  retract\_EV\_Hout = 0 )) }
			\nItemX{ act2 }{ SDCylinder :|  ((SDCylinder' = \{ a\mapsto b |  a\in  DOORS\cprod \{ DCYF,DCYR,DCYL\}  \land  b=MOVING\}  \land  open\_EV\_Hout = Hin) \lor  		\\\hspace*{1,4 cm}  (SDCylinder' = \{ a\mapsto b |  a\in  DOORS\cprod \{ DCYF,DCYR,DCYL\}  \land  b=STOP\}  \land  open\_EV\_Hout = 0) \lor 		\\\hspace*{1,4 cm}  (SDCylinder' = \{ a\mapsto b |  a\in  DOORS\cprod \{ DCYF,DCYR,DCYL\}  \land  b=MOVING\}  \land  close\_EV\_Hout = Hin) \lor 		\\\hspace*{1,4 cm}  (SDCylinder' = \{ a\mapsto b |  a\in  DOORS\cprod \{ DCYF,DCYR,DCYL\} \land  b=STOP\}  \land  close\_EV\_Hout = 0)) }
			\nItemX{ act3 }{ state :=  computing }
			\end{description}
		\EndAct
		\end{description}
	\EVT {Failure\_Detection\_Generic\_Monitoring}
	\EXTD {Failure\_Detection}
		\begin{description}
		\WhenGrd
			\begin{description}
			\nItemY{ grd1 }{ (\forall x,y,z\qdot  x\in 1\upto 3 \land  y\in 1\upto 3 \land  z\in 1\upto 3 \land  x\neq y \land  y\neq z \land  x\neq z \limp  		\\\hspace*{1,2 cm}  (handle(x)\neq handle(y) \land  handle(y)\neq handle(z) \land  handle(x)\neq handle(z)))		\\\hspace*{1,2 cm}  \lor 		\\\hspace*{1,2 cm}  (\forall x,y,z\qdot  x\in 1\upto 3 \land  y\in 1\upto 3 \land  z\in 1\upto 3 \land  x\neq y \land  y\neq z \land  x\neq z \limp  		\\\hspace*{1,2 cm}  (analogical\_switch(x)\neq analogical\_switch(y) \land  analogical\_switch(y)\neq analogical\_switch(z) \land  analogical\_switch(x)\neq analogical\_switch(z)))		\\\hspace*{1,2 cm}  \lor 		\\\hspace*{1,2 cm}  (\forall x,y,z\qdot  x\in 1\upto 3 \land  y\in 1\upto 3 \land  z\in 1\upto 3 \land  x\neq y \land  y\neq z \land  x\neq z \limp  		\\\hspace*{1,2 cm}  (gear\_extended(x)\neq gear\_extended(y) \land  gear\_extended(y)\neq gear\_extended(z) \land  gear\_extended(x)\neq gear\_extended(z)))		\\\hspace*{1,2 cm}  \lor  		\\\hspace*{1,2 cm}  (\forall x,y,z\qdot  x\in 1\upto 3 \land  y\in 1\upto 3 \land  z\in 1\upto 3 \land  x\neq y \land  y\neq z \land  x\neq z \limp  		\\\hspace*{1,2 cm}  (gear\_retracted(x)\neq gear\_retracted(y) \land  gear\_retracted(y)\neq gear\_retracted(z) \land  gear\_retracted(x)\neq gear\_retracted(z)))		\\\hspace*{1,2 cm}  \lor 		\\\hspace*{1,2 cm}  (\forall x,y,z\qdot  x\in 1\upto 3 \land  y\in 1\upto 3 \land  z\in 1\upto 3 \land  x\neq y \land  y\neq z \land  x\neq z \limp  		\\\hspace*{1,2 cm}  (gear\_shock\_absorber(x)\neq gear\_shock\_absorber(y) \land  gear\_shock\_absorber(y)\neq gear\_shock\_absorber(z) \land  gear\_shock\_absorber(x)\neq gear\_shock\_absorber(z)))		\\\hspace*{1,2 cm}  \lor 		\\\hspace*{1,2 cm}  (\forall x,y,z\qdot  x\in 1\upto 3 \land  y\in 1\upto 3 \land  z\in 1\upto 3 \land  x\neq y \land  y\neq z \land  x\neq z \limp  		\\\hspace*{1,2 cm}  (door\_open(x)\neq door\_open(y) \land  door\_open(y)\neq door\_open(z) \land  door\_open(x)\neq door\_open(z)))		\\\hspace*{1,2 cm}  \lor 		\\\hspace*{1,2 cm}  (\forall x,y,z\qdot  x\in 1\upto 3 \land  y\in 1\upto 3 \land  z\in 1\upto 3 \land  x\neq y \land  y\neq z \land  x\neq z \limp  		\\\hspace*{1,2 cm}  (door\_closed(x)\neq door\_closed(y) \land  door\_closed(y)\neq door\_closed(z) \land  door\_closed(x)\neq door\_closed(z)))		\\\hspace*{1,2 cm}  \lor 		\\\hspace*{1,2 cm}  (\forall x,y,z\qdot  x\in 1\upto 3 \land  y\in 1\upto 3 \land  z\in 1\upto 3 \land  x\neq y \land  y\neq z \land  x\neq z \limp  		\\\hspace*{1,2 cm}  (circuit\_pressurized(x)\neq circuit\_pressurized(y) \land  circuit\_pressurized(y)\neq circuit\_pressurized(z) \land  circuit\_pressurized(x)\neq circuit\_pressurized(z))) }{		\\\hspace*{1,4 cm}  Generic Monitoring uisng all sensors  } 
			\end{description}
		\ThenAct
			\begin{description}
			\nItemX{ act1 }{ anomaly :=  TRUE }
			\end{description}
		\EndAct
		\end{description}
	\EVT {Failure\_Detection\_Analogical\_Switch}
	\EXTD {Failure\_Detection}
		\begin{description}
		\WhenGrd
			\begin{description}
			\nItemY{ grd1 }{ analogical\_switch = \{ a\mapsto b |  a \in  1\upto 3 \land  b = open\}  		\\\hspace*{1,2 cm}  \lor 		\\\hspace*{1,2 cm}  analogical\_switch = \{ a\mapsto b |  a \in  1\upto 3 \land  b = closed\}  }{		\\\hspace*{1,4 cm}  Gears motion monitoring without considering time  } 
			\end{description}
		\ThenAct
			\begin{description}
			\nItemX{ act1 }{ anomaly :=  TRUE }
			\end{description}
		\EndAct
		\end{description}
	\EVT {Failure\_Detection\_Pressure\_Sensor}
	\EXTD {Failure\_Detection}
		\begin{description}
		\WhenGrd
			\begin{description}
			\nItemY{ grd1 }{ circuit\_pressurized \neq  \{ a\mapsto b |  a \in  1\upto 3 \land  b = TRUE\}  		\\\hspace*{1,2 cm}  \lor 		\\\hspace*{1,2 cm}  circuit\_pressurized \neq  \{ a\mapsto b |  a \in  1\upto 3 \land  b = FALSE\}   }{		\\\hspace*{1,4 cm}  Circuit pressurized motion monitoring without considering time  } 
			\end{description}
		\ThenAct
			\begin{description}
			\nItemX{ act1 }{ anomaly :=  TRUE }
			\end{description}
		\EndAct
		\end{description}
	\EVT {Failure\_Detection\_Doors}
	\EXTD {Failure\_Detection}
		\begin{description}
		\WhenGrd
			\begin{description}
			\nItemY{ grd1 }{ door\_closed \neq  \{ a\mapsto b |  a \in  1\upto 3 \land  b \in  DOORS \tfun  \{ FALSE\} \}  		\\\hspace*{1,2 cm}  \lor 		\\\hspace*{1,2 cm}  door\_open \neq  \{ a\mapsto b |  a \in  1\upto 3 \land  b \in  DOORS \tfun  \{ TRUE\} \} 		\\\hspace*{1,2 cm}  \lor 		\\\hspace*{1,2 cm}  door\_open \neq  \{ a\mapsto b |  a \in  1\upto 3 \land  b \in  DOORS \tfun  \{ FALSE\} \}   		\\\hspace*{1,2 cm}  \lor 		\\\hspace*{1,2 cm}  door\_closed \neq  \{ a\mapsto b |  a \in  1\upto 3 \land  b \in  DOORS \tfun  \{ TRUE\} \}   }{		\\\hspace*{1,4 cm}  Doors motion monitoring without considering time  } 
			\end{description}
		\ThenAct
			\begin{description}
			\nItemX{ act1 }{ anomaly :=  TRUE }
			\end{description}
		\EndAct
		\end{description}
	\EVT {Failure\_Detection\_Gears}
	\EXTD {Failure\_Detection}
		\begin{description}
		\WhenGrd
			\begin{description}
			\nItemY{ grd1 }{ gear\_retracted \neq  \{ a\mapsto b |  a \in  1\upto 3 \land  b \in  GEARS \tfun  \{ FALSE\} \}  		\\\hspace*{1,2 cm}  \lor 		\\\hspace*{1,2 cm}  gear\_retracted \neq  \{ a\mapsto b |  a \in  1\upto 3 \land  b \in  GEARS \tfun  \{ TRUE\} \} 		\\\hspace*{1,2 cm}  \lor 		\\\hspace*{1,2 cm}  gear\_extended \neq  \{ a\mapsto b |  a \in  1\upto 3 \land  b \in  GEARS \tfun  \{ FALSE\} \}   		\\\hspace*{1,2 cm}  \lor 		\\\hspace*{1,2 cm}  gear\_extended \neq  \{ a\mapsto b |  a \in  1\upto 3 \land  b \in  GEARS \tfun  \{ TRUE\} \}   }{		\\\hspace*{1,4 cm}  Gears motion monitoring without considering time  } 
			\end{description}
		\ThenAct
			\begin{description}
			\nItemX{ act1 }{ anomaly :=  TRUE }
			\end{description}
		\EndAct
		\end{description}
\END
\end{description}

\section{M8}
\label{sec:M8}

\begin{description}
\BTitle{M8}{27Jan2014}{10:44:59 AM}
\MACHINE{M8}\cmt{		\\\hspace*{1 cm}  Timing Requirements. }
\REFINES{M7}
\SEES{C1}
\VARIABLES
	\begin{description}
		\Item{ dstate }
		\Item{ lstate }
		\Item{ phase }
		\Item{ button }
		\Item{ p }
		\Item{ l }
		\Item{ i }
		\Item{ gstate }
		\Item{ handle }
		\Item{ analogical\_switch }
		\Item{ gear\_extended }
		\Item{ gear\_retracted }
		\Item{ gear\_shock\_absorber }
		\Item{ door\_closed }
		\Item{ door\_open }
		\Item{ circuit\_pressurized }
		\Item{ general\_EV }
		\Item{ close\_EV }
		\Item{ retract\_EV }
		\Item{ extend\_EV }
		\Item{ open\_EV }
		\Item{ gears\_locked\_down }
		\Item{ gears\_man }
		\Item{ anomaly }
		\Item{ general\_EV\_func }
		\Item{ close\_EV\_func }
		\Item{ retract\_EV\_func }
		\Item{ extend\_EV\_func }
		\Item{ open\_EV\_func }
		\Item{ gears\_locked\_down\_func }
		\Item{ gears\_man\_func }
		\Item{ anomaly\_func }
		\Item{ general\_EV\_Hout }
		\Item{ close\_EV\_Hout }
		\Item{ retract\_EV\_Hout }
		\Item{ extend\_EV\_Hout }
		\Item{ open\_EV\_Hout }
		\ItemY{ SDCylinder }{State of Door Cylinder}
		\ItemY{ SGCylinder }{State of Gear Cylinder}
		\ItemY{ A\_Switch\_Out }{State of Gear Cylinder}
		\Item{ state }
		\ItemY{ time }{current time}
		\ItemY{ at }{a future event activation set.}
		\ItemY{ index }{To take a function to index different sets for event activation set}
		\ItemY{ handleUp\_interval }{To keep an update time duration after handle up}
		\ItemY{ handleDown\_interval }{To keep an update time duration after handle down}
	\end{description}
\INVARIANTS
	\begin{description}
		\nItemY{ inv1 }{ time \in  \nat  }{ 		\\\hspace*{1,4 cm}  current updating time  }
		\nItemY{ inv2 }{ at \subseteq  \nat \cprod  \nat  }{ 		\\\hspace*{1,4 cm}  a set of times for activating event  }
		\nItemY{ inv3 }{ ran(at) \neq  \emptyset  \limp  time \leq  min(ran(at)) }{ 		\\\hspace*{1,4 cm}  if activation is a non empty set then the current time  will 
		\\\hspace*{1,2 cm}  be less than or equal to the minimum of activation set.  }
		\nItemY{ inv4 }{ index \in  \nat  }{ 		\\\hspace*{1,4 cm}  an index for event activation set to store multiple identical values }
		\nItemY{ inv5 }{ handleUp\_interval \in  \nat  }{ 		\\\hspace*{1,4 cm}  time interval after handle up }
		\nItemY{ inv6 }{ handleDown\_interval \in  \nat }{ 		\\\hspace*{1,4 cm}  time interval after handle down }
	\end{description}
\EVENTS
	\INITIALISATION
		\\\textit{extended}
		\begin{description}
		\BeginAct
			\begin{description}
			\nItemX{ act1 }{ button :=  DOWN }
			\nItemX{ act2 }{ phase :=  haltdown }
			\nItemXY{ act3 }{ dstate :|  (dstate'\in  DOORS \tfun  SDOORS \land  dstate'=\{ a\mapsto b|  a \in  DOORS \land  b=CLOSED\} ) }{ 		\\\hspace*{1,4 cm}  missing elements of the invariant }
			\nItemX{ act4 }{ lstate :=  \{ a\mapsto b| a\in DOORS\land  b=LOCKED\}  }
			\nItemX{ act5 }{ p :=  R }
			\nItemX{ act6 }{ l :=  R }
			\nItemX{ act7 }{ i :=  R }
			\nItemX{ act8 }{ gstate :|  (gstate' \in  GEARS \tfun  SGEARS \land  gstate'=\{ a\mapsto b |  a \in  GEARS  \land  b=EXTENDED\} ) }
			\nItemX{ act14 }{ handle :\in  1\upto 3 \tfun  \{ DOWN\}  }
			\nItemX{ act15 }{ analogical\_switch :\in  1\upto 3 \tfun  \{ open\}  }
			\nItemX{ act16 }{ gear\_extended :\in  1\upto 3 \tfun  (GEARS \tfun  \{ TRUE\} ) }
			\nItemX{ act17 }{ gear\_retracted :\in  1\upto 3 \tfun  (GEARS \tfun  \{ FALSE\} ) }
			\nItemX{ act18 }{ gear\_shock\_absorber :\in  1\upto 3 \tfun  \{ ground\}  }
			\nItemX{ act19 }{ door\_closed :\in  1\upto 3 \tfun  (DOORS \tfun  \{ TRUE\} ) }
			\nItemX{ act20 }{ door\_open :\in  1\upto 3 \tfun  (DOORS \tfun  \{ FALSE\} ) }
			\nItemX{ act21 }{ circuit\_pressurized :\in  1\upto 3 \tfun  \{ FALSE\}  }
			\nItemX{ act22 }{ general\_EV :=  FALSE }
			\nItemX{ act23 }{ close\_EV :=  TRUE }
			\nItemX{ act24 }{ retract\_EV :=  FALSE }
			\nItemX{ act25 }{ extend\_EV :=  TRUE }
			\nItemX{ act27 }{ open\_EV :=  FALSE }
			\nItemX{ act28 }{ gears\_locked\_down :=  TRUE }
			\nItemX{ act29 }{ gears\_man :=  FALSE }
			\nItemX{ act30 }{ anomaly :=  FALSE }
			\nItemX{ act31 }{ general\_EV\_func :\in  (1\upto 3 \tfun  POSITIONS) \cprod  (1\upto 3 \tfun  A\_Switch) \cprod  (1\upto 3 \tfun  (GEARS \tfun  BOOL)) \cprod  (1\upto 3 \tfun  (GEARS \tfun  BOOL)) \cprod  (1\upto 3 \tfun  GEAR\_ABSORBER) \cprod  (1\upto 3 \tfun  (DOORS \tfun  BOOL)) \cprod  (1\upto 3 \tfun  (DOORS \tfun  BOOL)) \cprod  (1\upto 3 \tfun  BOOL) \tfun  BOOL }
			\nItemX{ act32 }{ close\_EV\_func :\in  (1\upto 3 \tfun  POSITIONS) \cprod  (1\upto 3 \tfun  A\_Switch) \cprod  (1\upto 3 \tfun  (GEARS \tfun  BOOL)) \cprod  (1\upto 3 \tfun  (GEARS \tfun  BOOL)) \cprod  (1\upto 3 \tfun  GEAR\_ABSORBER) \cprod  (1\upto 3 \tfun  (DOORS \tfun  BOOL)) \cprod  (1\upto 3 \tfun  (DOORS \tfun  BOOL)) \cprod  (1\upto 3 \tfun  BOOL) \tfun  BOOL }
			\nItemX{ act33 }{ retract\_EV\_func :\in  (1\upto 3 \tfun  POSITIONS) \cprod  (1\upto 3 \tfun  A\_Switch) \cprod  (1\upto 3 \tfun  (GEARS \tfun  BOOL)) \cprod  (1\upto 3 \tfun  (GEARS \tfun  BOOL)) \cprod  (1\upto 3 \tfun  GEAR\_ABSORBER) \cprod  (1\upto 3 \tfun  (DOORS \tfun  BOOL)) \cprod  (1\upto 3 \tfun  (DOORS \tfun  BOOL)) \cprod  (1\upto 3 \tfun  BOOL) \tfun  BOOL }
			\nItemX{ act34 }{ extend\_EV\_func :\in  (1\upto 3 \tfun  POSITIONS) \cprod  (1\upto 3 \tfun  A\_Switch) \cprod  (1\upto 3 \tfun  (GEARS \tfun  BOOL)) \cprod  (1\upto 3 \tfun  (GEARS \tfun  BOOL)) \cprod  (1\upto 3 \tfun  GEAR\_ABSORBER) \cprod  (1\upto 3 \tfun  (DOORS \tfun  BOOL)) \cprod  (1\upto 3 \tfun  (DOORS \tfun  BOOL)) \cprod  (1\upto 3 \tfun  BOOL) \tfun  BOOL }
			\nItemX{ act35 }{ open\_EV\_func :\in  (1\upto 3 \tfun  POSITIONS) \cprod  (1\upto 3 \tfun  A\_Switch) \cprod  (1\upto 3 \tfun  (GEARS \tfun  BOOL)) \cprod  (1\upto 3 \tfun  (GEARS \tfun  BOOL)) \cprod  (1\upto 3 \tfun  GEAR\_ABSORBER) \cprod  (1\upto 3 \tfun  (DOORS \tfun  BOOL)) \cprod  (1\upto 3 \tfun  (DOORS \tfun  BOOL)) \cprod  (1\upto 3 \tfun  BOOL) \tfun  BOOL }
			\nItemX{ act36 }{ gears\_locked\_down\_func :\in  (1\upto 3 \tfun  POSITIONS) \cprod  (1\upto 3 \tfun  A\_Switch) \cprod  (1\upto 3 \tfun  (GEARS \tfun  BOOL)) \cprod  (1\upto 3 \tfun  (GEARS \tfun  BOOL)) \cprod  (1\upto 3 \tfun  GEAR\_ABSORBER) \cprod  (1\upto 3 \tfun  (DOORS \tfun  BOOL)) \cprod  (1\upto 3 \tfun  (DOORS \tfun  BOOL)) \cprod  (1\upto 3 \tfun  BOOL) \tfun  BOOL }
			\nItemX{ act37 }{ gears\_man\_func :\in  (1\upto 3 \tfun  POSITIONS) \cprod  (1\upto 3 \tfun  A\_Switch) \cprod  (1\upto 3 \tfun  (GEARS \tfun  BOOL)) \cprod  (1\upto 3 \tfun  (GEARS \tfun  BOOL)) \cprod  (1\upto 3 \tfun  GEAR\_ABSORBER) \cprod  (1\upto 3 \tfun  (DOORS \tfun  BOOL)) \cprod  (1\upto 3 \tfun  (DOORS \tfun  BOOL)) \cprod  (1\upto 3 \tfun  BOOL) \tfun  BOOL }
			\nItemX{ act38 }{ anomaly\_func :\in  (1\upto 3 \tfun  POSITIONS) \cprod  (1\upto 3 \tfun  A\_Switch) \cprod  (1\upto 3 \tfun  (GEARS \tfun  BOOL)) \cprod  (1\upto 3 \tfun  (GEARS \tfun  BOOL)) \cprod  (1\upto 3 \tfun  GEAR\_ABSORBER) \cprod  (1\upto 3 \tfun  (DOORS \tfun  BOOL)) \cprod  (1\upto 3 \tfun  (DOORS \tfun  BOOL)) \cprod  (1\upto 3 \tfun  BOOL) \tfun  BOOL }
			\nItemX{ act39 }{ A\_Switch\_Out :=  FALSE }
			\nItemX{ act40 }{ close\_EV\_Hout :=  0 }
			\nItemX{ act41 }{ retract\_EV\_Hout :=  0 }
			\nItemX{ act42 }{ extend\_EV\_Hout :=  0 }
			\nItemX{ act43 }{ open\_EV\_Hout :=  0 }
			\nItemX{ act44 }{ general\_EV\_Hout :=  0 }
			\nItemX{ act45 }{ SDCylinder :\in  DOORS \cprod  \{ DCYF,DCYR,DCYL\}  \tfun   \{ STOP\}  }
			\nItemX{ act46 }{ SGCylinder :\in  GEARS \cprod  \{ GCYF,GCYR,GCYL\}  \tfun  \{ STOP\}   }
			\nItemX{ act26 }{ state :=  computing }
			\nItem{ act47 }{ at :=  \emptyset  }
			\nItem{ act48 }{ time :=  0 }
			\nItem{ act49 }{ index :=  0 }
			\nItem{ act50 }{ handleUp\_interval :=  0 }
			\nItem{ act51 }{ handleDown\_interval :=  0 }
			\end{description}
		\EndAct
		\end{description}
	\EVT {opening\_doors\_DOWN}
	\EXTD {opening\_doors\_DOWN}
		\begin{description}
		\WhenGrd
			\begin{description}
			\nItemX{ grd1 }{ dstate[DOORS]= \{ CLOSED\}  }
			\nItemX{ grd5 }{ lstate[DOORS]=\{ UNLOCKED\}  }
			\nItemX{ grd7 }{ phase=movingdown }
			\nItemX{ grd8 }{ p=R }
			\nItemX{ grd9 }{ l=R }
			\nItemX{ grd10 }{ door\_open = \{ a\mapsto b |  a \in  1\upto 3 \land  b \in  DOORS \tfun  \{ FALSE\} \}   }
			\nItemX{ grd11 }{ door\_closed = \{ a\mapsto b |  a \in  1\upto 3 \land  b \in  DOORS \tfun  \{ FALSE\} \}   }
			\nItemX{ grd12 }{ \forall x\qdot x\in 1\upto 3 \limp  handle(x)=button }
			\nItemX{ grd3 }{ SDCylinder = \{ a\mapsto b |  a\in  DOORS\cprod CYLINDER \land  b=MOVING\}  }
			\nItemX{ grd13 }{ anomaly = FALSE }
			\end{description}
		\ThenAct
			\begin{description}
			\nItemX{ act1 }{ dstate :=  \{ a\mapsto b|  a \in  DOORS \land  b=OPEN\}  }
			\nItemX{ act2 }{ p:= E }
			\nItemX{ act3 }{ door\_open :\in  1\upto 3 \tfun  (DOORS \tfun  \{ TRUE\} ) }
			\nItemY{ act4 }{ at :=  at \bunion  \{ (index+1) \mapsto  (time+100)\}  }{ 		\\\hspace*{1,4 cm}  minimal interval for door open to gear extension }
			\nItem{ act5 }{ index :=  index+1 }
			\end{description}
		\EndAct
		\end{description}
	\EVT {opening\_doors\_UP}
	\EXTD {opening\_doors\_UP}
		\begin{description}
		\WhenGrd
			\begin{description}
			\nItemX{ grd1 }{ dstate[DOORS]= \{ CLOSED\}  }
			\nItemX{ grd4 }{ lstate[DOORS]=\{ UNLOCKED\}  }
			\nItemX{ grd5 }{ phase= movingup }
			\nItemX{ grd6 }{ p=E }
			\nItemX{ grd7 }{ l=E }
			\nItemX{ grd8 }{ door\_open = \{ a\mapsto b |  a \in  1\upto 3 \land  b \in  DOORS \tfun  \{ FALSE\} \}   }
			\nItemX{ grd9 }{ door\_closed = \{ a\mapsto b |  a \in  1\upto 3 \land  b \in  DOORS \tfun  \{ FALSE\} \}   }
			\nItemX{ grd10 }{ \forall x\qdot x\in 1\upto 3 \limp  handle(x)=button }
			\nItemX{ grd3 }{ SDCylinder = \{ a\mapsto b |  a\in  DOORS\cprod CYLINDER \land  b=MOVING\}  }
			\nItemX{ grd11 }{ anomaly = FALSE }
			\end{description}
		\ThenAct
			\begin{description}
			\nItemX{ act1 }{ dstate :=  \{ a\mapsto b|  a \in  DOORS \land  b=OPEN\}  }
			\nItemX{ act2 }{ p:= R }
			\nItemX{ act3 }{ door\_open:\in  1\upto 3 \tfun  (DOORS \tfun  \{ TRUE\} )  }
			\nItemY{ act4 }{ at :=  at \bunion  \{ (index+1) \mapsto  (time+100)\}  }{ 		\\\hspace*{1,4 cm}  minimal interval for door open to gear retraction }
			\nItem{ act5 }{ index :=  index+1 }
			\end{description}
		\EndAct
		\end{description}
	\EVT {closing\_doors\_UP}
	\EXTD {closing\_doors\_UP}
		\begin{description}
		\AnyPrm
			\begin{description}
			\ItemX{f }
			\end{description}
		\WhereGrd
			\begin{description}
			\nItemX{ grd1 }{ dstate[DOORS]=\{ OPEN\}  }
			\nItemX{ grd3 }{ f \in  DOORS \tfun  SDOORS }
			\nItemX{ grd4 }{ \forall e\qdot  e \in  DOORS \limp  f(e)=CLOSED }
			\nItemX{ grd5 }{ phase=movingup }
			\nItemX{ grd6 }{ p=R }
			\nItemX{ grd7 }{ gstate[GEARS]=\{ RETRACTED\}  }
			\nItemX{ grd8 }{ \forall x\qdot x\in 1\upto 3 \limp  handle(x)=button }
			\nItemX{ grd9 }{ anomaly = FALSE }
			\end{description}
		\ThenAct
			\begin{description}
			\nItemX{ act1 }{ dstate:= f }
			\end{description}
		\EndAct
		\end{description}
	\EVT {closing\_doors\_DOWN}
	\EXTD {closing\_doors\_DOWN}
		\begin{description}
		\AnyPrm
			\begin{description}
			\ItemX{f }
			\end{description}
		\WhereGrd
			\begin{description}
			\nItemX{ grd1 }{ dstate[DOORS]=\{ OPEN\}  }
			\nItemX{ grd3 }{ f \in  DOORS \tfun  SDOORS }
			\nItemX{ grd4 }{ \forall e\qdot  e \in  DOORS \limp  f(e)=CLOSED }
			\nItemX{ grd5 }{ phase=movingdown }
			\nItemX{ grd6 }{ p=E }
			\nItemX{ grd7 }{ gstate[GEARS]=\{ EXTENDED\}  }
			\nItemX{ grd8 }{ \forall x\qdot x\in 1\upto 3 \limp  handle(x)=button }
			\nItemX{ grd9 }{ anomaly = FALSE }
			\end{description}
		\ThenAct
			\begin{description}
			\nItemX{ act1 }{ dstate:= f }
			\end{description}
		\EndAct
		\end{description}
	\EVT {unlocking\_UP}
	\EXTD {unlocking\_UP}
		\begin{description}
		\WhenGrd
			\begin{description}
			\nItemX{ grd3 }{ lstate[DOORS]=\{ LOCKED\}  }
			\nItemX{ grd4 }{ phase=movingup }
			\nItemX{ grd5 }{ l=E }
			\nItemX{ grd6 }{ p=E }
			\nItemX{ grd7 }{ i=E }
			\nItemX{ grd8 }{ door\_open = \{ a\mapsto b |  a \in  1\upto 3 \land  b \in  DOORS \tfun  \{ FALSE\} \}   }
			\nItemX{ grd9 }{ door\_closed = \{ a\mapsto b |  a \in  1\upto 3 \land  b \in  DOORS \tfun  \{ TRUE\} \}   }
			\nItemX{ grd10 }{ \forall x\qdot x\in 1\upto 3 \limp  handle(x)=button }
			\nItemX{ grd11 }{ anomaly = FALSE }
			\end{description}
		\ThenAct
			\begin{description}
			\nItemX{ act1 }{ lstate:= \{ a\mapsto b| a\in DOORS \land  b=UNLOCKED\}  }
			\nItemX{ act2 }{ door\_closed :\in  1\upto 3 \tfun  (DOORS \tfun  \{ FALSE\} )  }
			\end{description}
		\EndAct
		\end{description}
	\EVT {locking\_UP}
	\EXTD {locking\_UP}
		\begin{description}
		\WhenGrd
			\begin{description}
			\nItemX{ grd3 }{ dstate[DOORS]=\{ CLOSED\}  }
			\nItemX{ grd4 }{ phase=movingup }
			\nItemX{ grd5 }{ lstate[DOORS]=\{ UNLOCKED\}  }
			\nItemX{ grd6 }{ p=R }
			\nItemX{ grd7 }{ l=E }
			\nItemX{ grd9 }{ door\_open = \{ a\mapsto b |  a \in  1\upto 3 \land  b \in  DOORS \tfun  \{ FALSE\} \}   }
			\nItemX{ grd10 }{ door\_closed = \{ a\mapsto b |  a \in  1\upto 3 \land  b \in  DOORS \tfun  \{ FALSE\} \}   }
			\nItemX{ grd11 }{ \forall x\qdot x\in 1\upto 3 \limp  handle(x)=button }
			\nItemX{ grd8 }{ SDCylinder = \{ a\mapsto b |  a\in  DOORS\cprod CYLINDER \land  b=STOP\}  }
			\nItemX{ grd12 }{ anomaly = FALSE }
			\end{description}
		\ThenAct
			\begin{description}
			\nItemX{ act1 }{ lstate:= \{ a\mapsto b| a\in DOORS \land  b=LOCKED\}  }
			\nItemX{ act3 }{ phase:= haltup }
			\nItemXY{ act4 }{ l:= R }{ 		\\\hspace*{1,4 cm}  added by D Mery }
			\nItemX{ act44 }{ door\_closed :\in  1\upto 3 \tfun  (DOORS \tfun  \{ TRUE\} )  }
			\nItemY{ act5 }{ at :=  at \bunion  \{ (index+1) \mapsto  (time+100)\}  }{ 		\\\hspace*{1,4 cm}  minimal interval for door closed to gear extension/retraction }
			\nItem{ act6 }{ index :=  index+1 }
			\end{description}
		\EndAct
		\end{description}
	\EVT {unlocking\_DOWN}
	\EXTD {unlocking\_DOWN}
		\begin{description}
		\WhenGrd
			\begin{description}
			\nItemX{ grd3 }{ lstate[DOORS]=\{ LOCKED\}  }
			\nItemX{ grd4 }{ phase=movingdown }
			\nItemX{ grd5 }{ l=R }
			\nItemX{ grd6 }{ p=R }
			\nItemX{ grd7 }{ i=R }
			\nItemX{ grd8 }{ door\_open = \{ a\mapsto b |  a \in  1\upto 3 \land  b \in  DOORS \tfun  \{ FALSE\} \}   }
			\nItemX{ grd9 }{ door\_closed = \{ a\mapsto b |  a \in  1\upto 3 \land  b \in  DOORS \tfun  \{ TRUE\} \}   }
			\nItemX{ grd10 }{ \forall x\qdot x\in 1\upto 3 \limp  handle(x)=button }
			\nItemX{ grd11 }{ anomaly = FALSE }
			\end{description}
		\ThenAct
			\begin{description}
			\nItemX{ act1 }{ lstate:= \{ a\mapsto b| a\in DOORS \land  b=UNLOCKED\}  }
			\nItemX{ act2 }{ door\_closed :\in  1\upto 3 \tfun  (DOORS \tfun  \{ FALSE\} ) }
			\end{description}
		\EndAct
		\end{description}
	\EVT {locking\_DOWN}
	\EXTD {locking\_DOWN}
		\begin{description}
		\WhenGrd
			\begin{description}
			\nItemX{ grd1 }{ dstate[DOORS]=\{ CLOSED\}  }
			\nItemX{ grd2 }{ phase=movingdown }
			\nItemX{ grd3 }{ lstate[DOORS]=\{ UNLOCKED\}  }
			\nItemX{ grd4 }{ p=E }
			\nItemX{ grd5 }{ l=R }
			\nItemX{ grd7 }{ door\_open = \{ a\mapsto b |  a \in  1\upto 3 \land  b \in  DOORS \tfun  \{ FALSE\} \}   }
			\nItemX{ grd8 }{ door\_closed = \{ a\mapsto b |  a \in  1\upto 3 \land  b \in  DOORS \tfun  \{ FALSE\} \}   }
			\nItemX{ grd9 }{ \forall x\qdot x\in 1\upto 3 \limp  handle(x)=button }
			\nItemX{ grd6 }{ SDCylinder = \{ a\mapsto b |  a\in  DOORS\cprod CYLINDER \land  b=STOP\}  }
			\nItemX{ grd10 }{ anomaly = FALSE }
			\end{description}
		\ThenAct
			\begin{description}
			\nItemX{ act1 }{ lstate:= \{ a\mapsto b| a\in DOORS \land  b = LOCKED\}  }
			\nItemX{ act3 }{ phase:= haltdown }
			\nItemX{ act4 }{ l:= E }
			\nItemX{ act5 }{ door\_closed :\in  1\upto 3 \tfun  (DOORS \tfun  \{ TRUE\} )  }
			\nItemY{ act6 }{ at :=  at \bunion  \{ (index+1) \mapsto  (time+100)\}  }{ 		\\\hspace*{1,4 cm}  minimal interval for door closed to extension/retraction }
			\nItem{ act7 }{ index :=  index+1 }
			\end{description}
		\EndAct
		\end{description}
	\EVT {PD1}
	\EXTD {PD1}
		\begin{description}
		\WhenGrd
			\begin{description}
			\nItemX{ grd1 }{ button=UP }
			\nItemX{ grd2 }{ phase=haltup }
			\nItemX{ grd3 }{ \forall x\qdot x\in 1\upto 3 \limp  handle(x)=DOWN }
			\end{description}
		\ThenAct
			\begin{description}
			\nItemX{ act1 }{ phase:= movingdown }
			\nItemX{ act2 }{ button:= DOWN }
			\nItemX{ act3 }{ l:= R }
			\nItemX{ act4 }{ p:= R }
			\nItemX{ act5 }{ i:= R }
			\end{description}
		\EndAct
		\end{description}
	\EVT {PU1}
	\EXTD {PU1}
		\begin{description}
		\WhenGrd
			\begin{description}
			\nItemX{ grd1 }{ button=DOWN }
			\nItemX{ grd2 }{ phase=haltdown }
			\nItemX{ grd3 }{ \forall x\qdot x\in 1\upto 3 \limp  handle(x)=UP }
			\end{description}
		\ThenAct
			\begin{description}
			\nItemX{ act1 }{ phase:= movingup }
			\nItemX{ act2 }{ button:= UP }
			\nItemX{ act3 }{ l:= E }
			\nItemX{ act4 }{ p:= E }
			\nItemX{ act5 }{ i:= E }
			\end{description}
		\EndAct
		\end{description}
	\EVT {PU2}
	\EXTD {PU2}
		\begin{description}
		\WhenGrd
			\begin{description}
			\nItemX{ grd1 }{ l=R }
			\nItemX{ grd2 }{ p=R }
			\nItemX{ grd3 }{ phase=movingdown }
			\nItemX{ grd4 }{ button=DOWN }
			\nItemX{ grd5 }{ i=R }
			\nItemX{ grd6 }{ lstate[DOORS]=\{ LOCKED\}  }
			\nItemX{ grd7 }{ door\_open = \{ a\mapsto b |  a \in  1\upto 3 \land  b \in  DOORS \tfun  \{ FALSE\} \}   }
			\nItemX{ grd8 }{ door\_closed = \{ a\mapsto b |  a \in  1\upto 3 \land  b \in  DOORS \tfun  \{ TRUE\} \}   }
			\nItemX{ grd9 }{ \forall x\qdot x\in 1\upto 3 \limp  handle(x)=UP }
			\end{description}
		\ThenAct
			\begin{description}
			\nItemX{ act1 }{ phase:= movingup }
			\nItemX{ act4 }{ button:=   UP }
			\nItemX{ act5 }{ l:= E }
			\nItemX{ act6 }{ p:= E }
			\nItemX{ act7 }{ i:= R }
			\end{description}
		\EndAct
		\end{description}
	\EVT {CompletePU2}
	\EXTD {CompletePU2}
		\begin{description}
		\WhenGrd
			\begin{description}
			\nItemX{ grd1 }{ phase=movingup }
			\nItemX{ grd2 }{ button=UP }
			\nItemX{ grd3 }{ l=E }
			\nItemX{ grd4 }{ p=E }
			\nItemX{ grd5 }{ i=R }
			\end{description}
		\ThenAct
			\begin{description}
			\nItemX{ act1 }{ phase:= haltup }
			\end{description}
		\EndAct
		\end{description}
	\EVT {PU3}
	\EXTD {PU3}
		\begin{description}
		\WhenGrd
			\begin{description}
			\nItemX{ grd1 }{ dstate[DOORS]=\{ CLOSED\}  }
			\nItemX{ grd2 }{ lstate[DOORS]=\{ UNLOCKED\}  }
			\nItemX{ grd3 }{ phase = movingdown }
			\nItemX{ grd4 }{ p=R }
			\nItemX{ grd5 }{ l=R }
			\nItemX{ grd6 }{ button=DOWN }
			\nItemX{ grd7 }{ door\_open = \{ a\mapsto b |  a \in  1\upto 3 \land  b \in  DOORS \tfun  \{ FALSE\} \}   }
			\nItemX{ grd8 }{ door\_closed = \{ a\mapsto b |  a \in  1\upto 3 \land  b \in  DOORS \tfun  \{ FALSE\} \}   }
			\nItemX{ grd9 }{ \forall x\qdot x\in 1\upto 3 \limp  handle(x)=UP }
			\end{description}
		\ThenAct
			\begin{description}
			\nItemX{ act1 }{ phase:= movingup }
			\nItemX{ act2 }{ p:= R }
			\nItemX{ act3 }{ l:= E }
			\nItemX{ act4 }{ button:= UP }
			\end{description}
		\EndAct
		\end{description}
	\EVT {PU4}
	\EXTD {PU4}
		\begin{description}
		\WhenGrd
			\begin{description}
			\nItemX{ grd1 }{ dstate[DOORS]=\{ OPEN\}  }
			\nItemX{ grd2 }{ phase=movingdown }
			\nItemX{ grd3 }{ p=E }
			\nItemX{ grd4 }{ button=DOWN }
			\nItemX{ grd7 }{ \forall x\qdot x\in 1\upto 3 \limp  handle(x)=UP }
			\end{description}
		\ThenAct
			\begin{description}
			\nItemX{ act1 }{ phase:= movingup }
			\nItemX{ act2 }{ p:= R }
			\nItemX{ act3 }{ button:= UP }
			\nItemX{ act4 }{ i:= E }
			\nItemX{ act5 }{ l:= E }
			\end{description}
		\EndAct
		\end{description}
	\EVT {PU5}
	\EXTD {PU5}
		\begin{description}
		\WhenGrd
			\begin{description}
			\nItemX{ grd1 }{ dstate[DOORS]=\{ CLOSED\}  }
			\nItemX{ grd2 }{ phase=movingdown }
			\nItemX{ grd3 }{ p=E }
			\nItemX{ grd4 }{ button=DOWN }
			\nItemX{ grd5 }{ lstate[DOORS]=\{ UNLOCKED\}  }
			\nItemX{ grd6 }{ door\_open = \{ a\mapsto b |  a \in  1\upto 3 \land  b \in  DOORS \tfun  \{ FALSE\} \}   }
			\nItemX{ grd7 }{ door\_closed = \{ a\mapsto b |  a \in  1\upto 3 \land  b \in  DOORS \tfun  \{ FALSE\} \}   }
			\nItemX{ grd8 }{ \forall x\qdot x\in 1\upto 3 \limp  handle(x)=UP }
			\end{description}
		\ThenAct
			\begin{description}
			\nItemX{ act1 }{ phase:= movingup }
			\nItemX{ act3 }{ button:= UP }
			\nItemX{ act4 }{ i:= E }
			\nItemX{ act5 }{ l:= E }
			\end{description}
		\EndAct
		\end{description}
	\EVT {PD2}
	\EXTD {PD2}
		\begin{description}
		\WhenGrd
			\begin{description}
			\nItemX{ grd1 }{ l=E }
			\nItemX{ grd2 }{ p=E }
			\nItemX{ grd3 }{ phase=movingup }
			\nItemX{ grd4 }{ i=E }
			\nItemX{ grd5 }{ lstate[DOORS]=\{ LOCKED\}  }
			\nItemX{ grd6 }{ \forall x\qdot x\in 1\upto 3 \limp  handle(x)=DOWN }
			\end{description}
		\ThenAct
			\begin{description}
			\nItemX{ act1 }{ phase:= movingdown }
			\nItemX{ act2 }{ button:= DOWN }
			\nItemX{ act3 }{ l:= R }
			\nItemX{ act4 }{ p:= R }
			\nItemX{ act5 }{ i:= E }
			\end{description}
		\EndAct
		\end{description}
	\EVT {CompletePD2}
	\EXTD {CompletePD2}
		\begin{description}
		\WhenGrd
			\begin{description}
			\nItemX{ grd1 }{ phase=movingdown }
			\nItemX{ grd2 }{ button=DOWN }
			\nItemX{ grd3 }{ l=R }
			\nItemX{ grd4 }{ p=R }
			\nItemX{ grd5 }{ i=E }
			\end{description}
		\ThenAct
			\begin{description}
			\nItemX{ act1 }{ phase:= haltdown }
			\end{description}
		\EndAct
		\end{description}
	\EVT {PD3}
	\EXTD {PD3}
		\begin{description}
		\WhenGrd
			\begin{description}
			\nItemX{ grd1 }{ dstate[DOORS]=\{ CLOSED\}  }
			\nItemX{ grd2 }{ lstate[DOORS]=\{ UNLOCKED\}  }
			\nItemX{ grd3 }{ phase=movingup }
			\nItemX{ grd4 }{ p=E }
			\nItemX{ grd5 }{ l=E }
			\nItemX{ grd6 }{ button=UP }
			\nItemX{ grd7 }{ door\_open = \{ a\mapsto b |  a \in  1\upto 3 \land  b \in  DOORS \tfun  \{ FALSE\} \}   }
			\nItemX{ grd8 }{ door\_closed = \{ a\mapsto b |  a \in  1\upto 3 \land  b \in  DOORS \tfun  \{ FALSE\} \}   }
			\nItemX{ grd9 }{ \forall x\qdot x\in 1\upto 3 \limp  handle(x)=DOWN }
			\end{description}
		\ThenAct
			\begin{description}
			\nItemX{ act1 }{ phase:= movingdown }
			\nItemX{ act2 }{ p:= E }
			\nItemX{ act3 }{ l:= R }
			\nItemX{ act4 }{ button:= DOWN }
			\end{description}
		\EndAct
		\end{description}
	\EVT {PD4}
	\EXTD {PD4}
		\begin{description}
		\WhenGrd
			\begin{description}
			\nItemX{ grd1 }{ dstate[DOORS]=\{ OPEN\}  }
			\nItemX{ grd2 }{ phase=movingup }
			\nItemX{ grd3 }{ p=R }
			\nItemX{ grd4 }{ button=UP }
			\nItemX{ grd6 }{ \forall x\qdot x\in 1\upto 3 \limp  handle(x)=DOWN }
			\end{description}
		\ThenAct
			\begin{description}
			\nItemX{ act1 }{ phase:= movingdown }
			\nItemX{ act2 }{ p:= E }
			\nItemX{ act3 }{ button:= DOWN }
			\nItemX{ act4 }{ i:= R }
			\nItemX{ act5 }{ l:= R }
			\end{description}
		\EndAct
		\end{description}
	\EVT {PD5}
	\EXTD {PD5}
		\begin{description}
		\WhenGrd
			\begin{description}
			\nItemX{ grd1 }{ dstate[DOORS]=\{ CLOSED\}  }
			\nItemX{ grd2 }{ phase=movingup }
			\nItemX{ grd3 }{ p=R }
			\nItemX{ grd4 }{ button=UP }
			\nItemX{ grd5 }{ lstate[DOORS]=\{ UNLOCKED\}  }
			\nItemX{ grd6 }{ door\_open = \{ a\mapsto b |  a \in  1\upto 3 \land  b \in  DOORS \tfun  \{ FALSE\} \}   }
			\nItemX{ grd7 }{ door\_closed = \{ a\mapsto b |  a \in  1\upto 3 \land  b \in  DOORS \tfun  \{ FALSE\} \}   }
			\nItemX{ grd8 }{ \forall x\qdot x\in 1\upto 3 \limp  handle(x)=DOWN }
			\end{description}
		\ThenAct
			\begin{description}
			\nItemX{ act1 }{ phase :=  movingdown }
			\nItemX{ act2 }{ button:= DOWN }
			\nItemX{ act3 }{ i:= R }
			\nItemX{ act4 }{ l:= R }
			\end{description}
		\EndAct
		\end{description}
	\EVT {retracting\_gears}
	\EXTD {retracting\_gears}
		\begin{description}
		\AnyPrm
			\begin{description}
			\Item{ind }
			\end{description}
		\WhereGrd
			\begin{description}
			\nItemX{ grd1 }{ dstate[DOORS]=\{ OPEN\}  }
			\nItemX{ grd2 }{ gstate[GEARS]=\{ EXTENDED\}  }
			\nItemX{ grd3 }{ p=R }
			\nItemX{ grd6 }{ gear\_extended = \{ a\mapsto b |  a \in  1\upto 3 \land  b \in  GEARS \tfun  \{ TRUE\} \}   }
			\nItemX{ grd7 }{ gear\_retracted = \{ a\mapsto b |  a \in  1\upto 3 \land  b \in  GEARS \tfun  \{ FALSE\} \}   }
			\nItemX{ grd8 }{ gear\_shock\_absorber = \{ a\mapsto b |  a \in  1\upto 3 \land  b = ground\}   }
			\nItemX{ grd9 }{ \forall x\qdot x\in 1\upto 3 \limp  handle(x)=button }
			\nItemX{ grd5 }{ SGCylinder = \{ a\mapsto b |  a\in  GEARS\cprod CYLINDER \land  b=MOVING\}  }
			\nItem{ grd10 }{ at \neq  \emptyset  }
			\nItem{ grd11 }{ time \in  ran(at) }
			\nItem{ grd12 }{ ind \in  dom(at) \land  ind\mapsto time \in  at  }
			\end{description}
		\ThenAct
			\begin{description}
			\nItemX{ act1 }{ gstate :=  \{ a\mapsto b| a \in  GEARS \land  b=RETRACTING\}  }
			\nItemX{ act2 }{ gear\_extended :\in  1\upto 3 \tfun  (GEARS \tfun  \{ FALSE\} )  }
			\nItemX{ act3 }{ gear\_shock\_absorber :=  \{ a\mapsto b |  a \in  1\upto 3 \land  b = flight\}   }
			\nItem{ act4 }{ at :=  at \setminus  \{ ind\mapsto time\}  }
			\end{description}
		\EndAct
		\end{description}
	\EVT {retraction}
	\EXTD {retraction}
		\begin{description}
		\WhenGrd
			\begin{description}
			\nItemX{ grd1 }{ dstate[DOORS]=\{ OPEN\}  }
			\nItemX{ grd2 }{ gstate[GEARS]=\{ RETRACTING\}  }
			\nItemX{ grd4 }{ gear\_extended = \{ a\mapsto b |  a \in  1\upto 3 \land  b \in  GEARS \tfun  \{ FALSE\} \}   }
			\nItemX{ grd5 }{ gear\_retracted = \{ a\mapsto b |  a \in  1\upto 3 \land  b \in  GEARS \tfun  \{ FALSE\} \}   }
			\nItemX{ grd6 }{ gear\_shock\_absorber = \{ a\mapsto b |  a \in  1\upto 3 \land  b = flight\}   }
			\nItemX{ grd7 }{ \forall x\qdot x\in 1\upto 3 \limp  handle(x)=button }
			\nItemX{ grd3 }{ SGCylinder = \{ a\mapsto b |  a\in  GEARS\cprod CYLINDER \land  b=STOP\}  }
			\end{description}
		\ThenAct
			\begin{description}
			\nItemX{ act1 }{ gstate:=   \{ a\mapsto b|  a \in  GEARS \land  b= RETRACTED\}  }
			\nItemX{ act2 }{ gear\_retracted :\in  1\upto 3 \tfun  (GEARS \tfun  \{ TRUE\} )  }
			\end{description}
		\EndAct
		\end{description}
	\EVT {extending\_gears}
	\EXTD {extending\_gears}
		\begin{description}
		\AnyPrm
			\begin{description}
			\Item{ind }
			\end{description}
		\WhereGrd
			\begin{description}
			\nItemX{ grd1 }{ dstate[DOORS]=\{ OPEN\}  }
			\nItemX{ grd2 }{ gstate[GEARS]=\{ RETRACTED\}  }
			\nItemX{ grd3 }{ p=E }
			\nItemX{ grd5 }{ gear\_retracted = \{ a\mapsto b |  a \in  1\upto 3 \land  b \in  GEARS \tfun  \{ TRUE\} \}   }
			\nItemX{ grd6 }{ gear\_extended = \{ a\mapsto b |  a \in  1\upto 3 \land  b \in  GEARS \tfun  \{ FALSE\} \}   }
			\nItemX{ grd7 }{ gear\_shock\_absorber = \{ a\mapsto b |  a \in  1\upto 3 \land  b = flight\}   }
			\nItemX{ grd8 }{ \forall x\qdot x\in 1\upto 3 \limp  handle(x)=button }
			\nItemX{ grd4 }{ SGCylinder = \{ a\mapsto b |  a\in  GEARS\cprod CYLINDER \land  b=MOVING\}  }
			\nItem{ grd9 }{ at \neq  \emptyset  }
			\nItem{ grd10 }{ time \in  ran(at) }
			\nItem{ grd11 }{ ind \in  dom(at) \land  ind\mapsto time \in  at  }
			\end{description}
		\ThenAct
			\begin{description}
			\nItemX{ act1 }{ gstate :=  \{ a\mapsto b| a \in  GEARS \land  b=EXTENDING\}  }
			\nItemX{ act2 }{ gear\_retracted :\in  1\upto 3 \tfun  (GEARS \tfun  \{ FALSE\} ) }
			\nItem{ act3 }{ at :=  at \setminus  \{ ind\mapsto time\}  }
			\end{description}
		\EndAct
		\end{description}
	\EVT {extension}
	\EXTD {extension}
		\begin{description}
		\WhenGrd
			\begin{description}
			\nItemX{ grd1 }{ dstate[DOORS]=\{ OPEN\}  }
			\nItemX{ grd2 }{ gstate[GEARS]=\{ EXTENDING\}  }
			\nItemX{ grd4 }{ gear\_retracted = \{ a\mapsto b |  a \in  1\upto 3 \land  b \in  GEARS \tfun  \{ FALSE\} \}   }
			\nItemX{ grd5 }{ gear\_extended = \{ a\mapsto b |  a \in  1\upto 3 \land  b \in  GEARS \tfun  \{ FALSE\} \}   }
			\nItemX{ grd6 }{ gear\_shock\_absorber = \{ a\mapsto b |  a \in  1\upto 3 \land  b = flight\}   }
			\nItemX{ grd7 }{ \forall x\qdot x\in 1\upto 3 \limp  handle(x)=button }
			\nItemX{ grd3 }{ SGCylinder = \{ a\mapsto b |  a\in  GEARS\cprod CYLINDER \land  b=STOP\}  }
			\end{description}
		\ThenAct
			\begin{description}
			\nItemX{ act1 }{ gstate :=  \{ a\mapsto b|  a \in  GEARS \land  b=EXTENDED\}  }
			\nItemX{ act2 }{ gear\_extended :\in  1\upto 3 \tfun  (GEARS \tfun  \{ TRUE\} )  }
			\nItemX{ act3 }{ gear\_shock\_absorber :=  \{ a\mapsto b |  a \in  1\upto 3 \land  b = ground\}   }
			\end{description}
		\EndAct
		\end{description}
	\EVT {HPD1}
	\EXTD {HPD1}
		\begin{description}
		\WhenGrd
			\begin{description}
			\nItemX{ grd3 }{ \forall x\qdot x\in 1\upto 3 \limp  handle(x)=UP }
			\end{description}
		\ThenAct
			\begin{description}
			\nItemX{ act2 }{ handle :\in  1\upto 3 \tfun  \{ DOWN\}  }
			\nItemY{ act3 }{ at :=  at \bunion  \{ (index+1) \mapsto  (time+160)\}  }{ 		\\\hspace*{1,4 cm}  analogical switch is seen open 160ms after handle position has changed  }
			\nItemY{ act4 }{ handleDown\_interval :=  time + 40000 }{ 		\\\hspace*{1,4 cm}  add a new time interval (current time + handle not changed interval)
		\\\hspace*{1,2 cm}  in the event activation set }
			\nItemY{ act5 }{ handleUp\_interval :=  0 }{ 		\\\hspace*{1,4 cm}  update the handle up interval as 0 }
			\nItemY{ act6 }{ index :=  index +1 }{ 		\\\hspace*{1,4 cm}  update the current index value  }
			\end{description}
		\EndAct
		\end{description}
	\EVT {HPU1}
	\EXTD {HPU1}
		\begin{description}
		\WhenGrd
			\begin{description}
			\nItemX{ grd3 }{ \forall x\qdot x\in 1\upto 3 \limp  handle(x)=DOWN }
			\end{description}
		\ThenAct
			\begin{description}
			\nItemX{ act2 }{ handle :\in  1\upto 3 \tfun  \{ UP\}  }
			\nItemY{ act3 }{ at :=  at \bunion  \{ (index+1) \mapsto  (time+160)\}   }{ 		\\\hspace*{1,4 cm}  analogical switch is seen open 160ms after handle position has changed }
			\nItemY{ act4 }{ handleUp\_interval :=  time + 40000 }{ 		\\\hspace*{1,4 cm}  add a new time interval (current time + handle not changed interval)
		\\\hspace*{1,2 cm}  in the event activation set }
			\nItemY{ act5 }{ handleDown\_interval :=  0 }{ 		\\\hspace*{1,4 cm}  update the handle down interval as 0 }
			\nItemY{ act6 }{ index :=  index + 1 }{ 		\\\hspace*{1,4 cm}  update the current index value  }
			\end{description}
		\EndAct
		\end{description}
	\EVT {Analogical\_switch\_closed}
	\EXTD {Analogical\_switch\_closed}
		\begin{description}
		\AnyPrm
			\begin{description}
			\ItemXY{in }{in port }
			\Item{ind }
			\end{description}
		\WhereGrd
			\begin{description}
			\nItemX{ grd1 }{ in = general\_EV }
			\nItemX{ grd2 }{ \forall x\qdot x\in 1\upto 3 \limp  (handle(x)=UP \lor  handle(x)=DOWN) }
			\nItem{ grd3 }{ at \neq  \emptyset  }
			\nItem{ grd4 }{ time \in  ran(at) }
			\nItem{ grd5 }{ ind \in  dom(at) \land  ind\mapsto time \in  at  }
			\end{description}
		\ThenAct
			\begin{description}
			\nItemX{ act3 }{ analogical\_switch :\in  1\upto 3 \tfun  \{ closed\}  }
			\nItemX{ act4 }{ A\_Switch\_Out :=  TRUE }
			\nItemY{ act5 }{ at :=  (at \bunion  \{ (index+1) \mapsto  (time+1200)\}  )\setminus  \{ ind\mapsto time\}   }{ 		\\\hspace*{1,4 cm}  from closed to open 1.2 sec. }
			\nItem{ act6 }{ index :=  index + 1 }
			\end{description}
		\EndAct
		\end{description}
	\EVT {Analogical\_switch\_open}
	\EXTD {Analogical\_switch\_open}
		\begin{description}
		\AnyPrm
			\begin{description}
			\ItemXY{in }{in port }
			\Item{ind }
			\end{description}
		\WhereGrd
			\begin{description}
			\nItemX{ grd1 }{ in = general\_EV }
			\nItemX{ grd2 }{ \forall x\qdot x\in 1\upto 3 \limp  (handle(x)=UP \lor  handle(x)=DOWN) }
			\nItem{ grd3 }{ at \neq  \emptyset  }
			\nItem{ grd4 }{ time \in  ran(at) }
			\nItem{ grd5 }{ ind \in  dom(at) \land  ind\mapsto time \in  at  }
			\end{description}
		\ThenAct
			\begin{description}
			\nItemX{ act3 }{ analogical\_switch :\in  1\upto 3 \tfun  \{ open\}  }
			\nItemX{ act4 }{ A\_Switch\_Out :=  FALSE }
			\nItemY{ act5 }{ at :=  (at \bunion  \{ (index+1) \mapsto  (time+800)\} )\setminus \{ ind\mapsto time\}   }{ 		\\\hspace*{1,4 cm}  from open to closed .8 sec. }
			\nItem{ act6 }{ index :=  index + 1 }
			\end{description}
		\EndAct
		\end{description}
	\EVT {Circuit\_pressurized\_OK}
	\EXTD {Circuit\_pressurized\_OK}
		\begin{description}
		\WhenGrd
			\begin{description}
			\nItemX{ grd1 }{ general\_EV\_Hout = Hin }
			\end{description}
		\ThenAct
			\begin{description}
			\nItemX{ act9 }{ circuit\_pressurized :\in  1\upto 3 \tfun  \{ TRUE\}  }
			\end{description}
		\EndAct
		\end{description}
	\EVT {Circuit\_pressurized\_notOK}
	\EXTD {Circuit\_pressurized\_notOK}
		\begin{description}
		\WhenGrd
			\begin{description}
			\nItemX{ grd1 }{ general\_EV\_Hout = 0 }
			\end{description}
		\ThenAct
			\begin{description}
			\nItemX{ act9 }{ circuit\_pressurized :\in  1\upto 3 \tfun  \{ FALSE\}  }
			\end{description}
		\EndAct
		\end{description}
	\EVT {Computing\_Module\_1\_2}
	\EXTD {Computing\_Module\_1\_2}
		\begin{description}
		\WhenGrd
			\begin{description}
			\nItemX{ grd1 }{ state=computing }
			\end{description}
		\ThenAct
			\begin{description}
			\nItemX{ act1 }{ general\_EV :=  general\_EV\_func(handle \mapsto  analogical\_switch \mapsto  gear\_extended \mapsto  gear\_retracted \mapsto  gear\_shock\_absorber \mapsto  door\_open \mapsto  door\_closed \mapsto  circuit\_pressurized) }
			\nItemX{ act2 }{ close\_EV :=  close\_EV\_func(handle \mapsto  analogical\_switch \mapsto  gear\_extended \mapsto  gear\_retracted \mapsto  gear\_shock\_absorber \mapsto  door\_open \mapsto  door\_closed \mapsto  circuit\_pressurized) }
			\nItemX{ act3 }{ retract\_EV :=  retract\_EV\_func(handle \mapsto  analogical\_switch \mapsto  gear\_extended \mapsto  gear\_retracted \mapsto  gear\_shock\_absorber \mapsto  door\_open \mapsto  door\_closed \mapsto  circuit\_pressurized) }
			\nItemX{ act4 }{ extend\_EV :=  extend\_EV\_func(handle \mapsto  analogical\_switch \mapsto  gear\_extended \mapsto  gear\_retracted \mapsto  gear\_shock\_absorber \mapsto  door\_open \mapsto  door\_closed \mapsto  circuit\_pressurized) }
			\nItemX{ act5 }{ open\_EV :=  open\_EV\_func(handle \mapsto  analogical\_switch \mapsto  gear\_extended \mapsto  gear\_retracted \mapsto  gear\_shock\_absorber \mapsto  door\_open \mapsto  door\_closed \mapsto  circuit\_pressurized) }
			\nItemX{ act6 }{ gears\_locked\_down :=  gears\_locked\_down\_func(handle \mapsto  analogical\_switch \mapsto  gear\_extended \mapsto  gear\_retracted \mapsto  gear\_shock\_absorber \mapsto  door\_open \mapsto  door\_closed \mapsto  circuit\_pressurized) }
			\nItemX{ act7 }{ gears\_man :=  gears\_man\_func(handle \mapsto  analogical\_switch \mapsto  gear\_extended \mapsto  gear\_retracted \mapsto  gear\_shock\_absorber \mapsto  door\_open \mapsto  door\_closed \mapsto  circuit\_pressurized) }
			\nItemX{ act8 }{ anomaly :=  anomaly\_func(handle \mapsto  analogical\_switch \mapsto  gear\_extended \mapsto  gear\_retracted \mapsto  gear\_shock\_absorber \mapsto  door\_open \mapsto  door\_closed \mapsto  circuit\_pressurized) }
			\nItemX{ act9 }{ state:= electroValve }
			\end{description}
		\EndAct
		\end{description}
	\EVT {Update\_Hout}\cmt{		\\\hspace*{2,8 cm}  Assign the value of Hout  }
	\EXTD {Update\_Hout}
		\begin{description}
		\WhenGrd
			\begin{description}
			\nItemX{ grd1 }{ state = electroValve }
			\end{description}
		\ThenAct
			\begin{description}
			\nItemXY{ act1 }{ general\_EV\_Hout :|  ((general\_EV = TRUE \land   general\_EV\_Hout' = Hin) \lor  (general\_EV = FALSE \land  general\_EV\_Hout' = 0)		\\\hspace*{1,2 cm}  \lor  (A\_Switch\_Out = TRUE \land   general\_EV\_Hout' = Hin) \lor  (A\_Switch\_Out = FALSE \land   general\_EV\_Hout' = 0)) }{ 		\\\hspace*{1,4 cm}  pass the current value of hydraulic input port (Hin) to hydraulic output port (Hout) }
			\nItemX{ act2 }{ close\_EV\_Hout :|  ((close\_EV = TRUE \land  close\_EV\_Hout' = Hin) \lor  (close\_EV = FALSE \land  close\_EV\_Hout' = 0)) }
			\nItemX{ act3 }{ open\_EV\_Hout :|  ((open\_EV = TRUE \land  open\_EV\_Hout' = Hin) \lor  (open\_EV = FALSE \land  open\_EV\_Hout' = 0)) }
			\nItemX{ act4 }{ extend\_EV\_Hout :|  ((extend\_EV = TRUE \land  extend\_EV\_Hout' = Hin) \lor  (extend\_EV = FALSE \land  extend\_EV\_Hout' = 0)) }
			\nItemX{ act5 }{ retract\_EV\_Hout :|  ((retract\_EV = TRUE \land  retract\_EV\_Hout' = Hin) \lor  (retract\_EV = FALSE \land  retract\_EV\_Hout' = 0)) }
			\nItemX{ act6 }{ state :=  cylinder }
			\nItemY{ act7 }{ at :=  at \bunion  		\\\hspace*{1,2 cm}  \{ (index+1) \mapsto  (time+2000)\}  \bunion  		\\\hspace*{1,2 cm}  \{ (index+2) \mapsto  (time+10000)\}  \bunion 		\\\hspace*{1,2 cm}  \{ (index+3) \mapsto  (time+500)\}  \bunion  		\\\hspace*{1,2 cm}  \{ (index+4) \mapsto  (time+2000)\}  \bunion 		\\\hspace*{1,2 cm}  \{ (index+5) \mapsto  (time+500)\}  \bunion  		\\\hspace*{1,2 cm}  \{ (index+6) \mapsto  (time+2000)\}  \bunion 		\\\hspace*{1,2 cm}  \{ (index+7) \mapsto  (time+500)\}  \bunion  		\\\hspace*{1,2 cm}  \{ (index+8) \mapsto  (time+10000)\}  \bunion 		\\\hspace*{1,2 cm}  \{ (index+9) \mapsto  (time+500)\}  \bunion  		\\\hspace*{1,2 cm}  \{ (index+10) \mapsto  (time+10000)\}  }{ 		\\\hspace*{1,4 cm}  general EV 2     (time is given in comments in sec. while in model these are in ms.)		\\\hspace*{1,2 cm}  general EV 10		\\\hspace*{1,2 cm}  openning EV 0.5		\\\hspace*{1,2 cm}  openning EV 2		\\\hspace*{1,2 cm}  closure EV 0.5		\\\hspace*{1,2 cm}  closure EV 2		\\\hspace*{1,2 cm}  retraction EV 0.5		\\\hspace*{1,2 cm}  retraction EV 10		\\\hspace*{1,2 cm}  extension 0.5		\\\hspace*{1,2 cm}  extension 10 }
			\nItem{ act8 }{ index :=  index + 10 }
			\end{description}
		\EndAct
		\end{description}
	\EVT {CylinderMovingOrStop}\cmt{		\\\hspace*{4,6 cm}  Cylinder Moving or Stop according to the output of hydraulic circuit }
	\EXTD {CylinderMovingOrStop}
		\begin{description}
		\WhenGrd
			\begin{description}
			\nItemX{ grd1 }{ state = cylinder }
			\end{description}
		\ThenAct
			\begin{description}
			\nItemX{ act1 }{ SGCylinder :|  ((SGCylinder' = \{ a\mapsto b |  a\in  GEARS\cprod \{ GCYF,GCYR,GCYL\}  \land  b=MOVING\}  \land  extend\_EV\_Hout = Hin ) \lor  		\\\hspace*{1,4 cm}  (SGCylinder' = \{ a\mapsto b |  a\in  GEARS\cprod \{ GCYF,GCYR,GCYL\}  \land  b=STOP\}  \land  extend\_EV\_Hout = 0 ) \lor 		\\\hspace*{1,4 cm}  (SGCylinder' = \{ a\mapsto b |  a\in  GEARS\cprod \{ GCYF,GCYR,GCYL\}  \land  b=MOVING\}  \land  retract\_EV\_Hout = Hin ) \lor 		\\\hspace*{1,4 cm}  (SGCylinder' = \{ a\mapsto b |  a\in  GEARS\cprod \{ GCYF,GCYR,GCYL\}  \land  b=STOP\}  \land  retract\_EV\_Hout = 0 )) }
			\nItemX{ act2 }{ SDCylinder :|  ((SDCylinder' = \{ a\mapsto b |  a\in  DOORS\cprod \{ DCYF,DCYR,DCYL\}  \land  b=MOVING\}  \land  open\_EV\_Hout = Hin) \lor  		\\\hspace*{1,4 cm}  (SDCylinder' = \{ a\mapsto b |  a\in  DOORS\cprod \{ DCYF,DCYR,DCYL\}  \land  b=STOP\}  \land  open\_EV\_Hout = 0) \lor 		\\\hspace*{1,4 cm}  (SDCylinder' = \{ a\mapsto b |  a\in  DOORS\cprod \{ DCYF,DCYR,DCYL\}  \land  b=MOVING\}  \land  close\_EV\_Hout = Hin) \lor 		\\\hspace*{1,4 cm}  (SDCylinder' = \{ a\mapsto b |  a\in  DOORS\cprod \{ DCYF,DCYR,DCYL\} \land  b=STOP\}  \land  close\_EV\_Hout = 0)) }
			\nItemX{ act3 }{ state :=  computing }
			\end{description}
		\EndAct
		\end{description}
	\EVT {Failure\_Detection\_Generic\_Monitoring}
	\EXTD {Failure\_Detection\_Generic\_Monitoring}
		\begin{description}
		\WhenGrd
			\begin{description}
			\nItemXY{ grd1 }{ (\forall x,y,z\qdot  x\in 1\upto 3 \land  y\in 1\upto 3 \land  z\in 1\upto 3 \land  x\neq y \land  y\neq z \land  x\neq z \limp  		\\\hspace*{1,2 cm}  (handle(x)\neq handle(y) \land  handle(y)\neq handle(z) \land  handle(x)\neq handle(z)))		\\\hspace*{1,2 cm}  \lor 		\\\hspace*{1,2 cm}  (\forall x,y,z\qdot  x\in 1\upto 3 \land  y\in 1\upto 3 \land  z\in 1\upto 3 \land  x\neq y \land  y\neq z \land  x\neq z \limp  		\\\hspace*{1,2 cm}  (analogical\_switch(x)\neq analogical\_switch(y) \land  analogical\_switch(y)\neq analogical\_switch(z) \land  analogical\_switch(x)\neq analogical\_switch(z)))		\\\hspace*{1,2 cm}  \lor 		\\\hspace*{1,2 cm}  (\forall x,y,z\qdot  x\in 1\upto 3 \land  y\in 1\upto 3 \land  z\in 1\upto 3 \land  x\neq y \land  y\neq z \land  x\neq z \limp  		\\\hspace*{1,2 cm}  (gear\_extended(x)\neq gear\_extended(y) \land  gear\_extended(y)\neq gear\_extended(z) \land  gear\_extended(x)\neq gear\_extended(z)))		\\\hspace*{1,2 cm}  \lor  		\\\hspace*{1,2 cm}  (\forall x,y,z\qdot  x\in 1\upto 3 \land  y\in 1\upto 3 \land  z\in 1\upto 3 \land  x\neq y \land  y\neq z \land  x\neq z \limp  		\\\hspace*{1,2 cm}  (gear\_retracted(x)\neq gear\_retracted(y) \land  gear\_retracted(y)\neq gear\_retracted(z) \land  gear\_retracted(x)\neq gear\_retracted(z)))		\\\hspace*{1,2 cm}  \lor 		\\\hspace*{1,2 cm}  (\forall x,y,z\qdot  x\in 1\upto 3 \land  y\in 1\upto 3 \land  z\in 1\upto 3 \land  x\neq y \land  y\neq z \land  x\neq z \limp  		\\\hspace*{1,2 cm}  (gear\_shock\_absorber(x)\neq gear\_shock\_absorber(y) \land  gear\_shock\_absorber(y)\neq gear\_shock\_absorber(z) \land  gear\_shock\_absorber(x)\neq gear\_shock\_absorber(z)))		\\\hspace*{1,2 cm}  \lor 		\\\hspace*{1,2 cm}  (\forall x,y,z\qdot  x\in 1\upto 3 \land  y\in 1\upto 3 \land  z\in 1\upto 3 \land  x\neq y \land  y\neq z \land  x\neq z \limp  		\\\hspace*{1,2 cm}  (door\_open(x)\neq door\_open(y) \land  door\_open(y)\neq door\_open(z) \land  door\_open(x)\neq door\_open(z)))		\\\hspace*{1,2 cm}  \lor 		\\\hspace*{1,2 cm}  (\forall x,y,z\qdot  x\in 1\upto 3 \land  y\in 1\upto 3 \land  z\in 1\upto 3 \land  x\neq y \land  y\neq z \land  x\neq z \limp  		\\\hspace*{1,2 cm}  (door\_closed(x)\neq door\_closed(y) \land  door\_closed(y)\neq door\_closed(z) \land  door\_closed(x)\neq door\_closed(z)))		\\\hspace*{1,2 cm}  \lor 		\\\hspace*{1,2 cm}  (\forall x,y,z\qdot  x\in 1\upto 3 \land  y\in 1\upto 3 \land  z\in 1\upto 3 \land  x\neq y \land  y\neq z \land  x\neq z \limp  		\\\hspace*{1,2 cm}  (circuit\_pressurized(x)\neq circuit\_pressurized(y) \land  circuit\_pressurized(y)\neq circuit\_pressurized(z) \land  circuit\_pressurized(x)\neq circuit\_pressurized(z))) }{		\\\hspace*{1,4 cm}  Generic Monitoring uisng all sensors  } 
			\end{description}
		\ThenAct
			\begin{description}
			\nItemX{ act1 }{ anomaly :=  TRUE }
			\end{description}
		\EndAct
		\end{description}
	\EVT {Failure\_Detection\_Analogical\_Switch}
	\EXTD {Failure\_Detection\_Analogical\_Switch}
		\begin{description}
		\AnyPrm
			\begin{description}
			\Item{ind }
			\end{description}
		\WhereGrd
			\begin{description}
			\nItemXY{ grd1 }{ analogical\_switch = \{ a\mapsto b |  a \in  1\upto 3 \land  b = open\}  		\\\hspace*{1,2 cm}  \lor 		\\\hspace*{1,2 cm}  analogical\_switch = \{ a\mapsto b |  a \in  1\upto 3 \land  b = closed\}  }{		\\\hspace*{1,4 cm}  Gears motion monitoring without considering time  } 
			\nItem{ grd2 }{ at \neq  \emptyset  }
			\nItem{ grd3 }{ time \in  ran(at) }
			\nItem{ grd4 }{ ind \in  dom(at) \land  ind\mapsto time \in  at  }
			\end{description}
		\ThenAct
			\begin{description}
			\nItemX{ act1 }{ anomaly :=  TRUE }
			\nItem{ act2 }{ at :=  at \setminus  \{ ind\mapsto time\}  }
			\end{description}
		\EndAct
		\end{description}
	\EVT {check\_handle\_delay}\cmt{		\\\hspace*{4,2 cm}  This event is used to set 280ms in the set "at" 
		\\\hspace*{4 cm}  for event activation to detect anomaly
		\\\hspace*{4 cm}  and detect that hanlde is not change from last 40 sec. }
		\begin{description}
		\WhenGrd
			\begin{description}
			\nItemY{ grd1 }{ time = handleUp\_interval
		\\\hspace*{1,2 cm}  \lor  
		\\\hspace*{1,2 cm}  time = handleDown\_interval }{		\\\hspace*{1,4 cm}  current time is either equal to handle up interval or equal to the handle down interval } 
			\end{description}
		\ThenAct
			\begin{description}
			\nItemY{ act1 }{ at :=  at \bunion  \{ (index + 1) \mapsto  (time + 280)\}  }{ 		\\\hspace*{1,4 cm}  To add a new interval to the event activation set }
			\nItemY{ act3 }{ index :=  index + 1 }{ 		\\\hspace*{1,4 cm}  update the current index value }
			\end{description}
		\EndAct
		\end{description}
	\EVT {Failure\_Detection\_Pressure\_Sensor}
	\EXTD {Failure\_Detection\_Pressure\_Sensor}
		\begin{description}
		\AnyPrm
			\begin{description}
			\Item{ind }
			\end{description}
		\WhereGrd
			\begin{description}
			\nItemXY{ grd1 }{ circuit\_pressurized \neq  \{ a\mapsto b |  a \in  1\upto 3 \land  b = TRUE\}  		\\\hspace*{1,2 cm}  \lor 		\\\hspace*{1,2 cm}  circuit\_pressurized \neq  \{ a\mapsto b |  a \in  1\upto 3 \land  b = FALSE\}   }{		\\\hspace*{1,4 cm}  Circuit pressurized motion monitoring without considering time  } 
			\nItem{ grd2 }{ at \neq  \emptyset  }
			\nItem{ grd3 }{ time \in  ran(at) }
			\nItem{ grd4 }{ ind \in  dom(at) \land  ind\mapsto time \in  at  }
			\end{description}
		\ThenAct
			\begin{description}
			\nItemX{ act1 }{ anomaly :=  TRUE }
			\nItem{ act2 }{ at :=  at \setminus  \{ ind\mapsto time\}  }
			\end{description}
		\EndAct
		\end{description}
	\EVT {Failure\_Detection\_Doors}
	\EXTD {Failure\_Detection\_Doors}
		\begin{description}
		\AnyPrm
			\begin{description}
			\Item{ind }
			\end{description}
		\WhereGrd
			\begin{description}
			\nItemXY{ grd1 }{ door\_closed \neq  \{ a\mapsto b |  a \in  1\upto 3 \land  b \in  DOORS \tfun  \{ FALSE\} \}  		\\\hspace*{1,2 cm}  \lor 		\\\hspace*{1,2 cm}  door\_open \neq  \{ a\mapsto b |  a \in  1\upto 3 \land  b \in  DOORS \tfun  \{ TRUE\} \} 		\\\hspace*{1,2 cm}  \lor 		\\\hspace*{1,2 cm}  door\_open \neq  \{ a\mapsto b |  a \in  1\upto 3 \land  b \in  DOORS \tfun  \{ FALSE\} \}   		\\\hspace*{1,2 cm}  \lor 		\\\hspace*{1,2 cm}  door\_closed \neq  \{ a\mapsto b |  a \in  1\upto 3 \land  b \in  DOORS \tfun  \{ TRUE\} \}   }{		\\\hspace*{1,4 cm}  Doors motion monitoring without considering time  } 
			\nItem{ grd2 }{ at \neq  \emptyset  }
			\nItem{ grd3 }{ time \in  ran(at) }
			\nItem{ grd4 }{ ind \in  dom(at) \land  ind\mapsto time \in  at  }
			\end{description}
		\ThenAct
			\begin{description}
			\nItemX{ act1 }{ anomaly :=  TRUE }
			\nItem{ act2 }{ at :=  at \setminus  \{ ind\mapsto time\}  }
			\end{description}
		\EndAct
		\end{description}
	\EVT {Failure\_Detection\_Gears}
	\EXTD {Failure\_Detection\_Gears}
		\begin{description}
		\AnyPrm
			\begin{description}
			\Item{ind }
			\end{description}
		\WhereGrd
			\begin{description}
			\nItemXY{ grd1 }{ gear\_retracted \neq  \{ a\mapsto b |  a \in  1\upto 3 \land  b \in  GEARS \tfun  \{ FALSE\} \}  		\\\hspace*{1,2 cm}  \lor 		\\\hspace*{1,2 cm}  gear\_retracted \neq  \{ a\mapsto b |  a \in  1\upto 3 \land  b \in  GEARS \tfun  \{ TRUE\} \} 		\\\hspace*{1,2 cm}  \lor 		\\\hspace*{1,2 cm}  gear\_extended \neq  \{ a\mapsto b |  a \in  1\upto 3 \land  b \in  GEARS \tfun  \{ FALSE\} \}   		\\\hspace*{1,2 cm}  \lor 		\\\hspace*{1,2 cm}  gear\_extended \neq  \{ a\mapsto b |  a \in  1\upto 3 \land  b \in  GEARS \tfun  \{ TRUE\} \}   }{		\\\hspace*{1,4 cm}  Gears motion monitoring without considering time  } 
			\nItem{ grd2 }{ at \neq  \emptyset  }
			\nItem{ grd3 }{ time \in  ran(at) }
			\nItem{ grd4 }{ ind \in  dom(at) \land  ind\mapsto time \in  at  }
			\end{description}
		\ThenAct
			\begin{description}
			\nItemX{ act1 }{ anomaly :=  TRUE }
			\nItem{ act2 }{ at :=  at \setminus  \{ ind\mapsto time\}  }
			\end{description}
		\EndAct
		\end{description}
	\EVT {tic\_tock}\cmt{		\\\hspace*{2,2 cm}  time progression  }
		\begin{description}
		\AnyPrm
			\begin{description}
			\Item{tm }
			\end{description}
		\WhereGrd
			\begin{description}
			\nItem{ grd1 }{ tm \in  \nat }
			\nItemY{ grd2 }{ tm >  time }{		\\\hspace*{1,4 cm}  to take a new value of time in the future } 
			\nItem{ grd3 }{ ran(at) \neq  \emptyset  \limp  tm \leq  min(ran(at)) }
			\end{description}
		\ThenAct
			\begin{description}
			\nItemY{ act1 }{ time :=  tm }{ 		\\\hspace*{1,4 cm}  assign a new value of time to the current time }
			\end{description}
		\EndAct
		\end{description}
\END
\end{description}

\section{M9}
\label{sec:M9}

\begin{description}
\BTitle{M9}{27Jan2014}{10:44:59 AM}
\MACHINE{M9}\cmt{		\\\hspace*{1 cm}  Pilot interface light implementation }
\REFINES{M8}
\SEES{C1, C2}
\VARIABLES
	\begin{description}
		\Item{ dstate }
		\Item{ lstate }
		\Item{ phase }
		\Item{ button }
		\Item{ p }
		\Item{ l }
		\Item{ i }
		\Item{ gstate }
		\Item{ handle }
		\Item{ analogical\_switch }
		\Item{ gear\_extended }
		\Item{ gear\_retracted }
		\Item{ gear\_shock\_absorber }
		\Item{ door\_closed }
		\Item{ door\_open }
		\Item{ circuit\_pressurized }
		\Item{ general\_EV }
		\Item{ close\_EV }
		\Item{ retract\_EV }
		\Item{ extend\_EV }
		\Item{ open\_EV }
		\Item{ gears\_locked\_down }
		\Item{ gears\_man }
		\Item{ anomaly }
		\Item{ general\_EV\_func }
		\Item{ close\_EV\_func }
		\Item{ retract\_EV\_func }
		\Item{ extend\_EV\_func }
		\Item{ open\_EV\_func }
		\Item{ gears\_locked\_down\_func }
		\Item{ gears\_man\_func }
		\Item{ anomaly\_func }
		\Item{ general\_EV\_Hout }
		\Item{ close\_EV\_Hout }
		\Item{ retract\_EV\_Hout }
		\Item{ extend\_EV\_Hout }
		\Item{ open\_EV\_Hout }
		\ItemY{ SDCylinder }{State of Door Cylinder}
		\ItemY{ SGCylinder }{State of Gear Cylinder}
		\ItemY{ A\_Switch\_Out }{State of Gear Cylinder}
		\Item{ state }
		\ItemY{ time }{current time}
		\ItemY{ at }{a future event activation set.}
		\ItemY{ index }{To take a function to index different sets for event activation set}
		\ItemY{ handleUp\_interval }{To keep an update time duration after handle up}
		\ItemY{ handleDown\_interval }{To keep an update time duration after handle down}
		\ItemY{ pilot\_interface\_light }{current pilot interface light}
	\end{description}
\INVARIANTS
	\begin{description}
		\nItemY{ inv1 }{ pilot\_interface\_light \in  colorSet \tfun  lightState }{ 		\\\hspace*{1,4 cm}  a function to map from colorset to light state }
	\end{description}
\EVENTS
	\INITIALISATION
		\\\textit{extended}
		\begin{description}
		\BeginAct
			\begin{description}
			\nItemX{ act1 }{ button :=  DOWN }
			\nItemX{ act2 }{ phase :=  haltdown }
			\nItemXY{ act3 }{ dstate :|  (dstate'\in  DOORS \tfun  SDOORS \land  dstate'=\{ a\mapsto b|  a \in  DOORS \land  b=CLOSED\} ) }{ 		\\\hspace*{1,4 cm}  missing elements of the invariant }
			\nItemX{ act4 }{ lstate :=  \{ a\mapsto b| a\in DOORS\land  b=LOCKED\}  }
			\nItemX{ act5 }{ p :=  R }
			\nItemX{ act6 }{ l :=  R }
			\nItemX{ act7 }{ i :=  R }
			\nItemX{ act8 }{ gstate :|  (gstate' \in  GEARS \tfun  SGEARS \land  gstate'=\{ a\mapsto b |  a \in  GEARS  \land  b=EXTENDED\} ) }
			\nItemX{ act14 }{ handle :\in  1\upto 3 \tfun  \{ DOWN\}  }
			\nItemX{ act15 }{ analogical\_switch :\in  1\upto 3 \tfun  \{ open\}  }
			\nItemX{ act16 }{ gear\_extended :\in  1\upto 3 \tfun  (GEARS \tfun  \{ TRUE\} ) }
			\nItemX{ act17 }{ gear\_retracted :\in  1\upto 3 \tfun  (GEARS \tfun  \{ FALSE\} ) }
			\nItemX{ act18 }{ gear\_shock\_absorber :\in  1\upto 3 \tfun  \{ ground\}  }
			\nItemX{ act19 }{ door\_closed :\in  1\upto 3 \tfun  (DOORS \tfun  \{ TRUE\} ) }
			\nItemX{ act20 }{ door\_open :\in  1\upto 3 \tfun  (DOORS \tfun  \{ FALSE\} ) }
			\nItemX{ act21 }{ circuit\_pressurized :\in  1\upto 3 \tfun  \{ FALSE\}  }
			\nItemX{ act22 }{ general\_EV :=  FALSE }
			\nItemX{ act23 }{ close\_EV :=  TRUE }
			\nItemX{ act24 }{ retract\_EV :=  FALSE }
			\nItemX{ act25 }{ extend\_EV :=  TRUE }
			\nItemX{ act27 }{ open\_EV :=  FALSE }
			\nItemX{ act28 }{ gears\_locked\_down :=  TRUE }
			\nItemX{ act29 }{ gears\_man :=  FALSE }
			\nItemX{ act30 }{ anomaly :=  FALSE }
			\nItemX{ act31 }{ general\_EV\_func :\in  (1\upto 3 \tfun  POSITIONS) \cprod  (1\upto 3 \tfun  A\_Switch) \cprod  (1\upto 3 \tfun  (GEARS \tfun  BOOL)) \cprod  (1\upto 3 \tfun  (GEARS \tfun  BOOL)) \cprod  (1\upto 3 \tfun  GEAR\_ABSORBER) \cprod  (1\upto 3 \tfun  (DOORS \tfun  BOOL)) \cprod  (1\upto 3 \tfun  (DOORS \tfun  BOOL)) \cprod  (1\upto 3 \tfun  BOOL) \tfun  BOOL }
			\nItemX{ act32 }{ close\_EV\_func :\in  (1\upto 3 \tfun  POSITIONS) \cprod  (1\upto 3 \tfun  A\_Switch) \cprod  (1\upto 3 \tfun  (GEARS \tfun  BOOL)) \cprod  (1\upto 3 \tfun  (GEARS \tfun  BOOL)) \cprod  (1\upto 3 \tfun  GEAR\_ABSORBER) \cprod  (1\upto 3 \tfun  (DOORS \tfun  BOOL)) \cprod  (1\upto 3 \tfun  (DOORS \tfun  BOOL)) \cprod  (1\upto 3 \tfun  BOOL) \tfun  BOOL }
			\nItemX{ act33 }{ retract\_EV\_func :\in  (1\upto 3 \tfun  POSITIONS) \cprod  (1\upto 3 \tfun  A\_Switch) \cprod  (1\upto 3 \tfun  (GEARS \tfun  BOOL)) \cprod  (1\upto 3 \tfun  (GEARS \tfun  BOOL)) \cprod  (1\upto 3 \tfun  GEAR\_ABSORBER) \cprod  (1\upto 3 \tfun  (DOORS \tfun  BOOL)) \cprod  (1\upto 3 \tfun  (DOORS \tfun  BOOL)) \cprod  (1\upto 3 \tfun  BOOL) \tfun  BOOL }
			\nItemX{ act34 }{ extend\_EV\_func :\in  (1\upto 3 \tfun  POSITIONS) \cprod  (1\upto 3 \tfun  A\_Switch) \cprod  (1\upto 3 \tfun  (GEARS \tfun  BOOL)) \cprod  (1\upto 3 \tfun  (GEARS \tfun  BOOL)) \cprod  (1\upto 3 \tfun  GEAR\_ABSORBER) \cprod  (1\upto 3 \tfun  (DOORS \tfun  BOOL)) \cprod  (1\upto 3 \tfun  (DOORS \tfun  BOOL)) \cprod  (1\upto 3 \tfun  BOOL) \tfun  BOOL }
			\nItemX{ act35 }{ open\_EV\_func :\in  (1\upto 3 \tfun  POSITIONS) \cprod  (1\upto 3 \tfun  A\_Switch) \cprod  (1\upto 3 \tfun  (GEARS \tfun  BOOL)) \cprod  (1\upto 3 \tfun  (GEARS \tfun  BOOL)) \cprod  (1\upto 3 \tfun  GEAR\_ABSORBER) \cprod  (1\upto 3 \tfun  (DOORS \tfun  BOOL)) \cprod  (1\upto 3 \tfun  (DOORS \tfun  BOOL)) \cprod  (1\upto 3 \tfun  BOOL) \tfun  BOOL }
			\nItemX{ act36 }{ gears\_locked\_down\_func :\in  (1\upto 3 \tfun  POSITIONS) \cprod  (1\upto 3 \tfun  A\_Switch) \cprod  (1\upto 3 \tfun  (GEARS \tfun  BOOL)) \cprod  (1\upto 3 \tfun  (GEARS \tfun  BOOL)) \cprod  (1\upto 3 \tfun  GEAR\_ABSORBER) \cprod  (1\upto 3 \tfun  (DOORS \tfun  BOOL)) \cprod  (1\upto 3 \tfun  (DOORS \tfun  BOOL)) \cprod  (1\upto 3 \tfun  BOOL) \tfun  BOOL }
			\nItemX{ act37 }{ gears\_man\_func :\in  (1\upto 3 \tfun  POSITIONS) \cprod  (1\upto 3 \tfun  A\_Switch) \cprod  (1\upto 3 \tfun  (GEARS \tfun  BOOL)) \cprod  (1\upto 3 \tfun  (GEARS \tfun  BOOL)) \cprod  (1\upto 3 \tfun  GEAR\_ABSORBER) \cprod  (1\upto 3 \tfun  (DOORS \tfun  BOOL)) \cprod  (1\upto 3 \tfun  (DOORS \tfun  BOOL)) \cprod  (1\upto 3 \tfun  BOOL) \tfun  BOOL }
			\nItemX{ act38 }{ anomaly\_func :\in  (1\upto 3 \tfun  POSITIONS) \cprod  (1\upto 3 \tfun  A\_Switch) \cprod  (1\upto 3 \tfun  (GEARS \tfun  BOOL)) \cprod  (1\upto 3 \tfun  (GEARS \tfun  BOOL)) \cprod  (1\upto 3 \tfun  GEAR\_ABSORBER) \cprod  (1\upto 3 \tfun  (DOORS \tfun  BOOL)) \cprod  (1\upto 3 \tfun  (DOORS \tfun  BOOL)) \cprod  (1\upto 3 \tfun  BOOL) \tfun  BOOL }
			\nItemX{ act39 }{ A\_Switch\_Out :=  FALSE }
			\nItemX{ act40 }{ close\_EV\_Hout :=  0 }
			\nItemX{ act41 }{ retract\_EV\_Hout :=  0 }
			\nItemX{ act42 }{ extend\_EV\_Hout :=  0 }
			\nItemX{ act43 }{ open\_EV\_Hout :=  0 }
			\nItemX{ act44 }{ general\_EV\_Hout :=  0 }
			\nItemX{ act45 }{ SDCylinder :\in  DOORS \cprod  \{ DCYF,DCYR,DCYL\}  \tfun   \{ STOP\}  }
			\nItemX{ act46 }{ SGCylinder :\in  GEARS \cprod  \{ GCYF,GCYR,GCYL\}  \tfun  \{ STOP\}   }
			\nItemX{ act26 }{ state :=  computing }
			\nItemX{ act47 }{ at :=  \emptyset  }
			\nItemX{ act48 }{ time :=  0 }
			\nItemX{ act49 }{ index :=  0 }
			\nItemX{ act50 }{ handleUp\_interval :=  0 }
			\nItemX{ act51 }{ handleDown\_interval :=  0 }
			\nItem{ act52 }{ pilot\_interface\_light :=  \{ Green\mapsto Off,Orange\mapsto Off,Red\mapsto Off\}  }
			\end{description}
		\EndAct
		\end{description}
	\EVT {opening\_doors\_DOWN}
	\EXTD {opening\_doors\_DOWN}
		\begin{description}
		\WhenGrd
			\begin{description}
			\nItemX{ grd1 }{ dstate[DOORS]= \{ CLOSED\}  }
			\nItemX{ grd5 }{ lstate[DOORS]=\{ UNLOCKED\}  }
			\nItemX{ grd7 }{ phase=movingdown }
			\nItemX{ grd8 }{ p=R }
			\nItemX{ grd9 }{ l=R }
			\nItemX{ grd10 }{ door\_open = \{ a\mapsto b |  a \in  1\upto 3 \land  b \in  DOORS \tfun  \{ FALSE\} \}   }
			\nItemX{ grd11 }{ door\_closed = \{ a\mapsto b |  a \in  1\upto 3 \land  b \in  DOORS \tfun  \{ FALSE\} \}   }
			\nItemX{ grd12 }{ \forall x\qdot x\in 1\upto 3 \limp  handle(x)=button }
			\nItemX{ grd3 }{ SDCylinder = \{ a\mapsto b |  a\in  DOORS\cprod CYLINDER \land  b=MOVING\}  }
			\nItemX{ grd13 }{ anomaly = FALSE }
			\end{description}
		\ThenAct
			\begin{description}
			\nItemX{ act1 }{ dstate :=  \{ a\mapsto b|  a \in  DOORS \land  b=OPEN\}  }
			\nItemX{ act2 }{ p:= E }
			\nItemX{ act3 }{ door\_open :\in  1\upto 3 \tfun  (DOORS \tfun  \{ TRUE\} ) }
			\nItemXY{ act4 }{ at :=  at \bunion  \{ (index+1) \mapsto  (time+100)\}  }{ 		\\\hspace*{1,4 cm}  minimal interval for door open to gear extension }
			\nItemX{ act5 }{ index :=  index+1 }
			\end{description}
		\EndAct
		\end{description}
	\EVT {opening\_doors\_UP}
	\EXTD {opening\_doors\_UP}
		\begin{description}
		\WhenGrd
			\begin{description}
			\nItemX{ grd1 }{ dstate[DOORS]= \{ CLOSED\}  }
			\nItemX{ grd4 }{ lstate[DOORS]=\{ UNLOCKED\}  }
			\nItemX{ grd5 }{ phase= movingup }
			\nItemX{ grd6 }{ p=E }
			\nItemX{ grd7 }{ l=E }
			\nItemX{ grd8 }{ door\_open = \{ a\mapsto b |  a \in  1\upto 3 \land  b \in  DOORS \tfun  \{ FALSE\} \}   }
			\nItemX{ grd9 }{ door\_closed = \{ a\mapsto b |  a \in  1\upto 3 \land  b \in  DOORS \tfun  \{ FALSE\} \}   }
			\nItemX{ grd10 }{ \forall x\qdot x\in 1\upto 3 \limp  handle(x)=button }
			\nItemX{ grd3 }{ SDCylinder = \{ a\mapsto b |  a\in  DOORS\cprod CYLINDER \land  b=MOVING\}  }
			\nItemX{ grd11 }{ anomaly = FALSE }
			\end{description}
		\ThenAct
			\begin{description}
			\nItemX{ act1 }{ dstate :=  \{ a\mapsto b|  a \in  DOORS \land  b=OPEN\}  }
			\nItemX{ act2 }{ p:= R }
			\nItemX{ act3 }{ door\_open:\in  1\upto 3 \tfun  (DOORS \tfun  \{ TRUE\} )  }
			\nItemXY{ act4 }{ at :=  at \bunion  \{ (index+1) \mapsto  (time+100)\}  }{ 		\\\hspace*{1,4 cm}  minimal interval for door open to gear retraction }
			\nItemX{ act5 }{ index :=  index+1 }
			\end{description}
		\EndAct
		\end{description}
	\EVT {closing\_doors\_UP}
	\EXTD {closing\_doors\_UP}
		\begin{description}
		\AnyPrm
			\begin{description}
			\ItemX{f }
			\end{description}
		\WhereGrd
			\begin{description}
			\nItemX{ grd1 }{ dstate[DOORS]=\{ OPEN\}  }
			\nItemX{ grd3 }{ f \in  DOORS \tfun  SDOORS }
			\nItemX{ grd4 }{ \forall e\qdot  e \in  DOORS \limp  f(e)=CLOSED }
			\nItemX{ grd5 }{ phase=movingup }
			\nItemX{ grd6 }{ p=R }
			\nItemX{ grd7 }{ gstate[GEARS]=\{ RETRACTED\}  }
			\nItemX{ grd8 }{ \forall x\qdot x\in 1\upto 3 \limp  handle(x)=button }
			\nItemX{ grd9 }{ anomaly = FALSE }
			\end{description}
		\ThenAct
			\begin{description}
			\nItemX{ act1 }{ dstate:= f }
			\end{description}
		\EndAct
		\end{description}
	\EVT {closing\_doors\_DOWN}
	\EXTD {closing\_doors\_DOWN}
		\begin{description}
		\AnyPrm
			\begin{description}
			\ItemX{f }
			\end{description}
		\WhereGrd
			\begin{description}
			\nItemX{ grd1 }{ dstate[DOORS]=\{ OPEN\}  }
			\nItemX{ grd3 }{ f \in  DOORS \tfun  SDOORS }
			\nItemX{ grd4 }{ \forall e\qdot  e \in  DOORS \limp  f(e)=CLOSED }
			\nItemX{ grd5 }{ phase=movingdown }
			\nItemX{ grd6 }{ p=E }
			\nItemX{ grd7 }{ gstate[GEARS]=\{ EXTENDED\}  }
			\nItemX{ grd8 }{ \forall x\qdot x\in 1\upto 3 \limp  handle(x)=button }
			\nItemX{ grd9 }{ anomaly = FALSE }
			\end{description}
		\ThenAct
			\begin{description}
			\nItemX{ act1 }{ dstate:= f }
			\end{description}
		\EndAct
		\end{description}
	\EVT {unlocking\_UP}
	\EXTD {unlocking\_UP}
		\begin{description}
		\WhenGrd
			\begin{description}
			\nItemX{ grd3 }{ lstate[DOORS]=\{ LOCKED\}  }
			\nItemX{ grd4 }{ phase=movingup }
			\nItemX{ grd5 }{ l=E }
			\nItemX{ grd6 }{ p=E }
			\nItemX{ grd7 }{ i=E }
			\nItemX{ grd8 }{ door\_open = \{ a\mapsto b |  a \in  1\upto 3 \land  b \in  DOORS \tfun  \{ FALSE\} \}   }
			\nItemX{ grd9 }{ door\_closed = \{ a\mapsto b |  a \in  1\upto 3 \land  b \in  DOORS \tfun  \{ TRUE\} \}   }
			\nItemX{ grd10 }{ \forall x\qdot x\in 1\upto 3 \limp  handle(x)=button }
			\nItemX{ grd11 }{ anomaly = FALSE }
			\end{description}
		\ThenAct
			\begin{description}
			\nItemX{ act1 }{ lstate:= \{ a\mapsto b| a\in DOORS \land  b=UNLOCKED\}  }
			\nItemX{ act2 }{ door\_closed :\in  1\upto 3 \tfun  (DOORS \tfun  \{ FALSE\} )  }
			\end{description}
		\EndAct
		\end{description}
	\EVT {locking\_UP}
	\EXTD {locking\_UP}
		\begin{description}
		\WhenGrd
			\begin{description}
			\nItemX{ grd3 }{ dstate[DOORS]=\{ CLOSED\}  }
			\nItemX{ grd4 }{ phase=movingup }
			\nItemX{ grd5 }{ lstate[DOORS]=\{ UNLOCKED\}  }
			\nItemX{ grd6 }{ p=R }
			\nItemX{ grd7 }{ l=E }
			\nItemX{ grd9 }{ door\_open = \{ a\mapsto b |  a \in  1\upto 3 \land  b \in  DOORS \tfun  \{ FALSE\} \}   }
			\nItemX{ grd10 }{ door\_closed = \{ a\mapsto b |  a \in  1\upto 3 \land  b \in  DOORS \tfun  \{ FALSE\} \}   }
			\nItemX{ grd11 }{ \forall x\qdot x\in 1\upto 3 \limp  handle(x)=button }
			\nItemX{ grd8 }{ SDCylinder = \{ a\mapsto b |  a\in  DOORS\cprod CYLINDER \land  b=STOP\}  }
			\nItemX{ grd12 }{ anomaly = FALSE }
			\end{description}
		\ThenAct
			\begin{description}
			\nItemX{ act1 }{ lstate:= \{ a\mapsto b| a\in DOORS \land  b=LOCKED\}  }
			\nItemX{ act3 }{ phase:= haltup }
			\nItemXY{ act4 }{ l:= R }{ 		\\\hspace*{1,4 cm}  added by D Mery }
			\nItemX{ act44 }{ door\_closed :\in  1\upto 3 \tfun  (DOORS \tfun  \{ TRUE\} )  }
			\nItemXY{ act5 }{ at :=  at \bunion  \{ (index+1) \mapsto  (time+100)\}  }{ 		\\\hspace*{1,4 cm}  minimal interval for door closed to gear extension/retraction }
			\nItemX{ act6 }{ index :=  index+1 }
			\end{description}
		\EndAct
		\end{description}
	\EVT {unlocking\_DOWN}
	\EXTD {unlocking\_DOWN}
		\begin{description}
		\WhenGrd
			\begin{description}
			\nItemX{ grd3 }{ lstate[DOORS]=\{ LOCKED\}  }
			\nItemX{ grd4 }{ phase=movingdown }
			\nItemX{ grd5 }{ l=R }
			\nItemX{ grd6 }{ p=R }
			\nItemX{ grd7 }{ i=R }
			\nItemX{ grd8 }{ door\_open = \{ a\mapsto b |  a \in  1\upto 3 \land  b \in  DOORS \tfun  \{ FALSE\} \}   }
			\nItemX{ grd9 }{ door\_closed = \{ a\mapsto b |  a \in  1\upto 3 \land  b \in  DOORS \tfun  \{ TRUE\} \}   }
			\nItemX{ grd10 }{ \forall x\qdot x\in 1\upto 3 \limp  handle(x)=button }
			\nItemX{ grd11 }{ anomaly = FALSE }
			\end{description}
		\ThenAct
			\begin{description}
			\nItemX{ act1 }{ lstate:= \{ a\mapsto b| a\in DOORS \land  b=UNLOCKED\}  }
			\nItemX{ act2 }{ door\_closed :\in  1\upto 3 \tfun  (DOORS \tfun  \{ FALSE\} ) }
			\end{description}
		\EndAct
		\end{description}
	\EVT {locking\_DOWN}
	\EXTD {locking\_DOWN}
		\begin{description}
		\WhenGrd
			\begin{description}
			\nItemX{ grd1 }{ dstate[DOORS]=\{ CLOSED\}  }
			\nItemX{ grd2 }{ phase=movingdown }
			\nItemX{ grd3 }{ lstate[DOORS]=\{ UNLOCKED\}  }
			\nItemX{ grd4 }{ p=E }
			\nItemX{ grd5 }{ l=R }
			\nItemX{ grd7 }{ door\_open = \{ a\mapsto b |  a \in  1\upto 3 \land  b \in  DOORS \tfun  \{ FALSE\} \}   }
			\nItemX{ grd8 }{ door\_closed = \{ a\mapsto b |  a \in  1\upto 3 \land  b \in  DOORS \tfun  \{ FALSE\} \}   }
			\nItemX{ grd9 }{ \forall x\qdot x\in 1\upto 3 \limp  handle(x)=button }
			\nItemX{ grd6 }{ SDCylinder = \{ a\mapsto b |  a\in  DOORS\cprod CYLINDER \land  b=STOP\}  }
			\nItemX{ grd10 }{ anomaly = FALSE }
			\end{description}
		\ThenAct
			\begin{description}
			\nItemX{ act1 }{ lstate:= \{ a\mapsto b| a\in DOORS \land  b = LOCKED\}  }
			\nItemX{ act3 }{ phase:= haltdown }
			\nItemX{ act4 }{ l:= E }
			\nItemX{ act5 }{ door\_closed :\in  1\upto 3 \tfun  (DOORS \tfun  \{ TRUE\} )  }
			\nItemXY{ act6 }{ at :=  at \bunion  \{ (index+1) \mapsto  (time+100)\}  }{ 		\\\hspace*{1,4 cm}  minimal interval for door closed to extension/retraction }
			\nItemX{ act7 }{ index :=  index+1 }
			\end{description}
		\EndAct
		\end{description}
	\EVT {PD1}
	\EXTD {PD1}
		\begin{description}
		\WhenGrd
			\begin{description}
			\nItemX{ grd1 }{ button=UP }
			\nItemX{ grd2 }{ phase=haltup }
			\nItemX{ grd3 }{ \forall x\qdot x\in 1\upto 3 \limp  handle(x)=DOWN }
			\end{description}
		\ThenAct
			\begin{description}
			\nItemX{ act1 }{ phase:= movingdown }
			\nItemX{ act2 }{ button:= DOWN }
			\nItemX{ act3 }{ l:= R }
			\nItemX{ act4 }{ p:= R }
			\nItemX{ act5 }{ i:= R }
			\end{description}
		\EndAct
		\end{description}
	\EVT {PU1}
	\EXTD {PU1}
		\begin{description}
		\WhenGrd
			\begin{description}
			\nItemX{ grd1 }{ button=DOWN }
			\nItemX{ grd2 }{ phase=haltdown }
			\nItemX{ grd3 }{ \forall x\qdot x\in 1\upto 3 \limp  handle(x)=UP }
			\end{description}
		\ThenAct
			\begin{description}
			\nItemX{ act1 }{ phase:= movingup }
			\nItemX{ act2 }{ button:= UP }
			\nItemX{ act3 }{ l:= E }
			\nItemX{ act4 }{ p:= E }
			\nItemX{ act5 }{ i:= E }
			\end{description}
		\EndAct
		\end{description}
	\EVT {PU2}
	\EXTD {PU2}
		\begin{description}
		\WhenGrd
			\begin{description}
			\nItemX{ grd1 }{ l=R }
			\nItemX{ grd2 }{ p=R }
			\nItemX{ grd3 }{ phase=movingdown }
			\nItemX{ grd4 }{ button=DOWN }
			\nItemX{ grd5 }{ i=R }
			\nItemX{ grd6 }{ lstate[DOORS]=\{ LOCKED\}  }
			\nItemX{ grd7 }{ door\_open = \{ a\mapsto b |  a \in  1\upto 3 \land  b \in  DOORS \tfun  \{ FALSE\} \}   }
			\nItemX{ grd8 }{ door\_closed = \{ a\mapsto b |  a \in  1\upto 3 \land  b \in  DOORS \tfun  \{ TRUE\} \}   }
			\nItemX{ grd9 }{ \forall x\qdot x\in 1\upto 3 \limp  handle(x)=UP }
			\end{description}
		\ThenAct
			\begin{description}
			\nItemX{ act1 }{ phase:= movingup }
			\nItemX{ act4 }{ button:=   UP }
			\nItemX{ act5 }{ l:= E }
			\nItemX{ act6 }{ p:= E }
			\nItemX{ act7 }{ i:= R }
			\end{description}
		\EndAct
		\end{description}
	\EVT {CompletePU2}
	\EXTD {CompletePU2}
		\begin{description}
		\WhenGrd
			\begin{description}
			\nItemX{ grd1 }{ phase=movingup }
			\nItemX{ grd2 }{ button=UP }
			\nItemX{ grd3 }{ l=E }
			\nItemX{ grd4 }{ p=E }
			\nItemX{ grd5 }{ i=R }
			\end{description}
		\ThenAct
			\begin{description}
			\nItemX{ act1 }{ phase:= haltup }
			\end{description}
		\EndAct
		\end{description}
	\EVT {PU3}
	\EXTD {PU3}
		\begin{description}
		\WhenGrd
			\begin{description}
			\nItemX{ grd1 }{ dstate[DOORS]=\{ CLOSED\}  }
			\nItemX{ grd2 }{ lstate[DOORS]=\{ UNLOCKED\}  }
			\nItemX{ grd3 }{ phase = movingdown }
			\nItemX{ grd4 }{ p=R }
			\nItemX{ grd5 }{ l=R }
			\nItemX{ grd6 }{ button=DOWN }
			\nItemX{ grd7 }{ door\_open = \{ a\mapsto b |  a \in  1\upto 3 \land  b \in  DOORS \tfun  \{ FALSE\} \}   }
			\nItemX{ grd8 }{ door\_closed = \{ a\mapsto b |  a \in  1\upto 3 \land  b \in  DOORS \tfun  \{ FALSE\} \}   }
			\nItemX{ grd9 }{ \forall x\qdot x\in 1\upto 3 \limp  handle(x)=UP }
			\end{description}
		\ThenAct
			\begin{description}
			\nItemX{ act1 }{ phase:= movingup }
			\nItemX{ act2 }{ p:= R }
			\nItemX{ act3 }{ l:= E }
			\nItemX{ act4 }{ button:= UP }
			\end{description}
		\EndAct
		\end{description}
	\EVT {PU4}
	\EXTD {PU4}
		\begin{description}
		\WhenGrd
			\begin{description}
			\nItemX{ grd1 }{ dstate[DOORS]=\{ OPEN\}  }
			\nItemX{ grd2 }{ phase=movingdown }
			\nItemX{ grd3 }{ p=E }
			\nItemX{ grd4 }{ button=DOWN }
			\nItemX{ grd7 }{ \forall x\qdot x\in 1\upto 3 \limp  handle(x)=UP }
			\end{description}
		\ThenAct
			\begin{description}
			\nItemX{ act1 }{ phase:= movingup }
			\nItemX{ act2 }{ p:= R }
			\nItemX{ act3 }{ button:= UP }
			\nItemX{ act4 }{ i:= E }
			\nItemX{ act5 }{ l:= E }
			\end{description}
		\EndAct
		\end{description}
	\EVT {PU5}
	\EXTD {PU5}
		\begin{description}
		\WhenGrd
			\begin{description}
			\nItemX{ grd1 }{ dstate[DOORS]=\{ CLOSED\}  }
			\nItemX{ grd2 }{ phase=movingdown }
			\nItemX{ grd3 }{ p=E }
			\nItemX{ grd4 }{ button=DOWN }
			\nItemX{ grd5 }{ lstate[DOORS]=\{ UNLOCKED\}  }
			\nItemX{ grd6 }{ door\_open = \{ a\mapsto b |  a \in  1\upto 3 \land  b \in  DOORS \tfun  \{ FALSE\} \}   }
			\nItemX{ grd7 }{ door\_closed = \{ a\mapsto b |  a \in  1\upto 3 \land  b \in  DOORS \tfun  \{ FALSE\} \}   }
			\nItemX{ grd8 }{ \forall x\qdot x\in 1\upto 3 \limp  handle(x)=UP }
			\end{description}
		\ThenAct
			\begin{description}
			\nItemX{ act1 }{ phase:= movingup }
			\nItemX{ act3 }{ button:= UP }
			\nItemX{ act4 }{ i:= E }
			\nItemX{ act5 }{ l:= E }
			\end{description}
		\EndAct
		\end{description}
	\EVT {PD2}
	\EXTD {PD2}
		\begin{description}
		\WhenGrd
			\begin{description}
			\nItemX{ grd1 }{ l=E }
			\nItemX{ grd2 }{ p=E }
			\nItemX{ grd3 }{ phase=movingup }
			\nItemX{ grd4 }{ i=E }
			\nItemX{ grd5 }{ lstate[DOORS]=\{ LOCKED\}  }
			\nItemX{ grd6 }{ \forall x\qdot x\in 1\upto 3 \limp  handle(x)=DOWN }
			\end{description}
		\ThenAct
			\begin{description}
			\nItemX{ act1 }{ phase:= movingdown }
			\nItemX{ act2 }{ button:= DOWN }
			\nItemX{ act3 }{ l:= R }
			\nItemX{ act4 }{ p:= R }
			\nItemX{ act5 }{ i:= E }
			\end{description}
		\EndAct
		\end{description}
	\EVT {CompletePD2}
	\EXTD {CompletePD2}
		\begin{description}
		\WhenGrd
			\begin{description}
			\nItemX{ grd1 }{ phase=movingdown }
			\nItemX{ grd2 }{ button=DOWN }
			\nItemX{ grd3 }{ l=R }
			\nItemX{ grd4 }{ p=R }
			\nItemX{ grd5 }{ i=E }
			\end{description}
		\ThenAct
			\begin{description}
			\nItemX{ act1 }{ phase:= haltdown }
			\end{description}
		\EndAct
		\end{description}
	\EVT {PD3}
	\EXTD {PD3}
		\begin{description}
		\WhenGrd
			\begin{description}
			\nItemX{ grd1 }{ dstate[DOORS]=\{ CLOSED\}  }
			\nItemX{ grd2 }{ lstate[DOORS]=\{ UNLOCKED\}  }
			\nItemX{ grd3 }{ phase=movingup }
			\nItemX{ grd4 }{ p=E }
			\nItemX{ grd5 }{ l=E }
			\nItemX{ grd6 }{ button=UP }
			\nItemX{ grd7 }{ door\_open = \{ a\mapsto b |  a \in  1\upto 3 \land  b \in  DOORS \tfun  \{ FALSE\} \}   }
			\nItemX{ grd8 }{ door\_closed = \{ a\mapsto b |  a \in  1\upto 3 \land  b \in  DOORS \tfun  \{ FALSE\} \}   }
			\nItemX{ grd9 }{ \forall x\qdot x\in 1\upto 3 \limp  handle(x)=DOWN }
			\end{description}
		\ThenAct
			\begin{description}
			\nItemX{ act1 }{ phase:= movingdown }
			\nItemX{ act2 }{ p:= E }
			\nItemX{ act3 }{ l:= R }
			\nItemX{ act4 }{ button:= DOWN }
			\end{description}
		\EndAct
		\end{description}
	\EVT {PD4}
	\EXTD {PD4}
		\begin{description}
		\WhenGrd
			\begin{description}
			\nItemX{ grd1 }{ dstate[DOORS]=\{ OPEN\}  }
			\nItemX{ grd2 }{ phase=movingup }
			\nItemX{ grd3 }{ p=R }
			\nItemX{ grd4 }{ button=UP }
			\nItemX{ grd6 }{ \forall x\qdot x\in 1\upto 3 \limp  handle(x)=DOWN }
			\end{description}
		\ThenAct
			\begin{description}
			\nItemX{ act1 }{ phase:= movingdown }
			\nItemX{ act2 }{ p:= E }
			\nItemX{ act3 }{ button:= DOWN }
			\nItemX{ act4 }{ i:= R }
			\nItemX{ act5 }{ l:= R }
			\end{description}
		\EndAct
		\end{description}
	\EVT {PD5}
	\EXTD {PD5}
		\begin{description}
		\WhenGrd
			\begin{description}
			\nItemX{ grd1 }{ dstate[DOORS]=\{ CLOSED\}  }
			\nItemX{ grd2 }{ phase=movingup }
			\nItemX{ grd3 }{ p=R }
			\nItemX{ grd4 }{ button=UP }
			\nItemX{ grd5 }{ lstate[DOORS]=\{ UNLOCKED\}  }
			\nItemX{ grd6 }{ door\_open = \{ a\mapsto b |  a \in  1\upto 3 \land  b \in  DOORS \tfun  \{ FALSE\} \}   }
			\nItemX{ grd7 }{ door\_closed = \{ a\mapsto b |  a \in  1\upto 3 \land  b \in  DOORS \tfun  \{ FALSE\} \}   }
			\nItemX{ grd8 }{ \forall x\qdot x\in 1\upto 3 \limp  handle(x)=DOWN }
			\end{description}
		\ThenAct
			\begin{description}
			\nItemX{ act1 }{ phase :=  movingdown }
			\nItemX{ act2 }{ button:= DOWN }
			\nItemX{ act3 }{ i:= R }
			\nItemX{ act4 }{ l:= R }
			\end{description}
		\EndAct
		\end{description}
	\EVT {retracting\_gears}
	\EXTD {retracting\_gears}
		\begin{description}
		\AnyPrm
			\begin{description}
			\ItemX{ind }
			\end{description}
		\WhereGrd
			\begin{description}
			\nItemX{ grd1 }{ dstate[DOORS]=\{ OPEN\}  }
			\nItemX{ grd2 }{ gstate[GEARS]=\{ EXTENDED\}  }
			\nItemX{ grd3 }{ p=R }
			\nItemX{ grd6 }{ gear\_extended = \{ a\mapsto b |  a \in  1\upto 3 \land  b \in  GEARS \tfun  \{ TRUE\} \}   }
			\nItemX{ grd7 }{ gear\_retracted = \{ a\mapsto b |  a \in  1\upto 3 \land  b \in  GEARS \tfun  \{ FALSE\} \}   }
			\nItemX{ grd8 }{ gear\_shock\_absorber = \{ a\mapsto b |  a \in  1\upto 3 \land  b = ground\}   }
			\nItemX{ grd9 }{ \forall x\qdot x\in 1\upto 3 \limp  handle(x)=button }
			\nItemX{ grd5 }{ SGCylinder = \{ a\mapsto b |  a\in  GEARS\cprod CYLINDER \land  b=MOVING\}  }
			\nItemX{ grd10 }{ at \neq  \emptyset  }
			\nItemX{ grd11 }{ time \in  ran(at) }
			\nItemX{ grd12 }{ ind \in  dom(at) \land  ind\mapsto time \in  at  }
			\end{description}
		\ThenAct
			\begin{description}
			\nItemX{ act1 }{ gstate :=  \{ a\mapsto b| a \in  GEARS \land  b=RETRACTING\}  }
			\nItemX{ act2 }{ gear\_extended :\in  1\upto 3 \tfun  (GEARS \tfun  \{ FALSE\} )  }
			\nItemX{ act3 }{ gear\_shock\_absorber :=  \{ a\mapsto b |  a \in  1\upto 3 \land  b = flight\}   }
			\nItemX{ act4 }{ at :=  at \setminus  \{ ind\mapsto time\}  }
			\end{description}
		\EndAct
		\end{description}
	\EVT {retraction}
	\EXTD {retraction}
		\begin{description}
		\WhenGrd
			\begin{description}
			\nItemX{ grd1 }{ dstate[DOORS]=\{ OPEN\}  }
			\nItemX{ grd2 }{ gstate[GEARS]=\{ RETRACTING\}  }
			\nItemX{ grd4 }{ gear\_extended = \{ a\mapsto b |  a \in  1\upto 3 \land  b \in  GEARS \tfun  \{ FALSE\} \}   }
			\nItemX{ grd5 }{ gear\_retracted = \{ a\mapsto b |  a \in  1\upto 3 \land  b \in  GEARS \tfun  \{ FALSE\} \}   }
			\nItemX{ grd6 }{ gear\_shock\_absorber = \{ a\mapsto b |  a \in  1\upto 3 \land  b = flight\}   }
			\nItemX{ grd7 }{ \forall x\qdot x\in 1\upto 3 \limp  handle(x)=button }
			\nItemX{ grd3 }{ SGCylinder = \{ a\mapsto b |  a\in  GEARS\cprod CYLINDER \land  b=STOP\}  }
			\end{description}
		\ThenAct
			\begin{description}
			\nItemX{ act1 }{ gstate:=   \{ a\mapsto b|  a \in  GEARS \land  b= RETRACTED\}  }
			\nItemX{ act2 }{ gear\_retracted :\in  1\upto 3 \tfun  (GEARS \tfun  \{ TRUE\} )  }
			\end{description}
		\EndAct
		\end{description}
	\EVT {extending\_gears}
	\EXTD {extending\_gears}
		\begin{description}
		\AnyPrm
			\begin{description}
			\ItemX{ind }
			\end{description}
		\WhereGrd
			\begin{description}
			\nItemX{ grd1 }{ dstate[DOORS]=\{ OPEN\}  }
			\nItemX{ grd2 }{ gstate[GEARS]=\{ RETRACTED\}  }
			\nItemX{ grd3 }{ p=E }
			\nItemX{ grd5 }{ gear\_retracted = \{ a\mapsto b |  a \in  1\upto 3 \land  b \in  GEARS \tfun  \{ TRUE\} \}   }
			\nItemX{ grd6 }{ gear\_extended = \{ a\mapsto b |  a \in  1\upto 3 \land  b \in  GEARS \tfun  \{ FALSE\} \}   }
			\nItemX{ grd7 }{ gear\_shock\_absorber = \{ a\mapsto b |  a \in  1\upto 3 \land  b = flight\}   }
			\nItemX{ grd8 }{ \forall x\qdot x\in 1\upto 3 \limp  handle(x)=button }
			\nItemX{ grd4 }{ SGCylinder = \{ a\mapsto b |  a\in  GEARS\cprod CYLINDER \land  b=MOVING\}  }
			\nItemX{ grd9 }{ at \neq  \emptyset  }
			\nItemX{ grd10 }{ time \in  ran(at) }
			\nItemX{ grd11 }{ ind \in  dom(at) \land  ind\mapsto time \in  at  }
			\end{description}
		\ThenAct
			\begin{description}
			\nItemX{ act1 }{ gstate :=  \{ a\mapsto b| a \in  GEARS \land  b=EXTENDING\}  }
			\nItemX{ act2 }{ gear\_retracted :\in  1\upto 3 \tfun  (GEARS \tfun  \{ FALSE\} ) }
			\nItemX{ act3 }{ at :=  at \setminus  \{ ind\mapsto time\}  }
			\end{description}
		\EndAct
		\end{description}
	\EVT {extension}
	\EXTD {extension}
		\begin{description}
		\WhenGrd
			\begin{description}
			\nItemX{ grd1 }{ dstate[DOORS]=\{ OPEN\}  }
			\nItemX{ grd2 }{ gstate[GEARS]=\{ EXTENDING\}  }
			\nItemX{ grd4 }{ gear\_retracted = \{ a\mapsto b |  a \in  1\upto 3 \land  b \in  GEARS \tfun  \{ FALSE\} \}   }
			\nItemX{ grd5 }{ gear\_extended = \{ a\mapsto b |  a \in  1\upto 3 \land  b \in  GEARS \tfun  \{ FALSE\} \}   }
			\nItemX{ grd6 }{ gear\_shock\_absorber = \{ a\mapsto b |  a \in  1\upto 3 \land  b = flight\}   }
			\nItemX{ grd7 }{ \forall x\qdot x\in 1\upto 3 \limp  handle(x)=button }
			\nItemX{ grd3 }{ SGCylinder = \{ a\mapsto b |  a\in  GEARS\cprod CYLINDER \land  b=STOP\}  }
			\end{description}
		\ThenAct
			\begin{description}
			\nItemX{ act1 }{ gstate :=  \{ a\mapsto b|  a \in  GEARS \land  b=EXTENDED\}  }
			\nItemX{ act2 }{ gear\_extended :\in  1\upto 3 \tfun  (GEARS \tfun  \{ TRUE\} )  }
			\nItemX{ act3 }{ gear\_shock\_absorber :=  \{ a\mapsto b |  a \in  1\upto 3 \land  b = ground\}   }
			\end{description}
		\EndAct
		\end{description}
	\EVT {HPD1}
	\EXTD {HPD1}
		\begin{description}
		\WhenGrd
			\begin{description}
			\nItemX{ grd3 }{ \forall x\qdot x\in 1\upto 3 \limp  handle(x)=UP }
			\end{description}
		\ThenAct
			\begin{description}
			\nItemX{ act2 }{ handle :\in  1\upto 3 \tfun  \{ DOWN\}  }
			\nItemXY{ act3 }{ at :=  at \bunion  \{ (index+1) \mapsto  (time+160)\}  }{ 		\\\hspace*{1,4 cm}  analogical switch is seen open 160ms after handle position has changed  }
			\nItemXY{ act4 }{ handleDown\_interval :=  time + 40000 }{ 		\\\hspace*{1,4 cm}  add a new time interval (current time + handle not changed interval)
		\\\hspace*{1,2 cm}  in the event activation set }
			\nItemXY{ act5 }{ handleUp\_interval :=  0 }{ 		\\\hspace*{1,4 cm}  update the handle up interval as 0 }
			\nItemXY{ act6 }{ index :=  index +1 }{ 		\\\hspace*{1,4 cm}  update the current index value  }
			\end{description}
		\EndAct
		\end{description}
	\EVT {HPU1}
	\EXTD {HPU1}
		\begin{description}
		\WhenGrd
			\begin{description}
			\nItemX{ grd3 }{ \forall x\qdot x\in 1\upto 3 \limp  handle(x)=DOWN }
			\end{description}
		\ThenAct
			\begin{description}
			\nItemX{ act2 }{ handle :\in  1\upto 3 \tfun  \{ UP\}  }
			\nItemXY{ act3 }{ at :=  at \bunion  \{ (index+1) \mapsto  (time+160)\}   }{ 		\\\hspace*{1,4 cm}  analogical switch is seen open 160ms after handle position has changed }
			\nItemXY{ act4 }{ handleUp\_interval :=  time + 40000 }{ 		\\\hspace*{1,4 cm}  add a new time interval (current time + handle not changed interval)
		\\\hspace*{1,2 cm}  in the event activation set }
			\nItemXY{ act5 }{ handleDown\_interval :=  0 }{ 		\\\hspace*{1,4 cm}  update the handle down interval as 0 }
			\nItemXY{ act6 }{ index :=  index + 1 }{ 		\\\hspace*{1,4 cm}  update the current index value  }
			\end{description}
		\EndAct
		\end{description}
	\EVT {Analogical\_switch\_closed}
	\EXTD {Analogical\_switch\_closed}
		\begin{description}
		\AnyPrm
			\begin{description}
			\ItemXY{in }{in port }
			\ItemX{ind }
			\end{description}
		\WhereGrd
			\begin{description}
			\nItemX{ grd1 }{ in = general\_EV }
			\nItemX{ grd2 }{ \forall x\qdot x\in 1\upto 3 \limp  (handle(x)=UP \lor  handle(x)=DOWN) }
			\nItemX{ grd3 }{ at \neq  \emptyset  }
			\nItemX{ grd4 }{ time \in  ran(at) }
			\nItemX{ grd5 }{ ind \in  dom(at) \land  ind\mapsto time \in  at  }
			\end{description}
		\ThenAct
			\begin{description}
			\nItemX{ act3 }{ analogical\_switch :\in  1\upto 3 \tfun  \{ closed\}  }
			\nItemX{ act4 }{ A\_Switch\_Out :=  TRUE }
			\nItemXY{ act5 }{ at :=  (at \bunion  \{ (index+1) \mapsto  (time+1200)\}  )\setminus  \{ ind\mapsto time\}   }{ 		\\\hspace*{1,4 cm}  from closed to open 1.2 sec. }
			\nItemX{ act6 }{ index :=  index + 1 }
			\end{description}
		\EndAct
		\end{description}
	\EVT {Analogical\_switch\_open}
	\EXTD {Analogical\_switch\_open}
		\begin{description}
		\AnyPrm
			\begin{description}
			\ItemXY{in }{in port }
			\ItemX{ind }
			\end{description}
		\WhereGrd
			\begin{description}
			\nItemX{ grd1 }{ in = general\_EV }
			\nItemX{ grd2 }{ \forall x\qdot x\in 1\upto 3 \limp  (handle(x)=UP \lor  handle(x)=DOWN) }
			\nItemX{ grd3 }{ at \neq  \emptyset  }
			\nItemX{ grd4 }{ time \in  ran(at) }
			\nItemX{ grd5 }{ ind \in  dom(at) \land  ind\mapsto time \in  at  }
			\end{description}
		\ThenAct
			\begin{description}
			\nItemX{ act3 }{ analogical\_switch :\in  1\upto 3 \tfun  \{ open\}  }
			\nItemX{ act4 }{ A\_Switch\_Out :=  FALSE }
			\nItemXY{ act5 }{ at :=  (at \bunion  \{ (index+1) \mapsto  (time+800)\} )\setminus \{ ind\mapsto time\}   }{ 		\\\hspace*{1,4 cm}  from open to closed .8 sec. }
			\nItemX{ act6 }{ index :=  index + 1 }
			\end{description}
		\EndAct
		\end{description}
	\EVT {Circuit\_pressurized\_OK}
	\EXTD {Circuit\_pressurized\_OK}
		\begin{description}
		\WhenGrd
			\begin{description}
			\nItemX{ grd1 }{ general\_EV\_Hout = Hin }
			\end{description}
		\ThenAct
			\begin{description}
			\nItemX{ act9 }{ circuit\_pressurized :\in  1\upto 3 \tfun  \{ TRUE\}  }
			\end{description}
		\EndAct
		\end{description}
	\EVT {Circuit\_pressurized\_notOK}
	\EXTD {Circuit\_pressurized\_notOK}
		\begin{description}
		\WhenGrd
			\begin{description}
			\nItemX{ grd1 }{ general\_EV\_Hout = 0 }
			\end{description}
		\ThenAct
			\begin{description}
			\nItemX{ act9 }{ circuit\_pressurized :\in  1\upto 3 \tfun  \{ FALSE\}  }
			\end{description}
		\EndAct
		\end{description}
	\EVT {Computing\_Module\_1\_2}
	\EXTD {Computing\_Module\_1\_2}
		\begin{description}
		\WhenGrd
			\begin{description}
			\nItemX{ grd1 }{ state=computing }
			\end{description}
		\ThenAct
			\begin{description}
			\nItemX{ act1 }{ general\_EV :=  general\_EV\_func(handle \mapsto  analogical\_switch \mapsto  gear\_extended \mapsto  gear\_retracted \mapsto  gear\_shock\_absorber \mapsto  door\_open \mapsto  door\_closed \mapsto  circuit\_pressurized) }
			\nItemX{ act2 }{ close\_EV :=  close\_EV\_func(handle \mapsto  analogical\_switch \mapsto  gear\_extended \mapsto  gear\_retracted \mapsto  gear\_shock\_absorber \mapsto  door\_open \mapsto  door\_closed \mapsto  circuit\_pressurized) }
			\nItemX{ act3 }{ retract\_EV :=  retract\_EV\_func(handle \mapsto  analogical\_switch \mapsto  gear\_extended \mapsto  gear\_retracted \mapsto  gear\_shock\_absorber \mapsto  door\_open \mapsto  door\_closed \mapsto  circuit\_pressurized) }
			\nItemX{ act4 }{ extend\_EV :=  extend\_EV\_func(handle \mapsto  analogical\_switch \mapsto  gear\_extended \mapsto  gear\_retracted \mapsto  gear\_shock\_absorber \mapsto  door\_open \mapsto  door\_closed \mapsto  circuit\_pressurized) }
			\nItemX{ act5 }{ open\_EV :=  open\_EV\_func(handle \mapsto  analogical\_switch \mapsto  gear\_extended \mapsto  gear\_retracted \mapsto  gear\_shock\_absorber \mapsto  door\_open \mapsto  door\_closed \mapsto  circuit\_pressurized) }
			\nItemX{ act6 }{ gears\_locked\_down :=  gears\_locked\_down\_func(handle \mapsto  analogical\_switch \mapsto  gear\_extended \mapsto  gear\_retracted \mapsto  gear\_shock\_absorber \mapsto  door\_open \mapsto  door\_closed \mapsto  circuit\_pressurized) }
			\nItemX{ act7 }{ gears\_man :=  gears\_man\_func(handle \mapsto  analogical\_switch \mapsto  gear\_extended \mapsto  gear\_retracted \mapsto  gear\_shock\_absorber \mapsto  door\_open \mapsto  door\_closed \mapsto  circuit\_pressurized) }
			\nItemX{ act8 }{ anomaly :=  anomaly\_func(handle \mapsto  analogical\_switch \mapsto  gear\_extended \mapsto  gear\_retracted \mapsto  gear\_shock\_absorber \mapsto  door\_open \mapsto  door\_closed \mapsto  circuit\_pressurized) }
			\nItemX{ act9 }{ state:= electroValve }
			\end{description}
		\EndAct
		\end{description}
	\EVT {Update\_Hout}\cmt{		\\\hspace*{2,8 cm}  Assign the value of Hout  }
	\EXTD {Update\_Hout}
		\begin{description}
		\WhenGrd
			\begin{description}
			\nItemX{ grd1 }{ state = electroValve }
			\end{description}
		\ThenAct
			\begin{description}
			\nItemXY{ act1 }{ general\_EV\_Hout :|  ((general\_EV = TRUE \land   general\_EV\_Hout' = Hin) \lor  (general\_EV = FALSE \land  general\_EV\_Hout' = 0)		\\\hspace*{1,2 cm}  \lor  (A\_Switch\_Out = TRUE \land   general\_EV\_Hout' = Hin) \lor  (A\_Switch\_Out = FALSE \land   general\_EV\_Hout' = 0)) }{ 		\\\hspace*{1,4 cm}  pass the current value of hydraulic input port (Hin) to hydraulic output port (Hout) }
			\nItemX{ act2 }{ close\_EV\_Hout :|  ((close\_EV = TRUE \land  close\_EV\_Hout' = Hin) \lor  (close\_EV = FALSE \land  close\_EV\_Hout' = 0)) }
			\nItemX{ act3 }{ open\_EV\_Hout :|  ((open\_EV = TRUE \land  open\_EV\_Hout' = Hin) \lor  (open\_EV = FALSE \land  open\_EV\_Hout' = 0)) }
			\nItemX{ act4 }{ extend\_EV\_Hout :|  ((extend\_EV = TRUE \land  extend\_EV\_Hout' = Hin) \lor  (extend\_EV = FALSE \land  extend\_EV\_Hout' = 0)) }
			\nItemX{ act5 }{ retract\_EV\_Hout :|  ((retract\_EV = TRUE \land  retract\_EV\_Hout' = Hin) \lor  (retract\_EV = FALSE \land  retract\_EV\_Hout' = 0)) }
			\nItemX{ act6 }{ state :=  cylinder }
			\nItemXY{ act7 }{ at :=  at \bunion  		\\\hspace*{1,2 cm}  \{ (index+1) \mapsto  (time+2000)\}  \bunion  		\\\hspace*{1,2 cm}  \{ (index+2) \mapsto  (time+10000)\}  \bunion 		\\\hspace*{1,2 cm}  \{ (index+3) \mapsto  (time+500)\}  \bunion  		\\\hspace*{1,2 cm}  \{ (index+4) \mapsto  (time+2000)\}  \bunion 		\\\hspace*{1,2 cm}  \{ (index+5) \mapsto  (time+500)\}  \bunion  		\\\hspace*{1,2 cm}  \{ (index+6) \mapsto  (time+2000)\}  \bunion 		\\\hspace*{1,2 cm}  \{ (index+7) \mapsto  (time+500)\}  \bunion  		\\\hspace*{1,2 cm}  \{ (index+8) \mapsto  (time+10000)\}  \bunion 		\\\hspace*{1,2 cm}  \{ (index+9) \mapsto  (time+500)\}  \bunion  		\\\hspace*{1,2 cm}  \{ (index+10) \mapsto  (time+10000)\}  }{ 		\\\hspace*{1,4 cm}  general EV 2     (time is given in comments in sec. while in model these are in ms.)		\\\hspace*{1,2 cm}  general EV 10		\\\hspace*{1,2 cm}  openning EV 0.5		\\\hspace*{1,2 cm}  openning EV 2		\\\hspace*{1,2 cm}  closure EV 0.5		\\\hspace*{1,2 cm}  closure EV 2		\\\hspace*{1,2 cm}  retraction EV 0.5		\\\hspace*{1,2 cm}  retraction EV 10		\\\hspace*{1,2 cm}  extension 0.5		\\\hspace*{1,2 cm}  extension 10 }
			\nItemX{ act8 }{ index :=  index + 10 }
			\end{description}
		\EndAct
		\end{description}
	\EVT {CylinderMovingOrStop}\cmt{		\\\hspace*{4,6 cm}  Cylinder Moving or Stop according to the output of hydraulic circuit }
	\EXTD {CylinderMovingOrStop}
		\begin{description}
		\WhenGrd
			\begin{description}
			\nItemX{ grd1 }{ state = cylinder }
			\end{description}
		\ThenAct
			\begin{description}
			\nItemX{ act1 }{ SGCylinder :|  ((SGCylinder' = \{ a\mapsto b |  a\in  GEARS\cprod \{ GCYF,GCYR,GCYL\}  \land  b=MOVING\}  \land  extend\_EV\_Hout = Hin ) \lor  		\\\hspace*{1,4 cm}  (SGCylinder' = \{ a\mapsto b |  a\in  GEARS\cprod \{ GCYF,GCYR,GCYL\}  \land  b=STOP\}  \land  extend\_EV\_Hout = 0 ) \lor 		\\\hspace*{1,4 cm}  (SGCylinder' = \{ a\mapsto b |  a\in  GEARS\cprod \{ GCYF,GCYR,GCYL\}  \land  b=MOVING\}  \land  retract\_EV\_Hout = Hin ) \lor 		\\\hspace*{1,4 cm}  (SGCylinder' = \{ a\mapsto b |  a\in  GEARS\cprod \{ GCYF,GCYR,GCYL\}  \land  b=STOP\}  \land  retract\_EV\_Hout = 0 )) }
			\nItemX{ act2 }{ SDCylinder :|  ((SDCylinder' = \{ a\mapsto b |  a\in  DOORS\cprod \{ DCYF,DCYR,DCYL\}  \land  b=MOVING\}  \land  open\_EV\_Hout = Hin) \lor  		\\\hspace*{1,4 cm}  (SDCylinder' = \{ a\mapsto b |  a\in  DOORS\cprod \{ DCYF,DCYR,DCYL\}  \land  b=STOP\}  \land  open\_EV\_Hout = 0) \lor 		\\\hspace*{1,4 cm}  (SDCylinder' = \{ a\mapsto b |  a\in  DOORS\cprod \{ DCYF,DCYR,DCYL\}  \land  b=MOVING\}  \land  close\_EV\_Hout = Hin) \lor 		\\\hspace*{1,4 cm}  (SDCylinder' = \{ a\mapsto b |  a\in  DOORS\cprod \{ DCYF,DCYR,DCYL\} \land  b=STOP\}  \land  close\_EV\_Hout = 0)) }
			\nItemX{ act3 }{ state :=  computing }
			\end{description}
		\EndAct
		\end{description}
	\EVT {Failure\_Detection\_Generic\_Monitoring}
	\EXTD {Failure\_Detection\_Generic\_Monitoring}
		\begin{description}
		\WhenGrd
			\begin{description}
			\nItemXY{ grd1 }{ (\forall x,y,z\qdot  x\in 1\upto 3 \land  y\in 1\upto 3 \land  z\in 1\upto 3 \land  x\neq y \land  y\neq z \land  x\neq z \limp  		\\\hspace*{1,2 cm}  (handle(x)\neq handle(y) \land  handle(y)\neq handle(z) \land  handle(x)\neq handle(z)))		\\\hspace*{1,2 cm}  \lor 		\\\hspace*{1,2 cm}  (\forall x,y,z\qdot  x\in 1\upto 3 \land  y\in 1\upto 3 \land  z\in 1\upto 3 \land  x\neq y \land  y\neq z \land  x\neq z \limp  		\\\hspace*{1,2 cm}  (analogical\_switch(x)\neq analogical\_switch(y) \land  analogical\_switch(y)\neq analogical\_switch(z) \land  analogical\_switch(x)\neq analogical\_switch(z)))		\\\hspace*{1,2 cm}  \lor 		\\\hspace*{1,2 cm}  (\forall x,y,z\qdot  x\in 1\upto 3 \land  y\in 1\upto 3 \land  z\in 1\upto 3 \land  x\neq y \land  y\neq z \land  x\neq z \limp  		\\\hspace*{1,2 cm}  (gear\_extended(x)\neq gear\_extended(y) \land  gear\_extended(y)\neq gear\_extended(z) \land  gear\_extended(x)\neq gear\_extended(z)))		\\\hspace*{1,2 cm}  \lor  		\\\hspace*{1,2 cm}  (\forall x,y,z\qdot  x\in 1\upto 3 \land  y\in 1\upto 3 \land  z\in 1\upto 3 \land  x\neq y \land  y\neq z \land  x\neq z \limp  		\\\hspace*{1,2 cm}  (gear\_retracted(x)\neq gear\_retracted(y) \land  gear\_retracted(y)\neq gear\_retracted(z) \land  gear\_retracted(x)\neq gear\_retracted(z)))		\\\hspace*{1,2 cm}  \lor 		\\\hspace*{1,2 cm}  (\forall x,y,z\qdot  x\in 1\upto 3 \land  y\in 1\upto 3 \land  z\in 1\upto 3 \land  x\neq y \land  y\neq z \land  x\neq z \limp  		\\\hspace*{1,2 cm}  (gear\_shock\_absorber(x)\neq gear\_shock\_absorber(y) \land  gear\_shock\_absorber(y)\neq gear\_shock\_absorber(z) \land  gear\_shock\_absorber(x)\neq gear\_shock\_absorber(z)))		\\\hspace*{1,2 cm}  \lor 		\\\hspace*{1,2 cm}  (\forall x,y,z\qdot  x\in 1\upto 3 \land  y\in 1\upto 3 \land  z\in 1\upto 3 \land  x\neq y \land  y\neq z \land  x\neq z \limp  		\\\hspace*{1,2 cm}  (door\_open(x)\neq door\_open(y) \land  door\_open(y)\neq door\_open(z) \land  door\_open(x)\neq door\_open(z)))		\\\hspace*{1,2 cm}  \lor 		\\\hspace*{1,2 cm}  (\forall x,y,z\qdot  x\in 1\upto 3 \land  y\in 1\upto 3 \land  z\in 1\upto 3 \land  x\neq y \land  y\neq z \land  x\neq z \limp  		\\\hspace*{1,2 cm}  (door\_closed(x)\neq door\_closed(y) \land  door\_closed(y)\neq door\_closed(z) \land  door\_closed(x)\neq door\_closed(z)))		\\\hspace*{1,2 cm}  \lor 		\\\hspace*{1,2 cm}  (\forall x,y,z\qdot  x\in 1\upto 3 \land  y\in 1\upto 3 \land  z\in 1\upto 3 \land  x\neq y \land  y\neq z \land  x\neq z \limp  		\\\hspace*{1,2 cm}  (circuit\_pressurized(x)\neq circuit\_pressurized(y) \land  circuit\_pressurized(y)\neq circuit\_pressurized(z) \land  circuit\_pressurized(x)\neq circuit\_pressurized(z))) }{		\\\hspace*{1,4 cm}  Generic Monitoring uisng all sensors  } 
			\end{description}
		\ThenAct
			\begin{description}
			\nItemX{ act1 }{ anomaly :=  TRUE }
			\end{description}
		\EndAct
		\end{description}
	\EVT {Failure\_Detection\_Analogical\_Switch}
	\EXTD {Failure\_Detection\_Analogical\_Switch}
		\begin{description}
		\AnyPrm
			\begin{description}
			\ItemX{ind }
			\end{description}
		\WhereGrd
			\begin{description}
			\nItemXY{ grd1 }{ analogical\_switch = \{ a\mapsto b |  a \in  1\upto 3 \land  b = open\}  		\\\hspace*{1,2 cm}  \lor 		\\\hspace*{1,2 cm}  analogical\_switch = \{ a\mapsto b |  a \in  1\upto 3 \land  b = closed\}  }{		\\\hspace*{1,4 cm}  Gears motion monitoring without considering time  } 
			\nItemX{ grd2 }{ at \neq  \emptyset  }
			\nItemX{ grd3 }{ time \in  ran(at) }
			\nItemX{ grd4 }{ ind \in  dom(at) \land  ind\mapsto time \in  at  }
			\end{description}
		\ThenAct
			\begin{description}
			\nItemX{ act1 }{ anomaly :=  TRUE }
			\nItemX{ act2 }{ at :=  at \setminus  \{ ind\mapsto time\}  }
			\end{description}
		\EndAct
		\end{description}
	\EVT {check\_handle\_delay}\cmt{		\\\hspace*{4,2 cm}  This event is used to set 280ms in the set "at" 
		\\\hspace*{4 cm}  for event activation to detect anomaly
		\\\hspace*{4 cm}  and detect that hanlde is not change from last 40 sec. }
	\EXTD {check\_handle\_delay}
		\begin{description}
		\WhenGrd
			\begin{description}
			\nItemXY{ grd1 }{ time = handleUp\_interval
		\\\hspace*{1,2 cm}  \lor  
		\\\hspace*{1,2 cm}  time = handleDown\_interval }{		\\\hspace*{1,4 cm}  current time is either equal to handle up interval or equal to the handle down interval } 
			\end{description}
		\ThenAct
			\begin{description}
			\nItemXY{ act1 }{ at :=  at \bunion  \{ (index + 1) \mapsto  (time + 280)\}  }{ 		\\\hspace*{1,4 cm}  To add a new interval to the event activation set }
			\nItemXY{ act3 }{ index :=  index + 1 }{ 		\\\hspace*{1,4 cm}  update the current index value }
			\end{description}
		\EndAct
		\end{description}
	\EVT {Failure\_Detection\_Pressure\_Sensor}
	\EXTD {Failure\_Detection\_Pressure\_Sensor}
		\begin{description}
		\AnyPrm
			\begin{description}
			\ItemX{ind }
			\end{description}
		\WhereGrd
			\begin{description}
			\nItemXY{ grd1 }{ circuit\_pressurized \neq  \{ a\mapsto b |  a \in  1\upto 3 \land  b = TRUE\}  		\\\hspace*{1,2 cm}  \lor 		\\\hspace*{1,2 cm}  circuit\_pressurized \neq  \{ a\mapsto b |  a \in  1\upto 3 \land  b = FALSE\}   }{		\\\hspace*{1,4 cm}  Circuit pressurized motion monitoring without considering time  } 
			\nItemX{ grd2 }{ at \neq  \emptyset  }
			\nItemX{ grd3 }{ time \in  ran(at) }
			\nItemX{ grd4 }{ ind \in  dom(at) \land  ind\mapsto time \in  at  }
			\end{description}
		\ThenAct
			\begin{description}
			\nItemX{ act1 }{ anomaly :=  TRUE }
			\nItemX{ act2 }{ at :=  at \setminus  \{ ind\mapsto time\}  }
			\end{description}
		\EndAct
		\end{description}
	\EVT {Failure\_Detection\_Doors}
	\EXTD {Failure\_Detection\_Doors}
		\begin{description}
		\AnyPrm
			\begin{description}
			\ItemX{ind }
			\end{description}
		\WhereGrd
			\begin{description}
			\nItemXY{ grd1 }{ door\_closed \neq  \{ a\mapsto b |  a \in  1\upto 3 \land  b \in  DOORS \tfun  \{ FALSE\} \}  		\\\hspace*{1,2 cm}  \lor 		\\\hspace*{1,2 cm}  door\_open \neq  \{ a\mapsto b |  a \in  1\upto 3 \land  b \in  DOORS \tfun  \{ TRUE\} \} 		\\\hspace*{1,2 cm}  \lor 		\\\hspace*{1,2 cm}  door\_open \neq  \{ a\mapsto b |  a \in  1\upto 3 \land  b \in  DOORS \tfun  \{ FALSE\} \}   		\\\hspace*{1,2 cm}  \lor 		\\\hspace*{1,2 cm}  door\_closed \neq  \{ a\mapsto b |  a \in  1\upto 3 \land  b \in  DOORS \tfun  \{ TRUE\} \}   }{		\\\hspace*{1,4 cm}  Doors motion monitoring without considering time  } 
			\nItemX{ grd2 }{ at \neq  \emptyset  }
			\nItemX{ grd3 }{ time \in  ran(at) }
			\nItemX{ grd4 }{ ind \in  dom(at) \land  ind\mapsto time \in  at  }
			\end{description}
		\ThenAct
			\begin{description}
			\nItemX{ act1 }{ anomaly :=  TRUE }
			\nItemX{ act2 }{ at :=  at \setminus  \{ ind\mapsto time\}  }
			\end{description}
		\EndAct
		\end{description}
	\EVT {Failure\_Detection\_Gears}
	\EXTD {Failure\_Detection\_Gears}
		\begin{description}
		\AnyPrm
			\begin{description}
			\ItemX{ind }
			\end{description}
		\WhereGrd
			\begin{description}
			\nItemXY{ grd1 }{ gear\_retracted \neq  \{ a\mapsto b |  a \in  1\upto 3 \land  b \in  GEARS \tfun  \{ FALSE\} \}  		\\\hspace*{1,2 cm}  \lor 		\\\hspace*{1,2 cm}  gear\_retracted \neq  \{ a\mapsto b |  a \in  1\upto 3 \land  b \in  GEARS \tfun  \{ TRUE\} \} 		\\\hspace*{1,2 cm}  \lor 		\\\hspace*{1,2 cm}  gear\_extended \neq  \{ a\mapsto b |  a \in  1\upto 3 \land  b \in  GEARS \tfun  \{ FALSE\} \}   		\\\hspace*{1,2 cm}  \lor 		\\\hspace*{1,2 cm}  gear\_extended \neq  \{ a\mapsto b |  a \in  1\upto 3 \land  b \in  GEARS \tfun  \{ TRUE\} \}   }{		\\\hspace*{1,4 cm}  Gears motion monitoring without considering time  } 
			\nItemX{ grd2 }{ at \neq  \emptyset  }
			\nItemX{ grd3 }{ time \in  ran(at) }
			\nItemX{ grd4 }{ ind \in  dom(at) \land  ind\mapsto time \in  at  }
			\end{description}
		\ThenAct
			\begin{description}
			\nItemX{ act1 }{ anomaly :=  TRUE }
			\nItemX{ act2 }{ at :=  at \setminus  \{ ind\mapsto time\}  }
			\end{description}
		\EndAct
		\end{description}
	\EVT {tic\_tock}\cmt{		\\\hspace*{2,2 cm}  time progression  }
	\EXTD {tic\_tock}
		\begin{description}
		\AnyPrm
			\begin{description}
			\ItemX{tm }
			\end{description}
		\WhereGrd
			\begin{description}
			\nItemX{ grd1 }{ tm \in  \nat }
			\nItemXY{ grd2 }{ tm >  time }{		\\\hspace*{1,4 cm}  to take a new value of time in the future } 
			\nItemX{ grd3 }{ ran(at) \neq  \emptyset  \limp  tm \leq  min(ran(at)) }
			\end{description}
		\ThenAct
			\begin{description}
			\nItemXY{ act1 }{ time :=  tm }{ 		\\\hspace*{1,4 cm}  assign a new value of time to the current time }
			\end{description}
		\EndAct
		\end{description}
	\EVT {pilot\_interface\_Green\_light\_On}\cmt{		\\\hspace*{6,6 cm}  green light is on when gears locked down is true  }
		\begin{description}
		\WhenGrd
			\begin{description}
			\nItemY{ grd1 }{ gears\_locked\_down = TRUE }{		\\\hspace*{1,4 cm}  gears locked down must be true  } 
			\end{description}
		\ThenAct
			\begin{description}
			\nItemY{ act1 }{ pilot\_interface\_light(Green) :=  On }{ 		\\\hspace*{1,4 cm}  To set on of Green light of pilot interface light  }
			\end{description}
		\EndAct
		\end{description}
	\EVT {pilot\_interface\_Orange\_light\_On}\cmt{		\\\hspace*{6,8 cm}  orange light is on when gears maneuvering is true }
		\begin{description}
		\WhenGrd
			\begin{description}
			\nItemY{ grd1 }{ gears\_man = TRUE }{		\\\hspace*{1,4 cm}  gears maneuvering must be true } 
			\end{description}
		\ThenAct
			\begin{description}
			\nItemY{ act1 }{ pilot\_interface\_light(Orange) :=  On }{ 		\\\hspace*{1,4 cm}  To set on of Orange light of pilot interface light  }
			\end{description}
		\EndAct
		\end{description}
	\EVT {pilot\_interface\_Red\_light\_On}\cmt{		\\\hspace*{6,2 cm}  red light is on when anomaly is detected (true) }
		\begin{description}
		\WhenGrd
			\begin{description}
			\nItemY{ grd1 }{ anomaly = TRUE }{		\\\hspace*{1,4 cm}  anomaly must be true } 
			\nItem{ grd2 }{ pilot\_interface\_light(Red) = Off }
			\end{description}
		\ThenAct
			\begin{description}
			\nItemY{ act1 }{ pilot\_interface\_light(Red) :=  On }{ 		\\\hspace*{1,4 cm}  To set on of Red light of pilot interface light  }
			\end{description}
		\EndAct
		\end{description}
	\EVT {pilot\_interface\_Green\_light\_Off}\cmt{		\\\hspace*{6,8 cm}  green light is off when gears locked down is false }
		\begin{description}
		\WhenGrd
			\begin{description}
			\nItemY{ grd1 }{ gears\_locked\_down = FALSE }{		\\\hspace*{1,4 cm}  gears locked down must be false } 
			\end{description}
		\ThenAct
			\begin{description}
			\nItemY{ act1 }{ pilot\_interface\_light(Green) :=  Off }{ 		\\\hspace*{1,4 cm}  To set off of Green light of pilot interface light  }
			\end{description}
		\EndAct
		\end{description}
	\EVT {pilot\_interface\_Orange\_light\_Off}\cmt{		\\\hspace*{7 cm}  orange light is off when gears maneuvering is false }
		\begin{description}
		\WhenGrd
			\begin{description}
			\nItemY{ grd1 }{ gears\_man = FALSE }{		\\\hspace*{1,4 cm}  gears maneuvering must be false } 
			\end{description}
		\ThenAct
			\begin{description}
			\nItemY{ act1 }{ pilot\_interface\_light(Orange) :=  Off }{ 		\\\hspace*{1,4 cm}  To set off of Orange light of pilot interface light  }
			\end{description}
		\EndAct
		\end{description}
	\EVT {pilot\_interface\_Red\_light\_Off}\cmt{		\\\hspace*{6,4 cm}  red light is off when anomaly is detected (false) }
		\begin{description}
		\WhenGrd
			\begin{description}
			\nItemY{ grd1 }{ anomaly = FALSE }{		\\\hspace*{1,4 cm}  anomaly must be false } 
			\nItem{ grd2 }{ pilot\_interface\_light(Red) = On }
			\end{description}
		\ThenAct
			\begin{description}
			\nItemY{ act1 }{ pilot\_interface\_light(Red) :=  Off }{ 		\\\hspace*{1,4 cm}  To set off of Red light of pilot interface light  }
			\end{description}
		\EndAct
		\end{description}
\END
\end{description}

\end{document}